\documentclass[a4paper,11pt,twocolumn,twoside]{article}

\usepackage{graphicx,natbib,aas_macros}
\usepackage{fancyhdr}

\tt

\title{Radio synchrotron emission: electron energy spectrum, supernovae, microquasars, 
active nuclei, cluster relics and halos; X-ray halos}
\author{Nimisha G. Kantharia \\
\normalsize Tata Institute of Fundamental
Research \\ \normalsize Post Bag 3, Ganeshkhind, Pune-411007, India   \\
\normalsize \it Email: nkprasadnetra@gmail.com \\
\normalsize \it URL: https://sites.google.com/view/ngkresearch/home}

\date{April 2019}

\begin{document}
\maketitle

\thispagestyle{empty}

\begin{abstract}
In this paper, transient high energy events and their hosts are discussed with focus
on signatures of radio synchrotron emission.
There are numerous outstanding questions ranging from the origin of the two phases of radio
emission in supernovae to the formation of conical jets launched at relativistic velocities
in active nuclei to the formation of radio hotspots
in FR~II sources to the formation of radio relics and halos in clusters of
galaxies to the origin of the relativistic electron population and its
energy spectrum which is responsible for the synchrotron emission to the composition
of the radio synchrotron-emitting plasma in these sources.  
Observations have helped build an empirical 
model of active nuclei but the origin of
the narrow line regions, broad line regions, dust, accretion disk in 
active nuclei and their connection to radio jets and lobes remains largely unknown.
This paper attempts to understand and resolve the 
above outstanding issues. 
The summary of a few inferences presented in the paper is:

\begin{itemize}

\item 
The matter before being ejected in a transient explosive event like a supernova has to be
rapidly accelerated in situ to at least the escape velocity. 
The energy distribution of the ejected cosmic rays including electrons has to be a normal 
distribution since the energising process will be statistical in nature.
The peak of the normal distribution has to be at or beyond the energy equivalent 
of the escape velocity for protons for a large fraction of matter to escape. 
This explains the ubiquitous curved+power law nature of the
observed radio synchrotron spectra and the cosmic ray energy spectra 
which peaks near 1 GeV. 
The dispersion of the E/m i.e. energy/mass distribution
for each species of particles will be different and will determine the range
in their random velocity component so that for a constant energy imparted per particle,
the light particles can acquire extremely high velocities.  

\item The prompt radio synchrotron emission detected soon after a supernova 
explosion is from light plasma composed of positrons and electrons which should 
escape the explosion site soon
after the neutrinos and radiate in the ambient circumstellar magnetic field.  
The delayed radio emission from supernova remnants can be explained by synchrotron 
emission from the relativistic proton-electron plasma mixed with and radiating in the
magnetic field frozen in the ejected thermal plasma.  
All matter has to be energised to escape velocities or more before expulsion. 
The main source of energy in both type I and II supernovae has to be a thermonuclear
explosion and the ejected matter will consist of relativistic neutrinos, 
relativistic positron-electron
plasma, non-thermal proton-electron plasma and thermal ionized plasma.  
Neutrinos and positrons can be generated in the thermonuclear reactions. 

\item The observable structure of an accreting rotating black hole has to consist of a 
quasi-spherical pseudosurface made of infalling matter which has been compressed
to degenerate matter densities near the event horizon,  a broad line 
region (BLR) accreted on the non-polar regions of the pseudosurface 
and accreted matter on the polar regions of the pseudosurface
which is enriched and episodically ejected due to energy injection from a
thermonuclear outburst which should occur when favourable physical conditions 
in the accreted matter are achieved.  
The ejected matter from the poles is observed as synchrotron jets and thermal
narrow line region (NLR).  
Dust is generated in the metal-rich clumps formed in the NLR. 
An accretion disk is formed in the non-polar regions from the excess matter
that collects beyond the BLR due to latitude-dependent accretion rates on a
rotating black hole which are minimum for the equatorial regions.  
The observed spectral lines appear at a redshift which is 
some combination of gravitational redshift, Doppler shift and galaxy redshift.  

\item The accreting black hole in all jetted sources is rotating and hosts an
accretion disk in the equatorial regions. 

\item Observations support a positron-electron composition for 
the fast radio jets launched from the polar regions of microquasars and active nuclei.
The light plasma will radiate in the magnetic field that
is frozen in the ejected thermal gas i.e. in the narrow line region. 

\item The fast positron-electron jet is launched from near the event horizon of
a fast spinning black hole and hence requires
relativistic escape velocities.  This explains the relativistic bulk expansion
velocities observed for radio jets and their ballistic nature which persists upto long
distances.  The radial ejection of jet plasma is from the polar region of 
the quasi-spherical pseudosurface which is not influenced by the
centrifugal potential i.e. is devoid of a broad line region and this 
region will be small in extent for a fast spinning black hole.  This 
readily explains the observed range of jet opening angles, conical 
shapes of jets and their high collimation.  The central
black holes in FR~II radio sources have to be the fastest spinning objects in 
the universe that we know.

\item The positron-electron plasma has to emit annihilation line photons
near 511 keV along the entire length of the jet.  Compton scattering of
these soft $\gamma-$ray photons to X-ray energies can explain the formation of
the thick X-ray beams detected along radio jets in several radio galaxies
(e.g. Cygnus~A, Pictor~A).  The annihilation photons can also explain the 
formation of hotspots and backflow lobes
in FR~II sources due to the radiation pressure they exert back on the plasmas
when the annihilation happens beyond the observable jet i.e. beyond
the periphery of thermal gas.  When the lobes flow back to the core in FR~II sources,
the large accretion disk around the black hole will deflect them by 
$\sim 90^{\circ}$ which explains the formation of winged radio structure.

\item The radio plasma in cluster radio halos and relics is 
commensurate with the relativistic positron-electron plasma which was 
ejected from the supermassive black hole in the central galaxy 
during its active phase and which has since diffused over a large region. 
The relics can indicate the orientation of the jet axis of the black hole during its
active phase indicating that a fraction of the  
plasma still retains its forward-directed motion. 
The formation of relics can be explained by radiation pressure exerted
back on the plasmas by pair annihilation photons formed beyond the extent of the
thermal X-ray halo.  

\end{itemize}

\end{abstract}

\tableofcontents

\section{Introduction}

Radio synchrotron emission which requires relativistic electrons and a magnetic field
is detected from most high energy phenomena in the universe underlining the
ubiquity of relativistic plasma and magnetic fields.  
Although the first extraterrestrial
detection of such radiation was less than a century ago, it has since been extensively 
detected from a large range of astronomical objects and has become one of the important
diagnostics available to us for the study of energetic phenomena.  With the advent
and development of multi-band telescopes,  it has also become possible to study the same
astrophysical object in different wavebands and hence obtain a complete observational 
picture which is a tremendous aid to the interpretation process.  This has helped 
shed light on several intriguing astrophysical systems.
However the large data quantities and involvement of many more scientists in research 
have also led to the emergence of several different interpretations of the same
observational results which has caused confusion.  

The paper discusses the strong radio synchrotron emitters namely supernovae, microquasars, active nuclei, 
cluster halos and relics.  While extensive observations have led to solid observational results
and have also allowed classification of these sources into different types,  
it has also led to multiple
explanations put forward to explain parts of the object so that we still do not have
a complete picture.  Explanation for several
key issues elude us so that we do not know all the sources of energy in active nuclei
the launch-site of jets, the origin of the relativistic
plasma responsible for the radio synchrotron emission, the connection between
the prompt short-lived radio emission from supernovae and long-lived radio emission from
supernova remnants,  the origin of radio halos and relics in clusters,
the reason for spectral curvature at low radio frequencies in most radio sources 
and so on.  Aiming to resolve these within the realm of known physics and 
respecting multiband observational results as inviolable boundary conditions, 
we research these problems in detail.  The results of this effort are presented in 
the paper.   

In the paper, black holes refer to compact objects with masses higher than neutron stars
so that their densities have surpassed the values at which degeneracy pressure of 
neutrons can be counter gravitational collapse.  
Since a neutron star is surmised to be typically
$3-4 R_s$ in radius, further contraction can be surmised to lead to matter being squeezed
within a sphere of radius $R_s$ or smaller.  Basically, all the mass is inside the
event horizon and can be supported either by the degeneracy pressure of quarks or
other particles.  We have no way of knowing whether the mass is distributed in a 
sphere of radius $R_s$ or $R_s/100$.  To summarise, we refer to that object
as a black hole wherein all the mass is compressed to within the event horizon -
the mass could range from stellar mass to billions of solar masses.
This is not to be confused with 
the mathematical black hole which is believed to be a singularity. 

The rest of this section presents short summaries of a few relevant points 
which were made in the context of quasars and nova outbursts but are applicable to many 
other astrophysical energetic systems like supernovae and active nuclei and
short summaries of known emission processes. 

 \subsection{Recap from \citet{2016arXiv160901593K,2017arXiv170909400K}}
\label{recap}
Detailed physical models 
which could explain the ultraviolet characteristics of quasars was suggested
in \citet{2016arXiv160901593K} and the multiwavelength characteristics of novae
in outburst and in quiescence was suggested in \citet{2017arXiv170909400K}. 
Several aspects of these models are applicable to other astrophysical events like 
supernova explosions and other classes of active nuclei and hence a short summary of 
the relevant points from these papers is presented here.  

  \subsubsection{Formation of a pseudosurface}
All accreting black holes should form a layer of degenerate matter at or beyond its
event horizon.  
As matter falls in towards the event horizon of the black hole, its
density will increase on being squeezed into an ever-decreasing volume
as the radial separation from the black hole decreases.  At some point outside the 
event horizon,
the matter can exceed the densities at which normal matter can exist
i.e. $> 10^5$ gm cm$^{-3}$.
Electrons will be stripped from the densely packed atoms and form the degenerate 
electron gas. 
At still higher densities, electrons and protons can fuse to form a neutron gas.
This is suggested in analogy to the degenerate matter that 
%is formed as a result of increasing densities and which has been shown to exist
has been shown to arrest further gravitational collapse in
white dwarfs and neutron stars.
The degenerate pressure of neutrons/electrons can then counter
the gravitational pull of the black hole and a quasi-stable layer of matter 
of a certain thickness can deposit outside the event horizon of the black hole.
This surface is referred to as the {\it pseudosurface} of the black hole.
Further accretion will deposit matter on the pseudosurface.  
Since black hole physics is well understood, it might be possible to derive conditions
under which such pseudosurfaces will form and their separation from the event horizon.
The black body component of the observed continuum emission from quasars
which peaks around blue/ultraviolet wavelengths is due to the hot pseudosurface.

   \subsubsection{Origin of relativistic particles}
\label{relpart}
Radio synchrotron emission is detected from some novae especially recurrent novae. 
This means that a population of relativistic electrons can exist 
and whose origin is 
often attributed to post-ejection shock acceleration.  In the shock acceleration 
scenario, one can then infer  
that the nova explosion only energises the heavy particles while the light charged
particles are electrically dragged with the heavy particles and are not relativistic 
when they leave the parent system.  In this scenario, electrons are accelerated by the 
forward shock set up by the explosion. 

As was suggested, a plausible origin for the relativistic electrons
is in the thermonuclear explosion by the same process that energises the heavy 
particles.  Under the simple assumption that the released energy is equally divided 
amongst all particles (both light and heavy) in the ejecta can lead
to electrons by virtue of their miniscule mass to acquire relativistic velocities
in addition to an outward velocity equal to the escape velocity from the white
dwarf.  The efficacy of this process will depend on the available average energy per
particle. 
This explanation does away with having to
find an independent mechanism to accelerate electrons to relativistic velocities.
There could exist a population of charged light particles which are bound to other 
light particles 
of opposite charge and which as a result might acquire higher expansion velocities
than the massive ejecta.  This population could leave the system before the main ejecta. 

   \subsubsection{Clump and dust formation}
Novae outbursts lead to an expanding ejecta and are characterised by 
detection of absorption and emission lines in
the optical bands with their velocity displacement and linewidths respectively, 
being determined by 
the expansion velocity of the line-forming gas.  The velocity shift of the 
absorption lines have been observed to 
systematically change with the outburst phase so that at least four major 
spectral line systems are 
chronologically identifiable in a nova outburst: pre-maximum, principal, diffuse enhanced 
and Orion lines.   The velocity displacement of the absorption lines 
increases from pre-maximum to Orion.  
Observations strongly support a single burst of explosion energy
in novae and hence it has been difficult to explain the higher velocity lines 
by subsequent faster ejections.   Dust is detected from the several nova ejecta
but its origin has remained unclear. 

These points were explained in \citet{2017arXiv170909400K}.
If similar energy is imparted to all particles in the ejecta, then the atoms of heavier
elements would lag as compared to hydrogen and helium in the ejecta of finite thickness.
The heavier atoms can then cluster due to mutual gravity and form clumps in the ejecta.  
These clumps, if optically thick, will be subject to the radiation pressure due to
the white dwarf radiation and can hence acquire a higher velocity than the optically thin
uniform density ejecta.   Absorption lines forming in these clumps will, hence, 
show a larger
velocity displacement compared to the lines forming in rest of the ejecta.  This scenario
explains the higher velocity displacements of the diffuse enhanced and Orion
absorption lines which are often due to transitions in heavy elements. 

Once clumps are formed as described above, it follows that the insides of 
the optically thick
clumps would be ideal sites for dust formation since they consist of heavy elements and
will be shielded from the harsh radiation field of the white dwarf.  It was suggested
that dust forms inside the clumps in the ejecta. 
The reasons for clump and subsequent dust formation in novae are sufficiently
general to be applicable to other systems where matter is ejected. 

   \subsubsection{Formation of accretion disk and bipolar outflow}
While observations strongly support a connection between an accretion disk and
bipolar outflows/jets in a system, the precise nature of the connection
has remained elusive.  One of
the theories that is commonly used to explain formation of an accretion disk is 
that the infalling particles carry angular momentum and hence form the disk to
lose their angular momentum before accreting on the central object. 
One of the theories of jet launching supports its ejection from the accretion disk. 

Distinct physical explanations pointed out 
in \citet{2017arXiv170909400K} are summarised here.  In this explanation, 
the angular momentum carried by the infalling particle is considered inconsequential.  
Instead it is pointed out that the incoming
particles should experience a latitude-dependent attractive force when falling 
onto a rotating
spherical massive object due to the combined effect of the latitude-independent 
gravitational and latitude-dependent centrifugal forces.  It was pointed out that
a particle falling in at the
poles of the rotating object will only be subject to the gravitational force whereas at
the equator, it will be subject to the same gravitational force but maximum opposing
centrifugal force so that the net attractive force at the equator will be lower
than at the poles.
The physical implication of this would be maximum accretion rates at the poles 
and minimum rates at the equator.  In case of spherical accretion
and uniform density of infalling matter, all the infalling matter 
would accumulate at the poles while only a fraction will accrete at the equators
with the remaining matter accumulating outside and forming
an accretion disk.  The radial extent of the accretion disk will be largest at the
equator and it will taper down towards the poles.  
Another outcome of the latitude-dependent accretion rates would be the formation 
of a prolate-shaped accreted envelope on the compact object which in the extreme
case can lead to bipolar ejections i.e. outflows. 

This explanation which follows from the nature of gravitational and centrifugal forces
and spherical accretion of matter by a rotating compact object can account for
several observations: (1) formation of an accretion disk in the equatorial plane of 
the accretor  
(2) formation of bipolar outflows/jets in several objects (3) the connection between
the accretion disk and bipolar outflows/jets.  The explanation also expects that
(1) there should no accretion disk or bipolar outflows around a non-rotating object,
(2) there should be a spherical ejection from a non-rotating object.  
Thus combining these with observations should help us better understand several
accreting astrophysical systems.

We will use these explanations when encountering a similar situation in
the systems under discussion in the paper.

 \subsection{Known Emission Processes}
In this section, synchrotron and inverse Compton processes are briefly introduced since
these frequently explain the observed radio, X-ray or $\gamma-$ray 
emission from many astrophysical systems.  

\subsubsection{Synchrotron emission}
It was suggested by \citet{1950PhRv...78..616A}
that the main physical process responsible for the observed radio emission 
from active nuclei was synchrotron which has stood the test of time 
with observations having proved this origin beyond doubt.  

Synchrotron emission refers to the radiation from relativistic
electrons accelerated in a magnetic field under the influence of the Lorentz force.
This is possible only
when the magnetic field $B$ and velocity $v$ of the electron are not parallel
so that Lorentz force $F = e (v \times B) \neq 0$ where $e$ is the electronic charge.
A large part of of the energy spectrum
of the radiating electrons is approximated by a power law with 
index $p$ i.e. $ N(E_e) dE_e = \kappa E_e^{-p} dE_e$ where $N(E_e)dE_e$ denotes the
number of electrons per unit volume with energies between $E_e$ to $E_e + dE_e$
and $\kappa$ is an indicator of the total number of electrons. 
The observed synchrotron spectrum due to such an electron population
will be a power law with an index $\alpha = (p-1)/2$ i.e. $S \propto \nu^{-\alpha}$.  

Synchrotron emissivity at frequency $\nu$ is defined as
\begin{equation}
J_\nu = A~\kappa~B^{(p+1)/2}~\nu^{-(p-1)/2}
\label{eqn1}
\end{equation}
where A is a constant.  Thus, the emissivity depends on the total number of radiating electrons
$\kappa$, magnetic field $B$ and the electron energy spectrum index $p$.  
An interesting feature of synchrotron emission which is particularly useful
in interpretation is that electrons of a given energy
emit maximally at a particular characteristic frequency $\nu_c$ (which is related to
the cyclotron frequency) in a given magnetic field.  The simplified formula that
quantifies this and is often used in literature is: 
\begin{equation}
\nu_c \sim 0.016~B_\perp~E_e^2 
\label{eqn2}
\end{equation}
where $\nu_c$ is in GHz, $B_\perp$ is in $\mu$G and $E_e$ is in GeV. 
This formula then allows us to calculate the energy of the electrons
which principally emit at some $\nu_c$, if the magnetic field
can be estimated from polarisation studies and equipartition arguments.
For example, the energy of the electrons which dominantly radiate 
between 0.1 to 10 GHz in a magnetic field of $10 \mu$G will be 0.8 to 8 GeV.
If the magnetic field is stronger at $100 \mu$G then electrons of lower energy 
i.e. between 0.25 to 2.5 GeV will radiate within the same band. 
Thus, Equation \ref{eqn2} allows us to generate the energy distribution of the 
parent population of electrons given a magnetic field.
It also means that if the magnetic field strength is spatially varying in an object, 
then the synchrotron radiation at the same frequency will be from electrons of
varying energies.  

Similarly a simplified equation is used in literature to estimate the
the synchrotron lifetime of a radiating electron:  
\begin{equation}
t_s \sim 1.06 \times 10^9 B_\perp^{-3/2} \nu_c^{-1/2}
\end{equation}
where $B_\perp$, $\nu_c$ are in $\mu$G and GHz whereas $t_s$ is in years.  
The lifetime of electrons in a magnetic field of $10 \mu$G and 100 $\mu$G emitting
at a radio frequency of 10 GHz will be $1.06\times10^7$ years and  
$1.06\times10^{5.5}$ years respectively.  Lifetime of electrons in the same fields
emitting at a X-ray frequency of $10^8$ GHz will be $1.06\times10^{3.5}$ years and
$1.06\times10^2$ years respectively.  This only showcases the shorter
lifetimes of  X-ray emitting electrons due to higher energy losses 
as compared to radio-emitting electrons. 
The estimated lifetimes of relativistic electrons then allows us to estimate the
distances they can traverse before they stop radiating. 
Thus, for a bulk speed of 0.1c, the X-ray emitting electron with a 
lifetime of $1.06\times10^2$ years can travel $1\times10^{14}$ km i.e. 3 pc
whereas the radio emitting electron with a lifetime of $1.06\times10^{5.5}$ years
can travel 10 kpc.  Obviously for larger bulk velocities 
the electrons will travel further from the core.  Lower magnetic fields will
cause lower energy losses and will also allow the electron to travel further
from the core.

The bulk velocity of the emitting electron (mass $m_e$) is often expressed in form of 
the bulk Lorentz factor $\Gamma$ as $E_e = (\Gamma - 1) m_e c^2 $ 
where standard MKS units are used.  
If energy is specified in GeV, then  $\Gamma = 1953.6 E_e + 1$. 
The bulk velocity $v_e$ is related to the bulk Lorentz factor $\Gamma$ as
$\Gamma = 1 / \sqrt{1-v_e^2/c^2}$ and is often expressed in terms of
$v_e = \beta c$.   

Synchrotron radiation is expected to be linearly polarised. 
The polarised emission can help estimate the
intensity of the ordered magnetic field from equipartition arguments.  
While the total intensity measures the 
total magnetic field (random and ordered), measurement of linearly polarised emission
allows the ordered component of the magnetic
field to be estimated.   Detailed treatment of synchrotron emission process can
be obtained from textbooks. 

\subsubsection{The Compton processes}
The mechanism wherein low energy (frequency) photons gain energy on being
scattered off relativistic electrons which lose energy in the process is referred
to as the inverse Compton process 
This process is inverse of the Compton process 
in which high energy photons lose energy on collision with electrons 
which in turn gain energy \citep{1923PhRv...21..483C}.  Both the processes 
are important in astronomical systems given favourable conditions.  
In the Compton process, the high energy photon (generally effective
with X-ray or $\gamma-$ray) loses
energy to electron whereas in the inverse Compton process, the electron loses
energy to a photon (any wavelength).  Inverse Compton process is often found
to be responsible for the formation of hard X-ray photons in active nuclei. 

The change in the wavelength of the photon, 
Compton scattered by an electron depends mainly on the scattering angle $\theta$. 
This Compton shift is estimated by the relation: 
\begin{equation}
\lambda^{'} - \lambda = h/(\gamma m_e c)~ (1 - cos\theta) 
\label{comp}
\end{equation}
where $h / m_e c = 2.43 \times 10^{-12}$ m is the de Broglie wavelength of 
the electron.  When $\theta = 90^{\circ}$, the shift in the wavelength of the 
high energy photon is equal to the de Broglie wavelength of the electron
while when $\theta = 180^{\circ}$ the shift is twice the de Broglie wavelength which
is the maximum value of the shift.
This process is likely to be most effective in situations involving
photons whose energy is close to the rest mass energy of electrons. 

The mean frequency of the photon energies which are inverse Compton scattered 
by relativistic electrons can be estimated from the equation:
\begin{equation}
<\nu_{IC}> = 4/3~\Gamma^2~\nu_0
\label{ic}
\end{equation}
where $\nu_0$ refers to the frequency of the low energy photon which is boosted to 
a higher frequency $\nu_{IC}$ due to collisions with relativistic electrons of
energy $\Gamma$.
For example, a photon of frequency $\nu_0=300$ GHz when scattered by 
an electron of energy $\Gamma=1000$ will result in a photon of frequency 
$\nu_{IC}=4\times10^8$ GHz which is an X-ray photon.  This has been found
to be an important process responsible for generation of X-ray photons
in active galaxies.
A special case of this process is when the radio photon
generated by the synchrotron process is boosted to X-ray energies by collisions with the 
same synchrotron-emitting electrons.  This is referred to as self synchrotron
Compton (SSC) process.   This process is found to be particularly active in bright
synchrotron emitting regions where both radio photons and energetic
electrons are available in plenty.  For example, a $\nu_0=10$ GHz radio photon will 
be radiated by an electron of energy 1.44 GeV ($\gamma=2813$) gyrating in a magnetic 
field of strength $300 \mu$G through the synchrotron process.  The 10 GHz photon 
can then collide with the electron of $\gamma=2813$ which can boost the radio
photon to X-ray energies i.e. $\nu_{IC}=10^8$ GHz. 
In SSC, the energy losses suffered by the relativistic
electrons will be due to both synchrotron and inverse Compton processes 
which can lead to reduction in their lifetimes. 

More detailed treatment of Compton processes can be obtained from textbooks. 

 \subsection{This work}
Keeping the above in mind, we study supernovae (and supernova remnants), 
microquasars, active nuclei and cluster relics, halos.  
The focus in this work is more on the synchrotron 
emitting sources and their properties although other wavebands are included when relevant
to the discussion.  Before we embark on study of these exotic objects, we discuss the
origin of the observed spectrum of cosmic ray energies and connect it to the observed 
synchrotron radio spectra.

\section{Cosmic rays}
\label{cosrays}
We start with a short summary of particle acceleration mechanisms from literature
and then proceed to put forth a more likely scenario for particle acceleration
that is supported by observations.  This builds on the brief summary
presented in the introduction. 

\subsection{Background on particle acceleration}
One of the mechanisms is Fermi second order acceleration in which the proton component of 
cosmic rays is accelerated in the interstellar medium
due to collisions with irregularities in the magnetic field \citep{1949PhRv...75.1169F}. 
%The magnetic field also confines the cosmic rays to the galaxy.  
While this mechanism results in a power law distribution for the energies of protons
and has hence been widely used, it runs into difficulties in accelerating 
heavy nuclei and electrons and the acceleration is a rather slow process with energy
increase being proportional to $v^2/c^2$  \citep{1949PhRv...75.1169F}.  
The mechanism requires a small pool of seed protons of energy $\ge 200$ MeV and it was suggested 
that such protons 
could be produced in the vicinity of magnetically active stars \citep{1949PhRv...75.1169F}.
It is important to keep in mind that this acceleration mechanism was suggested when
the origin of nova outbursts and supernova explosions were not fully understood
while active nuclei and black holes were yet to be identified and understood.  
Our knowledge, hence, about the immense energetics of such systems and its origin was 
fairly limited.   
In that situation, Fermi acceleration was revolutionary since it could explain the energies
of the detected cosmic rays especially protons and the observed power law spectrum although the 
process was slow.  Fermi's second order acceleration found it difficult to explain
the existence of relativistic electrons in cosmic rays because it was estimated that the losses
incurred by electrons in a collision generally exceed the gain so that the electron
does not gain energy \citep{1949PhRv...75.1169F}.  It is interesting to note that at that time, 
the ubiquity of relativistic electrons in the interstellar medium was not yet known 
and hence this inability was not considered a drawback of the mechanism.  It is
to be noted that the first detection of radio synchrotron emission from
outside the solar system \citep{1933Natur.132...66J}
had only been done in the early 1930s and systematic radio studies were only undertaken
in the 1940s.  The synchrotron origin for the low radio frequency emission was 
recognised in 1950 \citep{1950PhRv...78..616A}. 

In analogy to acceleration of protons due to repeated collisions with magnetic 
irregularities, a mechanism which instead of magnetic irregularities uses 
repeated shock crossings by the particle has been found to more satisfactorily explain
cosmic rays including electrons.   Referred to as shock acceleration, it
has been considered as a possible
mechanism for accelerating cosmic rays since 1960s.  Shock acceleration 
results in a power law distribution of energies of roughly the index observed in 
cosmic rays 
\citep{1978MNRAS.182..147B}.  The seed particle population was considered 
to be relativistic particles which were further accelerated 
by the shock front \citep{1978MNRAS.182..147B}.  In this process, acceleration is achieved
by the fast particles making repeated shock crossings and the confinement of the charged
particles to the shock front is achieved by scattering off Alfven waves which are also
generated by the particles \citep{1978MNRAS.182..147B}. 
This mechanism has come to be known as 
first order Fermi acceleration due to the increase in energy being proportional to $v/c$
which is faster than second order Fermi acceleration.
The first order Fermi acceleration is also referred to as diffusive shock 
acceleration (DSA) especially when the shock is supersonic.
DSA has been the favoured mechanism since
strong shock waves are expected in several astrophysical systems like supernovae
which are believed to be one of the major sites of particle acceleration.
Supernova remnants are also believed to be an important source of cosmic rays with the
energy being extracted from the expansion energy of the supernova remnant.
DSA is believed to impart similar velocities to all particles so that lighter
atoms/ions acquire lower energies than heavier atoms/ions.
The seed particles to the DSA mechanism are considered to be the pool of 
thermal particles. 
Currently DSA is the favoured mechanism for explaining the existence of 
cosmic rays in the universe.  Another acceleration mechanism which is often
used to explain the relativistic electrons in cluster halos is turbulence 
generated by a cluster merger.
It appears that while shock or turbulence-driven acceleration might be able to 
accelerate particles over
long timescales, the relativistic electrons and cosmic rays that are detected
immediately after energetic outbursts
like supernovae, novae etc has to be due to another, must faster physical process.

Now that we know about the huge energy output of events like supernova explosions
and jet launching, it seems advisable to revisit the particle acceleration mechanism.    
The easiest and obvious alternative to DSA is the scenario wherein energetic events 
like nova outbursts and supernova explosions 
adiabatically impart comparable energy to all the available particles 
(heavy and light particles)
before they are ejected from the explosion site as summarised in section \ref{relpart}
and \citet{2017arXiv170909400K}.  Since we already accept that heavy matter is energised
by the explosion energy, it appears contrived to assume that light particles are
left out.  
Although this explanation removes the need to require shocks for particle
acceleration, the role of the blast wave and shocks set up by the explosion 
and fast ejecta remains important in compressing matter and hence
enhancing matter densities and magnetic fields.  This, then, leads to simplifying the
discussion since there remains no need to invoke any extra physical process to accelerate
electrons to relativistic energies or search for a seed population of particles.
Observations support all particles leaving an explosion site with tremendous energy and
expending it in radiation, kinetic motion etc over a long timescale.
In fact, if we believe the energetic event is unable to accelerate electrons 
to relativistic energies then we have to explain why electrons are selectively
excluded from being energised whereas heavier particles are energised. 

In the following sections we discuss the observed cosmic ray spectrum. 

 \subsection{Observed cosmic ray energy spectrum}
\begin{figure}[t]
\centering
\includegraphics[width=6.0cm]{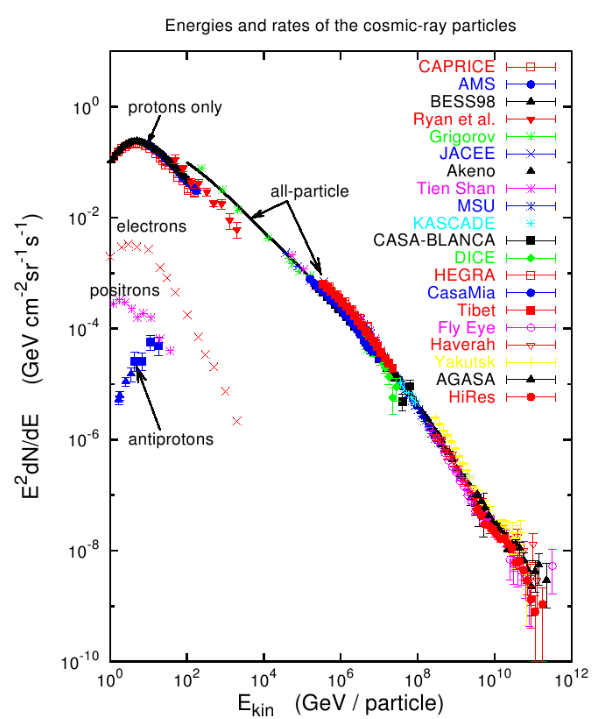}(a)
\includegraphics[width=6.0cm]{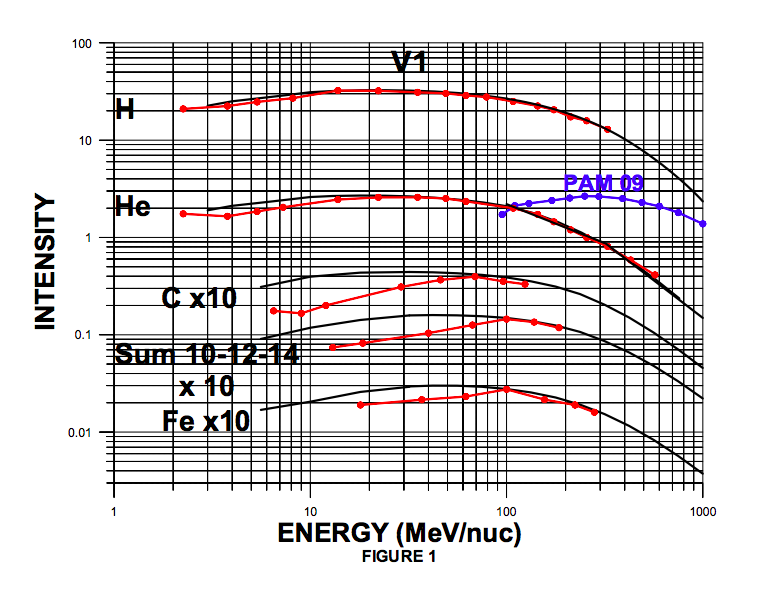}(b)
\caption{ \small (a) Figure shows the distribution of cosmic ray intensity with energy.  
Figure downloaded from the internet. 
(b) Figure reproduced from \citet{2015arXiv150801542W}. Figure shows the
distribution of intensity of low energy cosmic rays of H, He, C, (Ne+Mg+Si), Fe 
measured by Voyager 1 beyond the heliopause.  Notice the turnover observed in
the energy spectra of cosmic rays. }
\label{fig1}
\end{figure}

The observed cosmic ray energy distribution is shown in Figure \ref{fig1}.  
Figure \ref{fig1}a shows the combined spectrum of all cosmic ray particles (also
separately for electrons, protons, positrons) between energies of 1 GeV to a few times 
$10^{11}$ GeV.  A few gross features are recognisable from the spectrum: a turnover at low energies, 
a power law-like behaviour at higher energies and a spectral break inbetween often referred 
in literature as the `knee'.
The energy spectrum of most cosmic rays shows a turnover at energies $\sim 1$ GeV.  
Although the intensity of all species of cosmic rays is found to turnover at low
energies, the origin of this turnover is not well understood.  Some of it is attributed
to solar influence but considering that the turnover was even noted by the Voyager
spacecraft beyond the heliopause \citep{2015arXiv150801542W} as shown in Figure \ref{fig1}b 
indicates that the turnover has to be intrinsic to the distribution. 
It should be kept in mind that the global cosmic ray spectrum in Figure \ref{fig1}a 
contains
contribution from several independent cosmic ray sources - different types of sources
(e.g. supernovae and active nuclei) and
several same type of sources (e.g. contribution from several supernovae and
several active nuclei).   The most intense sources will dominate the cosmic
ray spectrum and depending on the energies of the cosmic rays generated by the different
sources, they can dominate different energy ranges.  The global integrated cosmic ray spectrum
shown in Figure \ref{fig1}a is observed to show two breaks at high energies, which is
interpreted in literature, to indicate two distinct dominant sources of contribution  
to the spectrum.  The break referred to as the `knee' occurs around few times 
$10^{15}$ eV and it is believed that the cosmic rays below this energy are 
accelerated in supernova explosions.  
The cosmic rays of energies greater than the knee are believed to be accelerated 
by active nuclei.  
The second break referred to as the `ankle' occurs between $10^{18}$ and $10^{19}$ eV.
Overall, the higher energy cosmic ray spectrum can be approximated by a 
power law of index $p$ between 2 and 3.

Figure \ref{fig1}b shows the energy spectrum for a few atomic species with
particle energies ranging from 0.001 GeV to about a GeV that was recorded by the Voyager
spacecraft beyond the heliopause \citep{2015arXiv150801542W}.  
The figure is a blow-up of the part of the spectrum where the intensity of lower
energy cosmic rays shows a turnover and it appears to be around 0.1 GeV which
is comparable for primary cosmic rays of hydrogen and iron although the lighter atoms
seem to show a flatter turnover than the heavier atoms \citep{2015arXiv150801542W}.  
The peaks in Figures \ref{fig1}a and b differ and these could be attributed 
to solar effects.

A typical nova outburst ejects about $\rm 10^{-5} M_\odot \sim 2 \times 10^{25}$ kg
of matter.  If this entire mass was composed of hydrogen then it would mean 
about $1.25\times 10^{52}$ protons and at least that many electrons are 
ejected in the process.  A typical nova outburst can release $\sim 10^{38}$ Joule 
of energy and if we assume that all of it is 
equally distributed amongst the particles, then it would impart an average energy
of $\sim 0.25 \times 10^{5}$ eV to each particle.  
Similarly, we can make order of magnitude estimates for a supernova explosion of
energy $10^{45}$ Joule which ejects half a solar mass of matter.  If the ejected mass
consists of protons then
it would eject $0.6 \times 10^{57}$ protons and an equal number of electrons.  
The energy acquired
by each particle then would be $0.5 \times 10^{7}$ eV.  This simple argument is 
included to
show the lower limit for the energy imparted to particles in nova or supernova
explosions.  In practise the ejecta will contain heavier atoms also and hence
the energy imparted to each particle will be higher.
In the next section we discuss the origin of the observed 
energy distribution of cosmic rays. 

 \subsection{Normal distribution of cosmic ray energies or universal spectrum}

\begin{figure}
\centering
\includegraphics[width=5cm]{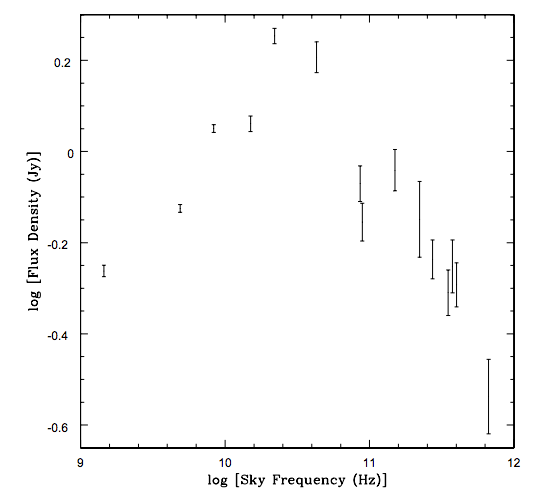} a
\includegraphics[width=4.5cm]{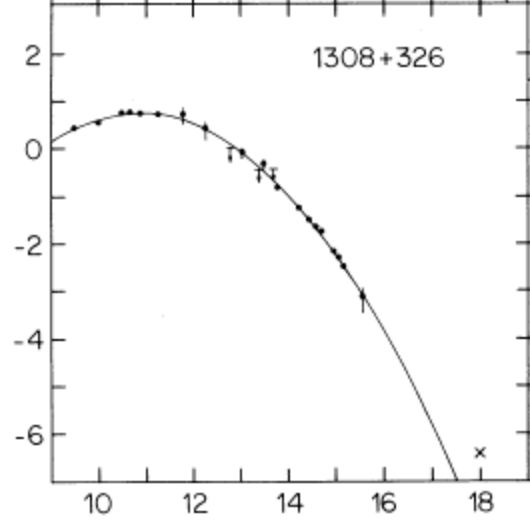} b
\caption{\small (a) The spectrum of the core of the radio galaxy Cygnus A.  Figure 
copied from \citet{1999AJ....118.2581C}.  Notice that the peak occurs near 30 GHz.
(b) Wide-band spectrum of the active nucleus 1308+326 copied 
from \citet{1986ApJ...308...78L}.  X-axis is in units of log(frequency).  Notice the
log parabolic nature of the spectrum. }
\label{fig2}
\end{figure}

\begin{figure}
\centering
\includegraphics[width=6cm]{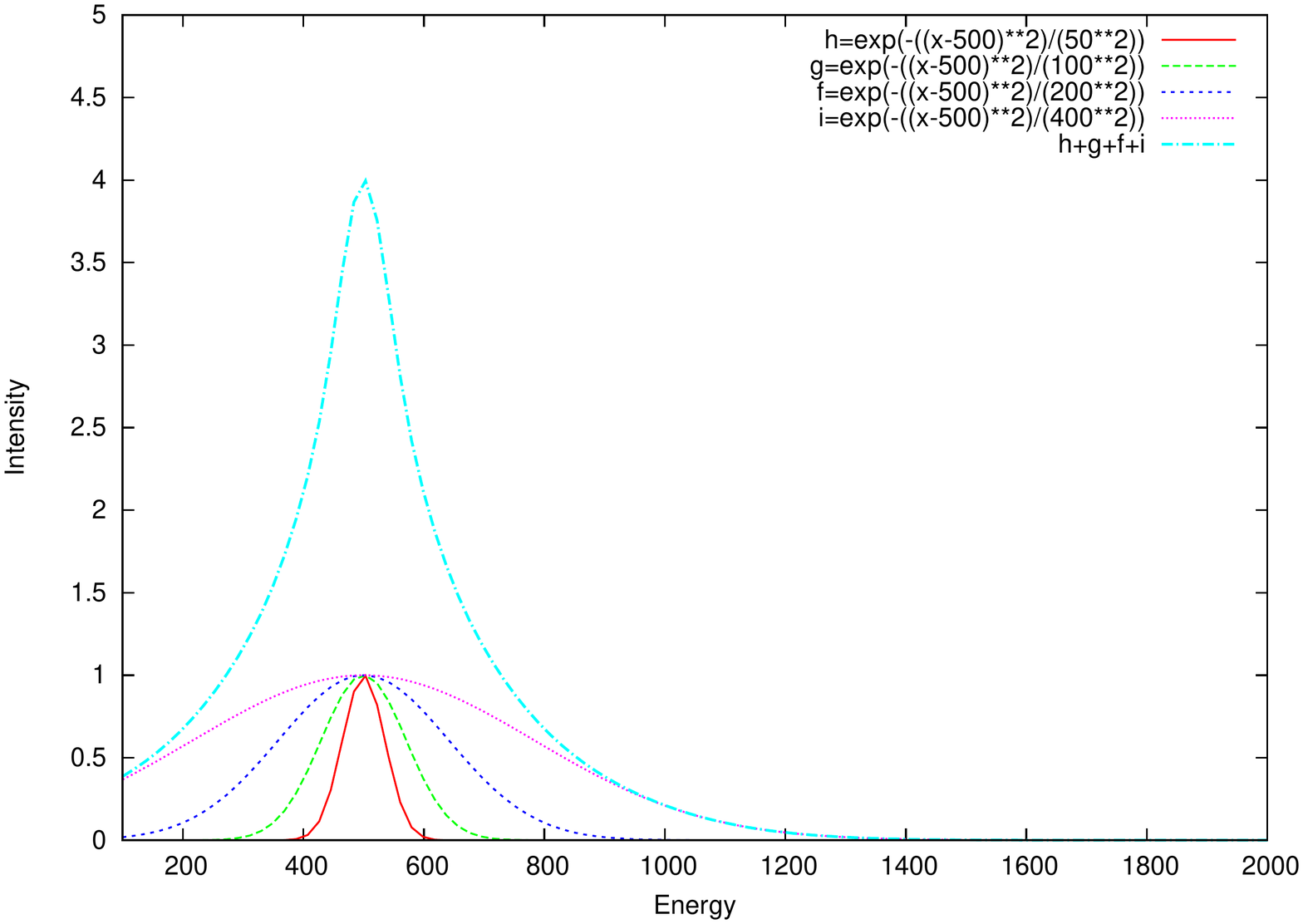}(a)
\includegraphics[width=6cm]{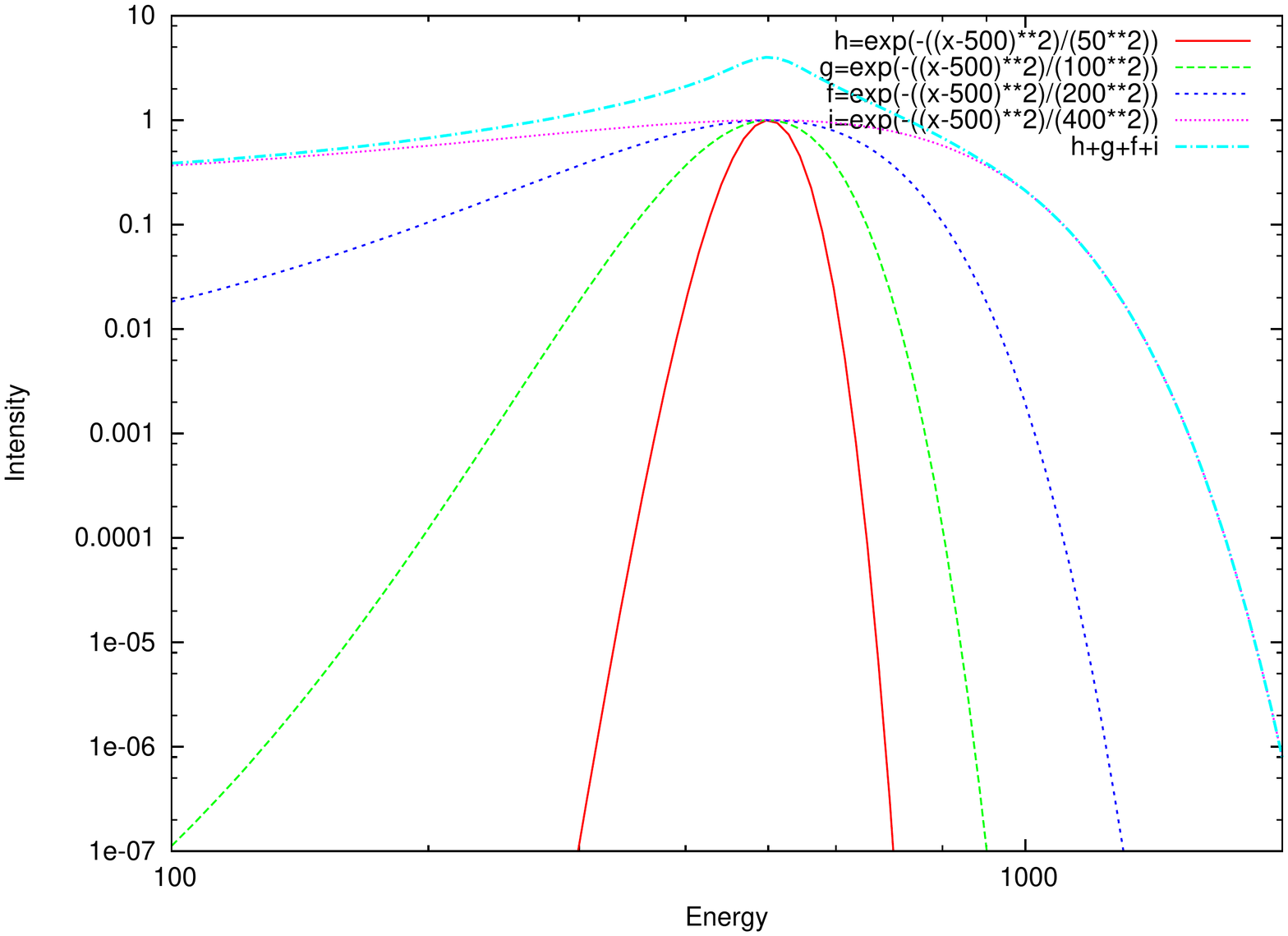}(b)
\includegraphics[width=6cm]{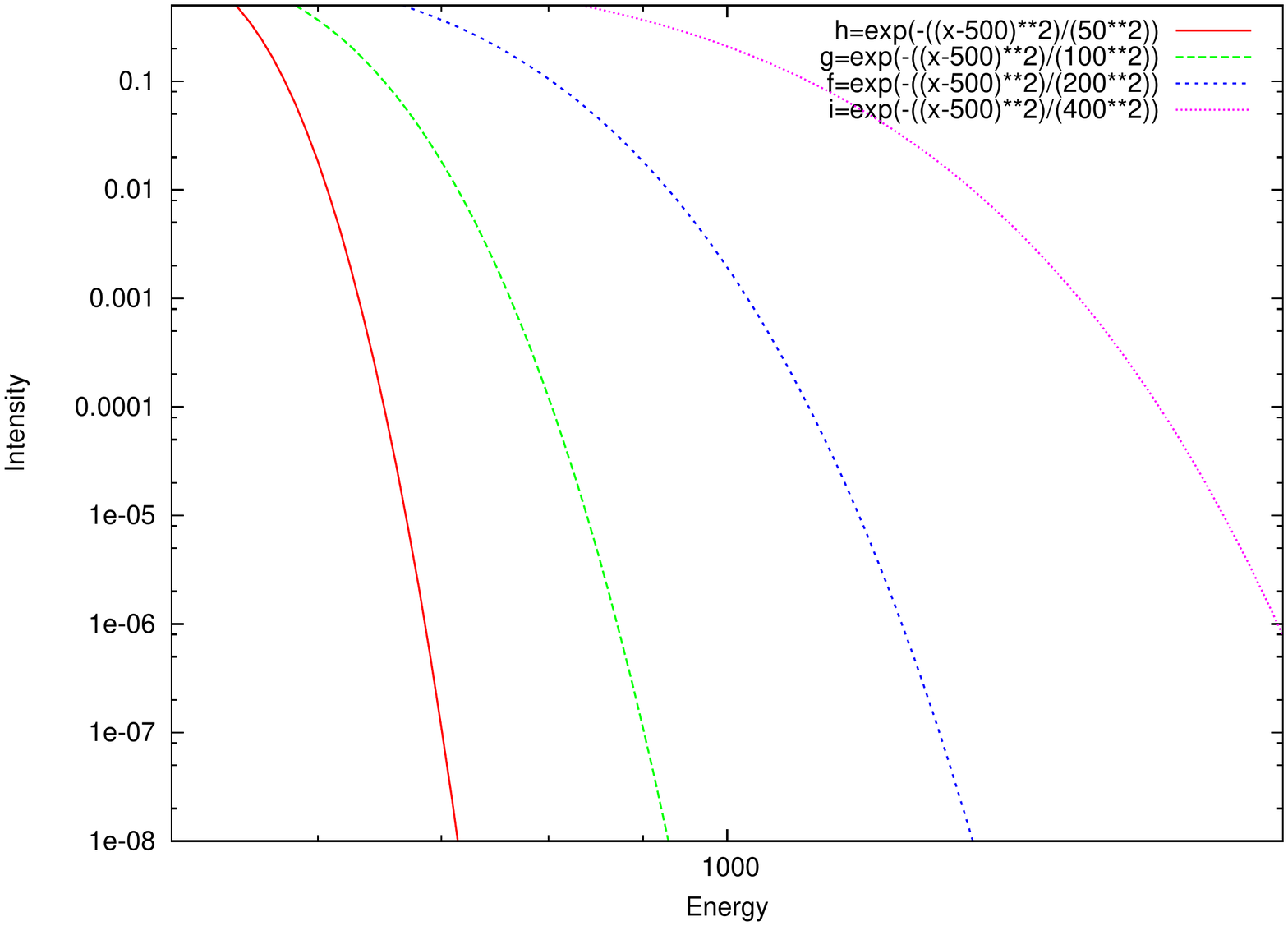}(c)
\includegraphics[width=6cm]{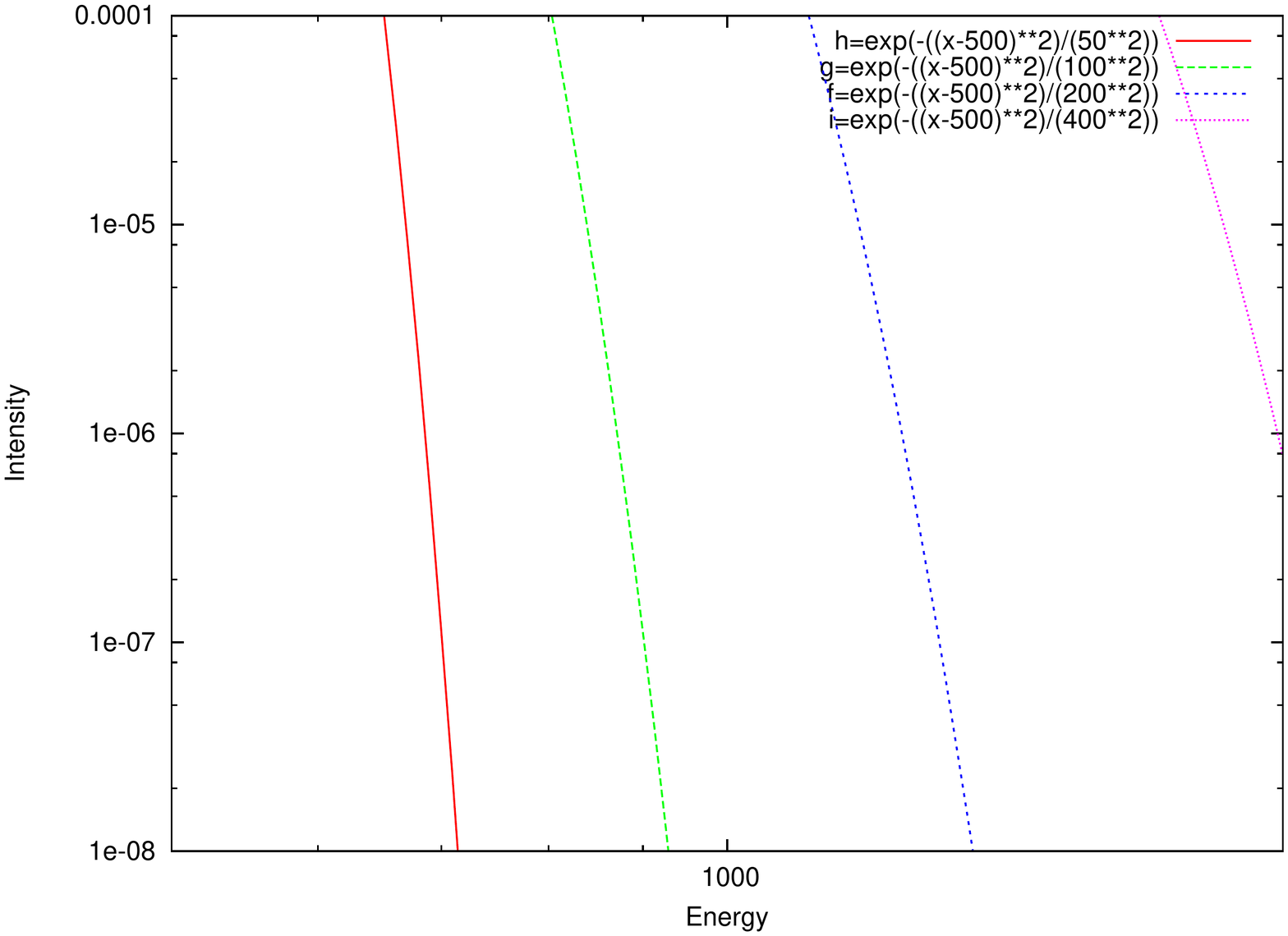}(d)
\caption{\small Indicative plots for the intensity of cosmic rays with their energy in
arbitrary units assuming that the injected distribution of energies is normal/Gaussian. 
(a) Linear scale.  (b) Logarthmic scale. (c) Logarithmic scale, the peak is omitted.  The
curves are power law for high energies and show a flattening at lower energies.  These
reproduce the observed curves fairly well.
(d) Logarithmic scale, the peak is omitted and the range of on the y-axis is limited to
four orders of magnitude.  Notice how the curves are better approximated with
a power law for the narrower energy range.}
\label{fig3}
\end{figure}

We keep in mind that the observed cosmic ray distribution shown in Figure \ref{fig1}(a) 
has a shape that can be approximated by a power law at high energies which peaks at
low energies (0.1  to 1 GeV) and the spectrum turns over at still 
lower energies.  
In literature, two explanations exist for the origin of the observed nature 
of the cosmic ray/electron 
energy distribution:  (1) the power law energy spectrum is due to the injection 
spectrum of electron energies expected from DSA and 
the low frequency turnover is due to solar modulation or a cutoff due to
some appropriate process (2) the observed energy
spectrum is intrinsic which is often referred to as the universal spectrum.  Since 
no satisfactory theory to explain the origin of a universal spectrum has emerged,
the first alternative has been favoured in literature.  Here we
explore the existence of processes that would support 
a universal spectrum.  Motivation for this arises from strong observational 
evidence for energising of all particles before they are ejected.
Fast neutrinos, which are amongst the first signatures of the explosion, are
detected from supernovae and hence require the neutrinos to be energised
at the site of the explosion before being expelled at relativistic
velocities.  It follows that the same should be true
for all other particles which are ejected following the explosive event. 
It appears that this argument convincingly rules out shock acceleration of 
electrons beyond the explosion site and strongly supports the same physical mechanism 
being responsible for the acceleration of neutrinos, electrons and 
heavy particles. 

The observed spectral energy distribution (SED) of blazars and the compact
cores of radio galaxies, especially in the radio band, is generally peaked
and curved  (see Figure \ref{fig2}) so that the power law behaviour 
is found to be valid in a narrow radio frequency range.  BL Lacs are classified as HBL
(high frequency BL Lac) if the spectrum peaks in X-rays and as 
LBL (low frequency BL Lac) if the spectrum peaks at radio 
frequencies.  The reasons used to explain the curved SED of blazars 
and compact radio cores range from an intrinsic curved injection
spectrum of relativistic electrons \citep[e.g.][]{2004A&A...413..489M}
to the effect of radiative losses and escape of high energy electrons 
\citep[e.g.][]{2002MNRAS.336..721K}.  The curved smooth nature of the 
observed spectrum from radio to ultraviolet bands of active nuclei espeically blazars is well-approximated 
by a log-parabolic function \citep{1965ApJS...10..331H,1986ApJ...308...78L}.  
The spectra of BL Lac objects
show a log-parabolic shape with a narrow range of curvature parameters 
\citep{2002babs.conf...63G} 
which could indicate that this shape is a general characteristic of blazars
\citep[e.g.][]{2004A&A...413..489M}.  
Comparing Figures \ref{fig1} and \ref{fig2} reveals that the shapes of both the energy
distribution of cosmic rays and radio spectra of radio cores/blazars can be
approximated by a power law component and a curved component.  In other words,
since the parent electron energy spectrum is curved,  the observed
radio synchrotron spectrum will also be curved when radiating in an
appropriate magnetic field.   

Being convinced that electrons and other primary cosmic rays are  
instaneously energised in explosive events before being ejected from the system,
we think of the energy distribution that these particles would 
acquire in such an event.   Since the number of particles that is being
energised is very large, it is
a statistical system for which the favoured distribution should be a
normal (or log-normal) distribution which 
which are maximum entropy distributions for a given mean and variance. 
Since the second law of thermodynamics tells us that
all processes lead to an increase in entropy of the system,  
the process of instantaneous energising of the cosmic rays has to 
result in a normal (or log-normal) distribution of particle energies. 
Another way of looking at it is from Gibbs entropy formula,
$S = -k~ \Sigma p_i$ln$(p_i)$ where $k$ is
the Boltzmann constant and $p_i$ is the probability that some microstate $i$ takes up an
energy $E_i$.  Gibbs formula applies to both equilibrium and non-equilibrium systems.
The most likely microstates (i.e. particle energies) are those which 
maximise S and hence the distribution, typically displayed by large statistical
systems, is normal (or log-normal).   

The particulate matter that acquires an outward component of velocity
which exceeds the escape velocity from the object will be
lost from the system.  One can think about the centroid
of the normal distribution to signify the escape velocity from the object
and the dispersion to signify the magnitude of the random velocity component i.e.
the difference in energy between the total imparted energy per particle and 
the required escape energy.  One can think about the escape velocity as the
radially outward directed component which can range from
sub-relativistic to relativisitic whereas the random velocity component
has to be relativistic if the plasma is to be capable of emitting synchrotron
radiation else no synchrotron radiation will be emitted. 

In Figure \ref{fig3}, four normal distributions having the same mean but
different variance are shown alongwith the result of adding
the four distributions. These can be 
taken to be indicative of the cosmic ray spectrum
so that the x-axis is an indicator of the cosmic ray energy and the y-axis is an
indicator of the intensity of the cosmic rays in arbitrary units. 
Figure \ref{fig3}a is in linear scale while rest of the panels are
on the logarithmic scale.   Figure \ref{fig3}b can be compared to the spectrum of
the radio core of Cygnus A (Figure \ref{fig2}).  The two lowermost 
panels are zoom-ins to demonstrate the peak and power-law parts of the gaussians.
The observed radio spectra of synchrotron sources also show such shapes. 
Only the particles whose energy exceed the escape velocity from the object
will be able to leave which means if the centroid is at the escape velocity
than only the distribution at higher energies will be observable.  The rest of
the particulate matter will remain on the system. 
Interestingly, the slope of the curves, which can be approximated by
power laws, for the four distributions within the
narrow range which highlights their linear nature plotted in 
the lowermost panel of Figure \ref{fig3}b appear to be comparable. 
The distributions with the same centroid 
but varying variance, plotted in Figure \ref{fig3} schematically 
depict the energy distributions which would be expected for
different species of particles.  The narrowest distribution should
correspond to the heaviest element and the broadest distribution correspond to
the lightest particle.  
This can be understood as follows: for comparable energy imparted to all particles, 
the heavier
particle will use up a larger fraction of its energy in escaping from the system 
i.e. $m v_{esc}^2/2$ due to its large mass $m$ while lighter particles like electrons 
will need to use only a small fraction of
its energy $ m_e v_{esc}^2/2$ to escape from the system and the remaining energy
will be in the random component which can be relativistic, making them efficient
synchrotron emitters.

As mentioned earlier, the global cosmic ray spectrum that we observe 
has to be a combination of multiple spectra from several different sources.  
In Figure \ref{fig3}a,b, the spectrum obtained by adding four
distinct distributions with the same mean has also been shown.  This spectrum closely
follows the distribution with the largest dispersion except near the peak which
is more pronounced compared to the individual distributions.  The nature of
the integrated spectrum can also account for
`knee`-like features.  This allows the inference that combination of multiple normal
distributions can explain the nature of the low energy turnover, `knee' 
and `ankle' features of the global cosmic ray spectra and supports the
existing hypothesis wherein the features delineate the distinct source of
cosmic rays.  

To summarize, it is suggested that the energy spectrum of relativistic 
electrons (and cosmic rays) 
that are energised in an explosive event like nova, supernova and other transient events
will follow a normal (or log-normal) distribution of energy with a mean and variance 
determined by the energetics of the event.  The distribution can be approximated by 
a power law at high energies and shows a smooth convex-shaped turnover at low energies.  
In the next section we 
discuss if the typical observed radio synchrotron spectrum of radio sources 
is consistent with the explanation given here.   

 \subsection{Observed radio synchrotron spectrum}
\begin{figure}
\centering
\includegraphics[width=6cm]{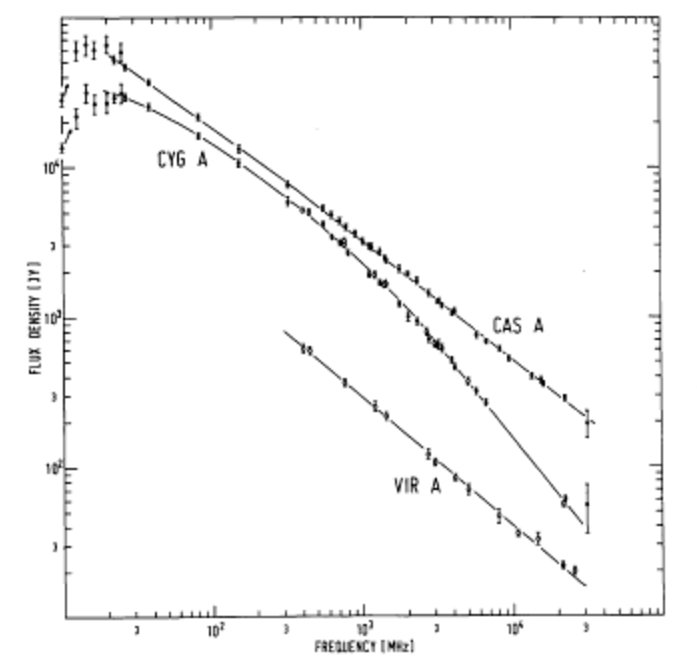}
\caption{\small Figure shows the observed radio spectrum of Cas~A and Cygnus~A copied from
\citet{1977A&A....61...99B}.  Note the turnover at low radio frequencies and the
convex-shaped spectrum of Cygnus~A.}
\label{fig4}
\end{figure}
In Figure \ref{fig4}, the observed integrated radio spectra of three strong radio sources:
the Galactic supernova remnant Cas~A and the radio galaxies Cygnus~A and Virgo~A are
shown.   The spectra of Cas~A and Cygnus~A show a turnover at frequencies lower than
about 30 MHz.  The spectrum of Cas~A at frequencies above 30 MHz and Virgo~A are 
well-approximated by a power law whereas the spectrum of Cygnus~A shows a convex 
shape beyond 30 MHz.  The behaviour of the spectrum of Cygnus~A 
is generally attributed to the effect of different components of the 
radio galaxy. 
It is interesting to note that the shape of the radio spectrum of Cygnus~A compares well
with the global cosmic ray spectrum shown in Figure \ref{fig1}a. 
The low frequency turnover in Cas~A is explained by free-free foreground absorption 
whereas it remains to be understood in the case of Cygnus~A.  In fact, the radio spectra
of several (all?) radio sources show a turnover at low frequencies and these
are generally attributed to free-free absorption, synchrotron self-absorption or
a low energy cutoff.  Moreover the different components of the radio source
also show such turnover (see Figure \ref{fig2}). 

From the discussion in the previous section, turnovers in the observed
spectral energy distribution are expected due to the very nature of the injection 
spectrum of the electron energies i.e a normal distribution.  
The observed differences in the peak frequencies 
will depend on the details of the energy injection and ageing. 
Thus, the core spectrum of Cygnus~A (Figure \ref{fig2}a) shows a peak around
30 GHz, whereas the integrated spectrum of Cygnus~A which includes other
radio components shows a peak around 30 MHz. 
If we assume that the turnover in the spectra of Cas~A and
Cygnus~A are due to the underlying electron energy spectrum then we can estimate
the energy of electrons that emit near 30 MHz given a magnetic field.
For a magnetic field of $100 \mu$G which is typical in supernova remnants and radio 
galaxies, the electrons emitting at 30 MHz would have energies of $\sim 140$ MeV.  
Interestingly, from Figure \ref{fig1}b, it is seen that several cosmic ray species 
are detected with a flat energy spectrum near 100 MeV. 

Since the effect of thermal free-free absorption, synchrotron self absorption and
the turnover due to the energy spectrum are similarly 
manifested in the radio spectrum, it is
difficult to disentangle the different contributions and verify the effect of the
curved injection spectrum.  Cases where the effect of free-free absorption
and synchrotron self-absorption is expected to be minimal need to be studied.  

\section{Supernovae and supernova remnants}
Supernova explosions are divided into two main types (type I and type II) based 
on the spectral lines that are detected during the phase of peak optical emission 
with type 1 lacking hydrogen lines and type II showing the presence of hydrogen lines
\citep{1941PASP...53..224M}.  Type Ia explosions are associated with a
white dwarf mass crossing the Chandrasekhar limit while type II 
explosion is associated with core-collapse 
in massive stars.  The origin of the energy release in Type II 
supernova explosion is believed to be the sudden gravitational 
contraction of the stellar core (implosion) when it runs out of fuel
and ignition of matter outside the core which blows away the outer parts of the star 
i.e. explosion \citep{1960ApJ...132..565H}.  The energy released when light elements
other than hydrogen ignite at a billion degrees K is about 
$5 \times 10^{17}$ ergs gm$^{-1}$
\citep{1960ApJ...132..565H}. If $0.5$ M$_\odot$ of matter were ignited in a supernova
explosion then it could lead to an instantaneous release of energy $\sim 10^{51}$ ergs 
which can be catastrophic for the star.  This energy is released within
1-100 seconds \citep{1960ApJ...132..565H}.  In case of hydrogen fusion, 
the energy release can also be large but is much slower 
\citep{1960ApJ...132..565H}.
Long-lived supernova remnants are left behind by supernova explosions of both types. 
It has also been possible to identify the explosion type from the
spectral composition of the supernova remnants. 

In the next few sections, we summarise the observed properties of these 
objects, describe the model which emerges and present a few case studies 
to demonstrate the validity of the model.

 \subsection{Summary of observed features}
Some of the observed properties of supernovae and supernova remnants are:
\begin{itemize}

\item The mean observed value of the peak B band absolute magnitude of supernovae
of type Ia is $-19.46$ magnitudes while it ranges from $-17$ to $-19.27$ magnitudes 
for supernovae of type II \citep[e.g.][]{2002AJ....123..745R}.   

\item Progenitors of type II supernovae suffer extensive mass loss 
before the explosion.  The mass loss could be continuous at typical
rates estimated to be $10^{-8}$ to $10^{-4}$ 
M$_\odot$yr$^{-1}$ and typical outflow velocities of 10 to 40 kms$^{-1}$
or the mass loss could be episodic. 
This matter forms the circumstellar medium around the star/supernova. 
%{\bf reference?}
Since there is no comparable mass loss prior to a type I supernova explosion,
no circumstellar medium is expected. 

\item Typical energy of $\sim 10^{51}$ ergs is released in a supernova explosion. 

\item Type I supernovae are observed to occur in all types of galaxies whereas type II
explosions are generally recorded in star forming galaxies. 

\item The peak luminosity of supernovae follows a direct relationship with $t_2$ 
which is the time taken by the peak luminosity to decline by 2 magnitudes.  
Longer $t_2$ is recorded for a higher peak luminosity of the supernova.  This behaviour 
is opposite to that seen in classical nova outbursts.  Since the peak luminosity
recorded for type 1a supernovae is confined to a small range, the $t_2$ is also
recorded within the narrow range of 30 to 40 days. 

\item Absorption lines and emission bands are detected in the supernova spectrum.
The spectrum eventually evolves to be pure nebular containing only 
emission lines.  These lines are used to estimate the expansion velocities. 

\item Initial expansion velocities in supernovae are generally measured to be 
upwards of $20000$ kms$^{-1}$ which 
are observed to decline to a few thousand kms$^{-1}$ (see Figure \ref{fig5}).  
The expansion velocities measured from Fe II lines are lower than 
those measured from hydrogen lines (see Figure \ref{fig5} for SN 1987A).  
Expansion velocities ranging from a couple to a few thousand kms$^{-1}$
are commonly recorded in supernova remnants.

\item The prompt radio emission, detected soon after the supernova explosion,
is due to the synchrotron process.  Prompt radio emission 
is commonly detected from type II supernovae while it has never been detected 
from type 1a supernovae. 

\item Prompt X-ray emission is detected from type II supernovae but has 
never been detected from a type 1a supernovae. 

\item The radio emission from supernova remnants is mainly due to the 
synchrotron process.  

\item The radio light curves of the prompt emission show a frequency-dependent 
detection epoch such that emission at the higher frequencies is 
detected before the emission at the lower frequencies.  The frequency-dependence is 
well-understood as being due to thermal free-free absorption by foreground or
mixed thermal gas, since the free-free optical
depth has a frequency dependence $\tau_\nu \propto \nu^{-2}$.  

\item The prompt radio emission from supernovae is, at best, weakly polarised.  
Strong linearly polarised radio emission is detected from supernova remnants.

\item Polarisation studies detect radially aligned magnetic fields in 
young supernova remnants whereas older remnants tend to 
show a tangential magnetic field \citep[e.g.][]{1976AuJPh..29..435D}. 

\item The magnetic fields in supernova remnants are estimated to be 
between 0.1 mG to 1 mG. 

\begin{figure}
\centering
\includegraphics[width=6cm]{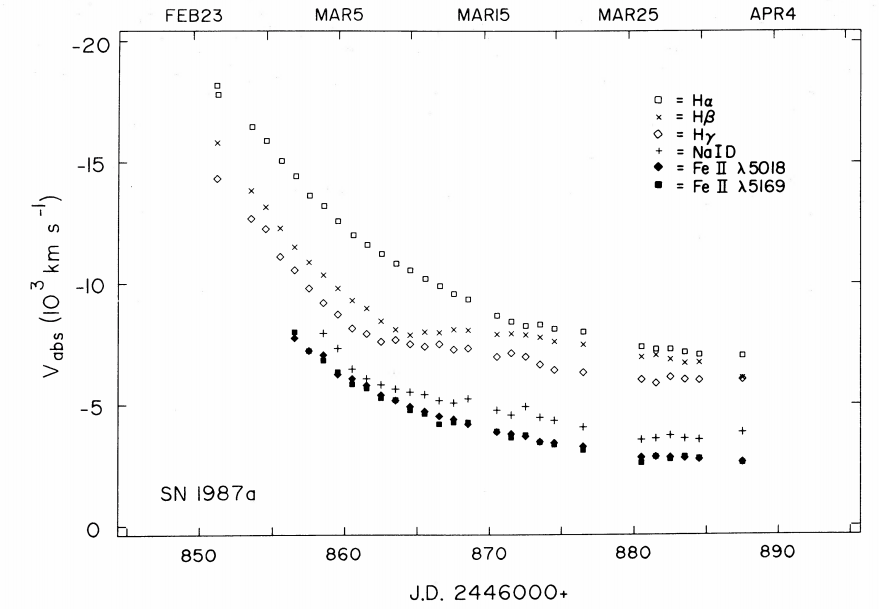}
\caption{\small Figure showing the evolution of the spectral line velocity
for SN 1987A copied from \citet{1987ApJ...320..589B}.  Notice the rapid decline in the expansion
velocity of the ejecta which is typical of several supernovae and the lower 
expansion velocities recorded at a given epoch for the heavier
elements like Fe II.  Even the higher level transitions in hydrogen (H$\beta$, H$\gamma$),
show lower velocity shifts than H$\alpha$ especially at early times and then tend towards
similar values at later epochs.  A velocity of $-18000$ kms$^{-1}$ was recorded for
H$\alpha$ in the first observation conducted a day or so after discovery with the wings
of the P Cygni profile extending upto $\pm 30000$ kms$^{-1}$ \citep{1987ApJ...320..589B}. }
\label{fig5}
\end{figure}

\item Many relatively young supernova remnant shells appear spherical 
supporting a predominantly spherically symmetric explosion in massive stars. 

\item Optical line emission is often confined to the filamentary structure
distributed across the supernova remnant whereas the continuous emission is 
observed to arise from a smoothly distributed emitting component. 

\item Forbidden line emission is detected from fast moving knots in 
supernova remnants which are often devoid of hydrogen and helium lines.  
H, He lines are observed from streaks and wisps of lower expansion
velocities, which are distributed in the remnant alongwith the knots 
\citep{1989agna.book.....O}. {\it The spectral nature of knots and streaks 
is reminiscent of the behaviour of the Orion and principal spectral lines 
observed from the nova ejecta.}

\item Asymmetric line profiles, such that the blue sides are 
stronger than the red parts, are commonly detected from supernova remnants.  
It is inferred that the red part of the line, which arises
in the distant part of the object, is fainter due to obscuration by dust
in the remnant \citep[e.g.][]{2017arXiv170100891M}. 

\begin{figure}[t]
\centering
\includegraphics[width=5cm]{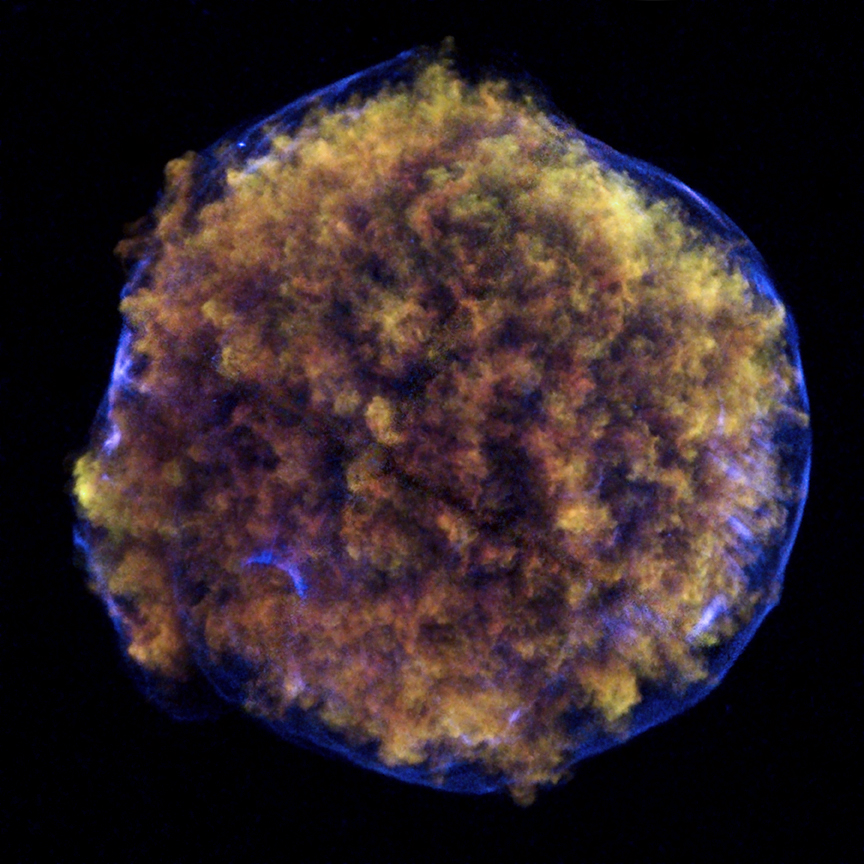} a
\includegraphics[width=5cm]{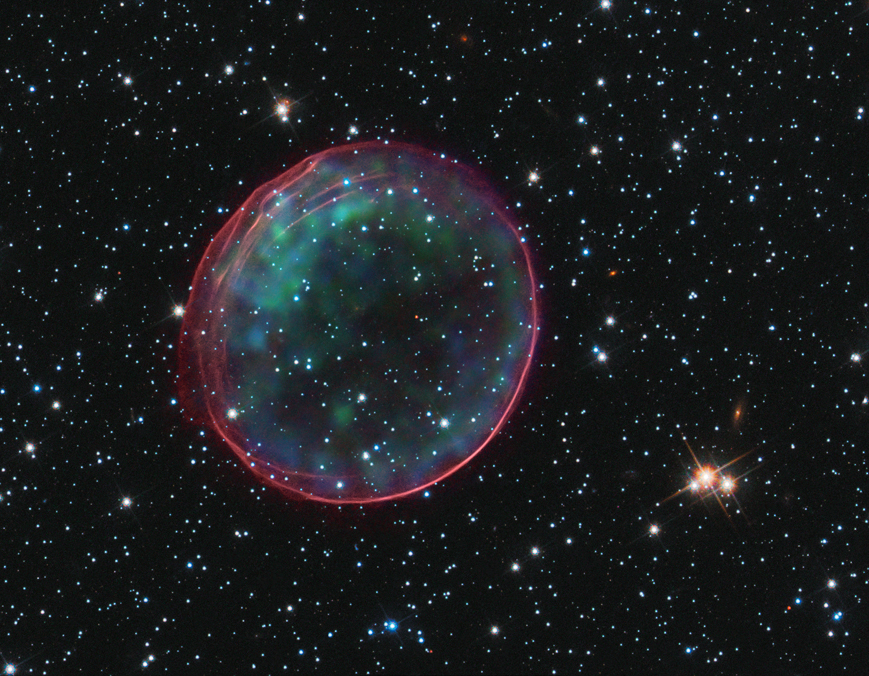} b
\caption{\small Images downloaded from http://chandra.harvard.edu/photo/.
(a) The X-ray image of the type 1a Galactic supernova
remnant Tycho. The outer thin ring in blue is hard X-ray emission whereas
the cloudy emission in yellow is low energy X-rays.  
(b) Composite X-ray and optical image of the supernova 
remnant SNR 0509-67.5 in the Large Magellanic Cloud (LMC). 
The thin outer ring in red is optical emission while the X-ray emission is the diffuse
emission within the ring.  In most supernova remnants, the outer ring is
observed in hard X-rays as shown for Tycho whereas in SNR 0509-67.5 the outer 
ring is detected in optical bands.   The formation process is likely to be the same 
for both though. }
\label{fig6}
\end{figure}

\item In several supernova remnants, a thin hard X-ray (or optical) rim 
of synchrotron emission has been observed encircling the main body of the remnant
from which thermal X-ray and multiband emission are detected
(e.g. see Figure \ref{fig6} where the type1a supernova remnants SNR 0509-67.5
in the Large Magellanic Cloud (LMC) and the Tycho remnant are shown).  
Such rims are predominantly detected around remnants
of type 1a supernova explosions e.g. Kepler, Tycho, SN 1006, G1.9+0.3.  One of
the rare type II supernova remnant which shows such a rim is Cas~A. 

\item The observed radio spectra of a large fraction of Galactic
supernova remnants show a low frequency turnover 
\citep[e.g.][]{1989ApJ...347..915K}.  The turnover is
generally attributed to thermal free-free absorption in the foreground ionized gas. 

\item The 20cm flux density of a few Galactic supernova remnants are listed
here.  The flux density of Tycho (1572 AD) was $\sim 40$ Jy in $1997$,  Kepler (1604 AD)
was $\sim 17$ Jy in $1988$,  Cas~A ($\sim1670$ AD) was around 2700 Jy in 1980 
and SN 1006 (1006 AD) was around 25 mJy in 2003. 

\item There exist supernova remnants from which the detected multiband 
emission especially X-rays is dominated by synchrotron emission. 
Faint thermal X-ray lines from small regions within the remnant are 
sometimes detected.  Occasionally emission is detected at TeV energies 
which is believed to be due to the inverse Compton process. 
A few examples of such remnants are  
G347.3-0.5 \citep{1999ApJ...525..357S}, G266.2-1.2 \citep{2001ApJ...548..814S},
G1.9+0.3 \citep{2008ApJ...680L..41R,2013ApJ...771L...9B}. 

\item G1.9+0.3 is the remnant of the youngest Galactic typa 1a supernova 
with an age less than 150 years and synchrotron radio and X-ray shells are
detected \citep{2008MNRAS.387L..54G}. 
This shell remnant is estimated to 
lie at a distance of about 8.5 kpc and the 100' diameter X-ray shell 
would correspond to about 2.2 pc in size \citep{2008ApJ...680L..41R}.

\end{itemize}

 \subsection{A comprehensive explanation}
We search for a comprehensive model which can coherently
explain the observational results on supernovae and supernova remnants. 
The model includes already known aspects of these objects from literature. 
Since nova and supernova explosions share several similar features 
e.g. sudden optical brightening, expulsion and fast expansion of
matter, dust formation, shell-like remnant, emission bands, the explanation also 
includes features of the nova model presented in \citet{2017arXiv170909400K}. 
Since the behaviour of a range of multi-band features have already been 
explained for novae in the above paper, we focus on mainly understanding
the behaviour of the detected synchrotron emission from supernovae and supernova
remnants and use this with multiband explanations to derive a detailed
explanation for the phenomena. 

We start with the known aspects of a supernova explosion which have
been consistently supported by observations.  In a type II explosion,
the core starts contracting to neutron star densities after running 
out of fuel and triggers a thermonuclear explosion in the surrounding
gas which ignites and blows out matter.  Typical energies of 
about $10^{51}$ ergs are inferred to have been released in the supernova explosion 
which can be explained by simultaneous ignition of about 0.5 M$_\odot$ of 
fuel with the energy being released within 10-100 seconds \citep{1960ApJ...132..565H}. 
The circumstellar medium around type II supernovae is dense due to
extensive mass loss suffered by the massive progenitor star whereas it
is tenuous or absent around type 1a supernovae which is an explosion in
a low mass star.  The type 1a explosion occurs when a white dwarf crosses the
Chandrasekhar mass limit which triggers a thermonuclear eruption which
blows the star apart.  The progenitor system for a type 1a
explosion is always a binary.  Following a supernova explosion, the
exploded star rapidly brightens by 15 magnitudes or more in the optical bands
and after reaching a maximum luminosity starts declining in brightness.  
Similar optical light curves are recorded for most type II supernovae which
are generally double-humped and for type I supernovae which are generally 
single-peaked.  There remain various unknowns such as the reason for this
optical light curve behaviour, origin of dust, origin and composition 
of the relativistic plasma,
the difference between prompt radio emission and remnant radio emission etc 
which we now discuss. 

The explosion releases copious amount of energy which is imparted to
the overlying matter - neutrinos, electrons, protons, ions, etc.  The thermonuclear
explosion and formation of neutron star releases neutrinos.  Neutrinos
constituted the first signal that reached us after the explosion in
SN 1987A in the LMC, indicating their high velocities.  The neutrinos had
to be energised at the explosion site before being expelled with 
relativistic velocities which far exceeded the escape velocity to explain
the immediacy with which neutrinos were detected.  It has been
implicitly accepted that the stellar matter (protons+ions) that is ejected at very high 
velocities is energised in the explosion which enables it to escape. 
However a distinct explanation 
for the acquisition of relativistic velocities by electrons such as DSA are
invoked in literature
without detailing why such a distinct treatment is required by electrons. 
Since there does not appear any reason for the special treatment of electrons,
we accept that electrons acquire energy in the same process that imparts 
energy to protons and ions
accelerating them beyond the escape velocity.  Since large energy is
pumped into matter made up of an extremely large number of particles, the
aggregate energy distribution of all particles and separately 
for each particle species has to be gaussian-shaped with a
mean and a dispersion as described earlier.  Thus, in the most simple 
scenario, it can be
assumed that the distribution for all particles is a gaussian centred
on the escape velocity from the star with a dispersion determined by
mass of the particles.  In other words, the dispersion
will quantify the random velocity component of the particles of a given
species.   Thus electrons by virtue of their smaller mass will have a
larger random velocity component and hence can have a much
wider gaussian distribution as compared to protons.   
This holds for comparable energy 
allocation to all particles which are expelled.  This explanation, as
discussed earlier, is supported by the roughly similar energy spectra
observed for all particles which turns over around 1 GeV. 
The gaussian-shape of the injected energy spectrum of electrons explains
the observed curved radio spectra of many supernova remnants. 

The supernova ejecta should also contain positrons 
generated by the thermonuclear explosion.  It is possible
that while a fraction might be annihilated close to the generation site,
there could be a large number which are energised to relativistic
velocities and escape alongwith an equal number of relativistic electrons,
thus forming a light positron-electron plasma.  There will also be a large
number of electrons which will be tied to protons/ions in the ejecta 
and these will form the heavier proton(ion)-electron plasma.
Thus a supernova explosion, can in principle energise and eject neutrinos,
a positron-electron plasma and a proton(ion)-electron plasma with decreasing
expansion velocities assuming comparable energy allocation to all particles. 
In all these cases, the energising of particles is completed before they
are expelled from the star at velocities in excess of the escape velocity
from the star.  No significant acceleration occurs after the ejection. 
Since all this is expected from simple physical arguments, we assume that
all explosions result in three energetic component ejecta. In reality,
it is possible that the details of the explosion and type of progenitor will
decide the ejected components.  This means that observational signatures
of all three components or any subset should be detectable from a supernova and
which should be determinable from interpreting observational
results on case-to-case basis.  

The observational signature of the neutrino component are neutrinos as was
detected in the case of SN 1987A in the LMC.  Neutrinos are ultra-light,
travel at ultra-relativistic
velocities, are all ejected within seconds of the explosion and hence their
detection depends heavily on having collected data at a critical juncture.  
Hence although neutrinos have long been expected from supernova explosions, till date 
they have only been detected from SN 1987A. 

Both the positron-electron and proton(ion)-electron plasma are relativistic
and hence can be a source of synchrotron radiation.  There should
be a single injection energy spectrum of the electrons, which are part of both 
plasma.  The only difference in the energies of the electrons that are
part of the two plasmas could be in the range of energies they encompass 
such that the positron-electron plasma which is lighter
is constituted from the higher velocity section of the parent distribution 
while the proton-electron plasma contains electrons of relatively lower
velocity.  However it would be difficult to confirm this observationally
and this is only assumed since it sounds reasonable. 
The detectable signals which could help differentiate between the two plasma
composition are as follows.  
The positron-electron plasma will be a source of the 511 keV annihilation
line.  In this process, a pair of positron-electron is expected to annhilate 
resulting in emission of two or three $\gamma-$ray photons of 
total energy of 1 MeV with the maximum
energy of a photon being 511 keV.  The proton(ion)-electron plasma will be
at a high temperature and can emit soft X-rays if at a million degrees
or optical continuum if around 10000 K.  This plasma will also be the
source for the several multiband absorption and emission lines that are detected
in the supernova spectrum.  Thus, in addition to synchrotron emission,
the two plasmas should also result in independent observational signatures.
While to the best of our knowledge, no annihilation line emission has been reported
%{\bf check}
from a supernova/supernova remnant indicating that the presence of a positron-electron
plasma is not conclusively established, the signatures of the 
proton(ion)-electron plasma are regularly detected from supernovae/supernova 
remnants proving its existence. 

We, hence, examine observational results to search for any distinct signatures
of the synchrotron emission from the two plasma of different compositions.  
This, then, brings into focus the two phases in which radio emission is detected
from supernovae of type II namely prompt emission soon after the explosion
and delayed emission in the remnant phase.  This observation, then, strongly
supports the existence of two components of relativistic plasma. 
We already know that the proton(ion)-electron plasma constitutes the massive
ejecta of the explosion and as noted above is responsible for the thermal
continuum and line emissions.  This ejecta is very dense and hence synchrotron
radio emission is not expected from this plasma in the early days after
the explosion due to excessive free-free mixed thermal optical depths of the 
plasma itself.  Synchrotron
emission will become detectable at later dates when the ejecta has
expanded sufficiently for the plasma to have become optically thin to
radio emission.  Thus, the proton(ion)-electron plasma is ruled out from
being the source of the prompt synchrotron emission.  This actually turns out to be
convenient since it clearly points to the existence of the positron-electron
plasma which has escaped from the explosion site before the proton-electron
plasma and after the neutrinos and can be responsible for the prompt
synchrotron emission from type II explosions.  Since the positron-electron
plasma precedes the proton-electron 
plasma, the radio emission is only obscured by the free-free optical
depth of the circumstellar medium which is much lower than that of the
dense plasma expelled by the star.  This, then, explains the frequency-dependent
detection of prompt radio emission from the positron-electron plasma.   
It also explains the high expansion velocities ($\sim 30000$ kms$^{-1}$) 
that have been inferred for the plasma giving rise to prompt radio emission 
from some supernovae such as SN 1987A.  {\it This discussion then conclusively
establishes the ejection and existence of positron-electron plasma which
is responsible for the prompt radio emission from explosions of type II. }
This raises the question of the existence of a positron-electron plasma
in case of type 1a explosions since prompt radio emission has never been
detected from this type.  Since the generation of a large quantity of positrons 
is in a thermonuclear reaction and type 1a explosions also undergo this
reaction, the positron-electron plasma is expected to have left the explosion
site.  However since the progenitor of type 1a are low mass stars which do
lose mass and hence do not form a dense circumstellar medium,  
the lack of prompt radio emission can be attributed 
to lack of an appropriate magnetic field in the circumstellar region
of the star.  This means the fast positron-electron plasma should be 
ejected in type 1a supernova but which will keep expanding at high
velocities suffering minor kinetic losses till it encounters an ambient magnetic
field when it will start radiating synchrotron emission.  
This plasma should also keep annihilating at some rate even as it expands
outwards.  It might, hence, be possible to detect the annihilation line
from supernova although it is possible that the lines might be too faint to
be detectable.  {\it To summarise:  The prompt radio emission detected
in supernovae of type II is from the positron-electron plasma radiating in
the ambient magnetic field. }
It is interesting that around several type 1a supernova remnants, a thin
ring of hard X-ray emission (or optical) is detected.  This well-defined
rim of emission could be a signature of the positron-electron plasma
and although this sounds speculative at first it is a well-founded inference. 
The rim is expanding ahead of the main remnant, is radiating in a
distinct magnetic field and does not contain any thermal component of
radiation - all signatures which support the positron-electron plasma
which left the explosion site but could not radiate due to the lack of
a magnetic field.  The X-ray and optical rim emission could be synchrotron
or could be the Compton-scattered positron-annihilation photons as the
soft $\gamma-$ray photons interact with the dense interstellar medium. 
The degeneracy in the origin needs to be further investigated. 
The radio emission that arises between the main remnant emission and
the outer rim in the rare type II supernova with a X-ray rim i.e. Cas~A
is synchrotron in origin but with distinct magnetic field properties.   
The reason such rims should not be expected in type II supernova remnants
is because the positron-electron plasma would have suffered large synchrotron
losses soon after ejection as they emit the prompt synchrotron radiation
detected at radio and X-ray bands.  Unless this emission is quenched
soon after, as was the case in SN 1987A, the plasma will continue
to lose energy and hence might not be as potent as in type 1a supernova
remnants.  However this needs to be observationally confirmed. 
{\it To summarise: from the above discussion it can be surmised that
the observational signatures of the positron-electron plasma are
(1) prompt synchrotron emission following a supernova explosion 
(2) positron-electron annihliation lines (3) the X-ray or optical
rims detected around supernova remnants. }

We now examine the signatures of protron(ion)-electron plasma. 
This as mentioned earlier contains most of mass that is ejected in
a supernova explosion.  In other words, this plasma is the 
supernova remnant which is detected in thermal continuum emission, 
spectral lines and radio synchrotron emission.  It consists of
ions of several elements, protons and relativistic electrons. 
This ejection of the outer parts of the star should have
also carried along the stellar magnetic field which should be
radially stretched as the remnant expands.  In fact, the detection
of a radial magnetic field in young supernova remnants as surmised
from radio polarisation studies, gives strong evidence to the frozen-in 
magnetic field in the proton(ion)-electron plasma.  The synchrotron
radio emission from this plasma should also show a frequency-dependent
detection due to the free-free optical depth of the thermal
gas mixed with the relativistic plasma.  This kind of behaviour
was actually detected in SN 1986J i.e. a prompt radio emission episode 
and a delayed episode of radio synchrotron
emission with both showing a frequency-dependent detection.
The radio synchrotron emission is detectable in several
remnants and easily resolved in nearby supernova remnants.
{\it To summarise: the radio synchrotron emission from the proton(ion)-electron
plasma gyrating in the magnetic field frozen in the plasma is detected
when the thermal free-free optical depth of the remnant declines leading
to a frequency-dependent detection.  The remnant will keep shining till the
relativistic electron population survives.  
The thermal continuum and spectral line emission also arise from the remnant.  }

Polarisation studies show that
the magnetic field which is radially oriented in young supernova remnants becomes
tangential in older remnants.  The radial behaviour is explained by field
lines frozen in the radially expanding ejecta.  As the remnant ages and slows
down, its interaction with the environment increases.  When the remnant
runs into dense matter, it can be compressed and the magnetic field lines will
also be quashed in the direction perpendicular to the radial vector i.e
becoming tangential.  Thus the behaviour of the field orientation 
in a supernova remnant is easily understood. 

It has been inferred from observations that clumps and dust are formed in supernova
remnants.  Since a massive ejecta composed of a range of elements is
expelled after the explosion, we examine the efficacy of the same
clump and dust formation scenarios that were remarkably useful in case of 
nova outbursts and was summarised in the introduction.   
The expansion velocities inferred from the spectral lines 
observed soon after the explosion show higher values for lines of 
hydrogen and lower velocities values for iron lines (see Figure \ref{fig5}).
This could be inferred to mean that in the supernova ejecta, the heavier elements 
lag behind the light elements like hydrogen and helium.  
The heavier atoms which are lagging behind can cluster 
together due to mutual gravity and form metal-rich clumps.   
The insides of the clumps can be shielded from the ultraviolet radiation 
field and hence provide favourable conditions for dust formation.  Thus, dust
detected in the supernova forms within these optically thick 
metal-rich clumps.
Since a hot neutron star is expected in a type II supernova, its radiation
field can exert radiation pressure on the optically thick clumps forcing
them to move faster.  This, then explains the high velocity metal-rich
knots that are detected in type II supernova remnants like Cas~A.  
Since the wispy hydrogen features are not optically thick, no radiation
pressure acts on it explaining their lower velocities.  Thus, the
explanations for both
dust formation and the excess velocity picked up by the optically thick
clumps can be similar to the explanations that work for the same 
phenomena in nova ejecta. 
Since no central hot star is left behind in a type 1a supernova explosion,
no difference in the expansion velocity of metal-rich knots and
hydrogen wisps is expected.  However the dust formation scenario in both types
of supernovae can remain the same.  

The discussion in this section can be summarised as:
\begin{itemize}

\item Relativistic electrons are energised alongwith the heavier matter 
before they are ejected in the supernova explosion.  Assuming all particles
are imparted comparable energies, the energy spectrum of each species of particles
will follow a normal (Gaussian) distribution with similar mean energies (greater
or equal to the escape velocity) but a 
varying dispersion dependent on the mass of the particle.  The heavy particle 
distribution will show a lower dispersion compared to the light particle distribution. 

\item Observations support three-component ejection from the
explosion site in the listed order due to decreasing expansion velocities:
neutrinos, a light particle ejecta consisting of positron-electron
plasma and a heavy ejecta consisting of proton(ion)-electron ejecta.

\item The prompt radio synchrotron emission detected soon after the explosion 
in type II supernovae is due to the positron-electron plasma
radiating in the circumstellar magnetic field.  
The radio synchrotron 
emission detected in the supernova remnant phase is from electrons 
within the main remnant shell (i.e. proton(ion)-electron plasma) that are 
radiating in the magnetic field frozen in the remnant.  

\item X-ray/optical rims have been
detected predominantly around remnants of type 1a supernovae and are suggested to
be due to emission from the positron-electron plasma.   
The emission could either be due to the synchrotron process or could be
the Compton scattering of the positron-electron annihilation photons as the
plasma strikes the surrounding higher density thermal matter. 

\item The magnetic field is observed to be predominantly radial in young 
supernova remnants.  This can be explained by stretching of the field
lines frozen in the matter in the radially expanding ejecta. 
Older remnants show a predominantly tangential field which would be expected when
the remnant interacts with the ambient dense medium leading to compression of the
radial field.

\item Clump formation in supernovae is supported by the detection of fast metal-rich
knots devoid of hydrogen.  These knots observing to be expanding faster than
hydrogen/helium-rich streaks are formed due to clustering of heavy atoms
which lag behind in the ejecta.  The knots will be optically thick and 
hence accelerated to higher velocities by the radiation pressure 
exerted by the photons from 
the central hot star.  The insides of the optically thick metal-rich knots 
will be well-shielded from the hard radiation field 
and hence conducive to dust formation.  This dust will eventually disperse
and increase in the dust content of the galaxy. 

\item Since the formation scenario for type 1a supernova remnants 
indicate that it should not be hosting a central hot object, this means that
metal-rich knots and the hydrogen-rich
wisps in the remnant should be moving with similar velocities.  
In other words, the detection of faster metal-rich
knots would mean there exists a hot compact object at the centre of the remnant.

\end{itemize}

We now discuss the observational results on a few 
supernovae and a supernova remnant in light of the above model. 

 \subsection{Case Studies}
We discuss the observational results on the Galactic
supernova remnant Cassiopeia~A and the 
extragalactic supernovae SN 1986J, SN 1987A and SN 1993J.  As in the previous sections,
the focus is on the synchrotron emission although other diagnostics are not excluded.
Cas~A has been traced to a type II supernova explosion.  The other
two well known Galactic supernova remnants namely Kepler and Tycho 
are believed to have been type Ia.
All three remnants show the presence of a hard synchrotron X-ray rim which surrounds
the main remnant and similar synchrotron radio characteristics.  
This supports the ejection of relativistic
plasma in all the remnants - the positron-electron plasma which is responsible
for the thin hard X-ray rim and proton(ion)-electron plasma which is responsible
for the radio emission coincident with the multi-band emission from the remnant
shell.   While prompt radio emission is likely to have been
detected from the Cas~A explosion, it would not have been detected from the
explosions which resulted in the Kepler and Tycho remnants. 

  \subsubsection{Type II SNR: Cas~A in Milky Way}
\begin{figure}
\centering
\includegraphics[width=5cm]{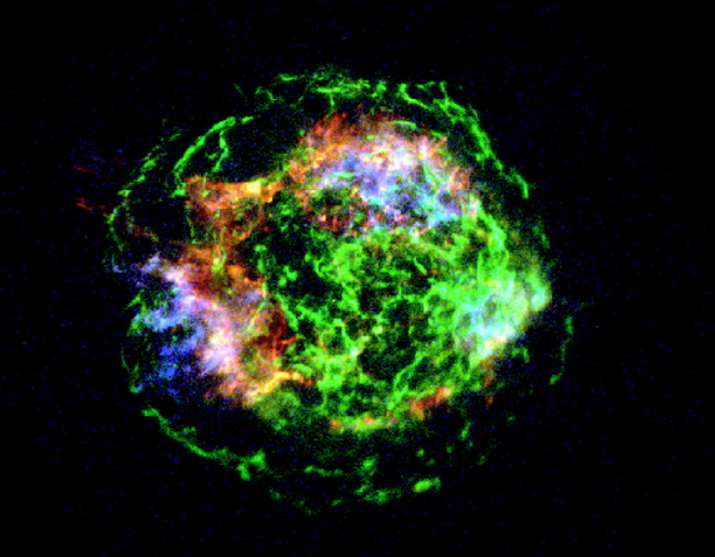} a
\includegraphics[width=5cm]{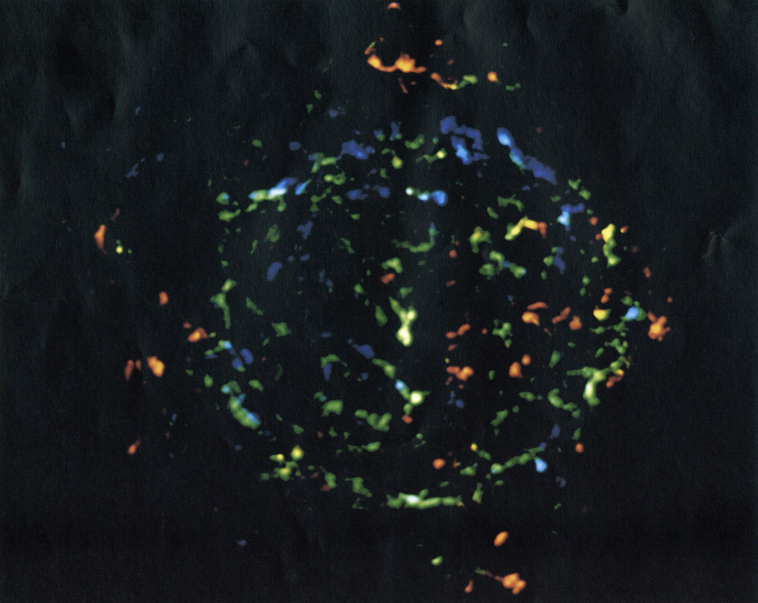} b
\includegraphics[width=5cm]{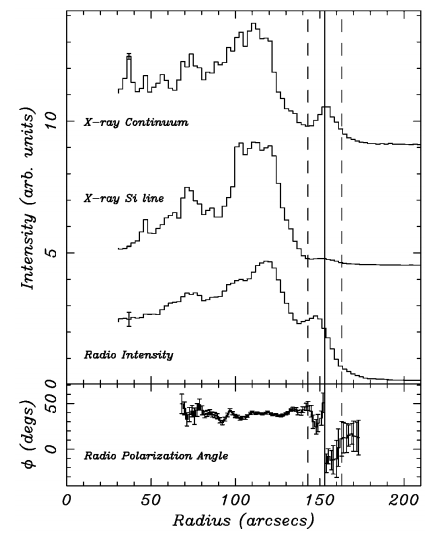} c
\caption{\small  (a) X-ray three-colour image of the supernova remnant Cassieopia~A
copied from \citet{2004ApJ...615L.117H}.  The colour coding is
Si He$\alpha$ (1.78-2.0 keV; red); Fe K (6.52-6.95 keV; blue) and 4.2-6.4 keV continuum 
(green).  Note the outer wispy ring in hard X-rays due to synchrotron emission.
(b) Figure showing radio spectral index of knots in Cas~A estimated in epoch 
1985 reproduced from \citet{1996ApJ...456..234A}. The red colour indicates a 
spectral index of $-0.9$ whereas blue indicates $0.75$ and the steeper knots lie 
outside the main ring of Cas~A. 
(c) This figure reproduced from \citet{2001ApJ...552L..39G} shows the radial variation
in brightness in different bands and in the radio polariation angle.  
Notice the outer parts are detected in
radio synchrotron and X-rays and the radio polarisation angle also changes.  
In the main remnant, the polarisation angle is constant and the
 magnetic field has been inferred to be mainly radial.  }
\label{fig7}
\end{figure}

Cas~A is the remnant of a type II supernova explosion located
in the Perseus arm at a distance of $\sim 3.5$ kpc 
that occurred about 350 years ago. 
The shell-like remnant of diameter $\sim 5'$ (linear size $\sim 5$ pc) 
left behind by the explosion is the brightest 
source in the radio sky.  The observable remnant consists of three main parts:
(1) a bright shell of radius $\sim 2'$ which emits thermal continuum emission,
multiband spectral lines and radio synchrotron emission.  This shell
contains most of the mass ejected in the explosion and refers to the
central bright region in Figure \ref{fig7}a; (2) faint radio synchrotron emission
which surrounds the bright shell and is referred to as the radio plateau in
literature; (3) a thin ring of wispy hard X-ray emission of radius 
$\sim 2.6'$ \citep{2001ApJ...552L..39G} (green ring in Figure \ref{fig7}a)  
which is coincident with the outermost part of the radio plateau.
The three top curves in Figure \ref{fig7}c, copied from
\citet{2001ApJ...552L..39G}, are the radial cut across the source in three tracers -
X-ray continuum, Si line emission and radio synchrotron emission.  
The main remnant is detected in all three tracers.  While the Si emission
is confined to the main remnant, the X-ray and radio synchrotron emission 
show a second peak beyond the main remnant
which corresponds to the X-ray rim and radio plateau respectively.
Figure \ref{fig7}b, copied from \citet{1996ApJ...456..234A}, shows the
distribution of radio spectral index between 1.4 and 5 GHz of the 
knotty emission in Cas~A.
The couple of distinct characteristics of the radio synchrotron emission 
from the main remnant and from the radio plateau are:
(1) the radio spectrum of knots in the radio plateau
is steeper with $\alpha > 0.8$ ($S \propto \nu^{-\alpha}$) 
compared to knots within the remnant 
\citep{1995ApJ...455L..59K,1996ApJ...456..234A} (also see Figure \ref{fig7}b).  
Steeper radio spectrum is also detected from a small region in the western part
of the main remnant which begs the inference that it is the foreground part
of the radio plateau (see Figure \ref{fig7}b). 
(2) The radio polarisation angle which is fairly constant within the main remnant 
and indicative of a radial magnetic field starts varying in the radio plateau and
then shows an abrupt jump which is coincident with the peak emission of the hard 
X-ray rim (lowermost curve in Figure \ref{fig7}c).  This behaviour
could be inferred to indicate the distinct nature of the magnetic field
in the remnant and in the radio plateau. 
The equipartition magnetic field in the knots, estimated from
radio data at 83 GHz ranges from 1 to 5 mG.  \citep{1999ApJ...518..284W}.  
The average magnetic field in Cas~A has been estimated to be $>0.5$ mG
from X-ray observations \citep{2003ApJ...584..758V}.  Assuming the X-ray 
rim in Cas~A is due to the synchrotron process magnetic field of 
0.1 mG has been estimated in the rim \citep{2003ApJ...584..758V}.

We try to understand this behaviour in light of the model put forward here. 
The X-ray rim and the radio synchrotron emission from outside
the main remnant lend support to their origin in the 
positron-electron ejecta which precedes the heavy proton(ion)-electron ejecta 
from the explosion.  The light ejecta radiates radio synchrotron emission in
the ambient magnetic field while the plasma in the main remnant emits in the
magnetic field frozen in itself.  This, then, explains the observed abrupt jump
in the radio polarisation angle (see Figure \ref{fig7}c).  However it
should be noted that the circumstellar magnetic field is also originally
the stellar field which was carried away in the mass loss suffered by 
the progenitor.  Since the wind velocities are much lower at $\sim 10-20$ kms$^{-1}$
compared to the explosion ejecta velocities, 
the circumstellar field will not display the strictly radial behaviour that
is exhibited by the remnant field.  
These, then explain the observed behaviour of the radio polarisation angle
in Cas~A shown in Figure \ref{fig7}c.  The parent
electron population within the main remnant and the outer radio plateau should be
the same i.e. electrons accelerated by the thermonuclear explosion in the supernova.
However different sections of the parent normal
distribution can contribute to the remnant and plateau so that 
the radiating electrons can be of varying range of energies.  These, when 
subject to distinct magnetic field strengths can result in electrons of
different energies i.e. different parts of the energy spectrum 
radiating at a give frequency so 
that the observed radio spectral indices are also different.
This means that emission at a given observing frequency can arise from
high energy electrons in the plateau for which a magnetic field of 0.1 mG has
been measured while it will arise from lower energy electrons in the remnant
wherein a magnetic field around 0.5 mG has been measured (see Equation \ref{eqn2}).
Since the electrons of lower energy will lie closer to the peak energy and
hence show a flatter energy spectrum than electrons of higher energy - this
could contribute to the steeper radio spectrum detected in the
radio plateau region outside the main remnant. 

It is believed that cosmic rays upto energies $\sim 10^{15}$ eV 
(1000 TeV) i.e. the knee in the cosmic ray energy spectrum are 
accelerated in supernovae or supernova remnants.  However detailed 
studies of supernova remnants have not been successful in supporting this conclusion. 
From a study of several supernova remnants, it was inferred that the high
energy cutoff of synchrotron-emitting electrons was around 80 TeV 
\citep{1999ApJ...525..368R}.  MAGIC observations have suggested an even lower
energy cutoff at $\sim 3.5$ TeV for the cosmic rays from Cas~A 
\citep{2017arXiv170900280G}.  
This is not surprising in light of the explanation provided in the paper wherein 
the acceleration of the cosmic rays is mainly effected by the energy released in the
thermonuclear explosion before matter is ejected.  This means even if particles
were accelerated to energies of 1000 TeV in the supernova explosion, 
these would have rapidly lost
energy and moved to lower energies after 350 years so that
a high energy cutoff near 3.5 TeV is observed by MAGIC in the cosmic ray
spectrum from Cas~A \citep{2017arXiv170900280G}.  This demonstrates that 
the issue of the maximum energy of
the particles accelerated by the supernova explosion remains unresolved
and we might be able to resolve this issue by targetting young powerful 
supernovae like SN 1986J for studies which aim at estimating the high energy 
cutoff of the cosmic ray spectrum studies instead of targetting supernova
remnants where the cosmic ray spectrum has aged.  Maybe such studies already
exist which the author has missed, in which case, we might be able to readily
be able to examine data on the same. 

Electrons of energy $\sim 80$ TeV gyrating in a magnetic field of 0.1 mG 
should be able to emit hard X-rays ($\nu \sim 10^{19}$ Hz) as observed in 
the outer rim but would have a short synchrotron lifetime of only 10 years.  
If the magnetic field
was much lower then, the electron energies required to explain the hard X-ray
emission assuming it is synchrotron in origin would be higher and these
could also survive longer.  For example,
if the magnetic field was $10 \mu$G, the electron energies emitting hard
X-rays ($\nu \sim 10^{19}$ Hz) would need to be 
250 TeV and their synchrotron lifetime would be $\sim 3000$ years.   
However, the two studies noted above seem to certainly rule out electrons of
energy beyond 80 TeV.  For electrons of energy 5 TeV to emit hard X-rays
would require a magnetic field of 25 mG.   This discussion is geared towards
understanding the hard X-ray rim that is detected around the main remnant
and the emission mechanism.  The above discussion suggests that the
only way the emission could be synchrotron
is if the electrons did not suffer any synchrotron or inverse Compton
losses during its expansion to the present location so that the electrons
retained the high energies they acquired from the explosion energy.  
However since the recent observation have not detected any high energy
electrons and the magnetic field is not observed to be so large as 25 mG,
it appears that the X-ray rim might owe its origin to a different physical
process.  Since the X-ray rim and radio plateau can be associated
with the positron-electron plasma, it is likely that the X-ray rim
is the positron annihiliation radiation at 511 keV that has been Compton-scattered
to hard X-ray frequencies by thermal electrons when the plasma has
encountered the ambient interstellar gas.  The low energy positron-electron plasma
continues to radiate by the synchrotron process in the radio plateau region. 

The detection of the hard X-ray rim  (which is more commonly detected
in supernova remnants of type 1a explosion) in Cas~A might suggest an episodic
mass loss by its progenitor star - a scenario similar to the more recent
type II supernova SN 1987A.  As in SN 1987A, the prompt radio emission in Cas~A
might have been short-lived, allowing the high energy positron-electron 
plasma to survive longer due to their expansion causing only kinetic losses. 
This is suggested due to predominant detection of X-ray rims around type 1a 
remnants wherein it is known that the circumstellar magnetic field is very
weak and hence the synchrotron losses suffered by the positron-electron
plasma are small and radiation from it is detected several hundred years
after the explosion.  

{\it The observational results on Cas~A can be briefly explained:  
following the explosion, a fast expanding
positron-electron ejecta would have emitted prompt radio and X-ray 
synchrotron emission 
and is currently responsible for the thin rim of hard X-ray emission
and radio plateau around the remnant.  A proton(ion)-electron plasma which
follows, is responsible
for synchrotron emission detected from the entire remnant and the multi-band
thermal emission including spectral lines. 
The magnetic field in the main remnant is radial indicating stretching of field
lines due to radial expansion of the remnant while it includes a larger
random component in the radio plateau which is indicative of the field 
in the circumstellar/interstellar medium.  }

  \subsubsection{Type II SN: SN1986J in NGC 891}
SN 1986J was detected in the galaxy NGC 891 (distance $\sim 10$ Mpc)
from radio continuum observations at 21~cm \citep{1986IAUC.4248....1V}.
At 10 Mpc,  $1''$ corresponds to about 50~pc. 
Although the supernova was identified in 1986, several observations
pointed to the explosion having occurred in 1983.2.  There existed a radio detection 
of the supernova in 1984 \citep{1987AJ.....94...61R}, 
the fitting of post-discovery radio light curves indicated an 
earlier explosion date than 1986 \citep{1987Natur.329..611C}
and the VLBI images taken in late 1986 \citep{1989ApJ...337L..85B} resolved
the radio emission also indicating an earlier date for the
explosion \citep{2002ApJ...581.1132B}.  After its identification,
SN 1986J has been intensively observed at radio wavelengths both in integrated
flux density and with the Very Long Baseline Interferometry (VLBI) which has 
resulted in images with a few milliarcsecond resolution.  
A peak radio emission of 128 mJy 
at 6 cm was recorded about 1200 days from the estimated date of explosion
\citep{2007ApJ...671.1959W}.  

\begin{figure}[t]
\centering
\includegraphics[width=6cm]{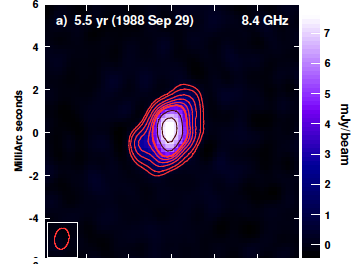}a
\includegraphics[width=6cm]{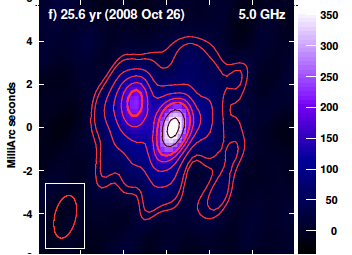}b
\caption{\small Figures showing SN 1986J at radio wavelengths and VLBI resolution are
copied from \citet{2017ApJ...839...10B}.  The figures show
SN 1986J at 8.4 GHz about 5.5 years (a) and at 5 GHz about 25.6 years (b) 
after the estimated date of explosion.
The central component in (a) expands to the north-east and
a new component emerges in the centre of the source and has exceeded the
brightness of the old component by 25 years as seen in (b).  Both the components
show a frequency-dependent onset. }
\label{fig8}
\end{figure}

The radio images of SN 1986J show a complex structure and the 
evolution of radio emission has been captured in detail by both
VLBI images and radio light curves allowing us a rare glimpse into the early time
behaviour of a supernova and its transition to a supernova remnant.   
In the VLBI images obtained in late 1986,
the source was inferred to be extended along a position angle of around
$150^{\circ}$ \citep{1989ApJ...337L..85B} which is similar to the 
major axis of the extended emission
observed about 5.5 years after the explosion in late 1988 as shown in 
Figure \ref{fig8}a copied from \citet{2017ApJ...839...10B}.
An equipartition magnetic field of 70 mG was estimated from the 
detected radio emission in late 1988 \citep{2002ApJ...581.1132B}.  
Even while the emission kept expanding about this axis, the central peak of
emission moved towards the north-east and was then referred to as the hotspot
in the shell.  Due to this, there was no central peak from 1999 to 2002 in 
the 5 GHz images.  A new component had appeared in the centre of the shell 
at 5 GHz in the image taken in 2005 and which got brighter
than the hotspot by 2008 \citep{2010ApJ...712.1057B}.
This component was first detected at higher radio frequencies which could
be due to frequency-dependent delayed detection \citep{2010ApJ...712.1057B}. 
As can be seen in Figure \ref{fig8}b, the radio
emission from SN 1986J had evolved to a quasi-spherical 
shell-like structure after 25 years which hosted a hotspot in the north-east and
the bright central component \citep{2002ApJ...581.1132B,2017ApJ...839...10B}.
The flux density of the new central component exceeded that of the shell 
hotspot about 25 years after the explosion \citep{2017ApJ...839...10B}.  
The radio emission from the shell hotspot and the central component 
measured about 14 years after the explosion had spectral 
indices ($S\propto\nu^{-\alpha}$) of $0.63$ and $0.76$ \citep{2017ApJ...851....7B}.  
The new central source got progressively brighter while the shell hotspot got
progressively fainter so that in images taken around 31.6 years after the 
explosion, the hotspot had faded and the central bright component was 
surrounded by a faint shell of radio emission \citep{2017ApJ...839...10B}. 
Several possible origins for the new central radio component in SN 1986J have
been examined and while it has been difficult to arrive at a conclusive origin, 
%while the actual process responsible remained unclear, 
it has been possible to infer 
that the emission was arising in the physical centre of the supernova 
shell \citep{2017ApJ...851....7B}.  

It is estimated that the hotspot plasma was expanding at $\sim 20000$ kms$^{-1}$
about 0.25 years after the explosion in 1983 which had slowed down to about 
6000 kms$^{-1}$ about 16 years later \citep{2002ApJ...581.1132B}.  
The expansion velocity of the central new component
is estimated to be $680^{+80}_{-380}$ kms$^{-1}$ assuming its expansion
since the explosion epoch of 1983.2 \citep{2017ApJ...839...10B}.  

The detection of two distinct radio components - the prompt component which
was detected soon after the discovery of the supernova which showed a
frequency-dependent detection with the peak emission at 5 and 1.4 GHz being
detected about 3 and 5 years after explosion \citep{1990ApJ...364..611W}  and
the new central component which also showed a frequency-dependent detection
\citep{2010ApJ...712.1057B} give strong support to the model suggested in the paper.
The late peaks recorded for the radio emission
from SN 1986J indicates the presence of extensive circumstellar medium
and light curve fitting has indeed shown that clumpy thermal medium is
required to explain the frequency-dependent behaviour of the
observed multifrequency light curves \citep{1990ApJ...364..611W}. 
The prompt emission was synchrotron emission from the positron-electron
plasma which was free-free absorbed by the circumstellar medium.  
The new component which became detectable at a later date and was inferred
to be located at the physical centre of the supernova was synchrotron
emission from the proton(ion)-electron plasma in the massive remnant. 
This emission was free-free absorbed by the thermal plasma in the
remnant which was several times denser than the circumstellar matter.  
The different velocities of the two components and the slightly
different radio spectral indices also support the separate origin 
for the radio emission from the hotspot and the central source. 
Thus, the emission from the positron-electron plasma is what we refer
to as the prompt radio emission from a supernova while the emission
from the proton-electron plasma is what we refer to as the emission
from the supernova remnant.  Thus, SN 1986J has provided us with
valuable data that has helped elucidate the above. 

The existence of the supernova remnant is verifiable 
from data at other bands and although the angular resolution is insufficient
to locate this remnant, the most likely location is in the centre of the
radio shell emitting synchrotron emission.  The emission detected at 
wavelengths of 1.2 mm and 3 mm from SN 1986J in late 1986 
was higher than the expected radio synchrotron emission as extrapolated 
from lower radio frequencies \citep{1989ESASP.290..387T}.  
An origin in a synchrotron nebula powered by a pulsar was suggested as a possible 
origin for this excess \citep{1989ESASP.290..387T}. 
The X-ray spectrum of SN 1986J taken in 2000 and 2003 show the presence of
iron, silicon, sulphur, oxygen, neon and magnesium with the abundance of
silicon and sulphur being super-solar \citep{2005xrrc.procE4.06H}.  The 
X-ray luminosity in the energy range 0.5-2.5 keV 
is observed to decline rapidly \citep{2005xrrc.procE4.06H}.
The optical spectrum of SN 1986J obtained in 1986 contained narrow H$\alpha$
lines of widths $\sim 700$ kms$^{-1}$ \citep{1987AJ.....94...61R} whereas the
forbidden lines of oxygen were detected with larger widths ranging from
1000 to 2000 kms$^{-1}$ \citep{1991ApJ...372..531L}.  
Narrow faint lines of neutral helium were also detected.  In 1989,
the narrow hydrogen lines became fainter while the forbidden
lines of oxygen remained unchanged
\citep{1991ApJ...372..531L}.  Further observations of the forbidden oxygen
lines showed that all the profiles displayed a double-peaked structure and that 
both peaks were blue-shifted
wrt to the galaxy redshift by $-1000$ and $-3500$ kms$^{-1}$ 
\citep{2008ApJ...684.1170M}.   While the strengths of the H$\alpha$, 
[O II] and [O III] had declined considerably by 2007 with the change in the hydrogen
line being maximum,  the [O I] lines seemed to have evolved the least
\citep{2008ApJ...684.1170M}.  While all this data is significant in
enhancing our knowledge of the remnant, here, this information is
mainly used to infer that the massive remnant which emits thermal
continuum and spectral lines was emitting, expanding and evolving
alongwith the radio synchrotron emission. 
Infrared emission inferred to be from warm dust at a temperature of
a couple hundred degrees Kelvin is
detectable after more than two decades \citep{2016ApJ...833..231T}.  
Since dust is expected to form in the main remnant of the explosion which is rich
in metals as discussed earlier, 
these observations also give evidence to the presence of the supernova remnant
and which should be coincident with the central radio source. 
If we are able to resolve the emission from SN 1986J
at several bands before it fades, then in future we should detect multiband
thermal continuum emission including spectral lines coincident with the central 
radio synchrotron emitting component.   While this is obvious from the
radio evolution of the source, obtaining independent evidence from multi-band
data should put our understanding on a firm footing allowing us to explore
further.

We can convert the radio flux densities measured for SN 1986J at a distance of 10 Mpc
to a fictitious supernova located 10 kpc from us.  This entails multiplying
the observed flux densities by a factor of a million and can be useful
in comparing with the order of magnitude flux densities of Galactic 
supernova remnants like Cas~A, Kepler and Tycho remnants.  
The peak flux density of 128 mJy noted at 5 GHz around 1200 days after
the explosion in SN 1986J would be recorded as a source of flux density
of 128000 Jy if the supernova had exploded at a distance of 10 kpc.  
In 2012 i.e. 29.6 years after the explosion a flux density of 1.82 mJy was
recorded at 5 GHz \citep{2017ApJ...851....7B}.  If the supernova
was located at a distance of 10 kpc then we would have recorded a
flux density of 1820 Jy.  The flux density of the remnant is evolving 
and is comparable to that of Galactic remnants. 
 
{\it To summarise:  
The two distinct radio components sequentially detected in SN 1986J with
both components showing a frequency-dependent delay in detection
due to optical depth effects lends
strong support to the existence of a fast light ejecta composed of 
positrons and electrons which is responsible for the prompt radio
emission and a slower massive ejecta composed of protons/ions and
electrons which is the radio component that makes a late appearance
and is the supernova remnant.  
The optical depth of the circumstellar medium delays the detection
of the prompt radio emission while the optical depth of the massive
ejecta itself delays the detection of the radio emission from the remnant. 
SN 1986J is hence a useful example demonstrating the origin of the
prompt and delayed radio emission from the supernova phase and
the supernova remnant phase respectively. 
}

  \subsubsection{Type II SN: SN1987A in LMC}

\begin{figure}[t]
\centering
\includegraphics[width=5cm]{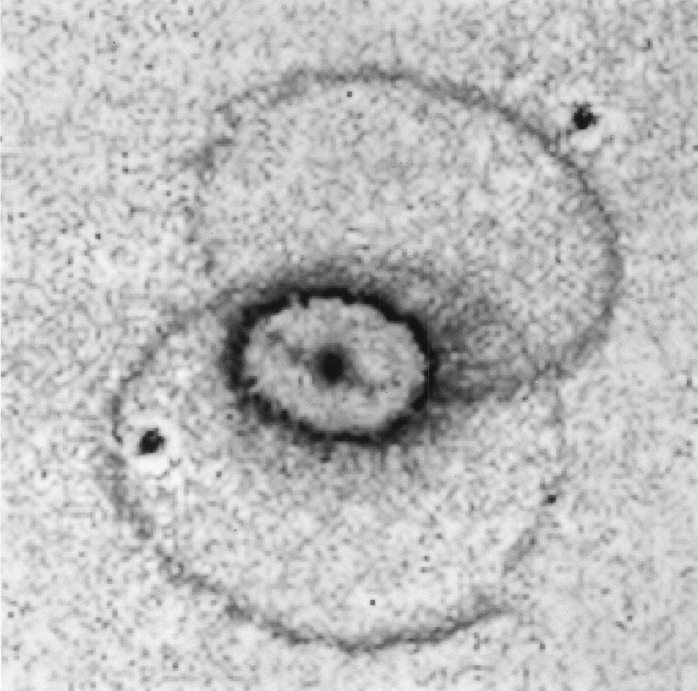}(a)
\includegraphics[width=6cm]{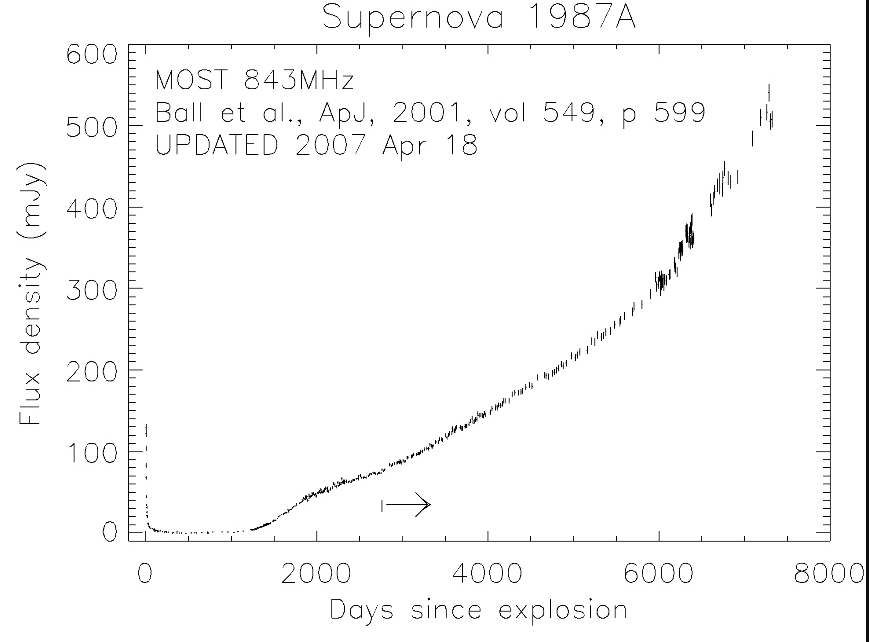}(b)
\caption{\small (a) Figure copied from \citet{1995ApJ...452..680B} showing the triple ring 
system around the SN 1987A.  The semi-major axis of the inner equatorial ring is 0.81''
(0.19 pc) and semi-minor axis is 0.57'' (0.14 pc) while the semi-major axes of the
northern outer ring and southern outer rings are 1.77'' (0.43 pc)  and 1.84'' (0.45 pc) 
\citep{1995ApJ...452..680B}. (b) Figure downloaded from
http://www.physics.usyd.edu.au/astrop/SN1987A and presented in
Ball et al (2001).  The radio light curve shows the prompt radio emission which was
rapidly quenched and the second longer-lasting phase of radio emission 
detected from SN 1987A at 843 MHz.}
\label{fig9}
\end{figure}

A supernova explosion of type II was recorded on February 23, 1987 in the
LMC, first surmised from the detection of neutrinos of energies a
few tens of MeV in a span of 6-13 seconds
\citep{1987PhRvL..58.1490H,1987PhRvL..58.1494B} which preceded
the detection of the explosion at electromagnetic bands.  
Till date, this appears to be the only
supernova explosion from which neutrinos have been detected.  Neutrinos are
expected in supernova explosions since they are 
released in nuclear fusion reactions and when protons combine with electrons
to form the degenerate neutron core.  Being one of the nearest
explosions, since LMC is at a distance of $\sim 50$ kpc, this supernova 
has since been extensively observed with high linear resolution.  
At this distance, $1''$ corresponds to about 0.25 pc.  SN 1987A has allowed
us to study several evolutionary aspects of type II supernova explosions and
also provided important input to nature of mass loss in the progenitor 
star.  Some of the peculiar observed characteristics of SN 1987A are summarised
below with comments included in italics: 

\begin{itemize}

\item Prompt radio emission was detected at frequencies between 840 MHz and
2.4 GHz with the MOST telescope with the first detection
on 25 February 1987 i.e. 2 days after the
detection of neutrinos \citep{1987Natur.327...38T}.  The frequency-dependent
delayed detection and peaks in the multi-frequency light curve were
observed within a week even for the lowest observed frequency 
\citep{1987Natur.327...38T} and the emission at all observed
frequencies had disappeared by September 1987 \citep{1995ApJ...453..864B}.  
A light curve at 843 MHz which includes the prompt
emission soon after the detection and the more persistent radio emission detected
later is shown in Figure \ref{fig9}b.  
Generally it takes the order of 100 days or more for the prompt radio emission 
near GHz frequencies from type II supernovae
to become detectable - for example the light curve of SN 1993J in 
Figure \ref{fig12} - 
and it takes a few years for the emission to fade.  Thus, by type II standards,
the prompt radio emission from SN 1987A was extremely prompt, 
fainter and faded very quickly. 

\item Narrow ultraviolet and optical lines were detected from SN 1987A 
when it was observed soon after
the explosion.  These lines were believed to be forming in the circumstellar medium
\citep{1989ApJ...336..429F, 1989A&A...217...31W}.  

\item An unexpected result from optical imaging was the detection of a triple ring
system (see Figure \ref{fig9}) within $\sim 5''$ of SN 1987A 
\citep{1989ApJ...347L..61C, 1990ApJ...362L..13W, 1995ApJ...452..680B}.
The rings had 
to be due to mass loss by the progenitor system prior to the explosion otherwise
they would require superluminal expansion, if ejected in the explosion. 
{\it The equatorial nature of the three rings indicates that the progenitor
suffered extensive mass loss from its equatorial regions.  }

\item The inner ring is likely expanding at a rate of $\rm \le 50 kms^{-1}$ as
surmised from optical emission lines of [O III] detected in December 1987 
\citep{1989A&A...217...31W}.

\begin{figure}[t]
\centering
\includegraphics[width=6cm]{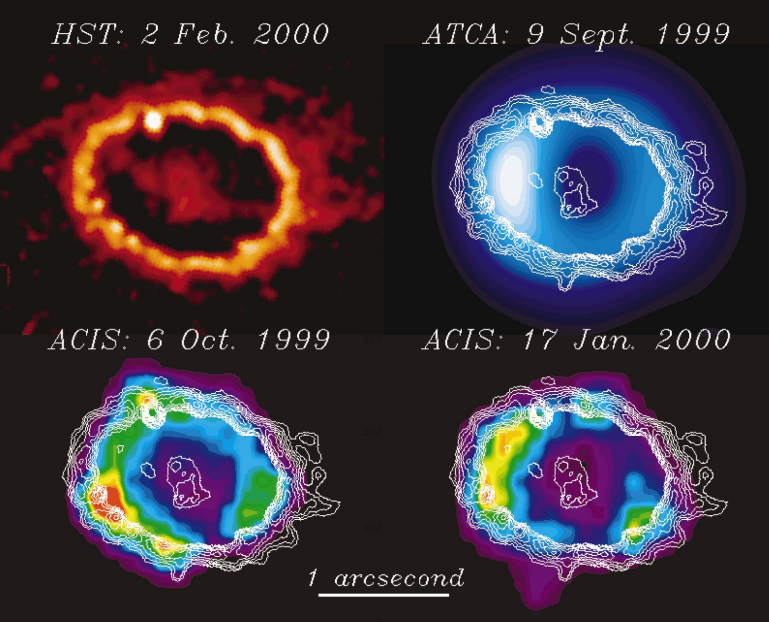}(a)
\includegraphics[width=6cm]{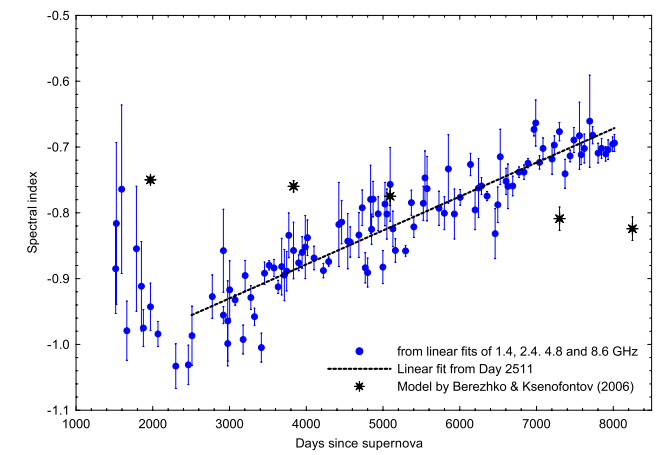}(b)
\caption{\small (a) Figure showing multi-band emission from SN 1987A reproduced from \citet{2000ApJ...543L.149B}.
The bulk ejecta (which we refer to as the supernova remnant) 
from the supernova explosion has to be the compact emission
located in the central part of the ring.  The positron-electron ejecta moving with much larger
velocities has reached the optical ring when it begins to radiate radio and 
X-ray synchrotron emission in the magnetic field of the ring.
(b) Figure reproduced from \citet{2010ApJ...710.1515Z} showing the spectral index
variation in SN 1987A from day 2511.  Notice how the spectrum gets flatter with time. }
\label{fig10}
\end{figure}

\item The radio emission in the second brightening phase was first 
detected at the low frequency of 843 MHz \citep{1995ApJ...453..864B} and
then at 4.8 GHz \citep{1992Natur.355..147S} indicating that the radio detection
in this second phase of the radio emission was not governed by free-free absorption.
{\it This contrasts with the case in SN 1986J wherein the emergence
of both radio components (shell hotspot and central source) was delayed due to
free-free absorption.  The observed behaviour of the radio emission in
SN 1987A helps one infer
that the second phase was not from the proton-electron plasma in the remnant 
which then means that the emission was still from the positron-electron plasma.   } 

\item The second brightening at 4.8 GHz was detected 
in July 1990, between days 1140 and 1200 after the explosion 
\citep{1992Natur.355..147S,1993PASAu..10..331S,1995ApJ...453..864B}. 
This radio emission when resolved in images taken in 1992 \citep{1997ApJ...479..845G}
was along a ring located towards the inner edge of 
the equatorial optical ring (see Figure \ref{fig10}a).  X-ray emission 
was also detected from the same region 
\citep{1994A&A...281L..45B} (see Figure \ref{fig10}a).
Both radio and X-ray emission have been since been rising
\citep{2014IAUS..296...15S, 2016ApJ...829...40F}.  Optical hotspots started
appearing in the inner equatorial ring from around 1996 \citep{1996IAUC.6368....1G}.

\item Assuming the second phase of radio emission 
was associated with the forward shock which had reached the optical ring, an 
expansion velocity of 29200 kms$^{-1}$ was estimated for the shock 
\citep{1993PASAu..10..331S}.
From the expansion of the radio ring it was then 
surmised  that the shock had slowed down to about 3500 kms$^{-1}$
after entering the inner equatorial ring \citep{2005coex.conf...89S}.
{\it We need to replace the shock by the presence of relativistic positron-electron
plasma which by virtue of being light can expand with high velocities
and retain the same till it encounters dense matter. }

\item Wide P Cygni profiles were detected for the Balmer lines of hydrogen with
the velocity shift of the H$\alpha$ absorption line 
being $-18000$ kms$^{-1}$ in the first observation taken on 25 February 1987 
(see Figure \ref{fig5})
while the wings of the P Cygni profile extended to $\pm 30000$ kms$^{-1}$
\citep{1987ApJ...320..589B,1988AJ.....95.1087P}.  Fe II 5169A was detected at a 
velocity shift of $-11355$ kms$^{-1}$ around day 4 \citep{1988AJ.....95.1087P}. 
The velocity shift of the Balmer lines of hydrogen stabilised around $-4000$ to
$-5000$ kms$^{-1}$ around day 80 while those of Fe II stabilised 
around $-2000$ kms$^{-1}$ \citep{1988AJ.....95.1087P} (see Figure \ref{fig5}).
{\it Since these high velocity lines have to emerge from the massive remnant of
the explosion, their velocities signify the expansion velocity of the 
massive remnant.  The initially recorded expansion velocities of 18000 kms$^{-1}$
or more had slowed down to 4000-5000 kms$^{-1}$ in 80 days.  This alongwith
similar behaviour recorded in other supernovae inform us of the rapid slowing
down of the expansion of the massive remnant unlike the light positron-electron
plasma. }

\item The HST optical image of the inner equatorial ring in Figure \ref{fig10}a
shows a bright ring with hotspots and a blob in the central regions.  The 
optical hotspots progressively move outwards \citep{2015ApJ...806L..19F} as the 
diameter of the radio and X-ray emission rings increase. 
The central blob of emission is not detected at low radio frequencies or X-rays. 
{\it The central blob is the massive remnant of the explosion which is
the source of the broad absorption and emission lines and which is
optically thick to radio synchrotron emission. }

\item The central blob detected in SN 1987A is expanding as can be seen 
in Figure \ref{fig11}.  This nebula has evolved to a structure stretched 
along north-south with a similarly elongated cavity in its centre.  
The nebula which was fairly compact in early days has expanded and 
encountered the southern part of the inner equatorial ring.  The central
blob as seen in the optical image given in Figure \ref{fig10}a, 
shows wispy extensions stretching out in the north-east and south-west which
appear to be along the segments of the northern ring that intersects the
inner equatorial ring (see Figure \ref{fig9}a).  The wispy extensions
are not detected in later images shown in Figure \ref{fig11}b.  
{\it The central blob is the massive metal-rich remnant of the explosion
which is also expanding.  The wispy extensions to the central blob seem 
to be the gas excited in the northern ring when the positron-electron
plasma passed through it leading to its temporary brightening.  This emission has 
since faded.  Since all the rings are equatorial, the progenitor has 
suffered excess mass loss in the equatorial planes which might explain
the prolate-shaped remnant assuming projection effects are not misleading. }

\item Synchrotron emission from the ring is detected at a short
wavelength of 0.87 mm while only the central blob
is detected at still shorter wavelengths and which has been inferred to be emission 
due to dust in the blob  \citep{2014ApJ...782L...2I}. 
{\it This observation lends further support to the central blob being
the metal-rich remnant of the explosion where dust is being formed and
which is still optically thick to radio synchrotron from mixed 
relativistic plasma.}

\item The radio spectral index measured in 1992 was 0.97 which in 2000 had 
reduced to 0.88
($S\propto\nu^{-\alpha}$) \citep{2002PASA...19..207M}.  The spectral index has
continued to decrease as seen in Figure \ref{fig10}b taken from 
\citet{2010ApJ...710.1515Z}. {\it Note that this behaviour, 
which is opposite of what is expected from an ageing electron 
energy spectrum, is exhibited
by the radio emission coincident with the inner equatorial ring.  No new
radio component from the central remnant has yet appeared.}

\item No central compact object has been detected from SN 1987A.
%{\bf reference}.

\begin{figure}[t]
\centering
\mbox{
\includegraphics[width=4cm]{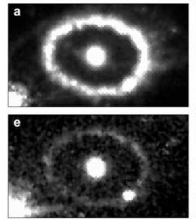}a
\includegraphics[width=4cm]{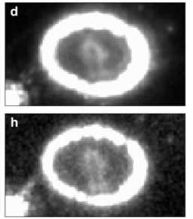}b}
\caption{\small From \citet{2011Natur.474..484L}.  Figures show the R 
(top panels) and B band (lower panels) evolution of SN 1987A.  
%The light curve shows that the emission increased around day 5500.  
The left panels show SN 1987A on 24 September 1994 (day 2770) and the right 
side panels are from 9 April 2009 (day 8101).  
Notice how the circumstellar ring brightens with time and the central ejecta 
increases in size and evolves to an elliptical ring.   It is suggested
that the electron-positron plasma is the cause of the synchrotron emission from
the ring while in future radio synchrotron emission might be detectable from the
non-thermal electron-proton plasma in the central component. }
\label{fig11}
\end{figure}

\end{itemize}

The triple ring structure around SN 1987A indicates that the progenitor
lost mass through episodic ejections instead of undergoing continuous mass loss. 
These two types of mass loss by massive stars before they explode as supernovae
has been confirmed from studies of massive supergiant stars. 
For example, it was inferred from the study of
two massive red supergiant stars that Betegeuse 
was undergoing a steady mass loss whereas VY CMa was experiencing 
episodic mass loss \citep{2009AJ....137.3558S}. 

While it is difficult to explain the equatorial nature of the triple ring
system, its structure can be trivially explained if 
the progenitor of SN 1987A was in a loose binary with an orbital radius 
of the order of half a parsec or so.  In this case, the distinct centres of the
three rings detected around SN1987A would be due to three episodes of
mass loss suffered by the progenitor when it was at three different locations 
in the orbit.  The centres of the three rings then outline the orbital path
of SN 1987A.  From the observed structure,
(assuming the orbital motion was not affected by the explosion)
it appears that the star is moving northwards towards the centre of
the northern ring.  
The rings are fairly symmetric indicating mass loss from a single object
in the centre of the rings.  The southern ring is observed to be 
the largest and can be inferred to indicate the first episode of mass loss.
Since the size of the northern ring is similar, one can infer that both the mass
loss episodes happened in the same orbit.  
%The star should have large proper motion for this to happen.   
The separation between the centres of the two large rings is about 0.4 parsec 
which means that if the star was moving with 
a velocity of 100 km/s (this value has no observational basis from data on SN 1987A), 
it would have taken about 3900 years to traverse
that distance.  In the 3900 years that the star took between 
the two mass loss episodes, the southern
ring would be 0.04 parsecs in size if it was expanding at a rate of
10 km/s (the rings are likely to have been ejected with a higher velocity
and also probably expanding at a higher velocity).  
The star has then completed its orbit in which both the rings have
expanded to a radius of $\sim 0.4$ parsecs.  During another orbit when the progenitor is
between the two rings, it suffers another another mass loss episode
which results in the inner ring of diametre 0.19 pc.   No further mass loss
seems to have occured before the progenitor exploded 
as a supernova.  The supernova still appears to be at the centre
of the last mass loss episode which suggests that the expansion velocities
of matter ejected in the last episode of mass loss is much larger 
than the space motion of the star.
If a companion star to SN 1987A forming a loose binary can be identified 
then it would give a boost to this explanation.  Such large orbits are not
exceptional since orbital periods have been observed to range 
from less than an hour to hundreds of thousands years.  For example, while
cataclysmic variables have short orbital periods, there exist long period double 
(or triple) binaries like Alpha Centauri AB and Proxima Centauri.  
Proxima Centauri has an orbital period of 550000 years around Alpha Centauri AB 
with the major axis being about 0.2 light years.
Alpha Centauri A and B form a tight binary with an orbital period of 
about 80 years and a separation of about 11 AU.  

The triple ring system around SN 1987A are not spherical shells but
equatorial rings with the supernova being located at the centre of the
inner ring. 
The reasons for the mass loss being confined to equatorial rings and
not spherical are not clear and require further investigation.  
The rings have to have been ejected with high radial velocity larger than
the orbital motion of the star, so that the rings  
continued to expand around the ejection position of the star and were not
affected by the motion of the stars.  
All the three rings since due to equatorial mass loss should lie in the same
plane.  This can then account for the enhanced optical emission which is detected
in the overlap region between the inner ring and northern part of the southern 
ring (see Figure \ref{fig9}a).
{\it To summarise - the triple equatorial ring system around SN 1987A, 
due to episodic mass loss from the progenitor, can be explained if
the star was in a loose binary and underwent three distinct episodes of 
mass loss from the equatorial regions, at much higher ejection velocities
than the orbital motion and from three different positions in 
its orbit.  The reason for equatorial mass loss instead of spherical mass loss 
from the progenitor star is not clear and needs to be investigated. }

The radio properties of SN 1987A can be easily explained by the 
fast ejection 
of a positron-electron plasma which is responsible for the short-lived prompt
radio emission which was detected with a frequency-dependent delay
and the second phase of radio emission coincident with the inner equatorial ring
but detected with a frequency-dependent delay.  That the prompt radio
emission from SN 1987A was detected almost immediately after the explosion
indicating low free-free thermal optical depths 
and was also extinguished much sooner than is generally observed in
type II supernovae probably due to the lack of an appropriate magnetic
field, argue for the lack of a distributed circumstellar medium around the supernova.  
As has been inferred from the triple ring structure, the progenitor 
lost mass predominantly in episodic ejections so that the
distributed circumstellar matter around SN 1987A could be absent.
However the detection of prompt emission soon after the explosion
indicates that there does exist some distributed circumstellar medium
in the immediate vicinity of the star. 

Since the progenitor star, for some inexplicable reason, lost mass in
episodic ejections only from the equatorial regions, it is intriguing
on what happened to the outer parts of the star in the non-equatorial regions. 
The detection of a central nebula which has evolved to be elongated perpendicular
to the equator suggests that the star did retain more mass in the 
non-equatorial regions which was blasted away in the supernova explosion.
The nebula seems to have expanded to 0.14 pc from the explosion site
around day 8101 (Figure \ref{fig11}b) which means that the remnant
has been expanding with a mean velocity of about 6200 kms$^{-1}$ since
the explosion.
However before drawing any firm conclusions it will be useful to wait 
for more observational results on the central remnant to become available.

No radio synchrotron emission has been detected from the central nebula. 
This could indicate that the nebula is still optically thick to radio frequencies.
Since this supernova is continously being monitored, we should
be able to detect the frequency-dependent switching on of the 
radio synchrotron emission from the central remnant in the future. 

{\it To summarise - all the synchrotron signatures of SN 1987A from the time of
explosion to the bright radio ring are due to the fast positron-electron
plasma set forth by the explosion.  This means the synchrotron radiating 
plasma should also be radiating the positron annihilation line at 511 keV as 
it expands away from the explosion site.  Deep observations of this
line immediately after the explosion and when the prompt radio emission
becomes detectable might help confirm this.  
The central nebula which contains most of the mass ejected
in the explosion i.e. is the supernova remnant.  No radio synchrotron emission
has been detected from the proton(ion)-electron plasma in the main remnant.
This is likely due to the large optical thickness of the thermal material
to radio emission and one can expect it to become transparent in the near
future allowing detection of radio synchrotron emission from the remnant. }

  \subsubsection{Type II SN: SN1993J in M~81 }
\begin{figure}
\centering
\includegraphics[width=7cm]{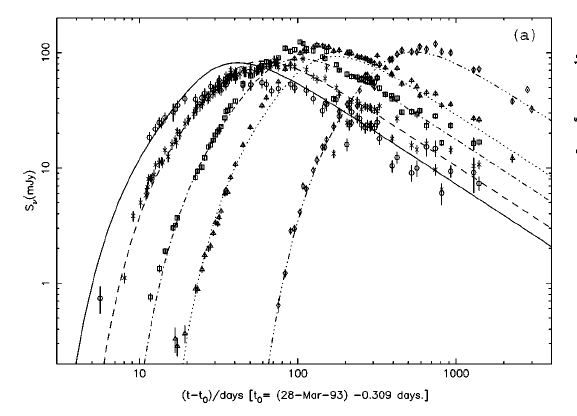}(a)
\includegraphics[width=6cm]{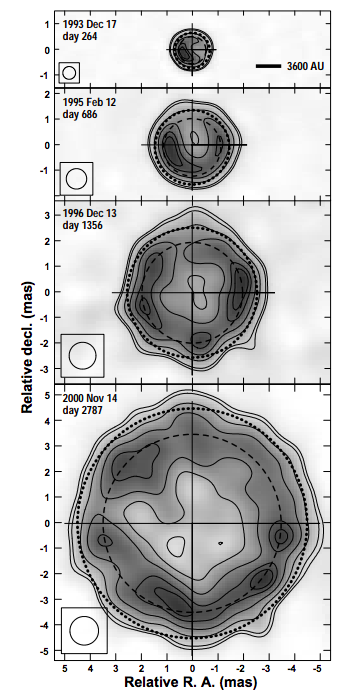}(b)
\caption{\small (a) The figure showing the radio light curves of SN 1993J is reproduced
from \citet{2002ARA&A..40..387W}.  Light curves at 22.5 GHz (open circles), 14.9 GHz (stars),
8.4 GHz (open squares), 4.9 GHz (open triangles), 1.5 GHz (open diamonds) are shown.  The onset
of the highest frequency emission precedes the lower frequencies.
(b) The figure showing a few VLBI images at 8.4 GHz of SN 1993J for different 
epochs is reproduced from \citet{2002ApJ...581..404B}.   Notice the shell which should
be composed of only positron-electron plasma is
similar to a supernova remnant and no central radio component has been detected
yet.  The image epochs shown in (b) are included in the light curve in (a). }
\label{fig12}
\end{figure}
SN 1993J was a supernova of type II which was recorded on 28 March 1993
in the galaxy M 81 located at a distance of about 4 Mpc 
\citep{1993IAUC.5731....1R}.  At this
distance, 1 arcsec corresponds to about 20 pc.  The optical (BVRI) light
curves of SN 1993J showed a double-peaked structure \citep{1993PASJ...45L..59V}
which is commonly observed in type II supernovae.  This supernova has
been extensively observed at radio wavelengths including high resolution VLBI imaging.  
The prompt radio emission near 25 GHz was detected around day 5 while
it was detected at 1.4 GHz around day 75 \citep{2007ApJ...671.1959W}. 
The prompt radio emission from SN 1993J has persisted for more than a decade
after its frequency-dependent detection at radio bands (see Figure \ref{fig12}a).  
The radio emission of SN 1993J at 6 cm peaked around day 132 at $\sim 117$ mJy
whereas it peaked at 20 cm around day 739 at 119 mJy
\citep{2007ApJ...671.1959W}.  The flux density of SN 1993J at 1.7 GHz measured
with the VLBI in 2010 had reduced to 1.7 mJy \citep{2010evn..confE..57B}.
If the supernova had exploded at a distance of 10 kpc from us, then the recorded
peak flux density at 1.4 GHz would have been 19000 Jy while the flux density
at 1.7 GHz in 2010 would have been 270 Jy. 
The radio emission at 8.4 GHz was resolved with VLBI into a shell-like
structure on day 175 \citep{2003ApJ...597..374B}.  Since then the radio shell has
been showing near-symmetrical expansion (see Figure \ref{fig12}b) and its
radio synchrotron structure is similar to relatively older 
shell supernova remnants like Cas~A and Tycho.  This is not surprising
since the spherical radio shell indicates that these supernova explosions 
have been spherically symmetric and ejected a spherical shell of relativistic 
plasma which has evolved to the observed symmetric radio shell. 
The synchrotron radio emission from the older
supernova remnants should be emitted by the proton-electron plasma while in SN 1993J,
the positron-electron plasma should be emitting.
The symmetric shell in SN 1993J also helps us infer that the  
progenitor star suffered mass loss in the form of a uniform continuous wind
so that the circumstellar medium and magnetic field are fairly uniform around SN 1993J.  
In fact, if one visualises the positron-electron plasma being incapable
of sweeping up matter, then the detailed structure of the prompt radio emission 
in the shell should trace the structure of the underlying circumstellar magnetic field. 
Interestingly, the rate of decline of radio emission 
increases from $t^{-0.7}$ to $t^{-2.7}$ beyond about day 3100 
with little change in the radio spectral index which remains around $\sim -0.8$ 
\citep{2007ApJ...671.1959W}.  However \citep{2002ApJ...581..404B} find that
the radio spectral index flattens from a 
mean value of $\sim 0.85$ upto day 1000 to $\sim 0.63$ observed upto 3164 days 
\citep{2002ApJ...581..404B}.  More recent data on this supernova, if available,
should be able to resolve this issue.  Interestingly it was found that the
behaviour of the X-ray light curve was similar to the radio light curve so
that the X-ray emission also showed a faster decline after day 3100
\citep{2007ApJ...671.1959W}.  The shell in SN 1993J was estimated to
be expanding with velocities $\sim 22500$ kms$^{-1}$ early on 
which reduced to 17200 kms$^{-1}$ around day 30, further slowed
down to $\sim 8900$ kms$^{-1}$ around day 1600 and to 8700 kms$^{-1}$
around day 3164 \citep{2002ApJ...581..404B}.  Only faintly
polarised radio signal is detected from SN 1993J \citep{2003ApJ...597..374B}. 
{\it To summarise:  prompt radio emission from SN 1993J is resolved into 
a spherical shell indicating
the ejection of a positron-electron plasma radiating in a uniform circumstellar
magnetic field.  The fast expanding plasma has slowed down to expansion
velocities less than 8700 kms$^{-1}$. }

A radio source due to synchrotron radiation from a proton-electron
plasma, coincident with the massive remnant of the explosion is expected
at the centre of the radio shell. 
Unlike SN 1986J, no new synchrotron radio source in the centre of 
the shell in SN 1993J has been detected \citep{2003ApJ...597..374B,2010evn..confE..57B}.  
This does not seem to be abnormal since while a radio source 
has emerged in the centre of SN 1986J, it is yet to emerge in SN 1987A.  
The detection of the central radio source depends sensitively on the
properties of the ejected mass especially the drop in the free-free optical depth
which can follow different evolutionary timescales for different supernovae. 
Since the radio shell in SN 1993J has an extent of only a few milliarcsecs, it has
not been possible to ascertain the presence or otherwise of a central
optical nebula i.e. a remnant as was possible in SN 1987A.  
The existence of the remnant has to be surmised from lower 
angular resolution spectral observations at optical, X-ray and infrared bands.  
The evolution of the optical spectral features of SN 1993J in the first few 
years ranged from detection of broad hydrogen lines which weakened over time
and the eventual appearance of the nebular spectrum  
\citep{2000AJ....120.1487M}.
It is noted that from around day 300 or so, a box-like profile was recorded
for H$\alpha$ and a few other elements \citep{2000AJ....120.1487M}.
After day 670 or so, it was noticed that a narrow emission feature
had appeared on top of the box-like profiles \citep{2000AJ....120.1487M}.
While the nature of the optical features is intriguing 
and hence have been difficult to interprete, we
can certainly infer that the broad optical features arose in the massive remnant.
Infrared emission from warm dust is detected from SN 1993J more than two decades
after the explosion \citep{2016ApJ...833..231T}.  
The formation of dust in SN 1993J is also deduced from the dimming of the red
part of the spectral lines of [O III] and from which dust of mass 0.08 to 0.15 M$_\odot$ 
was inferred to have formed by 2009 \citep{2017MNRAS.465.4044B}.  
It was found 
that the dust mass associated with supernova remnants of type II increases
as they evolve and that Cas~A had formed dust of mass $\sim 1.1$ M$_\odot$ 
about 330 years after explosion \citep{2017MNRAS.465.4044B}. 
{\it To summarise: In SN 1993J, the positron-electron plasma that the explosion
set forth has been radiating radio synchrotron emission in a spherically 
symmetric shell-like structure
supporting both a spherical explosion and a uniform density circumstellar medium.
Detection of wide optical lines and dust in SN 1993J
supports the existence of a supernova remnant which in analogy to SN 1986J and
SN 1987A should be in the centre of the observed radio shell.  
No radio synchrotron source is detected in the centre of the shell indicating
that the massive remnant of the explosion is still optically thick to radio
frequencies.  }

\subsubsection{Summary}
Most type II supernovae are observed to peak at 6 cm within a few hundred 
days after the explosion whereas type 1b and 1c are observed to
peak within 50 days with typical observed peak radio
luminosity being between $10^{26}$ to $10^{29}$ erg~s$^{-1}$~Hz$^{-1}$. 
\citep[see Table 3 in][]{2002ARA&A..40..387W}.  

In literature, a supernova explosion is explained by the scenario in which 
the core starts contracting into a degenerate neutron star and triggers a 
thermonuclear explosion in the gas outside the core which energises and 
expels matter from the star.  This scenario has been strongly supported by 
observations.  The first signature of the explosion has been observed to be 
fast neutrinos.  We sketch the further evolution of the supernova especially 
from the radio signatures.  We point out that
that observations support the ejection of a fast
positron-electron plasma followed by a relatively slower proton(ion)-electron plasma
in addition to the neutrinos.  In fact, most of the radio observations and
several multi-band features are easily explained once the inevitability of 
both the plasma being ejected following the explosion and 
the expected observational implications of the same are appreciated. 
It is also possible that both the plasmas are expelled at comparable
velocities but the latter suffers larger kinetic losses and quickly slows down. 
The positron-electron plasma gyrates and radiates in the circumstellar magnetic
field and is responsible for the prompt radio emission from a supernova
which becomes detectable once the foreground densities in the circumstellar
gas have sufficiently declined. 
The absence of a circumstellar field will lead to absence of prompt radio emission.
Although the relativistic electrons in the massive remnant will be
radiating in the field frozen in the remant, we detect this only when 
the initially large free-free optical depths of the thermal
gas mixed with the radio plasma have sufficiently declined.  This emission 
should be what we refer to as the supernova remnant. 
The frequency-dependent detection of radio synchrotron emission from 
both the plasma, as was observed in SN 1986J, unequivocally supports 
the above interpretation.  
The existence of a faster positron-electron plasma and a slower remnant is also supported
by the observations of SN 1987A wherein an expanding central remnant is detected
in dust and in optical bands while the fast plasma is responsible for  
the prompt radio emission and the currently detectable
emission from the inner equatorial ring.  The central remnant in SN 1987A
continues to be optically thick to radio frequencies but eventually 
radio synchrotron emission 
should become detectable from it.  In fact, frequency-dependent
detection of radio emission from a 
central component in SN 1987A, SN 1993J and other such supernovae should conclusively
support this explanation.
The existence of the two types of plasma ejected by a supernova explosion is
also supported by the detection of 
a thin outer rim in X-rays (or optical) and radio synchrotron emission around several old
supernova remnants like Cas~A, Kepler etc in addition to radio 
synchrotron emission from the massive remnant.  The rim emission is from
the positron-electron plasma while
the radio emission from the remnant is from proton-electron plasma radiating
in the field frozen in the remnant.  The rim emission appears to be due to
Compton scattering of the annihilation photons to lower energies. 

The above discussion also highlights some testable inferences regarding the
differences in the supernovae of type Ia and type II:
(1) The optical light curves of type II supernovae are generally observed
to be double-peaked whereas those of type 1a are single-peaked. 
This behaviour can be inferred to indicate that the second peak in type II
supernovae is due to the central hot star left behind by the explosion. 
Since no central
star is expected in type 1a supernovae, the second peak is also missing. 
(2) The detected higher expansion velocities of metal-rich knots 
as compared to the hydrogen/helium-rich wisps in supernova remnants are due
to the effect of the radiation pressure exerted on the optically thick
metal-rich knots by photons from the central hot star.  
This feature should, hence, only be detected in remnants
which do host a hot object at its centre. 
Since no central star is expected to be left behind in a type 1a supernova explosion,
all the compact features in the remnant should be expanding with similar
velocities.  This, then also provides an observational method to identify the 
existence of a central hot object in a remnant. 

\section{Accreting rotating black holes}
In this section we discuss objects which owe several of their characteristics to 
energy input from phenomena near black holes.  Black holes 
are objects wherein matter densities are so large
(larger than in neutron stars and white dwarfs) that matter is compressed into a 
volume much smaller than a sphere of Schwarzchild radius ($R_s$) defined by the mass of 
the object.  The fictitious surface of a black hole is referred to as the event
horizon.  Since the velocity required by matter or light to escape from 
the region inside the event horizon
is greater than the velocity of light, nothing can escape from the black hole and 
anything which falls through the event horizon of the black hole is trapped forever.   
For a static black hole, the event horizon is the sphere of radius $R_s$.
For a rotating black hole, the event horizon is prolate-shaped and
the sphere of radius $R_s$ is referred to as the ergosurface with
the region between the event horizon and ergosurface referred to as the ergosphere.
The escape velocity from the erogsphere is equal to velocity of light and
hence electromagnetic radiation can escape from it as is observationally
supported by the spectral lines showing a gravitational redshift greater than 0.5 
which are regularly detected in the quasar spectra and which have to necessarily
arise from a region with a separation less than $R_s$
from the black hole \citep{2016arXiv160901593K}.
The gravitational potential  ($\phi = Gm M_{BH}/R_s$) experienced by a particle of mass $m$ 
at the Schwarzchild radius $R_s$ will be same for a stellar mass or a supermassive 
black hole - this is because the ratio $M_{BH}/R_s=1/3~$M$_\odot$~km$^{-1}$ 
is the same for black holes of all masses.  In other words, as the mass of the
black hole increases, its Schwarzchild radius also increases so that the ratio
$M_{BH}/R_s$ is a constant. 
An important physical implication of this is that it 
should lead to similarity and uniformity in the physical properties of
phenomena in the immediate vicinity of the black hole irrespective of its mass
and the observed differences will be due to the differing linear dimensions 
of the systems. 

Some of the well known black hole-powered objects are microquasars and active galactic 
nuclei (active galaxies).  
Microquasars are stellar mass black hole binaries and are mostly detectable in 
the near universe while phenomena associated with supermassive black holes  
can be probed to cosmological distances.  Microquasars should exist in 
distant galaxies but their small size and hence lower energy phenomena
might not allow their detection from the distant universe except occasionally
during high energy transient flares.
A supermassive black hole e.g. $\sim 10^8$ M$_\odot$  will have 
$R_s \sim 3\times10^8$ km, whereas a stellar mass black hole e.g. $\sim 20$ M$_\odot$ will
have $R_s \sim 60$ km.  Such widely different linear scales will lead to differences in 
global accretion rate, total energy budget, angular sizes and hence detectability. 
The phenomena associated with black holes 
emit over almost the entire electromagnetic spectrum from $\gamma-$rays to
radio bands.  Synchrotron emission is detected in many such high 
energy phenomena indicating that such processes always generate
relativistic electrons and that magnetic fields are ubiquitous. 
Radio synchrotron emission is commonly detected from many active nuclei, 
microquasars and is the only diagnostic, currently available to us, 
for studying relics and halos in clusters. 
These strong synchrotron radio emitters are discussed in this section.
We follow the inferences from \citep{2016arXiv160901593K} and
\citep{2017arXiv170909400K} especially regarding the origin of 
(1) relativistic particles,
(2) accretion disk and bipolar jets and (3) degenerate matter surface 
around the event horizon of a black hole.  These were summarised in the introduction.

We begin the discussion with microquasars.  Although these were identified much later
than active nuclei, the extensive and unique observational results that exist on 
microquasars shed significant insight on the physical processes that occur near 
accreting black holes.
We follow it up with a discussion on active nuclei and 
on the origin of radio relics and lobes in clusters of galaxies.  
A physical model that derives from observational results is included for the 
aforementioned
objects and is used in understanding particular examples.  As in earlier papers,
the aim throughout has been to look for consistent and cohesive explanations 
of observational results within the framework of known physics instead of
contrived and exotic ones in unknown physics.  It should be 
added that it has been possible to conduct 
this study due to the careful observational results and inferences 
by several scientists. 

\subsection{Microquasars}

Microquasars are binary stellar systems consisting of an accreting compact object
like a black hole and a gaseous companion.  The name seems to capture the essence of their
observational properties in the sense that these are stellar-size versions of a quasar
or a radio-loud active nucleus.  
Microquasars display stellar-like compact appearance but are 
much brighter in the optical and
X-ray bands, are variable, host radio jets and are not located at the centres of galaxies.
Another category of such point-like sources which
are extremely bright in X-rays ($L_X > 10^{39}$ erg s$^{-1}$ in a band like $0.3-8$ keV)
and are not located at the centres of galaxies are the ultraluminous X-ray sources (ULX). 
It is believed that ULX are binaries which either host an accreting intermediate mass 
black hole ($10^2 - 10^5$ M$_\odot$) 
%{\bf reference} 
or an accreting stellar mass 
black hole.  Since the known microquasars are fainter than ULX in the X-rays,
these have been considered as different class of objects.  
ULX are rare with none having been detected in the Milky Way and generally only
one ULX being detected in most galaxies.  There do exist exceptions such as the ring 
galaxies Cartwheel and NGC 922 which host several ULX sources mostly located
in the star forming ring or other star forming regions in the galaxy 
\citep{2004A&A...426..787W,2012ApJ...747..150P}. 

Microquasars were identified and named as such when 
radio jets and lobes were detected around
the Galactic source 1E1740.7-2942 \citep{1992Natur.358..215M}.  
Astrophysicists have long appreciated the fact that if we can better understand the 
physical mechanisms active in these nearby objects, we might be able to improve 
our understanding of extragalactic quasars and active nuclei.  
More than 40 microquasars have been identified in our Galaxy \citep{2012arXiv1206.1041M}.
A few notable Galactic microquasars are SS 433, GRS 1915+105, Cygnus X-3 and Scorpius X-1.
We briefly include a few interesting observational results on these microquasars and then
discuss the results on SS 433 in detail.
Several microquasars have been monitored at radio bands \citep[e.g.][]{2017ASPC..510..492T} 
and other wavebands over long periods which has resulted
in a database of simultaneous multi-frequency observations.
Radio jet blobs are
observed to expand at relativistic velocities ranging from superluminal in GRS 1915+105
\citep{1994Natur.371...46M} to $0.45c$ in Scorpius X-1 \citep{2001ApJ...553L..27F} 
to $0.25c$ in SS 433 \citep{1979ApJ...233L..63M} and which if
interpreted as the escape velocity of the blob from the compact 
object ($v_{esc} = c / \sqrt n$ where
$n$ indicates the radial separation of the launching region from the
black hole in units of $R_s$) would mean that 
these were launched from regions separated by less than $16 R_s$ from the compact object.  
In Table \ref{vesc}, the separation from the black hole for different jet
velocities (assuming they represent the escape velocity from the black hole)
and the gravitational redshift that should be shown by a line photon
ejected from that separation are listed. 

\begin{table}
\centering
\caption{The connected nature of the separation from the black hole $R=nR_s$,
escape velocity required by matter to overcome the gravity of the black hole 
$V_{esc} = c/\sqrt(n)$, 
the gravitational redshift $z_g=1/(2n)$ that line photons will suffer when emitted
from R is shown here.  In the last column the intrinsic redshift that will be shown
by the line photon emerging at R is listed.  Within the ergosphere, $V_{esc}=c$.}
\begin{tabular}{l|l|l|l}
\hline
V$_{esc}$  &  $R$  & $z_g$ & $z_{in}$ \\
\hline
0.1c  & $100R_s$ & 0.005 & 0.005 \\
0.25c & $16R_s$ & 0.03125 & 0.03226 \\
0.5c  & $4R_s$  & 0.125 & 0.143 \\
0.75c & $1.78R_s$  & 0.281 & 0.391 \\
0.9c  & $1.23R_s$ & 0.4065 & 0.6849 \\
0.99c & $1.02R_s$ & 0.49 & 0.9608 \\
\hline
\end{tabular}
\label{vesc}
\end{table}

\subsubsection{Summary of observed features}
The observed features of microquasars can be summarised as follows: 
\begin{itemize}

\item A bright star-like object displaced from the centre of the galaxy which is bright
in multiple wavelengths including radio.

\item Multi-band variability observed on timescales of minutes to hours. 

\item Presence of a radio core, relativistically expanding radio jets, fast radio 
variability and 
radio flares often showing some form of correlation with X-ray and $\gamma-$ray emission.   

\item Microquasars are seen to be either in a radio quiescent state wherein they
maintain a constant flux density or a radio flare state wherein they rapidly brighten.
Such flaring is also observed in other bands. 

\item Microquasars display a hard spectral state in
which hard X-rays due to non-thermal processes dominate and a soft spectral state wherein 
soft X-rays due to a thermal process dominate the X-ray emission. 

\item Most microquasars show a correlation between the radio and X-ray states.  A microquasar
in a hard X-ray state shows persistent radio synchrotron emission from the jets which seem 
to be disrupted in the soft X-ray state e.g. Cygnus X-1.  However the opposite is also
at times observed to occur. 

\item High resolution milliarcsec radio images of microquasars show a sequence of 
events - the core gets brighter and 
increases in size, a component detaches from the core and the core gets fainter, the 
component starts expanding away from the core at relativistic velocities, sometimes such a
phenomenon is also noted on the opposite side of the core either simultaneously
or with a delay, the component(s) expand, diffuse out over some timescale and 
eventually become undetectable. 
The sequence of events is not periodic but is often observed to repeat on timescales 
ranging from hours to days to years. 

\item The range of observed properties of microquasars: jet sizes range from 
10 to $5\times10^5$ astronomical units, 
the jet radio luminosity
ranges from $10^{24}$ to $10^{26}$ W \citep[e.g.][]{2001IAUS..205..264S},
the jet bulk velocities appear to range from $0.25c$ to superluminal.   
While the radio jets in some microquasars are long-lived, in others they
rapidly fade. 

\begin{figure}
\centering
\includegraphics[width=8cm]{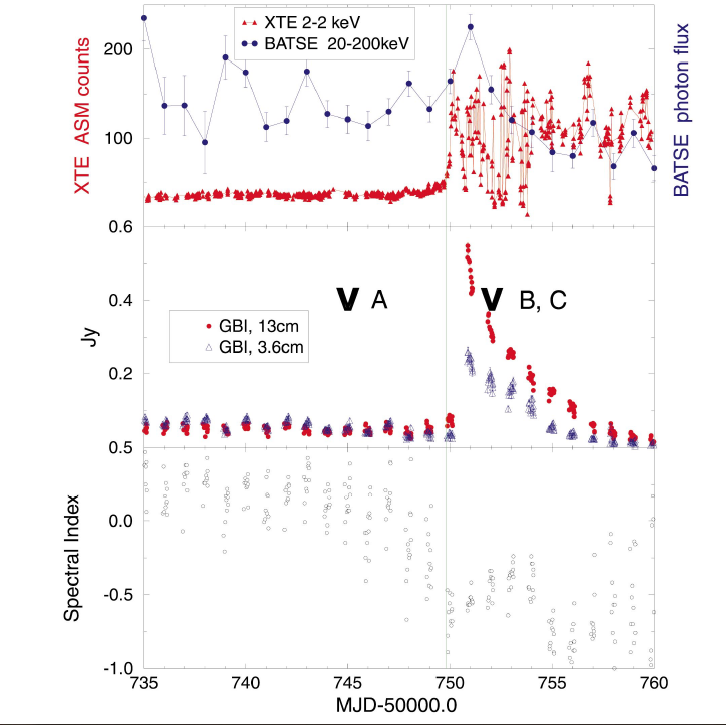}
\caption{\small Figure showing light curves of the microquasar GRS1915+105 reproduced
from \citet{2000ApJ...543..373D}.  The X-ray (top) and radio (middle)
light curves range from 14 October to 8 November 1997 during which a flare was recorded.
The variation in the radio spectral index is shown in the bottom panel and the change
from a flat spectrum to a steep spectrum during the flare can be seen.  }
\label{fig13}
\end{figure}

\item The flat radio spectrum of the core in GRS 1915+105 is 
always resolved into a compact jet in high resolution observations and the radio emission
consists of a compact jet (AU-scale) and a large scale jet (several 100 AU-scale) 
with superluminal motion being registered in the large scale jet  
\citep{2000ApJ...543..373D}. 
Coeval radio and X-ray observations of GRS 1915+105 have provided some interesting 
results as noted in \citet{2000ApJ...543..373D} such as
(a) a plateau/quiescent state is observed in which the radio spectrum is flat and the radio
source is compact (few AU) with typical radio flux densities 10-100 mJy and a
constant flux of X-ray emission in the 2-12 keV band and (b) a flare state is observed
in which the radio emission brightens upto a Jansky at a wavelength of 13 cm 
within a day, the flat spectrum transitions to an optically thin synchrotron spectrum and 
the X-rays between 2-12 keV brighten and show fast variability (see Figure \ref{fig13}).  
The source in the flare state reverts to the radio plateau phase in a few days 
(see Figure \ref{fig13}). 
Their study of GRS 1915+105 prompted \citet{2000ApJ...543..373D}
to suggest that the unresolved flat spectrum core is
a compact, quasi-continuous synchrotron jet and that both a steady jet and 
episodic ejections are observed from GRS 1915+105.  

\begin{figure}
\centering
\includegraphics[width=9cm]{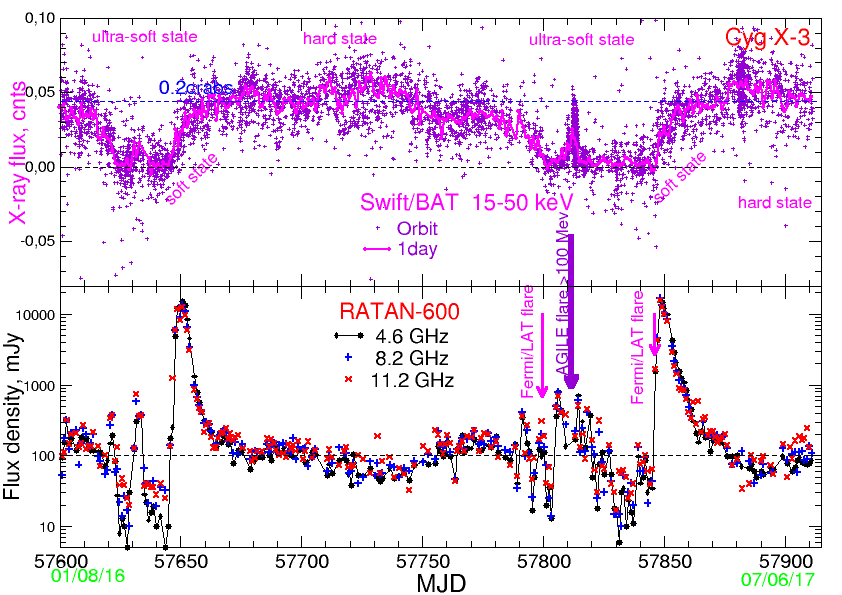}
\caption{\small Figure, showing the evolution of radio and X-ray emission of the microquasar
Cygnus X-3, reproduced from \citet{2017ATel10126....1T}.  
A giant radio flare is often seen to occur at
the end of the ultra-soft X-ray phase which is characterised by weak hard X-ray
emission and constant soft X-ray emission.  In the radio flares, 
the mean quiescent flux density of $\sim 100$ mJy
increases to $\sim 20$ Jy.  The excellent coincidence of a giant radio flare and the
$\gamma-$ray flare is also marked in the figure. }
\label{fig14}
\end{figure}

\item From VLBI observations of Cygnus X-3,  it is inferred that
the microquasar has milliarcsecond jets expanding at $0.25c$
with an opening angle of $80^{\circ}$ and is radio variable \citep{1988ApJ...331..494M}. 
In Cygnus X-3, the correlation
of radio flares \citep{2016ATel.9444....1T,2016ATel.9501....1T} with
$\gamma-$ray flares \citep{2016ATel.9502....1C} have been well established
(see Figure \ref{fig14}) with the $\gamma-$ray flares preceding the radio flares
by a few days.  It has been noted that the radio emission dips below
the mean value just before a radio flare occurs so that the radio dip is 
a precursor of a flare \citep{1994AJ....108..179W} (see Figure \ref{fig14}).  
Frequent radio flares are also recorded from SS 433 \citep[e.g.][]{2016ATel.9481....1T}
although these are fainter than in Cygnus X-3.

\item The microquasar Circinus X-1 displays curved radio jets \citep{1993MNRAS.261..593S}.

\item Scorpius X-1 shows ejection of blobs on opposite sides of the radio core which
then move out and disappear over a day or so and another pair emerges from the core 
with the north-east component generally being brighter \citep{2001ApJ...553L..27F}.  
The ejected blobs move outwards with constant but distinct velocities ranging from $0.31c$ to
$0.57c$ \citep{2001ApJ...553L..27F}.
Radio variability on timescales of hours is noted which is explained by the VLBI images
which show the blobs forming in the core as can be surmised by its brightening, 
the blob being detached and moving away from the core and 
eventually disappearing \citep{2001ApJ...553L..27F}.

\item The electron-positron annihilation line at 511 keV has been observed from
the microquasars - 1E 1740.7-2942 
\citep{1991ApJ...383L..45B,1991ApJ...383L..49S},  Cyg X-1 \citep{1989ApJ...343L..57L},
V404 Cygni \citep{2016Natur.531..341S} and Nova Muscae in which 
the annihilation line persisted for at least 10h 
\citep{1992ApJ...389L..75S,1992ApJ...389L..79G}. 

\item The microquasars 1E1740.7-2942 and GRS 1758-258 located close to the Galactic centre,
are persistent soft $\gamma-$ray ($30-500$ keV) sources
and host double-sided radio jets extending upto a few parsecs
but no optical or near-infrared counterparts have been detected 
\citep[e.g.][]{1994ApJS...92..369M}. 

\end{itemize}

 \subsubsection{A comprehensive explanation}
\label{mqsr}
Here we use the inferences that can be drawn from multi-band observational data 
on microquasars and existing laws of physics to present a simple physical model 
for microquasars.  The model builds on the already known aspects of microquasars
so that a consistent explanation for the myriad observed phenomena emerges. 

The most likely compact object which would be responsible for microquasars
appears to be black holes as in active nuclei.  One of the most important reasons
for this is the mass and hence size of the system.  A neutron star is generally
expected to be less than 2-3 M$_\odot$ in mass with a radius of 
about 10 km i.e. around 3$R_s$. 
However if it a stellar black hole, it can be $10-30$ M$_\odot$ or maybe more
and $R_s$ will range from 30 to 90 km or more.  The larger surface area in the higher 
gravitational potential at $R_s$ as compared to neutron stars
can lead to more energetic and hence detectable phenomena.  
The detection of radio jets expanding with velocities in excess of $0.45c$ which if
equal to the ecsape velocity from the compact object could mean
that the ejection happens from within $5 R_s$ of the compact object.  If the
compact object was a neutron star which are known to have an extremely large magnetic
field, it is likely that either the field or some effect of the magnetic field 
would have been detectable
on the core and jet emission.  However no such effect seems to be present as seen from
the uncanny similarity between the microquasar jets and jets in active nuclei. 
Hence it appears more likely that the 
compact object in all microquasars is a massive stellar-mass accreting black hole and
we assume the same.   The companion is a gas-rich star from which the black hole
is spherically accreting matter.

The jets in microquasars are bipolar indicating the presence of an accretion disk in
the equatorial regions of the black hole.  This then indicates a rotating black hole
which has led to a latitude-dependent accretion rate so that over time an accretion disk has formed.
A bright soft X-ray and optical core is often detected in microquasars indicating the presence 
of a hot degenerate matter surface, emitting black body radiation, which has
formed outside the event horizon of the black hole from the infalling matter.  
We refer to this surface as the pseudosurface of the black hole as was done for the
supermassive black hole in quasars. 
Thermal emission is generally detected in soft X-ray to infrared wavelengths.  
Any form of instability in the pseudosurface of the black hole 
should be detectable as variability in the thermal continuum from X-rays 
to infrared wavelengths.  This instability would not lead to 
variability in the radio bands. 
The relatively high accretion rates in the polar regions of the rotating
black hole should lead to accumulation,
compression and heating of matter in a small circular polar region. 
The process is similar to the accreting white dwarf in novae except the accretion
on white dwarfs appears to be predominantly over the entire surface 
due to their slower rotation.  Taking cue from novae, 
when temperatures of 100 million degrees or higher are achieved in 
the lower layers of the accreted fuel in the polar regions of the pseudosurface, 
these should ignite in an explosive thermonuclear reaction.
If hydrogen ignites, then energy released is expected to be $6\times10^{18}$ ergs gm$^{-1}$
and if elements heavier than hydrogen like carbon and oxygen ignite when temperatures
beyond billion degrees K are reached, then energy released is expected to be 
$5\times10^{17}$ ergs gm$^{-1}$ \citep{1960ApJ...132..565H}.  
While the energy release in the fusion of heavier elements can be achieved
within 1-100 seconds, it takes longer for hydrogen fusion \citep{1960ApJ...132..565H}. 
Thermonuclear explosions have long been identified as a process which can 
instantaneously release huge quantities of energy 
\citep{1939PhRv...55..434B,1960ApJ...132..565H,1965stst.conf..297M} 
and has successfully explained the outbursts in classical, recurrent novae
and supernovae.  When physical conditions of temperatures and densities 
in the accreted matter on the pseudosurface of a black hole which has been accreting normal
matter from its companion become similar to that on the accreting white dwarf in a nova then
a thermonuclear explosion has to be the most likely physical process that should occur. 
We suggest that is precisely what occurs in the polar regions of the accreting black hole.
The thermonuclear process can instantaneously inject energy on very short
timescales and hence energise and eject matter at high velocities. 
This, then, is similar to the ejecta that erupts from the nova 
following the thermonuclear explosion and eventually forms a
shell around the star.  The difference in the rotation speeds of the compact object
i.e. the fast rotation of the black hole leads to extreme bipolarity of the ejected matter
in form of collimated jets in microquasars.  This is due to the variation in the
accretion rates across the pseudosurface and the presence of an accretion disk.
This process then trivially explains the formation of 
relativistic synchrotron jets and optical line forming regions both of which are
directed outwards along the polar axis and observed in SS 433 to be expanding with 
velocities of $0.25c$.  The ejected matter moves away from the compact object along jets. 
As long as the black hole keeps accreting matter, thermonuclear explosions will
keep occuring which will release energy and propel matter outwards along the jet.  
The local conditions should determine the frequency of explosions and eruptions from 
the polar regions for a given system and which  
explains the continuing ejection of blobs/knots of radio emission which is
observable with VLBI in several microquasars (e.g. Scorpius X-1, GRS 1915+105) 
On formation, these knots of emission might be obscured 
due to free-free absorption or synchrotron self-absorption which eventually decreases
as it is ejected and expands leading to frequency-dependent detection with higher
radio frequencies being detected earlier as is often observed (see Figure \ref{fig15}b). 
It also explains the behaviour shown by GRS 1915+105 wherein in quiescence the
radio emission shows faint emission with a flat spectrum which evolves to bright 
optically thin
synchrotron emission with a steep spectrum during a flare (see Figure \ref{fig13}). 
Similar frequency-dependent behaviour is regularly observed in supernovae and novae.
From similar arguments, the accreted matter in the non-polar regions should also
ignite although the lower accretion rates might lead to infrequent eruptions. 
This might explain the occasionally observed extended radio emission in the
equatorial regions of SS 433.  {\it To summarise, the energy source of jets
and optical line forming gas observed in the polar regions of a microquasar
is a thermonuclear outburst in the accreted matter on the pseudosurface of 
the black hole.}  Variability in the radio synchrotron emission will result
from the varying jet characteristics such as a new knot/blob being energised
and ejected.  This can also explain variability in the hard synchrotron X-rays and $\gamma-$
rays when the latter are generated by the inverse Compton process.

The axis of the rotating black hole often shows precession, so that the 
jet direction changes (e.g SS 433), which can be understood
as being due to the gravitational torque exerted by the companion star as they orbit
each other.   

The separation of the line forming matter from the black hole will determine the
gravitational redshift ($z_g$) component in the spectral line redshift.  If it is
located several hundred $R_s$ away, then there will be a negligible contribution
of gravitational redshift to the observed redshift of the line.  However if
the photon emerges very close to the black hole then the contribution can be significant. 
The expansion velocities when estimated from the proper motion of blobs in the jets
of microquasars are found to be relativistic.  Assuming these are comparable
to the escape velocity from the black hole,  it would mean that the observed
redshift of the line 
photons emitted from the jet launching site should include a gravitational redshift
component.  In Table \ref{vesc}, the $z_g$ (and the intrinsic redshift $z_{in}$) 
of a line forming at different separations from the black hole are listed.  
As listed in Table \ref{vesc},  a jet which is launched from a separation of $4R_s$ 
will need to be ejected with at least $0.5c$ to be able to escape the black hole 
while a spectral line forming at $4R_s$ will experience a gravitational redshift
of 0.125 (intrinsic redshift = 0.143) due to the strong gravitational potential 
of the black hole.  This means that the spectral line from the microquasar
would be observed at a redshift of about 0.7. 

The frequent recurrence of radio blob/knot ejection in microquasars
over hours/days indicates frequently recurring thermonuclear
outbursts in the accreted matter on the black hole in the X-ray binary.
The behaviour is reminiscent of dwarf nova outbursts which
recur over timescales of days and last over timescales of days so that 
the optical brightening is short-lived.  
It was suggested that these small amplitude outbursts in dwarf novae
are also caused by thermonuclear
explosions which are not sufficiently energetic to expel matter and only succeed in
expanding the accreted envelope which eventually contracts \citep{2017arXiv170909400K}.  
This is expected since the accreted matter on the white dwarf (or black hole) which is
already heated to high temperatures can be repeatedly ignited leading to
an inflation of the outer envelope till nucleosynthesis leads to formation of iron
or a sufficiently energetic explosion occurs which succeeds in ejecting the envelope.
This also means that for the few microquasars that we observe with radio jets and cores,
there might exist many more which are like dwarf novae in that the thermonuclear outburst
is not sufficiently energetic to expel matter and the thermonuclear reaction 
enriches the accreted matter but seldom expels it. 

{\it 
This discussion then concludes that (1) the main source of black body continuum emission
from the microquasar is the hot pseudosurface of the black hole.  This is
mainly detected at wavelengths ranging from soft X-rays to infrared, (2) the main transient 
source of energy in microquasars is a thermonuclear outburst which leads to
jet launching consisting of relativistic plasma and optical line emitting
gas, (3) the main source of synchrotron continuum emission is the relativistic
plasma ejected from the pseudosurface following the thermonuclear explosion.  This 
emission dominates at hard X-ray and radio wavelengths, (4) the redshift of the line photons
emerging from the jet launching location should also include a gravitational redshift
component, (5) in some microquasars, frequent thermonuclear explosions 
can enrich the accreted matter instead of ejecting it.    }

In the next section we use this model to understand the detailed observational
results on SS 433.   

 \subsubsection{Case Studies}
We discuss, in detail, the case of SS 433 which is very well observed and
has provided us important insight into phenomena that occur 
close to accreting black holes. 
We then discuss the case of the short $\gamma-$ray burst GRB 170817A which 
was detected in the galaxy NGC 4993 on
17 August 2017 and sparked off wide-ranging observations, simulations
and interpretation.  It was interpreted to be a signal from the merger of 
two neutron stars - we investigate if
the observational results can also be explained by a better understood physical
process since GRBs are fairly common in the universe. 
In an earlier paper \citep{2016arXiv160901593K}, the similarity between quasar
spectra and the afterglow spectra of several GRBs was pointed out 
and it was suggested that GRBs are 
energetic transient events on quasars.  We continue this line of argument which can be
generalised to suggest that GRBs are transient events on a black hole 
and examine if GRB 170817A could have occurred on a microquasar in NGC 4993.
It is to be noted that the only fast source of huge quantities of energy that has been
well established from theory and observations, is a thermonuclear outburst 
when a large quantity of matter is simultaneously ignited.  

  \paragraph{SS 433 in Milky Way}:
SS 433 is a microquasar 
at a distance of about 5 kpc, is luminous in the optical (V=14.2 magnitudes)
and has been estimated to have a large line-of-sight extinction of $A_v = 8$ magnitudes
\citep{1984ARA&A..22..507M}.  The star-like object which had been independently detected in 
the optical, radio and X-ray bands was shown to correspond to the same system and it
was suggested that it might be connected to the supernova remnant W 50 
\citep{1978Natur.276...44C}.  The similarity of SS 433 to Circinus X-1 was pointed
out \citep{1978Natur.276...44C}. 
\begin{figure}[t]
\centering
\includegraphics[width=7cm]{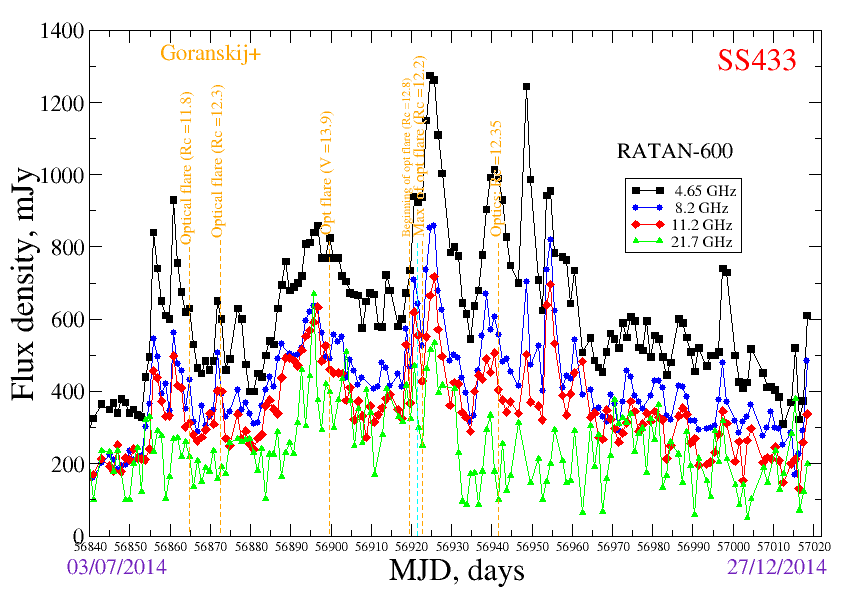}a
\includegraphics[width=6cm]{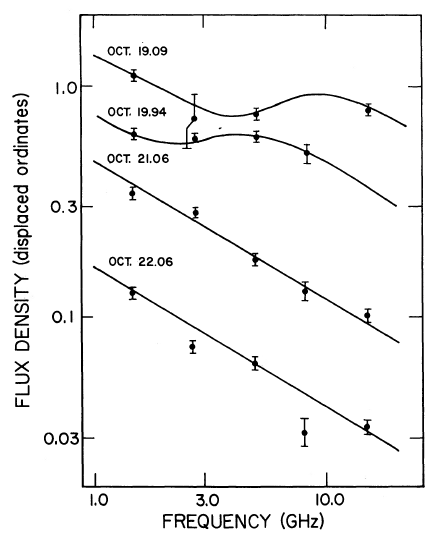}b
\caption{\small (a) The evolution of radio flux density of SS~433 shows the occurrence
of multiple radio flares and coincident optical flares.  Figure reproduced from
\citet{2014ATel.6492....1T}. (b) Figure copied from \citet{1982ApJ...260..220S} 
showing the varying radio spectrum of SS~433 during a flare with changes
recorded within a day. }
\label{fig15}
\end{figure}

Some of the other observed properties of SS 433 can be summarised to be:
\begin{itemize}

\item Multi-band variability is recorded in SS 433.

\item The orbital period of the binary is measured to be around 13 days 
\citep{1980ApJ...235L.131C}. 

\item  It was noted that three optical spectral features from SS 433 were found to move 
in velocity space by about 25000 kms$^{-1}$ in 28 days \citep{1979ApJ...230L..41M}.  
Further observations showed that there existed two sets of hydrogen Balmer lines and He I lines  with
one pair showing large and changing redshift and the other pair showing large and changing
blueshift with the variations being periodic over a 164 day duration such that the
maximum positive and negative velocities were 50000 kms$^{-1}$ and $-35000$ kms$^{-1}$
respectively and these were symmetric about a redshift of $\sim 0.04$ ($\sim 12000$ kms$^{-1}$)
\citep{1979ApJ...233L..63M}.  These were explained by Doppler shifted lines arising 
in oppositely directed precessing jets/clouds expanding at velocities of $\sim 0.26c$ 
oriented at an angle of about 65 degrees to the line of sight 
with this being referred to as the kinematic model for SS~433 
\citep{1979Natur.279..701A,1979MNRAS.187P..13F,1979A&A....76L...3M,1979ApJ...233L..63M}. 

\item Broad emission lines of widths $\sim 4000$ kms$^{-1}$
were detected from SS~433 \citep{1978Natur.276...44C}. 
The spectrum consisted of unshifted wide Balmer lines of hydrogen whose equivalent widths
are variable, He I, He II and several other lines \citep{1984ARA&A..22..507M}. 

\item Occasionally the shifted features are absent from the spectra for several days and
they also undergo fast intensity changes with the changes in the blue and
red-shifted components being quasi-simultaneous \citep{1984ARA&A..22..507M}.  It was 
observed 
in data from 1983 that the moving line velocities increased to 0.3c and
then disappeared from the spectrum for some time \citep{1984ApJ...281..313M}. 

\item Radio emission is detected from SS~433 \citep{1978Natur.276..571R,1979AJ.....84.1037S}
and is variable (see Figure \ref{fig15}).
The radio spectrum of SS433 between 1 and 10 GHz changes within a day during 
a radio flare \citep{1982ApJ...260..220S} (see Figure \ref{fig15}b).
On many days the spectrum is a power law,  on other days it
is absorbed at the lowest frequencies and on still other days it shows
enhanced emission at the higher frequencies so that
the spectrum becomes flatter (see Figure \ref{fig15}b).

\item The radio source consists of double-sided milliarcsec 
jets (see Figure \ref{fig16}a) which 
are observed to precess \citep{1981ApJ...246L.141H}.  The
entire source is located at the centre of the large supernova remnant W~50 and
is believed to be the compact object formed in the supernova explosion whose 
remnant is W~50. 
Extended radio lobe-like structures are seen surrounding a spherically symmetric
W~50 along the jet axis of SS~433 and are often referred to as the radio ears of W~50
(see Figure \ref{fig16}b).
%{\bf reference}
\begin{figure}[t]
\centering
\includegraphics[width=6cm]{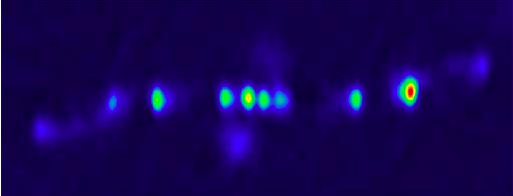}a
\includegraphics[width=6cm]{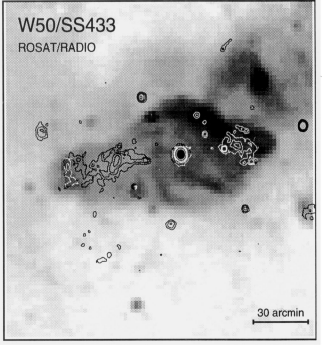}b
\caption{\small (a) VLBA image of SS 433 (Credit: Mioduszewski et al., NRAO/AUI/NSF).  
Knots in the precessing jet are noticeable as is the emission
extending outwards from the equatorial regions.  
(b) The ROSAT X-ray contours superposed on radio emission at 11 cm in grey scale copied
from \citet{1996A&A...312..306B}.  The entire structure hosting radio jets 
shown in (a) is contained within the central bright X-ray core in (b).  } 
\label{fig16}
\end{figure}

\item Combining the data from radio synchrotron and optical line emitting gas 
has helped constrain most of the parameters of the system.  
The jets precess on a cone of half-angle $20^{\circ}$ over a period of about
163 days and the central axis of this cone is inclined by $79^{\circ}$ 
to our sightline 
%{\bf reference}.  
{\it Since both the relativistic plasma and optical line emitting gas support this, it
tells us that the thermal gas in which the lines form and the relativistic
plasma which is observed as the radio jet have to part of the same relativistic
ejection from the microquasar.}  From the width of the emission lines
and assuming a conical geometry, the optical jets are found to be
highly collimated with an opening angle of $1^{\circ}-1.4^{\circ}$ 
\citep{1987SvAL...13..200B}. 

\item The optical line 
emitting regions occupies a smaller volume than the radio emitting region 
The relative narrowness of the moving optical lines indicate highly collimated jets. 
%{\bf references}.  
 
\item The first X-ray imaging observations indicated that 90\% of
the soft X-ray (1-3 keV) emission was coincident with the optical object while the rest
arose in two extended jet/lobe-like regions about $\sim 30'$ long and aligned along the major 
axis of W 50 and which established the connection between SS 433 and 
W 50 \citep{1980Natur.287..806S,1983ApJ...273..688W}.  The X-ray emission in the jets/lobes 
is much softer than from the central
source \citep{1983ApJ...273..688W}.  The position angle of the radio jets 
on arcsecond scales \citep[e.g.][]{1981ApJ...246L.141H},
of the X-ray lobes on several arcminute scales \citep[e.g.][]{1983ApJ...273..688W}
and the major axis of W 50 on several arcminute scales are all measured
to be around $100^{\circ}$ (see Figure \ref{fig16}).   
The east-west extend of the radio source in W50 is about 2 degrees which at a distance
of 5 kpc is about 175 pc.  

\item Variability of short timescales $\sim 2-8$ minutes are detected at 
radio, infrared and X-ray band \citep{2005MNRAS.362..957C}.

\item Radio and optical flares in SS~433 are observed to be coincident
\citep[e.g.][]{2014ATel.6492....1T} as shown in Figure \ref{fig15}a. 

\item The evolution of the radio knots in the two jets are independent 
\citep{1984IAUS..110..289S}.  It was suggested that the knotty
radio emission in jets is related to flare activity in SS~433 
\citep{1984IAUS..110..289S}.

\item The radio spectrum of SS~433 shows a turnover at low frequencies $\sim 150$ MHz
\citep[e.g.][]{2018arXiv180203406B}.

\item No radio emission is detected from the core of SS~433 in high resolution images 
which has been inferred to be due to synchrotron self-absorption and
free-free absorption \citep{1999NewAR..43..553P}.

\item Spatially resolved $\gamma-$ray photons upto 25 TeV have been 
detected from the lobes of SS 433 near the X-ray peaks \citep{2018arXiv181001892H}. 
The emission from radio bands to TeV energies
are all compatible with the same population of electrons in a magnetic
field of about 16 $\mu$G \citep{2018arXiv181001892H}. 

\end{itemize}

SS~433 is the only source wherein the discrete plasma blobs emitting
radio synchrotron and ionized gas emitting moving optical lines 
have been observed to be moving outwards with the same relativistic velocity 
i.e. 0.26c.  In most other microquasars, only the radio blobs in the jet are observed to
be expanding relativistically.  
The black hole in the SS~433 will accrete matter which 
rapidly accumulates on the poles of its pseudosurface due to the highest possible
accretion rates.  When favourable physical conditions are achieved, 
a thermonuclear explosion in the accreted matter on the poles of the
pseudosurface will occur, energising and ejecting matter.  Since the radio blobs
are seen to expand away from the black hole, it tells us that the matter 
has been ejected with velocities equal to or exceeding the escape velocity.
For simplicity, we assume that the matter is ejected at the escape
velocity which implies ejection from a region separated by about $14.8 R_s$ from
the black hole.  The detection of both radio synchrotron blobs and optical lines
expanding at 0.26c gives a rare insight into the properties of matter 
that is launched from the polar regions in the vicinity of the black hole.   
Such ejections consisting of both relativistic plasma and thermal ionized matter are 
commonly observed in supernova explosions whereas ionized matter is always ejected
in classical nova outbursts.  As was 
pointed out in the discussion on supernovae, the energetic ejection 
of matter following the thermonuclear explosion consists of three components -
neutrinos, a fast positron-electron ejecta and a slower proton(ion)-electron ejecta
and observational results strongly supported the existence of the two plasma
with different compositions.  The same should be expected as a result of
the thermonuclear explosion on the black hole pseudosurface.  SS~433 has
provided crucial evidence on the ejection of the proton(ion)-electron plasma
in which the optical lines form.  Since the optical line
forming gas in SS~433 is observed only close to the black hole and not
along the entire length of the radio jet, it could signal its rapid slowing down
after ejection at 0.26c.
This, then would indicate that the knotty radio emission that travels at
0.26c a long way from the core of SS~433 is due to the fast positron-electron 
plasma that was also ejected.  Although the annihiliation line 
has not been detected in SS~433,
its detection from other microquasars (e.g. 1E1740.7-2942)
lends strong support to the presence of positrons in microquasars. 
The few microquasars which are
detected as persistent soft $\gamma-$ray sources also support this inference. 
{\it To summarise, the accreting black hole in SS~433 undergoes repeated
thermonuclear outbursts which expels a proton(ion)-electron and positron-electron
plasma at 0.26c.  The proton(ion)-electron plasma from which optical lines
emerge rapidly slows down whereas the positron-electron plasma continues
at high velocities over long distances.  }

The expected gravitational redshift that a photon emerging from a region
separated from the black hole by 
$14.8 R_s$ should show is 0.034.  The moving optical lines from
SS~433 are observed to be varying about a constant redshift of about 0.04
i.e. $\sim 12000$ kms$^{-1}$.  
If we assume this is the intrinsic redshift due to a gravitational redshift component,
then it translates to a gravitational redshift of 0.0384.
The comparable values of the observed redshift (0.0384) and 
the expected gravitational redshift (0.034) from escape velocity arguments
give independent support to
ejection of matter from a separation  $\le 14.8 R_s$ in addition to
explaining the origin of the observed redshift of 0.04 over which the moving optical 
lines are observed to be symmetric.  This region located close 
to the black hole will have a finite thickness
in which the gravitational redshift will change.  Even 
small changes in the properties of this
region can lead to detectable changes in the gravitational redshift
component and escape velocity of the matter.  For example,
if the launching region moved closer to the black hole, the
escape velocity required would increase.  If at this location, the 
explosion energy is unable to accelerate matter to the escape velocity, then
the ejection would not occur which can explain the 
occasional disappearance of the optical lines.  If ejection does happen
from this region then it explains the occasional appearance of the moving
lines at velocities in excess of 0.26c.  Thus a gravitational redshift
origin for the excess redshift of 0.04 is able to explain several observations
of SS~433.  Interestingly soon after the detection of 
the shifting optical lines, there were suggestions that the observed velocities
of the shifting lines are a combination of Doppler shifts and gravitational redshift
\citep[e.g.][]{1979A&A....76L...3M}.  However this proposed gravitational redshift 
component was not connected to the redshift over which the lines are symmetric 
i.e. 0.04. 

The coincidence of radio flares with X-ray and/or $\gamma-$rays can be expected
if the emission is due to processes involving relativistic electrons.
No correlation would be observed if the emissions involve completely independent
processes.  For example, the radio flares and soft X-ray will not show
any correlation if the former is due to synchrotron process at the poles and
the latter is due to thermal emission from the pseudosurface unless there was
a connection between the two, say through the thermonuclear explosion which
heats up the entire pseudosurface and also accelerates the electrons to
relativistic energies.

The two poles are separated by the accretion disk and will hence 
not undergo simultaneous ejection events.
Since both poles will have high
accretion rates compared to other parts of the pseudosurface, the thermonuclear
blasts and ejections will be frequent but quasi-simultaneous. 
This explains the independent behaviour of the jets that has been observed in SS 433.
The similar shifts in the velocity of the
optical lines indicate that the ejection from both poles happen from a launching region 
separated from the black hole by about $14.8 R_s$. 

{\it To summarise: The ejected proton(ion)-electron plasma is detected in SS~433 in
the form of Doppler shifted optical lines.   These lines and radio synchrotron blobs
are observed to expand with a relativistic velocity of 0.26c which is shown
to be the escape velocity at a separation of $14.8 R_s$ from the black hole. 
It can be inferred that matter is launched from this separation. 
The line photons emerging at this distance would show a gravitational redshift
of about 0.034 (intrinsic redshift will be 0.035) 
which is close to the observed value of about 0.04
lending strong support to this explanation.  The observed flares are when a new component
of relativistic plasma is ejected. }

  \paragraph{GRB 170817A in NGC 4993}:
GRB 170817A which was detected by Fermi-GBM on 17 August 2017 became one of the 
most famous short $\gamma-$ray burst (GRB) due to
its linking to a source of gravitational waves GW 170817 which had been
recorded a couple seconds prior to it by LIGO \citep{2017ApJ...848L..13A}.  
The GRB became more famous when a bright optical transient source AT 2017go 
was detected in the S0 galaxy NGC 4993 within 11 hours of the GRB 
within the angular region defined by the large observing beams 
of the two instruments \citep{2017Sci...358.1556C}.  It was suggested 
that all these were signals from the same source and since the
source of gravitational waves which was localised to a region of area 31 square degrees,
was linked to merging of two neutron stars at a luminosity distance of 40 Mpc as
deduced from templates of gravitational wave signals, the
electromagnetic signals also got linked to the same source \citep{2017ApJ...848L..12A}.  

The electromagnetic signals associated with GRB 170817A consisted of three components -
(1) a relatively faint short $\gamma-$ray burst,  (2) bright thermal emission 
at ultraviolet, optical, and infrared bands which is detected a few days after 
the GRB and fades by day 110.  This emission component is known in literature
as a kilonova and (3) a synchrotron component which dominates radio and
X-ray bands at all epochs from around day 9,  peaks in both bands around
day 160 and then declines.  A constant spectral index of $\sim 0.585$ fits the
synchrotron spectrum from X-ray to radio bands at all epochs \citep{2018ApJ...856L..18M}.  
Figure 1 in \citet{2018ApJ...856L..18M} shows the wide band spectral energy distribution 
between day 9 and day 160 where the distinct nature of the kilonova and the synchrotron
emission are identifiable.  The X-ray luminosity on day 160 was estimated to be 
$\sim 5\times10^{39}$ ergs$^{-1}$ \citep{2018arXiv180806617T} which is similar
to ULX sources.  
A fit to only the post-peak X-ray data i.e. $ > 160$ days results in a spectral 
index of $0.8 \pm 0.4$, a fit to only the radio
data results in a spectral index of $0.4 \pm 0.3$ while the fit to the
entire spectrum results in a spectral index $0.585\pm0.005$ \citep{2018arXiv180806617T}. 
No further $\gamma-$ray signals were detected from this source after the initial
GRB which was about $2-6$ order of magnitudes fainter than other short GRB at
known redshifts \citep{2017ApJ...848L..13A}. 
The luminosity of most GRBs upto $z=4.5$ in the energy range 1 keV to 10 MeV
is observed to be between 
$10^{51}$ and $10^{54}$ erg s$^{-1}$ while the luminosity of GRB 170817A 
was only $\sim 10^{47}$ erg s$^{-1}$.  
  
Since the $\gamma-$ray pulse of GRBs is frequently observed to be
followed by multiband emission referred to as the afterglow 
which ranges from strong thermal continuum to wide multi-redshifted 
absorption lines to radio synchrotron emission, the
association of the GRB with the optical transient in NGC 4993 has a sound basis.
%Here we examine the observations to better understand
%the physical system in which this transient event occurred and we limit
%ourselves to systems whose existence is well established through extensive observations.
Since the observed spectra of GRBs show similarity to quasar spectra 
in terms of the presence of wide emission and multi-redshifted absorption lines, 
it was suggested that GRBs can be explosive events on quasars \citep{2016arXiv160901593K}. 
We pursue this line of reasoning and further investigate if the connection of GRBs to 
active nuclei or more generally to events near black holes 
can be conclusively established from observational results. 

A quasar consists of a black hole with the infalling matter forming its 
pseudo-surface around which a broad line region forms in the non-polar regions -
all of this very close to the event horizon of the black hole.  Quasars are
generally bright in the ultraviolet and optical bands and are the most luminous of
active nuclei.  However it is possible that a quasar with its structure intact
has evolved to dimness.  If a transient energetic event occurs on the dim quasar then 
it can brighten the system for some time.   This can, then, explain a GRB as an
energetic event on a dim quasar.  One possible origin of this energetic event was
suggested as a thermonuclear outburst on the pseudosurface of a dim quasar which will
emit $\gamma-$rays till it lasts.  If a large mass is ignited simultaneously then
huge quantities of energy can be injected into the system in a short time and this can
heat the pseudosurface which starts emitting as a black body especially in the
X-ray to infrared bands.  This can also lead to ejecting overlying matter.  
Thus explosions on dark quasars are a strong contender as a possible origin site of GRBs and
are backed by observational evidence in form of the broad lines and extremely
high energy outputs.  This argument
can be extended to accreting black holes of lower masses - intermediate mass black holes
or stellar mass black holes.   The surface area involved
in the explosion and luminosity of the system will depend on the mass of the black hole
so that it will be largest in case of a supermassive black hole and smallest in a
stellar mass black hole.  

We test the validity of the above explanation to observations of GRB 170817A. 
The short GRB in NGC 4993 was localised to a region displaced from the centre 
of the galaxy 
%(see Figure \ref{grb}).  
This then leaves the option of the explosion having been on a microquasar hosting
a stellar mass black hole in the galaxy or on a intermediate mass black hole.
Since very little is known about intermediate mass black holes in galaxy disks,
we confine this discussion to investigating the origin of the GRB on a dim microquasar
by finding viable explanations for the three emission components that were
detected from GRB 170817A.  A burst of $\gamma-$rays is expected from
a thermonuclear outburst and should exist for the duration of the burst.  Whether
we detect the $\gamma-$rays of energy released in the explosion or enhanced
energies depends on whether the photons reach us directly or are inverse
Compton scattered to higher energies.  This can vary from case to case.  Anyway
the burst of $\gamma-$ray photons from GRBs can be explained in this origin.
The thermonuclear outburst which happens on the pseudosurface can impart some
of the released energy to the highly conducting electrons in the pseudosurface,
and hence enhance its temperature.  The thermal emission that followed the 
$\gamma-$rays i.e. the kilonova emission from GRB 170817A can be explained by emission 
from the hot pseudosurface which was heated by the thermonuclear explosion.  
The emission lasted till the pseudosurface cooled down.  The
relativistic plasma responsible for the synchrotron emission would have
been accelerated by the thermonuclear energy before being ejected.  For a rotating
black hole, matter would be accreted directly at the poles whereas an accretion
disk would have formed in the non-polar regions.  Thus the explosion would
be able to easily eject energised matter from the poles, thus explaining the 
long-lived synchrotron afterglow from GRB 170817A along jets.  
In short, the short GRB being due to a thermonuclear explosion on the pseudosurface 
of a black hole is able to explain all the observed features of GRB 170817A
and hence supports the origin of GRB 170817A on a microquasar. 

The radio emission of the microquasar in quiescence is likely to be undetectable
at a distance of 40 Mpc.  For example, the quiscence radio emission of SS 433 
which is estimated to be at a distance of 5 kpc is around 300 mJy at 5 GHz.
Emission of similar strength from the microquasar in which GRB 170817A appears to
have occurred at a distance of 40 Mpc would be 0.005 $\mu$Jy. 
If radio flares of 5 Jy occurred in this source then these would be 
recorded by us as signals of strength 0.08 $\mu$Jy. 
Thus, it is not surprising that the source became detectable only during
a major energy outburst. 

{\it To summarise: All the three observationally recorded electromagnetic components
of GRB 170817A can be explained by a thermonuclear outburst in the accreted
matter on the pseudosurface of the black hole in a microquasar.
This provides strong reasons to believe that the GRB 170817A and the afterglow
occurred on a microquasar. }

\subsection{Active Galaxies}

Active galaxies are those galaxies which host an accreting supermassive black hole
in their centres and exhibit multi-band features whose origin can be traced
to the active nucleus. 
Active galactic nuclei generally show the following observable
characteristics: (1)  bright non-stellar multi-component continuum emission 
detectable in all wavebands (2) a non-stellar line spectrum often
consisting of broad permitted emission lines and/or narrow permitted and forbidden
emission lines (3) single-sided or double-sided radio and/or multi-band jets extending
out from the nuclear region  (4) radio lobes and/or hotspots located 
on either side of the nucleus and (5) multi-band variability.   
The general picture of an active nucleus in literature comprises of a central 
accreting supermassive black hole hosting an accretion disk from which jets 
are launched that form the radio lobes and hotspots.  However the detailed 
physical processes responsible for energy production, injection, jet ejection,
broad line forming region, variability etc are not known with certainty. 
In this section, we search for explanations for all these processes within
the purview of known physics so that the working of the entire physical system 
becomes apparent. 
Radio galaxies extend over regions ranging in size from kpc to hundreds of kpc 
to Mpc while the radio structure in Seyferts, gigahertz-peaked 
sources (GPS) and compact steep spectrum (CSS) sources are generally 
comparable or smaller than galactic dimensions.   
The host galaxies of radio galaxies, GPS, CSS sources are ellipticals
whereas those of Seyferts are generally disk galaxies.  There remains ambiguity regarding
the host galaxy of quasars and existing observations indicate the prevalence of both
elliptical and disk galaxies.

In the early studies, many elliptical galaxies were found to emit nuclear radio emission 
of lower luminosity than radio galaxies
but stronger than the radio emission detected from the nuclei of spiral galaxies 
\citep{1969ApJ...157..481R}.  Moreover it was noted that radio emission was predominantly 
detected from elliptical galaxies whose absolute photographic magnitudes were 
brighter than $-20$ magnitudes and a correlation between the presence of 
radio emission and presence of optical emission lines was noticed \citep{1969ApJ...157..481R}.
It was noted that the probability of nuclear activity increased when the galaxy 
was a member of 
a group or cluster \citep[e.g.][]{1974MNRAS.168..307S} with the probability being 
maximum for galaxies in pairs \citep{1981IAUS...94..167H}. 
The central source in interacting spiral galaxies was observed to be stronger by a 
factor of 2 to 3 than in isolated spiral galaxies \citep{1980A&A....89L...1H}.  
It was shown that the nuclear emission in spiral galaxies was
more extended and displayed a steeper spectrum ($\alpha \sim 0.7$; $S \propto \nu^{-\alpha}$)
compared to the nuclear source in lenticulars and ellipticals
($\alpha \sim 0.1$) \citep{1974IAUS...58..257E,1978ApJ...221..456C,1981IAUS...94..167H}.  
A good correlation between the presence of a strong
nuclear radio source and extended radio emission in elliptical galaxies was measured
which was inferred to indicate that the nucleus played a role in the extended emission 
\citep{1974IAUS...58..257E}.
Enhanced nuclear activity in galaxies was attributed to increased gas 
accretion by the nuclear black hole \citep{1978ApJ...221..456C}.  

\begin{figure}[t]
\centering
\includegraphics[width=7cm]{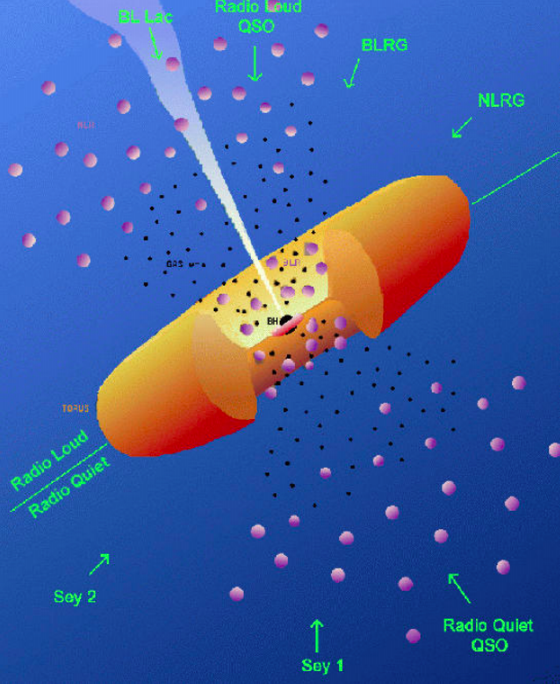}
\caption{\small Unification of active galaxies summarised in a schematic by 
\citet{1995PASP..107..803U}.
Figure copied from http://pulsar.sternwarte.uni-erlangen.de/wilms/teach/xray2.old/xray20233.html.
The pink filled circles represent narrow line regions whereas the black dots represent
broad line regions.  The orange structure is the dust torus whereas the white coloured
outflow from the black hole is the radio jet.  The dark pink band which cuts across
the black hole is the accretion disk.  The orientation-based unification model
for the different types of active galaxies is labelled.  }
\label{fig17}
\end{figure}

Active nuclei have been classified based on observational features such as (1) radio 
morphology/power e.g. FR I and FR II classes or (2) 
observed widths of spectral lines e.g. Narrow Line Radio Galaxy(NLRG)/Seyfert 2 and 
Broad Line Radio Galaxy (BLRG)/Seyfert 1 or 
(3) on the type of host galaxy e.g. Seyfert if disk galaxy host and radio galaxy if 
an elliptical host or (4) on the radio detectability e.g. radio-quiet
and radio-loud galaxy or (5) on the light output from the central core 
e.g. quasars, BLRG, Seyfert 1 which have bright cores and  NRLG, Seyfert 2 
which have a faint core or (6) size of the radio source e.g radio galaxy for extended
radio structures and Gigahertz Peaked Sources (GPS), Compact Steep Spectrum sources (CSS) for
small powerful radio sources.  It has been useful to have all these results 
derived from meticulous observations which have enhanced our understanding of active 
galaxies and also added to their complexity.   
Several studies have been aimed at unifying these active galaxies based on attributing
many of the observed differences to the orientation parameter. 
The schematic in Figure \ref{fig17} by \citet{1995PASP..107..803U}   
summarises the widely accepted unification model for active nuclei as
surmised from observations. 
This orientation model (see Figure \ref{fig17}) explains a host of observations indicating
that the different parts of the active nucleus have been appropriately 
located wrt to the black hole and sightline. 
In the orientation model shown in Figure \ref{fig17}, the supermassive
black hole is surrounded by an equatorial accretion disk which in turn is surrounded
by a dust torus.  The broad and narrow lines are believed to arise in clouds distributed within
biconical polar regions around the black hole with the narrow line clouds being extended to
larger separations from the black hole.  The narrow lines form in 
the narrow line region (NLR) while the broad lines form in the broad line
region (BLR).  In this model, the active nuclei observed close to
the polar axis of the black hole are classed as radio-loud blazars if radio-bright and
radio-quiet quasars if radio-faint while rest of the objects are 
explained by increasing polar angles with NLRG and Seyfert 2 being observed through the torus. 
Although this observational model has been able to explain several observables,
there remain outstanding questions like the origin and reasons for the surmised
location of the various line forming regions around the black hole, origin of
the intense photoionizing continuum, and whether orientation is the only varying
parameter between the different active nuclei.  

The continuum light output from the central parts of active galaxies especially in case of
quasars, BLRG and Seyfert 1 galaxies, is a significant fraction
of the galaxy luminosity and at times as in case of quasars, can render the light
from the extended galaxy undetectable. 
The wideband continuum spectrum of the nuclear emission of active galaxies and even
normal elliptical galaxies tend to show an
`ultraviolet/blue bump'.  In many cases, it is found that the wideband 
nuclear spectral energy distribution can be approximated by a 
combination of a power law,  a black body spectrum and a thermal spectrum.   

It was estimated that $10^3-10^6$ M$_\odot$ of warm ionized gas detected in [O II] 3727A is 
present in the central few hundred parsecs of elliptical and lenticular galaxies  and
many times this gas appeared to be arranged in the form of a rotating disk 
\citep[e.g.][]{1960ApJ...132..325O,1984PASP...96..287C,1986AJ.....91.1062P}.  
The rotating disk is found to show varying orientation with the minor axis which has 
prompted an external origin explanation for the gas \citep[e.g.][]{1984PASP...96..287C}.  
More recent observations 
of large samples of elliptical and S0 galaxies collectively referred to as early type galaxies 
(ETG) in literature find that this gas is often kinematically misaligned wrt to the stars
\citep[e.g.][]{2011MNRAS.417..882D} for which a gravitational torque as the cause
has been suggested \citep{2016arXiv160604242K}.  
About $\sim 50 \%$ of ETGs show
the presence of ionized gas in the central parts. Dust has been detected in ellipticals 
from which mass of $10^7-10^8$ M$_\odot$ have been estimated for the cold interstellar gas 
in the central parts of elliptical galaxies 
\citep[e.g.][]{1987ApJ...312L..11J}.  More recent studies estimate even larger
gas masses in centres of elliptical galaxies.  These studies have conclusively
established the existence of a significant quantity of gas 
in the centres of ETGs.  

It is interesting to note that the existence of a supermassive black hole at centres
of galaxies was not hypothesized till 1960s.  In 1950s, it was believed that
the centres of strong radio sources like Cygnus~A and Virgo~A, which were just being
optically identified, could be due to colliding nebulae
of gas or colliding galaxies \citep[e.g.][]{1954ApJ...119..215B}.
The existence of a very massive object at the centres of galaxies was
slowly being appreciated and it was speculated that the strengths and multiple redshifts of
lines observed in a quasar spectrum could be explained due to a gravitational redshift component
if the line photons arose very close to a supermassive object \citep{1964ApJ...140....1G}. 
The existence of supermassive black holes/dead quasars at centres of galaxies 
was discussed by \citet{1969Natur.223..690L} whereas
\citet{1971MNRAS.152..461L} discussed the existence of a supermassive black hole in the
centre of our Galaxy.
Backed by several observations, it is now widely accepted that there exists a
supermassive accreting black hole at the centre of active galaxies and which hosts an
accretion disk in the equatorial regions. 
Several studies have discussed the structure, accretion rates, physical conditions,
viscosity, angular momentum transfer, jet formation, corona, outflows etc in accretion
disks near black holes \citep[e.g.][]{1969Natur.223..690L,1973A&A....24..337S,
1976MNRAS.176..465B,1978AcA....28...91P,
1981ARA&A..19..137P, 1996ApJ...469..784S,1999ApJ...523..203S}. 

We start by summarising observational results on active galaxies which includes
quasars, BLRG, NLRG, FR I, FR II, GPS, CSS, Seyfert 1 and 2 and follow it
with a discussion on the physical model which emerges from the observations
and present a few case studies to demonstrate the validity of the model. 
While efforts have been made to keep the discussion concise
and succinct, the volume of observational results in literature and our
understanding of these intriguing and amazing objects is so vast, evolving and
wide-ranging as reflected in literature that it is a formidable task.  
It is hoped that the discussion is sufficiently clear and tight that the inferences 
are obvious. 

 \subsubsection{Summary of observed features}

\begin{itemize}

\item An active galactic nucleus is generally detectable in all wavebands.
The brightest nuclear emission is detected from quasars, BLRG and Seyfert 1 galaxies.   
The spectral energy distribution (SED) follows a power law in infrared and optical wavelengths,
becomes flat in blue and ultraviolet wavelengths and then starts declining around
2500 A with the decline becomes steeper at wavelengths lower than 1200 A so that
at X-ray bands, the same power law appears to be applicable
\citep[e.g.][]{1982ApJ...254...22M}.  The spectra of most nuclei show an excess in the
blue or ultraviolet emission which is referred to as the blue bump in literature.  Several
normal elliptical galaxies also show this blue bump 
%{\bf reference} 
in their SED. 

\item The multi-band continuum emission from the active nucleus is often variable. 
Variability in the X-ray, ultraviolet and optical bands is often found to be either
simultaneous or the lower frequency emission follows the higher frequency emission 
(e.g. NGC 5548; Clavel et al. 1992) and
this has been well studied in Seyfert galaxies.  However occasionally a flare confined
to the ultraviolet band is also recorded from these galaxies 
(e.g. NGC 5548;  Clavel et al. 1992).  Similar phenomena are noted in NGC 4151
(e.g. Perola et al. 1986).  

\item Optical/ultraviolet emission lines of a wide range of excitation are 
generally detected from the active nucleus suggesting a non-stellar hard exciting source. 
Narrow (FWHM of few hundred kms$^{-1}$) emission lines are detectable from almost all 
active nuclei.  These consist of low and high ionization lines.  Many narrow lines 
are asymmetric with a wider wing on the blue side compared to the red side.  Broad 
(FWZM of thousands of kms$^{-1}$) permitted emission lines are detected from quasars, BLRG 
and Seyfert 1 galaxies.  No broad forbidden lines are detected.  Broad lines are
often variable in both amplitude and width which is correlated with the variability 
in the thermal continuum emission
whereas no variability is detected in the narrow lines \citep[e.g.][]{1989agna.book.....O}.
This property has been used in reverberation studies of Seyfert 1 galaxies wherein the delay
in correlated variability in the continuum and broad lines are used to
determine the size of the BLR and the separation of the BLR from the continuum
emitting region.
%{\bf reference}.  
This method has resulted in deriving a size of a few light days for the BLR
in Seyfert 1 galaxies \citep[e.g.][]{1986ApJ...305..175G}.

\item The radio structure associated with the active nucleus in Seyferts is 
fainter and more compact than radio galaxies and is generally confined within
the optical galaxy. 

\item Extended Lyman$\alpha$ emission coincident with the extended radio structure 
has been detected towards several radio galaxies 
\citep[e.g. 3C~294][]{1990ApJ...365..487M}.
Also the velocities recorded for the Lyman$\alpha$ on either side of the core
are observed to be blueshifted and redshifted - for example in 3C~294, the Lyman$\alpha$
emission coincident with the northern lobe is blueshifted by about 900
kms$^{-1}$ whereas the emission coincident with the southern lobe is redshifted
by about 500 kms$^{-1}$ \citep{1990ApJ...365..487M}.
The Lyman$\alpha$ emission often shows asymmetric distribution on the two
sides of the core. {\it These observations wherein the radio emission and line emission
are of similar extents and the latter show expansion on either side of the core
support the ejection of both the ionized gas and radio plasma from the 
vicinity of the black hole.}

\item An approximate linear correlation exists between the radio power at 1.4 GHz and 
the luminosity of the [O III]5007A line for Seyfert galaxies \citep{1978A&A....64..433D}.
A correlation between the radio power and optical continuum luminosity
is also observed for Seyfert galaxies \citep[e.g.][]{1987ApJ...313..651E}.
\citet{1980ApJ...240..429W} noted that radio emission showed remarkable similarity with
the distribution of the optical line emitting gas in a few Seyfert galaxies and suggested that
these galaxies also hosted a double radio source.

\item In the Seyfert 2 galaxy NGC 5929, double-peaked [O III] profiles
are detected in addition to a double-lobed radio structure both separated by
200-300 pc from the nucleus \citep{1986MNRAS.222..189W}.
Correlation between discrete components in [O III] and radio emission are 
detectable with both diagnostics extended along similar position angles
\citep{1986MNRAS.222..189W}. 
%{\bf are the two peaks red and blue shifted wrt galaxy velocity?}
Such double-peaked optical emission lines are often detected from Seyfert galaxies 
with the peaks displaced by a few hundred kms$^{-1}$ on either side of the rest frequency
i.e. consisting of blue and redshifted components. 
%{\bf reference}

\item The typical widths of high excitation forbidden lines like [O III] from
the NLR close to the nucleus is around 300-500 kms$^{-1}$.  Such high excitation
lines are also detected at distances of several kpc from the nucleus but
are fainter, significantly narrower at
$<45$ kms$^{-1}$ and the velocity gradient within the region is small.
Such a region has been referred to as the
extended NLR i.e. ENLR \citep{1987MNRAS.228..671U}. 

\item The range of excitation observed in the emission lines from an active nucleus
is much larger than that from a HII region or a planetary nebula and which appears to have
a photoionization origin \citep[e.g.][]{1989agna.book.....O}. 

\item Soft X-ray emission especially upto 2 keV is seen to show good correlation 
with the narrow line region traced by forbidden lines of oxygen [O III]5007A
and H$\alpha$+[N II]6548,6583 in terms
of morphology and extension with the extension being along the radio jet axis
in the Seyfert galaxy NGC 1068 \citep{2001ApJ...556....6Y}. 

\item Radio continuum emission is detected from several components of active galaxies -
compact core, collimated jets, extended lobes and compact hotspots 
and the emission is due to the synchrotron process. 

\item Collimated narrow single-sided jets and symmetric two-sided lobes and 
hotspots are commonly detected in radio 
images of powerful quasars and FR II galaxies \citep{1974MNRAS.167P..31F}.  The
central core, symmetric and often bent jets and lobes are commonly detected from 
FR I galaxies.   

\item Narrow single-sided X-ray jets are detected close to the core and
the emission process has been found to be synchrotron. 
%{\bf reference}
However  
X-ray beams which are extended from the nucleus till the hotspots but laterally more diffuse 
then radio jets 
have been detected in a few FR II radio galaxies like Cygnus~A and Pictor~A.  The emission
process for these X-rays remains unknown. 

\item The radio power of the active galaxies in the 3CR catalogue of strong radio sources 
is a strong function of redshift such that it increases with redshift. 
%{\bf reference?}

\item Radio jet collimation has been found to be inversely correlated with luminosity such
that most radio luminous (FR II) sources host highly collimated jets 
(jet opening angles $\le 3^{\circ}$) whereas
lower luminosity (FR I) sources show larger jet opening angles ($\sim 5^{\circ}-15^{\circ}$). 
The bulk velocity of the jet in FR II is measured to be higher than in FR I sources.  

\item 
Some jets show a limb-brightened morphology starting from close to the black hole as seen for
M 87 in Figure \ref{fig18}.
The jet in M87 was the first jet to be discovered (Curtis 1918) 
%{\bf reference} 
and is detected across the electromagnetic spectrum - radio to TeV $\gamma-$rays 
%{\bf reference}.  
VLBI images at 1.3 mm have spatially resolved
the base of the jet in M 87 indicating that it has been launched from
a region separated from the black hole by $\le 5.5 R_s$  \citep{2012Sci...338..355D}.   
\begin{figure}
\centering
\includegraphics[width=7cm]{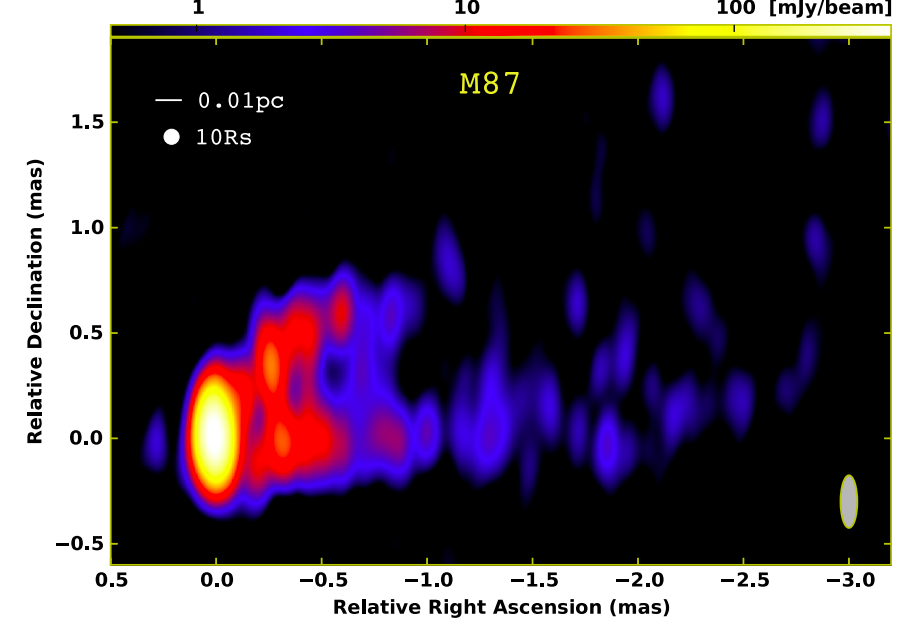}
\caption{\small 
The central region of M 87 at 86 GHz
and high resolution. Figure copied from \citet{2016ApJ...817..131H}.  
Notice the limb-brightening in the jet 
and the presence of a faint counter-jet close to the core.  }
\label{fig18}
\end{figure}

\begin{figure}
\centering
\includegraphics[width=6cm]{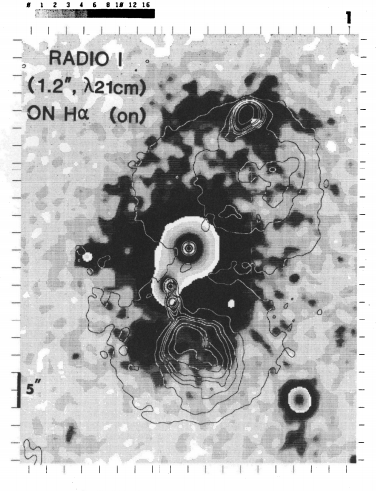}
\caption{\small Figure reproduced from \citet{1985ApJ...290..496V},  
showing the radio continuum emission in contours and optical H$\alpha$
emission in grey scale for the central galaxy in Coma cluster - 3C 277.3. 
Notice the co-extensive nature of the two diagnostics.  }
\label{fig19}
\end{figure}

\item Optical emission extended along the radio jets was detected beyond the host 
galaxy in Centaurus A, Fornax A and NGC 7720 \citep{1964ApJ...140...35M}.

\item The narrow line emission in many radio galaxies has been mapped and found to extend over
a region comparable to the radio structure as shown in Figure \ref{fig19}
for the radio galaxy 3C 277.3 located at the centre of the Coma cluster
\citep{1981ApJ...247L...5M, 1985ApJ...290..496V}.
The optical line brightness was observed to be correlated with
radio continuum intensity in 3C 277.3 and the optical lines were
red-shifted in the north and blue-shifted in the south of the
nuclear source indicating expansion  \citep{1981ApJ...247L...5M}.  From such studies, 
an interaction between the radio plasma and the optical emission line gas
has been inferred.
Additionally, it has also been inferred that  
a common energy source is responsible for the optical lines and radio continuum 
\citep{1989ApJ...336..702B}.
A tighter correlation between radio jets and narrow emission lines is found
in steep spectrum extended radio galaxies \citep{1989ApJ...336..702B}.
Extended emission line gas has more commonly been
observed in steep spectrum radio-loud quasars as compared to flat spectrum radio loud or
radio-quiet quasars \citep{1985ApJ...293..120B,1987ApJ...316..584S}.
The median size of the optical line emitting region in radio galaxies is found to 
be about 10 kpc
\citep{1989ApJ...336..681B}.  The extent of the emission line region is found to be
smaller for radio-quiet active galaxies. 
%{\bf reference}

\begin{figure}
\centering
\includegraphics[width=6cm]{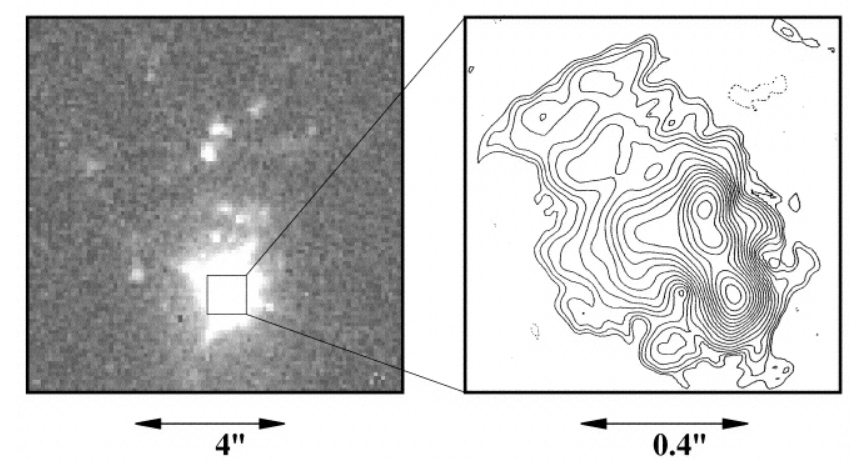}(a)
\includegraphics[width=6cm]{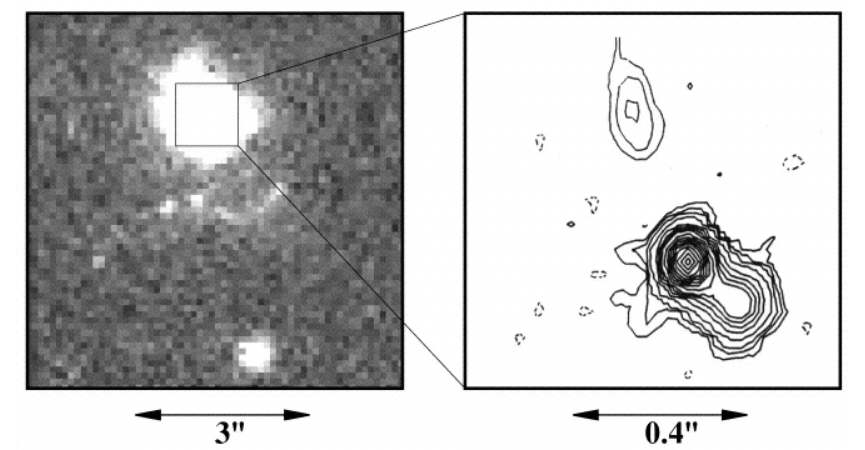}(b)
\caption{\small Figures showing the distribution of redshifted narrow band 
forbidden line emission (left) and
radio continuum emission (right) of compact steep spectrum sources, reproduced from 
\citet{2000AJ....120.2284A}.  (a) 3C 48 
(b) 3C 147.  Notice that [O III] emission is extended along the radio axis and
is more extended than the radio continuum emission for both the sources.  }
\label{fig20}
\end{figure}

\begin{figure}
\centering
\includegraphics[width=7cm]{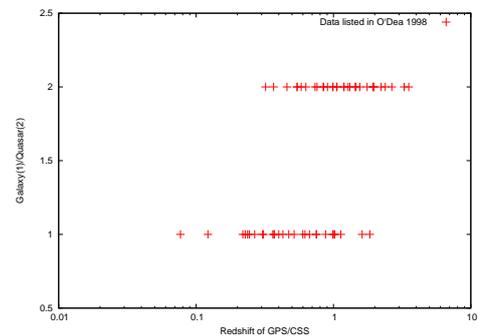}
\caption{\small Figure showing the redshift distribution of GPS, CSS galaxies and quasars.
1 refers to a source classified as a galaxy and 2 to a quasar classification. }
\label{fig21}
\end{figure}

\item From studies
of a large sample of radio galaxies, it is found that extended optical emission line gas 
on scales of 40-100 kpc is commonly associated with powerful radio 
galaxies \citep{1988ApJS...68..643B}.
No correlation is observed between
broad line emission and radio power \citep{1989ApJ...336..702B}.

\item Optical line emitting gas is detected 
in about 90\% of the 53 observed 3CR radio 
galaxies with $0.05 < z < 2$ and is distributed
over scales ranging from sub-parsec to about 300 kpc \citep{1995ApJS...99...27M}. 
The emission line regions are found to be aligned with the radio jets/lobes
for sources located at $z>0.3$ while for sources at $z>0.6$  the optical/ultraviolet 
continuum emission also shows an alignment with the radio axis 
\citep{1995ApJS...99...27M}. 

\item Emission lines are detected near the radio jets especially 
at the outer edges of bends in radio jets and near knots and hotspots 
with the spectrum resembling the higher excitation spectrum of Seyfert 2 nuclei
and not HII regions \citep[e.g.][]{1984ARA&A..22..319B}. Typical electron densities estimated
from the optical emission lines are $\sim 10^2-10^3$ cm$^{-3}$
for temperatures of $\sim 20000$ K \citep[e.g.][]{1984ARA&A..22..319B}.  The typical bulk
velocities are estimated to be a few hundred kms$^{-1}$ with detected line widths
typically being less than 1000 kms$^{-1}$.

\item The observed forbidden line ratios ([O II]/[O III]) detected from the
nuclear region and the extended region along
the radio jet in radio galaxies appear to be well explained by the same ionizing 
source.  The ionizing source is inferred to be a black body of temperature 
around $1.3\times10^5$ K  
i.e. an ionizing radiation field consisting of mean photon 
energies of 30-40 eV which rules out a stellar continuum origin \citep{1987MNRAS.227...97R}. 

\item Optical line emitting gas is also detected along the radio axis in GPS
and CSS sources (see Figure \ref{fig20}).
The radio emission in GPS and CSS sources shows relatively lower linear polarisation
and higher rotation measures 
%{\bf reference}.  
The optical brightness of the central
source in GPS and CSS sources is dominated by emission lines
\citep{2000AJ....120.2284A}.  Broad ($\sim 2000$ kms$^{-1}$) and flat-topped [O III] profiles
are observed from CSS nuclei \citep{1994ApJS...91..491G}.
A correlation between the narrow line luminosity and the total
radio power is also noted in GPS and CSS sources
\citep[e.g.][]{1994ApJS...91..491G,1995PASP..107..205G}.

\item GPS and CSS constitute about 10\% and 30\% respectively of the radio source
population and their host galaxies resemble the elliptical hosts of radio galaxies
\citep[e.g.][]{1998PASP..110..493O}. 

\item The emission line redshifts of GPS and CSS are $\ge 0.1$ as shown in Figure \ref{fig21}.

\item Jets are observed to be one-sided close to the core in both FR I and FR II galaxies. 
Double-sided jets are detected on kpc scales in FR I galaxies but continue to
be single-sided along their entire length in FR II sources so that the jet-counterjet
brightness ratios are large \citep[e.g.][]{1994AJ....108..766B,1995PNAS...9211413L}. 
The non-detection of a counter-jet is attributed to Doppler beaming owing to relativistic 
bulk jet speeds indicating that such speeds continue till the hotspot in FR II sources.  
The jet-counterjet asymmetry can be explained by relativistic jet velocities
$\beta$ between 0.65 and 0.8 for jet viewing angle between $45^\circ$ and $75^\circ$
\citep{1997MNRAS.286..425W}.  These studies do not support significant intrinsic asymmetry
in the jets.  

\item Many jets show the same opening angle at parsec and kiloparsec scales (e.g. NGC 6251
in Figure \ref{fig22}) which would be indicative of a free jet \citep{1980ARA&A..18..165M}. 
However larger opening angles are noted in milliarcsec jets imaged with VLBA/VLBI
e.g. Virgo~A, \citep{1999Natur.401..891J} and Cygnus~A \citep{2016A&A...585A..33B}
as compared to the opening angles measured at parsec/kiloparsec scales. 

\item FR I radio galaxies and radio-quiet active galaxies with similar host optical magnitude 
show similar narrow emission line luminosities \citep{1995ApJ...451...88B}. 
It is also found that the host galaxy and nuclear properties of the class of 
radio sources which are known as 
FR 0 in literature are comparable to FR I except that extended radio structure is
only associated with FR I \citep{2019MNRAS.482.2294B}.
%Observational results
%support the radio emission from FR 0 due to mildly relativistic jets while disfavouring
%a young source argument or varying jet properties \citep{2019MNRAS.482.2294B}. 
The ultraviolet continuum from the active nucleus in FR I sources is fainter than in
FR II sources. 
{\it shift this to explanation: These observational results would argue for a decrease in both
bulk velocity and 
the random relativistic velocities of the electrons going from FR I to FR 0 to radio-quiet
active galaxies  which in turn could argue for an increasing separation of the launching
region from the black hole and hence decreasing escape velocity i.e. lower energy
thermonuclear bursts which lead to lower energy ejections. }
 
\item Optical and X-ray synchrotron emission from several jets close to the core have been observed.
Additionally optical emission with polarisation properties similar to the radio
synchrotron emission has been found to be coincident with radio hotspots and 
radio lobes in a few sources e.g. 3C 285 \citep{1977A&A....59L..15T}, 3C 33 
\citep{1978ApJ...222L..55S,1986Natur.319..459M}. 

\item Excess soft X-ray emission (0.1 - 2 keV) over a power law is detected in most 
Seyfert 1 galaxies
which is also found to be variable \citep{1989MNRAS.240..833T}.  This excess 
is  well-fit by a multi-temperature black body with temperatures ranging from 40 to 150 eV
which indicates that the emitting region is closer than $3 R_s$ 
\citep{1999ApL&C..39..125L,1997A&AS..126..525P}.    These galaxies also show 
a ultraviolet excess.  
The hard X-ray emission (2-10 keV) is well described
by a power law so that the entire X-ray spectra for many Seyfert 1 is well described by the
combination of a power law and a black body spectrum.

\item In some FR I sources, the X-ray jet is bright close to the core and which
coincides with a faint radio jet whereas further from the core, the X-ray jet fades
and the radio jet brightens \citep[e.g. 3C 31][]{2008MNRAS.386..657L}.  

\item An anti-correlation is observed between the hot spot prominence and jet bending 
such that when the hot spot is more prominent, the jet is straight and vice versa 
\citep{1994AJ....108..766B}.   

\item Radio polarisation studies find that the magnetic field lines are along the jet 
in luminous sources such as FR II galaxies.  
In fainter sources like FR I galaxies, the magnetic field lines are along the jet
close to the central active nucleus but perpendicular to the jet at larger distances from the 
nucleus (see Figure \ref{fig22}).  It is observed that the fraction of the jet length over which
the magnetic field is parallel to the jet is determined by the radio core luminosity such
that a stronger radio core shows a parallel field over a longer jet length than a fainter
radio core \citep{1982IAUS...97..121B}.  Moreover jet-counterjet brightness asymmetry is more
pronounced in the radio sources with a powerful core i.e. ones which show a parallel
magnetic field configuration \citep{1982IAUS...97..121B}.  

\begin{figure}
\centering
\includegraphics[width=8cm]{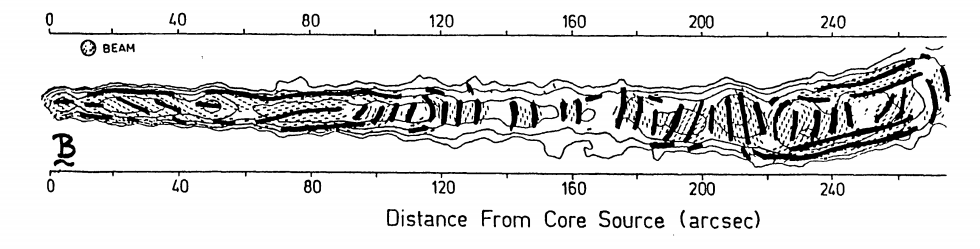}
\caption{\small Figure showing the typical magnetic field orientation in a FR~I galaxy.
The figure showing NGC~6251 is reproduced from \citet{1982IAUS...97..121B}.  
Note how the magnetic field is along the jet
in the inner parts and flips to being transverse to the jet at larger distances 
when the jet starts flaring up. }
\label{fig22}
\end{figure}

\item Quasars, BLRG and Seyfert 1 galaxies show the presence of wide emission lines
in their nuclear spectra. 
The ratio of the strengths of nuclear broad H$\alpha$ to H$\beta$ is around 3.5 
in Seyfert 1 whereas it is around 6 in BLRG  \citep{1989agna.book.....O}.  It is also 
observed that the emission lines of Fe II are absent or significantly fainter in the
nuclear spectrum of BLRG than in Seyfert 1 galaxies \citep{1989agna.book.....O}. 
It is also observed that lobe-dominated radio sources show weaker [Fe II] emission in their 
core spectrum compared to core-dominated sources \citep{1984ARA&A..22..319B}. 

\item The spectra of some Seyfert 2 galaxies show the presence 
of broad lines in polarised light \citep[e.g. NGC 1068;][]{1985ApJ...297..621A}
while the spectra of narrow lined Seyfert 2 nuclei are observed to transform
to an intermediate 1.5 type or
type 1 spectrum and then revert back to type 2 over time.
The galaxy NGC 1566 has been observed to transit from a Seyfert 2 to a
Seyfert 1 type i.e. its core brightens and broad permitted lines of hydrogen
appear with no change in the narrow lines over a period of about 1300 days
\citep{1985ApJ...288..205A,1986ApJ...308...23A}.
Energy of the order of $\sim 10^{51}$ ergs has been estimated to be released
but an origin in supernova explosions have been ruled out \citep{1986ApJ...308...23A}.
Radio emission and narrow line [O III] emission have been detected from NGC 1566. 

\item The broad lines of higher ionization are observed to be wider and more 
variable than the broad lines of low ionization in Seyfert 1 and quasars.
This has been understood in terms of stratification in the BLR with
the high ionization region lying closer to the ionizing source and hence directly
subject to variations in the ionizing source 
\citep[e.g.][]{2006pces.conf...89P}. 

\item It was inferred that the radial extent of the
BLR in the Seyfert galaxy Arp 151 increased by a factor of two between 2008 and 2011
when the continuum varied \citep{2018ApJ...856..108P}.  It was inferred that  
the inner boundary of the BLR remained the same as measured from reverberation mapping
in data taken over seven years \citep{2018ApJ...856..108P}.

\item Quasar spectra show broad emission lines at the highest detected redshift
and absorption features at several lower redshifts.  The presence of multiple
redshifts in the quasar spectra has been explained by a varying 
gravitational redshift component in radially distributed line forming
zones in the BLR \citep{2016arXiv160901593K}.  Similar behaviour 
but of much smaller redshift magnitude has also been observed in other active nuclei.

\item Broad emission lines like C IV from the BLR are often observed
to be asymmetric with a steeper profile on the red side compared
to the blue side.  The velocity of the high ionization lines like 
C IV in quasars is systematically blue-shifted by $\sim 600$ kms$^{-1}$ 
wrt to the lower ionization lines like Mg II \citep{1982ApJ...263...79G}.
{\it These two observations might be connected
if the strong gravitational potential in which the C IV line is formed is responsible 
for the line asymmetry which
in turn leads to the measured centroid being biased towards the blue.} 

\item The radio and $\gamma-$ray emission especially between 0.1 and 100 GeV
show a tight correlation in active nuclei especially blazars.  
The correlation which is observed to be strongest for BL Lac objects declines at 
energies $>100 $ GeV \citep{2017A&A...606A.138L}.
Recently a high-energy neutrino IceCube-170922A with energy surmised to be
about 290 TeV has been detected from a flaring blazar TXS 0506+056 \citep{2018arXiv180708816I}. 
While neutrinos of lower energy have been detected from the sun and 
the supernova SN 1987A, this is the first detection of neutrinos 
associated with an active nucleus which is in the active
phase and also emitting $\gamma-$rays.  

\item Quasars could be divided into two groups based on the detection of Fe II emission
lines: (1) when Fe II emission lines are not detected, 
narrow lines are stronger, [O III]/H$\beta$ ratio is large, hydrogen lines are broad and
complex and the radio emission is extended with a double-lobed structure and a steep spectrum.
(2) When Fe II emission is strong, narrow lines are weak, the hydrogen
lines are narrower and smoother with relatively lower ratios of [O III]/H$\beta$ 
and the radio structure is
compact with a flat spectrum \citep{1979ApJ...228L..55M,1984ApJ...281..535B}. 

\item The optical continuum of quasars (also Seyfert 1 and `N' 
type radio galaxies) was observed to be either variable (change by $\le 0.5$ magnitude 
within a year or so) or violently variable (change by a magnitude 
or so even within a night), with the latter being referred to as optically
violently variable (OVV) quasars \citep[e.g.][]{1969CoKon..65..485P,1971MNRAS.152...79C}.  
The OVV quasars are also characterised by flat radio spectra and radio variability at
high frequencies.  The variability of these sources is not periodic. 
In more recent terminology, the OVV quasars are referred to as flat spectrum radio
quasars (FSRQ). 

\item Quasar host galaxies range from disk galaxies to elliptical galaxies
with typical sizes between 2 kpc and 15 kpc and absolute
R band magnitudes within $-21$ and $-24$ magnitudes \citep{2013arXiv1302.1366K}. 
Majority of quasars are radio-quiet.

\item The spectra of several active compact sources like blazars show a smooth log 
parabolic/convex shape from 1.4 GHz to about 1400 A with peaks
occurring between 10 and 1000 GHz \citep[e.g.][]{1986ApJ...308...78L}
(also see Figure \ref{fig2}). 

\item Multi-phase interstellar mediums is detected in almost all elliptical and 
lenticular galaxies and dust has also been detected 
\citep{1994A&AS..105..341G}.

\item The infrared emission in Seyfert 1 galaxies is observed to be confined
to a core source ($<0.1$ pc) whereas the radio emission is detected from a
core source and an extended source of several hundred pcs to a few kpcs 
\citep{1978A&A....64..433D}.  The typical radio power of Seyfert galaxies at 21 cm is
$10^{21} - 10^{22}$ W Hz$^{-1}$ \citep{1978A&A....64..433D}.  
A small fraction of Seyferts host a flat spectrum compact radio source
while some Seyferts show the presence of a radio halo of several tens of kpc
\citep{1978A&A....64..433D}.

\begin{figure}
\centering
\includegraphics[width=6cm]{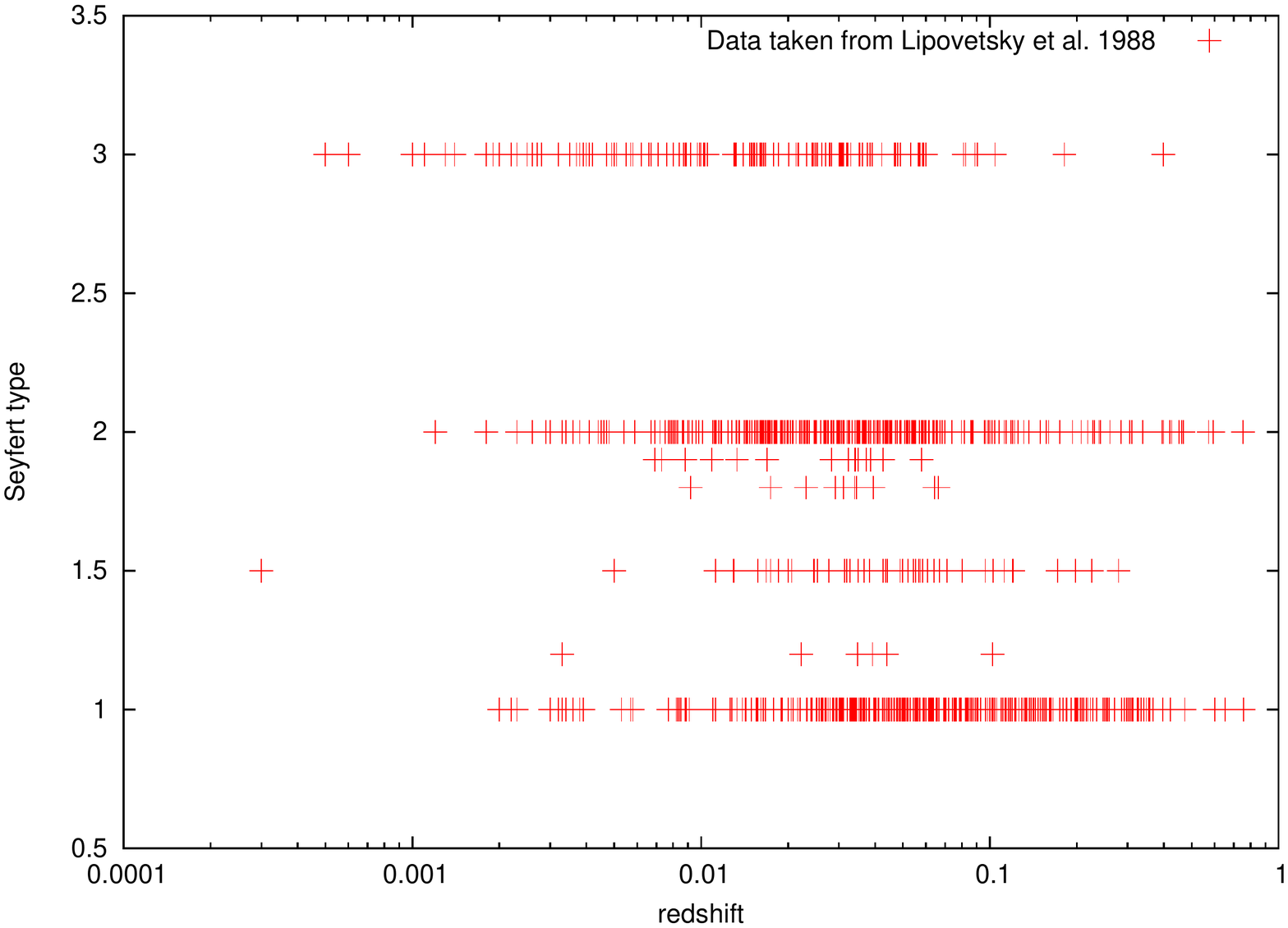} a
\includegraphics[width=6cm]{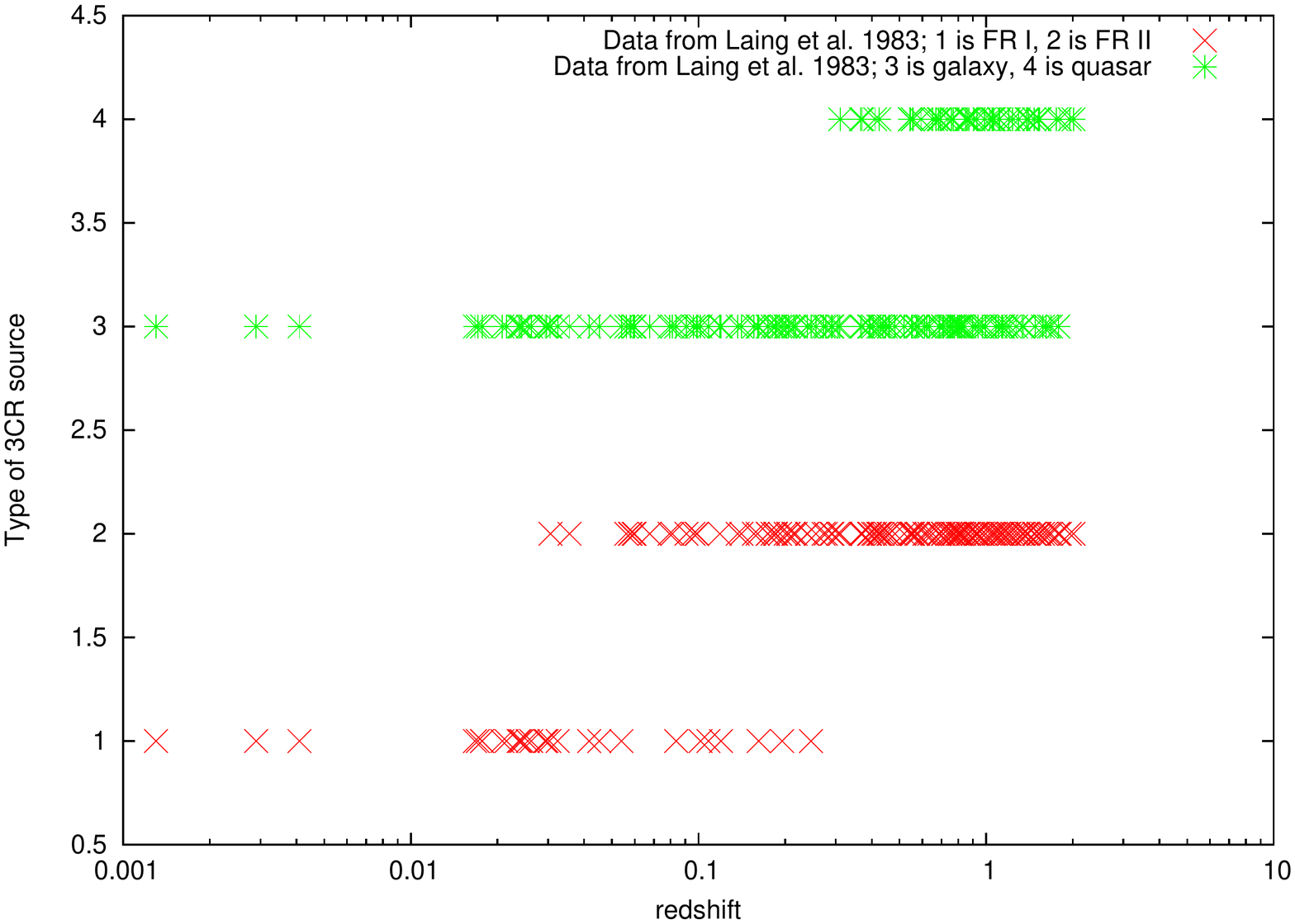} b
\includegraphics[width=6cm]{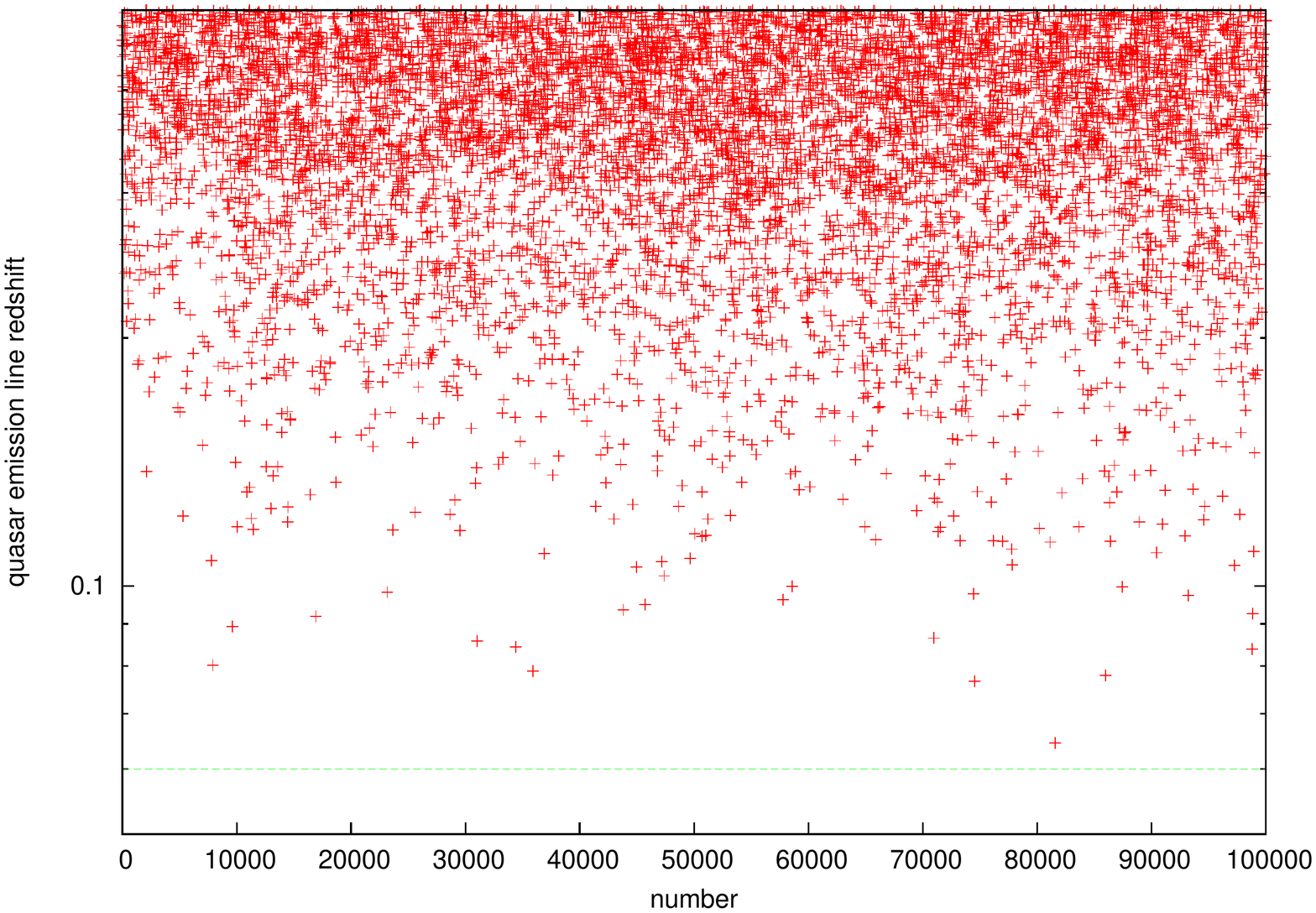} c
\caption{\small (a)
Redshift distribution of Seyfert galaxies from the catalogue compiled
by \citet{1988SoSAO..55....5L}.  The number on the y-axis between 1 and 2 denotes the type of
the active nucleus.  Ignore 3.  
(b) The redshift distribution of more than 100 3CR radio sources classified into
FR~I (1) and FR~II (2) (red crosses) and classified into galaxy (3) and quasars (4) 
(green stars) as listed in \citet{1983MNRAS.204..151L}.  FR~II galaxies are
not detected at low redshifts unlike FR~I galaxies.  Also quasars are detected
at much higher redshifts than galaxies.  
Such a gradation is indicative of the increasing influence of a gravitational redshift
component which is maximum for quasars. 
(c) Distribution of the smallest redshifts ($z \le 0.5$)  detected for the 
quasars in the seventh data release of the SDSS quasar catalogue 
\citep{2010AJ....139.2360S}.}
\label{fig23}
\end{figure}

\item The sequence of active nuclei in terms of increasing 
emission line redshifts is Seyfert 2, Seyfert 1, FR I/NLRG, FR II/BLRG
galaxies, GPS/CSS, quasars.  For this sequence, 
there is a rise in the minimum redshift that an object in the specific class
is detected such that it is lowest for Seyferts and highest for quasars 
as shown in Figure \ref{fig23}.

\item H$\alpha$ and [N II] emission lines of varying relative
intensities are detected from the central regions of
all spiral galaxies which has helped classify them into LINERs 
\citep{1980A&A....87..152H} and HII nuclei.  It is found that
80\% Sa, Sb galaxies are LINERs while
20\% of Sc galaxies are LINERs \citep{1989agna.book.....O}.
LINER gas is also detected in a large fraction of early type
galaxies \citep[e.g.][]{2010A&A...519A..40A,2011MNRAS.417..882D}.

\item The angle between the radio jet axis and optical polarisation angle is found to be
different for Seyfert 1 and Seyfert 2 galaxies such that the two are predominantly 
parallel in Seyfert 1 galaxies and predominantly perpendicular in Seyfert 2 galaxies
\citep{1983Natur.303..158A}.   

\item Similar polarisation fraction and angle are detected for permitted 
emission lines and continuum emission from the nucleus of the 
Seyfert 2 galaxy NGC 1068  
whereas the forbidden lines are 
weakly polarised with a different position angle \citep{1976ApJ...206L...5A}. 
It is inferred that the continuum and allowed emission lines
formed in the nuclear region and hence were similarly scattered by dust in a cloud
of size  $\sim 1''$ leading to similar polarisation properties whereas 
the forbidden lines formed outside the nuclear region and hence were not scattered 
by the dust \citep{1976ApJ...206L...5A}.  

\item In NGC 1068, broad permitted lines (FWZI $\sim 7500$ kms$^{-1}$) 
were detected in polarised light \citep{1985ApJ...297..621A}.
The broad lines were redshifted by $\sim 600$ kms$^{-1}$ wrt 
the narrow lines \citep{1985ApJ...297..621A}.
Such behaviour is not detected in all Seyfert 2 type galaxies.

\item A few inferences from HI absorption studies of radio-loud nuclei  
\citet{2018arXiv180701475M}: 
(1) HI absorption has been detected from several active nuclei 
(2) HI absorption is detected from the radio core and/or jets and/or lobes.
(3) Blueshifted HI absorption features are more widely detected than redshifted 
features.  Blueshifted features are interpreted as outflows whereas redshifted
features are interepreted as gas flows from the accretion disk into the nucleus.  
%mass outflow rates estimated from HI are larger than estimated for the ionized gas. 
(4) The HI detected in the central regions of active nuclei is deduced to be in form
of a circumnuclear disk.  It is found to form a thick disk of small radial
extent in high power sources and a thin, extended disk in low power sources. 
(5) The circumnuclear HI disk merges with the galactic disks in Seyfert
galaxies. 

\item Some of the main inferences from reverberation mapping 
\citep{1972ApJ...171..467B, 1982ApJ...255..419B} of the BLR in Seyfert 1 galaxies as 
summarised by \citet{2006pces.conf...89P} 
are that (1) The variability in the continuum emission in the ultraviolet to optical
bands are in phase to within a day or so.  (2) Variability in the continuum first induces
variability in the highest ionization lines - He II 1640A, C IV 1549A, Ly$\alpha$ and
then in lower ionization lines - H$\beta$, C III] 1909A lines.  
The lags increase with the luminosity of the nuclear source in Seyfert 1.

\item The NLR is distributed in a conical
region with the black hole at the apex as seen in the [O III] image of 
Circinus in Figure \ref{fig24} copied
from \citet{1994Msngr..78...20M}.  The conical region is often referred to as the
ionization cone.
The radio jets and lobes are also extended along the same axis.
This Seyfert 2 galaxy is gas-rich and hosts a large warped HI disk and a starburst.
The radio jets are spread over a smaller conical angle than the NLR. 

\begin{figure}
\centering
\includegraphics[width=7cm]{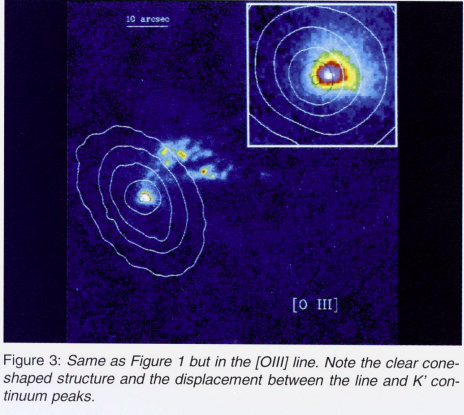}
\caption{\small [O III] image of the Seyfert 2 galaxy Circinus copied from
\citet{1994Msngr..78...20M}.  The conical clumpy outflow in [O III] 
which defines the NLR of the active nucleus is clearly seen.  The inset shows a zoom-in into
the central regions of the image where the cone is even better defined.  
The contours indicate the optical continuum morphology of the galaxy. 
Radio jets and lobes are detected along the [O III] cone and extend out to
several kpc from the nucleus.   }
\label{fig24}
\end{figure}

\item Broad Balmer lines have been detected in the polarised spectra of a few NLRGs
and are also detected in total flux when deep spectra are obtained 
\citep[e.g.][]{1997ApJ...482L..37O,1999AJ....118.1963C}. These observations
indicate that some NLRG also host a BLR.  When resolved, the broad lines
in NLRG appear to be arising in a polar biconical region 
around the core like in Figure \ref{fig24}.  
The biconical region has been inferred to be a reflection nebula around the
core which scatters the broad lines from the BLR into the sightline 
\citep[e.g.][]{1999AJ....118.1963C}.  

\item The highest emission line redshift at which Seyfert 1 galaxies are detected is
larger than Seyfert 2 galaxies \citep{1974ApJ...192..581K}. The highest
emission line redshift at which FR II radio 
galaxies are detected is higher than FR I radio sources. 

\item The spectrum of the NLR consist of 
emission lines ranging from neutral atoms to multiply-ionized
ions.  There exists a correlation between the lines from neutral 
and singly-ionized ions indicating their origin in a common zone and between  
the different high ionization lines which indicate their origin in a common
zone - this then suggests the existence of at least 
two separate ionization zones in the NLR \citep{1978ApJ...223...56K}. 
Similar line widths are recorded for the narrow emission lines from 
Seyfert 2 and NLRG.  Moreover it has been noted that 
the velocities of the low and high ionization narrow lines from Seyfert 2 nuclei
differ by a few hundred kms$^{-1}$ \citep{1978ApJ...223...56K}.  

\item Continuum emission of Seyfert 2 galaxies 
contain a significant contribution from stellar light in addition
to the active nucleus \citep{1978ApJ...223...56K}. 

\item Strong multiple permitted Fe II lines of widths $\le$ Balmer lines 
of hydrogen are frequently detected in Seyfert 1 galaxies but
rarely in Seyfert 2 galaxies \citep{1977ApJ...215..733O,1978ApJ...223...56K}.  
The forbidden [Fe II] lines are generally weak \citep{1978ApJS...38..187P}.  
The intensity ratio [O III]/H$\beta$ is generally 
smaller in Seyfert 1 compared to Seyfert 2 galaxies 
\citep{1974ApJ...192..581K}.  Forbidden lines of widths 300-800 kms$^{-1}$ 
are detected in both Seyfert 1 and 2 galaxies  \citep{1977ApJ...215..733O} 
but appear  to be stronger in Seyfert 2 \citep{1978ApJ...223...56K}.
Iron lines are generally strong
in radio-quiet quasars but absent or weak in radio galaxies.  

\item Using statistical arguments, it was shown that the origin of the large line widths in 
Seyfert 1 galaxies was not rotation \citep{1977ApJ...215..733O}.  

\item In a sample of 55 Seyfert galaxies (19 Sy 1 and 36 Sy 2) 
observed in the radio continuum at 3.6 cm only one was not detected 
\citep{2001ApJS..132..199S}. 
Radio emission from the active core is commonly detected from Seyfert galaxies
and in a few Seyferts, the radio emission is extended with double-sided conical jets.  
It is noticed that the position angle of the radio axis of the Seyfert
is seldom perpendicular to the galaxy major axis \citep{2000ApJ...537..152K}
{\it This could indicate large conical angles of ejection due to a low spin
the black hole. }

\item The nuclear radio power in Seyferts ranges from $10^{38}$ to $10^{40}$ ergs$^{-1}$,
in normal spirals with strong nuclear emission from $10^{36}$ to 
$10^{38}$ ergs$^{-1}$ and in other spirals is  $< 10^{36}$ ergs$^{-1}$
\citep{1973A&A....29..263V}.  The nuclear/core radio power in several early
type galaxies including LINERs is in the range $10^{36}$ to
$10^{38}$ ergs$^{-1}$, for FR 0 sources ranges from 
$10^{38}$ to $10^{40}$ ergs$^{-1}$ and for the core
of FR I sources ranges from $10^{38}$ to $10^{42}$ ergs$^{-1}$.
%{\bf reference?}
The X-ray luminosity of active nuclei are generally
estimated to range from $10^{42}$ to $10^{46}$ erg~s$^{-1}$.
The bolometric luminosites of active nuclei are estimated to be in the
range $10^{44}$ to $10^{48}$ erg~s$^{-1}$ \citep[e.g.][]{2002ApJ...579..530W}.

\item The properties of some Seyfert nuclei have been observed to switch 
between type 1 and type 2 in terms of the high nuclear continuum luminosity and
presence of broad lines in the spectrum transition to a fainter nuclear continuum
luminosity and disappearance of the broad lines  e.g. NGC~1566, Mrk~590.  

\item The jet opening angle in blazars is estimated to be $1^{\circ}-2^\circ$
\citep{2009A&A...507L..33P}.  The jet opening angle in Cygnus~A is measured to 
be $1.6^{\circ}$ far from the core 
and about $10^\circ$ close to the core \citep{2016A&A...585A..33B}.
The jet opening angles are modified by the angle made by the jet with the
sightline. 

\item  In the FR II BLRG, Pictor~A, a single radio jet is detected whereas
double-sided X-ray beams which are thicker than the radio jet extend from the
core to the hotspots \citep{2016MNRAS.455.3526H} (see Figure \ref{fig34}).  
The X-ray beam in the 
east terminates in a X-ray hotspot which is distinct from the bright radio hotspot.
The opening angle of the jet is $\sim 3^{\circ}$ \citep{2016MNRAS.455.3526H}.
The radio spectrum of Pictor~A turns over around 100 MHz \citep{1997A&A...328...12P}.
X-ray beams are also detected in the FR II galaxy Cygnus~A.

\item A large fraction of the extragalactic $\gamma-$ray sources detected by
the Fermi LAT are blazars with only $\sim 3\%$ not being associated with blazars
but with nearby FR I galaxies \citep{2012arXiv1205.1686G}. 

\item The $\gamma-$ray flares detected in radio galaxies by Fermi LAT
have generally been associated with the detection of a new radio component
e.g. FR~II BLRG 3C~111 \citep{2012ApJ...751L...3G} and 
and FR~I BLRG 3C~120 \citep{2015ApJ...808..162C}.  In both the sources,  
the $\gamma-$ray flaring is observed to be associated with the 
brightening of the radio core followed by the ejection of
a new radio blob along the jet and subsequent dimming of the core 
\citep{2012ApJ...751L...3G,2015ApJ...808..162C}.
This kind of behaviour has been observed in microquasars. 
However all radio flaring events do not seem to result in $\gamma-$ray flares
as observed from the periodic radio flaring in 3C~120 \citep{2015ApJ...808..162C}. 

\item Hard X-rays have been detected from radio lobes of 
Fornax~A which have been attributed to inverse Compton scattering
of the CMB photons by relativistic electrons \citep{1995ApJ...449L.149F,1995ApJ...453L..13K}.
Soft to hard X-rays have been detected from the radio lobes of several other galaxies
like 3C~219 \citep{1999A&A...342...57B}, 3C~452 \citep{2002ApJ...580L.111I},
Pictor~A \citep{2003ApJ...586..123G,2005MNRAS.363..649H}, 3C~98 \citep{2005ApJ...632..781I} 
with most of these galaxies showing a FR~II radio morphology and the coincident
X-ray emission being attributed to inverse Compton scattering of the CMB photons.
X-ray emission is observed to be coincident with at least one of the radio lobes 
in most of the FR~II radio sources. 

\item The soft X-ray emission which is coincident with radio lobes
in FR~I galaxies and especially those located in groups and clusters of galaxies
is often observed to be fainter than the rest of the diffuse X-ray emission that
forms a halo around
the radio galaxy.  These regions are referred to as X-ray cavities in literature 
and are believed to have formed due to the radio plasma in the lobes pushing 
out the thermal plasma.  The X-ray cavities are often surrounded by denser
and cooler gas. 

\end{itemize}

 \subsubsection{A comprehensive explanation}
We derive a model for active nuclei which well explains their wide-ranging observed
properties.  We begin with a summary of the physical model for
a quasar as described by \citet{2016arXiv160901593K} which was mainly derived
from results at ultraviolet bands and then expand and refine it to 
encompass the observed multi-band features (especially radio) of active 
galaxies.  The similarity in the observed radio features of a microquasar and
a radio galaxy is remarkable and underlines the uniformity of physical processes
near an accreting black hole irrespective of its mass.  It then means that 
the model put forward for microquasars
should be applicable to active nuclei in its entirety with observed 
differences being confined to effects of the very different
linear dimensions associated with the two systems which will lead to wide differences in 
the energy budget, variability amplitudes and timescales,
mass ejected and linear scales.  
Since microquasars showcase significant properties of
active black holes, discussion on the same was included before active nuclei.
We begin the section by a short recap of the unique model that had emerged from 
the behaviour of quasar spectra and then expand it. 
The results which are relevant to active nuclei and which have already been 
justified in case of novae in \citet{2017arXiv170909400K} and 
the effect of gravitational torques on gas distribution and rotation in galaxies 
justified in \citet{2016arXiv160604242K} are referred to where required. 
This is followed by a few case studies which highlights the efficacy of the model. 

  \paragraph{Recap: Quasar model}
The ultraviolet spectra of quasars consist of wide high-redshift 
permitted emission lines and a host of multi-redshifted absorption lines blueshifted
wrt to the emission lines.  Most of the redshifts involved are significantly higher
than other active galaxies. 
Since this sort of spectrum is unique to quasars and observed in quasars at 
all redshifts, \citet{2016arXiv160901593K} explored an intrinsic origin for the multiple
redshifts.  The physical process that was predicted by the general theory of relativity
wherein the line photon suffers a redshift when arising in a strong gravitational potential
was found to well explain the observed behaviour of the redshifts.  
It was reasoned that if an intrinsic redshift component contributed
to the redshift of the observed lines then the lowest redshift instead
of the highest redshift detected in the quasar spectrum would indicate the cosmological 
redshift and the difference redshift would quantify the largest value of the intrinsic
component.  
Data also showed that the lowest redshift detected in a quasar spectrum
increased with quasar redshift (emission line redshift) which is exactly what would
be expected if the quasar redshift included an intrinsic component in addition
to the increasing cosmological component.  
The gravitational redshifts of the emission lines for the sample of quasars 
listed in \citet{2013ApJ...779..161S} were determined to be between 
0.016 (photon emitted from $31.25 R_s$) and 0.5535 (photon emitted from $0.9 R_s$) 
with a median of 0.25 (the photon would be emitted from $2R_s$).  
Gravitational redshift is always $\le 1$ and gravitational redshift larger
than 0.5 would mean that the photon arises within the ergosphere and gives irrefutable
evidence to the existence of rotating black holes and escaping of a photon
from within the ergosphere i.e. escape velocity being equal to velocity of light. 
When the intrinsic redshift component due to the gravitational redshift is 
removed from the observed emission line redshift, quasar 
redshifts are significantly reduced. 
In fact, since the absorption lines of Mg II and Fe II occur at the lowest redshifts 
detected in a quasar spectrum, these provide the distribution of the quasar redshifts.
The existence of gravitational redshift component in the quasar spectral lines has
motivated a unique model for the active nucleus in quasars as
was shown in \citet{2016arXiv160901593K} which we summarise below. 
This also begs a critical revisit to our understanding based on the existence 
of quasars at high redshifts and their extreme luminosities in addition to
the existence of superluminal motions.  

It was reasoned that matter falling in towards the supermassive black hole should
progressively increase in densities as it approaches the black hole (or any
other compact object) and this can lead to significant consequences.   If 
the infalling densities became $> 10^5$ gm/cc outside the event 
horizon of the black hole than matter could enter the degenerate state.
This was briefly explained in Section \ref{recap}.  A more detailed explanation
is presented here due to its importance in active nuclei.  In analogy to
white dwarfs and neutron stars, it was suggested that the degenerate pressure of electrons
or neutrons (or other particles) would prevent the collapse of these 
layers of matter into the black hole.
Thus, layers of degenerate matter can be expected to deposit 
outside the event horizon of the black hole whose outer surface would then constitute 
the pseudosurface of the black hole which forms a quasi-stable configuration
close to the black hole.  Emission from this hot pseudosurface
would then be the electromagnetic signal that arises closest to the 
black hole that we can detect.  For 
simplicity, the pseudosurface can be assumed to be spherical - however an
uneven asymmetric latitude-dependent surface cannot be ruled out and hence
it is probably more appropriate to describe it as quasi-spherical. 
This pseudosurface can form anywhere outside the event horizon of the black hole.
Observations show that it is formed close to the event horizon in quasars. 
Since this layer is formed in the strong gravitational potential of the black hole
it can be gravitationally heated to high temperatures with the high conductivity
of the degenerate matter ensuring
that it quickly uniformises through the layer.  This hot pseudosurface can be 
a source of continuum emission contributing black body radiation 
to the spectral energy distribution of the active nucleus and explaining 
the blue bump or ultraviolet excess which is observed from active nuclei and nuclei
of quiescent elliptical galaxies.  Further accretion will lead to the infalling  
normal matter collecting on this degenerate pseudosurface. 
The matter accreted immediately outside the pseudosurface of the rotating
black hole in the non-polar regions gives rise 
to the broad emission lines.  The dense matter layers beyond the emission line
layer give rise to the absorption lines detected in the quasar spectrum. 
Such stratification, very close to the black hole, leads to line photons arising in a
rapidly varying strong gravitational potential and hence appearing at 
a range of redshifts in the quasar spectrum.  Thus, the continuum emission
and multi-redshifted spectral lines in quasar spectra dictate a unique
model for the distribution of matter around the supermassive black hole as was
pointed out in \citet{2016arXiv160901593K}. 

It was also suggested that the structure of nuclei of active galaxies other than
quasars are also likely to be similar with the observational differences 
likely due to some differing parameters like 
the separation of the line forming matter from the supermassive black holes
since the gravitational redshift critically depends on it and the quanta of
this component in the spectral lines other active nuclei is lower than in quasars. 

\paragraph{Model for active galaxies}
With the basic model for an active nucleus being determined from the data on the
most extreme of the active nuclei i.e. quasars, we elaborate on this further by
studying multifrequency (especially radio) data on all types of active galaxies and 
examining the myriad of physical processes that can be active in these.
Moreover the explanation for microquasars will apply to active nuclei and hence
the model discussed in Section \ref{mqsr} is also relevant to this discussion. 
We begin by outlining and describing the expected physical processes and
their observational implications, followed
by a description of the model that can explain an active nucleus of a galaxy 
and then elaborate on the effect of varying physical parameters of the black hole/accretion
on the observed properties.  We then describe the match between observations and
the model structure of the active galaxy and end with case studies to demonstrate
the comprehensive nature of the model.  We hope to have convinced readers of the
validity of the model by then and the experts will notice that this has raised
many other interesting research problems.

\noindent
{\bf Expected physical structure \& processes:}
The physical system that we are trying to understand consists of a rotating supermassive
black hole immersed in a gaseous medium i.e. the interstellar medium
of the host galaxy.  Since the black hole dimensions are very small compared to
the scale height of the galaxy, it is reasonable to assume spherical gas inflow.
The black hole accretes matter with 
%surface density increasing by 1/r$^2$ and 
volume densities increasing as 1/r$^3$ where $r$ indicates the separation of
the infalling matter from the black hole. 
Outside the event horizon, it can be postulated that the mass densities 
become so large that matter transforms to the degenerate state and   
its degeneracy pressure provides support against
collapse into the black hole.  A quasi-stable layer 
of degenerate matter accumulates outside the event horizon forming the pseudosurface
of the black hole.  The local accretion rates vary with latitude due to latitude-dependent 
centrifugal force which leads to a prolate-shaped event horizon of a rotating black hole 
and could lead to the formation of a prolate-shaped pseudosurface.  
Subsequent to the formation of the pseudosurface, infalling matter deposits on it and
does not fall into the black hole.
All the infalling matter in the polar region deposits directly onto the pseudosurface. 
In the non-polar regions, rotation leads to relatively lower accretion rates so that
some matter will deposit onto the pseudosurface and the remaining infalling matter
will accumulate outside in the form of an accretion disk.  Eventually, 
the infalling matter will collect in the accretion disk with the same accretion rate 
as at the poles and then flow onto the pseudosurface at a lower rate.
Thus in rotating black holes, an accretion disk will always be present beyond the 
layer of normal matter that accumulates on the pseudosurface in
the non-polar regions.  The layer of matter deposited on the non-polar pseudosurface
and the accretion disk is likely to be dragged along by the rotating black hole.  
The detection of molecular gas arranged in a rotating disk morphology whose axis is coincident 
with the radio jet axis in Centaurus~A 
%{\bf reference}
(https://www.eso.org/public/news/eso1222/),
Hydra~A and Abell~262 \citep{2019arXiv190209164O} could be signatures of
an accretion disk around the supermassive black hole.  
The linear extent molecular gas disk is few kiloparsecs in both the cases.  
Such disks should be present in all accreting rotating black holes and they should
be detectable in molecular lines and dust emission. 
The distinct environment of
the collected matter in the polar regions and the non-polar regions is to be noted. 
In the polar regions, matter accumulates at the highest possible accretion rates
for the system on the pseudosurface and is confined on the inner side by the 
pseudosurface but free on the outer side.  In the non-polar regions, matter accumulation
is slower and the layer of normal matter which accumulates on the pseudosurface
is sandwiched between the accretion disk on the outer side and
pseudosurface on the inner side.   {\it Thus the structure of all rotating accreting 
black holes is expected to consist of (1) a quasi-spherical pseudosurface composed of 
degenerate matter (2) 
normal matter collected on the pseudosurface (3) a non-polar accretion disk. }
The detailed physical properties of such a structure will depend on the specifics of the 
system e.g. a supermassive black hole will form an accretion disk which is larger
in linear dimensions than a stellar mass black hole. 

Further reflection on the differences in the accumulation of matter at the poles and
in the non-polar regions allows us to understand the evolution of the accreted
matter.
The rotating matter accumulated in the non-polar parts
on the pseudosurface are blocked on both sides and hence a likely evolutionary
path is conversion to the degenerate state as densities increase due to
compression and hence thickening of the degenerate matter layer.  This 
accreted matter can be ejected from the system only if
sufficient energy is pumped into it so that it can overcome rotation and 
shatter the accretion disk which if not impossible should have a lower probability
than retention of all the accreted matter.  Thus, the evolution of infalling 
matter in non-polar regions can be summarised to be: accumulation in
the accretion disk, flowing onto the pseudosurface, getting compressed to higher
densities and thickening of the non-polar pseudosurface or occasional ejections. 
In contrast to this, the matter accreted in the polar regions of the pseudosurface is  
likely to follow a different evolutionary path wherein some of the accreted matter
is ejected following an energy pulse.  In this case, the only condition to
be met is that the injected energy is sufficient to eject the matter
at escape velocities.  This then boils down to the following scenario which has already
been shown to be valid in the case of the accreting white dwarf in a nova binary and
on the accreting black hole in X-ray binaries.  Continuous accretion of normal matter
at the poles will lead to compression of the lower accumulated layers and 
hence their heating.  Once the temperature of the lower layers exceeds $10^8$ K,
an explosive thermonuclear reaction can occur which can inject a pulse of
energy to the accreted matter.  We already know from other astrophysical systems
and physics that this type of energy release is huge (depending on the quantity
of matter that is ignited) and fast.  This energy can accelerate the matter to
escape velocities and since the matter is not
confined on the outside, matter can explode out of the system.
Thus, the polar regions can be envisaged to be frequently going through a cycle of matter
accretion, compression, ignition and ejection.   This ensures that 
the infalling matter in the polar region always forms a thin layer on the pseudosurface
before being expelled and hence the degenerate matter layer would not thicken
with time.  However there are likely to be 
a fraction of the explosions which are not of sufficient energy to accelerate
matter to escape velocities and hence this matter will accumulate on the
polar region till the next explosion occurs.   Another outcome of the thermonuclear
explosions will be enrichment of the accreted matter.  Thermonuclear explosions can
be expected to occur over the entire pseudosurface since the only requirement
is that matter is compressed and heated to high temperatures which is expected
in matter accreted by a compact object. 
Although the explosions might not be energetic enough to eject matter in the 
non-polar regions, these will lead to enrichment of matter.  Thus all matter
that has experienced a thermonuclear reaction will be enriched till iron is formed
after which the exothermic nature of the thermonuclear reaction will cease.  
This means that the matter in
the non-polar regions which is rarely ejected will be continuously enriched with time 
while in the polar regions, the matter that is not ejected will be enriched. 
The enriched matter can accumulate on the
pseudosurface at the poles and thicken the pseudosurface.   {\it The expected
physical processes in the accreted matter on the pseudosurface, which underline
the quasi-stable nature of structure close to the black hole, will be 
(1) existence of a layer of normal matter on the non-polar pseudosurface accreted
through the accretion disk.  Repeated thermonuclear explosions in the accreted
matter in the non-polar regions which enrich the accreted matter and
hasten its eventual evolution 
to the degenerate state which thickens the non-polar pseudosurface.  
The accumulated matter in the non-polar regions forms the broad line region (BLR).
(2) Repeated thermonuclear explosions in the accreted matter
at the poles powering episodic outbursts of matter.  The matter ejected
from the poles forms the narrow line regions (NLR).  In some cases,
the explosions lead to enrichment of matter and thickening of the polar 
pseudosurface.  }

The above discussion has already led to a model for the active nucleus/galaxy which
we summarise below.

\noindent
{\bf Summarising the model:}
Now that the basic physical structure of a rotating accreting black hole and physical
processes in the accreted matter have been outlined, we can describe the physical
model for an active nucleus/galaxy and connect it to observational results before
elaborating on the different parts. 

We recall the main observable signatures of an active nucleus: (1) luminous multiband 
continuum emission from an unresolved core, (2) occasional presence of broad 
permitted emission lines arising in the BLR, 
(3) ubiquitous narrow permitted and forbidden emission lines  
arising in the distributed NLR, 
(4) occasional presence of radio core, jets, lobes, hotspots (5) multi-band variability. 

In literature, the origin of the black body component that is required to
explain the observed SED of the core has been attributed 
to the accretion disk since black holes do not have a surface and
observations require a hot emitting body. 
%and lack of a better candidate in the vicinity of the black hole. 
Now with the inevitable formation of a hot dense pseudosurface around the black hole
as described above, it has to be the source of the black body component. 
The hot pseudosurface will also be
the source of hard ionizing photons required to explain the high excitation line spectra
observed from active nuclei.  The line spectra will be formed in the accreted gas either
in the matter accumulated on the non-polar pseudosurface or in the matter
ejected from the polar regions. 
These points mean that
the existence of a hot pseudosurface which is expected from physical arguments
outlined above, is well supported by observational results.  
Continuum emission can also be contributed by
the hot normal matter accreted onto the non-polar pseudosurface.
The nature of the observed line emission from the BLR and NLR 
are expected to be different.  The matter in the polar regions 
is accreted, heated, enriched, ejected and will be eventually spread over a large region
whereas in the non-polar regions, the matter is confined to a small region near
the black hole, compressed, heated, enriched but seldom ejected.  
The polar line forming region 
will be the NLR which has been surmised from observations to exist from near
the core to large distances along the polar direction whereas the non-polar 
line forming region will be the BLR which has been surmised from 
observations to be compact.
The co-existence of the extended NLR and radio jet/lobes/hotspots along the 
polar axis (generally the minor axis of the galaxy which is generally the rotation axis
of the black hole) supports their origin in matter ejected from the poles of the 
black hole following energy injection from a thermonuclear explosion in the accreted matter. 
The accreted matter in the non-polar regions which is sandwiched between the
pseudosurface and the accretion disk is compact and rotates with the black hole. 
Since this region lies close to the black hole, it is responsible for 
the broad emission lines with the large widths being due to the large and 
varying gravitational potential within the BLR.  The large
gravitational potential in which the photon is formed leads to a shift in its
wavelength (gravitational redshift component) and 
the finite thickness of the line forming region leads to broadening of the lines.  
Both the factors depend critically on the separation of this region i.e. BLR from
the black hole.  If the BLR forms very far from the black hole then the spectral
lines will not show a gravitational redshift component nor will they be 
gravitationally broadened. 
This helps us infer that all the active 
nuclei which show the presence of a broad line in their spectra, have formed
a BLR sufficiently close to the black hole for the line photons to include
a contribution from a gravitational redshift.  The active nuclei from which
only narrow lines are detected can indicate that the BLR is located far from 
the black hole so that the influence of the gravitational potential is
negligible.  However it is also possible that the sightline has missed intercepting 
the BLR in these cases.  Careful examination of observational results should
help us disentangle the two types of cases which can lead to non-detection of broad
lines. 

\begin{figure}
\includegraphics[width=8cm]{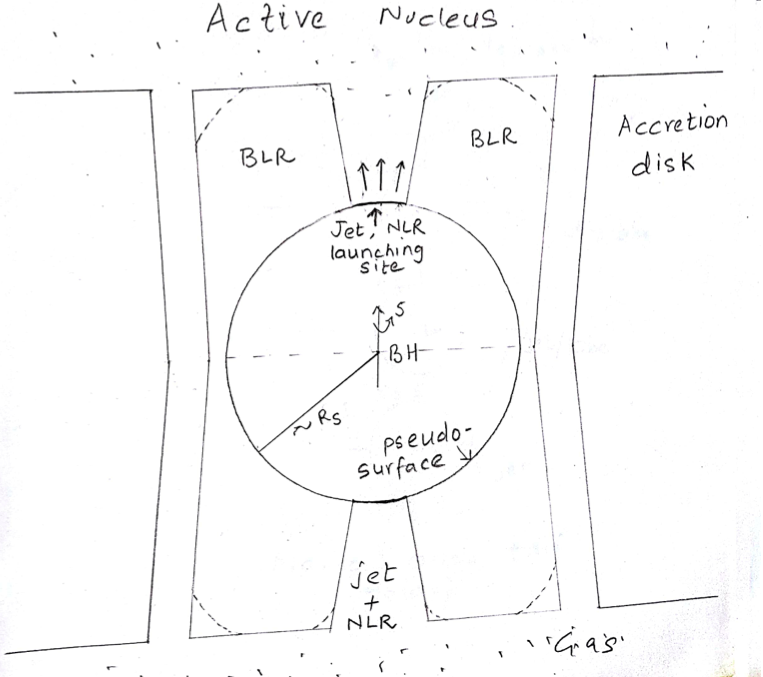}
\caption{\small A zoom-in of the nuclear region of an accreting
black hole, showing the structure as supported by observations and physics.  
The BLR is hour-glass shaped with a hole in the polar regions from where the jet
and NLR are launched.  The accretion disk contains the excess matter which accumulates
outside the pseudosurface due to the reduced non-polar accretion rates in a rotating
black hole.  The pseudosurface can also be prolate-shaped for a fast rotating black hole.}
%{\bf modify the pseudosurface to a prolate-shape instead of spherical.}}
\label{fig25}
\end{figure}

Thus the structure of the active nucleus or more generally the distribution of
matter around an accreting 
rotating black hole is illustrated in the schematic in Figure \ref{fig25} as 
expected from known physics and known physical processes. 
This structure trivially explains several
observed properties of an active nucleus.  {\it To summarise: 
(1) a layer of hot dense degenerate matter accumulates around the black hole 
forming its pseudosurface.  The pseudosurface emits black body radiation which is the
photoionizing radiation field exciting the line forming regions, (2) matter accreted
in the non-polar regions of the pseudosurface forms a BLR  
from which broad emission and absorption lines emerge, 
(3) episodic thermonuclear explosions in the accreted matter at the poles
launch bipolar radio synchrotron jets and NLR from which narrow 
forbidden lines emerge. }

\noindent
{\bf Spin, mass, pseudosurface of black hole:}
In general relativity, the concept of an innermost stable circular orbit (ISCO) has been
introduced.
% which is relevant for black holes. 
Matter located inside the ISCO is unstable and plunges into the black hole.
The radius of the ISCO for a non-rotating black hole has been estimated to be 
$3 R_s$ whereas for a rotating black hole, the ISCO is the equatorial radius  
of the event horizon which can be as small as $R_s/2$ for a maximally rotating black hole. 
In the quasar sample that was examined in Kantharia (2016), the median gravitational
redshift shown by quasars was $z_g=0.25$ indicating that
the emission line photon in half the quasars arises at a separation $< 2 R_s$ from the black
hole which then provides another unambiguous piece of evidence to spinning black holes in
quasars. 
The bipolar radio jets and NLR detected in many active galaxies 
also argue for a spinning black hole and hence it appears safe to assume that all
active galaxies host a supermassive rotating accreting black hole.
To recap, at least three different observations support the existence of
spinning black holes in active nuclei:  (1) a gravitational redshift that is
$> 0.5$ which indicates the existence of an ergosphere  
(2) existence of matter within the ISCO of a non-rotating black hole as
surmised from the gravitational redshift component in spectral lines 
(3) presence of bipolar radio jets and NLR in active nuclei. 

For an accreting spinning supermassive black hole, the fundamental variables are
mass (or Schwarzchild radius) and spin rate of the black hole and the
infalling matter densities.  Table \ref{spin} shows the nature of 
dependence of the observed properties of active nuclei
on the spin and mass of the black hole for the same infalling matter 
densities.  The black hole spin emerges as the single-most 
important parameter which can acccount for a range of observed properties of active 
nuclei.  Higher spin rate of a black hole should lead to lower accretion rates in
the non-polar regions due to the larger magnitude of the centrifugal force 
($F_c \propto r_i \omega^2$ where $\omega$ is the angular velocity of the black hole
and $r_i$ is the separation of the surface at latitude $i$ from the polar axis)
which reduces the effective gravitational force acting on the infalling matter. 
It will also lead to the BLR \& accretion disk enclosing almost the entire
pseudosurface due to the modified
gravitational force being effective to larger latitudes so that broad
lines from the BLR should be detectable from such nuclei over most sightlines. 
The pseudosurface and BLR can be radially thinner, hotter and more 
variable - all due to their proximity to the black hole. 
On the other hand, the accretion disk around such black holes will be radially
extended due to the lower accretion rates on the pseudosurface. 
The higher spin of the black hole will lead to reducing the extent of 
the polar region (while the BLR encloses most of the pseudosurface) 
wherein the effect of the centrifugal force ($r_i \rightarrow 0$ in the polar region) 
is negligible 
so that the infalling matter is subject to the highest possible gravitational force.
Infalling matter will accumulate directly in this region instead through 
an accretion disk and can be compressed and heated.
When the temperature exceeds 100 million degrees K, a thermonuclear outburst
can eject the accreted matter i.e. jets and NLR from this polar region.  
Since the extent of this region will determine the jet opening angle, faster
rotating black holes will have smaller jet opening angles.   This ejection
can happen from close to the black hole since matter accumulates
in a small region and matter has to 
be ejected with relativistic velocities to be able to leave the system.
In a faster spinning black hole, energy will be focussed within a smaller opening
angle thus explaining the highly relativistic velocities observed for narrow jets.
{\it To summarise, black hole spin is responsible for the latitude-extent of the 
non-polar BLR, accretion disk and opening angle of polar jets amongst other features. 
A range in the spin rate of black holes explains the range of radio jet opening angles that
are observed in active nuclei and the detectability of broad lines. }

The main effect of a larger mass, given other parameters are the same, is 
to increase the linear dimensions of the
system with one of the important outcomes being increased detectability of the system. 
Mass differences can lead to black hole powered objects being classed into 
different types e.g. microquasars host stellar mass black holes and quasars host
supermassive black holes with the former being randomly distributed in a galaxy
while the latter always located at the centre of the galaxy. 
The main effect of larger infalling densities given other parameters
are the same, would be to enhance the accretion rates and hence 
thicken the pseudosurface and/or
the BLR in the non-polar regions and lead to more frequent thermonuclear
outbursts in the polar regions.  In such a case, the photons from the BLR 
will progressively arise from an increasing distance from the black hole as the
pseudosurface thickens. 
The frequent thermonuclear explosions in the polar regions, if not energetic
enough to expel matter, can lead to a radio-quiet active nucleus
and if energetic enough to expel matter, can lead to frequent episodes of matter ejection
and hence a radio-loud nucleus with distributed optical line emitting gas i.e. NLR. 
We now examine in detail, the properties 
listed in Table \ref{spin} starting with the non-polar regions. 

\begin{table*}
\small
\centering
\caption{ The expected behaviour of physical parameters for a spinning, accreting
black hole which can be surmised from physics are listed in this table.
All cases assume the same infalling matter densities.  A black hole
of the same mass is considered for examining
the spin-dependent properties and a black hole of the same spin is considered for
examining the mass-dependent properties.  L refers to 'largest' of the property
while S means the smallest and O
means that there is no effect on that parameter.  Accretion rate is considered
in terms of total mass accreted so that around a more massive black hole the
accretion rate will be larger by virtue of a larger surface area (i.e. larger $R_s$). }
\begin{tabular}{l|l||c|c||c|c}
\hline
Region & Physical parameter & \multicolumn{2}{|c|}{BH of same mass} & \multicolumn{2}{|c}{BH of same spin}\\
\hline
&  & High spin BH  & Low spin BH  & High M$_{BH}$ & Low M$_{BH}$ \\
\hline
\multicolumn{6}{c}{\it Non-Polar region properties} \\
%    & Radio intensity & O & O & &\\
& & & & & \\
& Global accretion rate$^1$ & S & L & O & O \\
& & & & & \\
 Pseudosurface  & Radial thickness &  S & L & S & L \\
(degenerate matter    & Latitude thickness  & L  & S  & O & O \\
(layer)    & Separation from BH & S & L & O & O \\
%& Luminosity    & L & S & L  & S \\
   & Temperature & L & S & O & O \\
    &  UV luminosity    & L & S & L & S \\
 & Variability & L & S &  & \\
& & & & & \\
Broad line Region & Radial extent & S & L & S$^2$ & L$^2$ \\
    & Latitude thickness  & L & S & O  & O \\
    & Ionization,excitation & L & S & L & S \\
    & Variability & L & S &  & \\
& & & & & \\
 Accretion disk & Radial extent & L & S & S & L\\
    & Latitude thickness  & L  & S & O & O\\
& & & & & \\
Examples$^3$ & & quasar & -  & quasar & - \\
 & & BLRG &  NLRG & BLRG &  NLRG  \\
& & Seyfert 1 &  Seyfert 2 & Seyfert 1 & Seyfert 2\\
% &  &  & & &\\
\hline
\multicolumn{6}{c}{\it Polar region properties} \\
& & & & & \\
& Global accretion rate$^4$ & S & L & L & S \\
& & & & & \\
Pseudosurface  & Thickness$^5$  & S  & L & O & O\\
(degenerate matter    & Separation from BH$^5$ & S & L & O & O \\
layer)    & Emitting region & S & L & O & O \\
%      &  UV luminosity  & S & L & L & S\\
& & & & & \\
Accreted matter layer & Thickness & L & S & O & O \\
  & Circular size  & S & L & L  & S \\
  & Separation from BH$^5$ & S & L & O & O \\
& & & & & \\
Thermonuclear outburst$^6$ & Frequency  & L  & S & O & O\\
in accreted matter         & Energy directivity & L & S & O & O \\
& & & & & \\
Expulsion of matter$^6$ & Recurrence rate & L  & S & O & O \\
    & Ejection speeds & L & S & O & O \\
%   & Efficiency  & L  & S & O & O \\
    & Jet opening angle & S & L & O & O \\
    & Jet lengths & L & S & O & O \\
    & NLR length  & L  & S & O & O \\
    & Random relativistic speeds & L & S & O & O \\
    & Radio synchrotron & L & S & L & S \\
Examples$^3$ &  & FR II$^{7,8}$  & FR I & Radio galaxy$^8$ & Sy,LINERs\\
         &  & GPS, CSS$^{7,8}$  & Seyferts & Radio qsr$^8$ & Radio-faint qsr \\
\hline
Host galaxy$^9$ &  & Elliptical  & Spiral & Elliptical & Spiral \\
\hline
\end{tabular}

$^1$ {\small Varying accretion rates due to latitude dependent centrifugal potential;
largest at poles.}
$^2$ {\small Larger mass of the black hole increases
the surface area of the outer event horizon leading to the listed outcomes.}
$^3$ {\small The examples are compared in pairs. }
$^4$ {\small Varying accretion rates caused by different linear sizes of the polar
region involved.} 
$^5$ {\small This follows from the frequent outbursts so that most
accreted matter is ejected.}
$^6$ {\small Achieving conditions conducive
to thermonuclear explosion should be easier in a small region with high accretion rates. }
$^7$ {\small FR II and FR I are comparable in size and
GPS, CSS are comparable in size to the radio structure in Seyferts}
$^8$ {\small These examples are made on basis of the radio loudness property.}
$^9$ {\small This is based on observational results.}

\label{spin}
\end{table*}

It can be inferred from the highly luminous nuclear continuum 
(e.g. quasars) and broad emission lines
of high ionization and a significant gravitational redshift component
(e.g. quasars) that the pseudosurface and BLR in that active nucleus have formed 
in close proximity to a fast spinning black hole. 
Similarly we can infer from the radio jets observed to be expanding with highly 
relativistic velocities and launched within small opening angles 
(e.g. FR II sources) that the ballistic
ejection in that active nucleus has happened from close proximity to 
a fast spinning black hole where the escape velocity is relativistic.
The main observable attributes of the non-polar regions
are the black body radiation from the pseudosurface and continuum and 
broad lines from the BLR whereas that of the polar regions are the radio jets
and optical line forming NLR. 
The polar region involved in the thermonuclear explosion gets larger and the latitude extent
of the BLR smaller for black holes of lower spin rates.  The 
jet is launched within a larger opening angle and the efficiency of
energy transfer, collimation and subsequent processes decline so that the jet
is relativistic close to the core from where it is launched 
but quickly becomes sub-relativistic (e.g. FR~I). 
Since the radio structure in FR~II and FR~I sources extend upto several hundred
kiloparsecs but show distinct jet behaviour (opening angle and persistence
of ballistic nature), they represent the high and low spin cases respectively 
in high mass black holes.  The dimensions of the
radio structure of GPS,CSS sources is comparable to Seyferts although the
radio strengths are very different.  These, then, might represent the high and low spin
systems of a lower mass black hole with other parameters contributing to rest of 
their observed differences. 

The ejection of matter occurs from the polar region of the quasi-spherical
pseudosurface which will have a finite curvature.  For simplicity if we assume
that the pseudosurface is a spherical surface then 
radial ejection of matter from the poles of such a surface, 
irrespective of the black hole spin and mass, will always lead to a conical outflow.
This means that the observed conical radio jets and NLR are expected when the ejection
is from the pseudosurface and no other explanation is necessary.  In other words, the
conical outflows provide independent evidence to the ejection of jets and NLR from the
polar pseudosurface.  The spin of the black hole will vary the cone angle such
that a fast spinning black hole will launch the jet within a small conical angle 
while a slower black hole will launch the jet in a larger conical angle.  Thus
the conical angle which is viewed in two dimensions and which we refer to as the 
jet opening angle is a function of the black hole spin.  The observed conical
angle will be enhanced by effects like a precessing spin axis of the black hole. 

The synchrotron emission from radio jets in several cases is observed to 
consist of a discrete clumpy component and a uniform component.  While the bulk expansion
of the clumpy component is measured to be relativistic from proper motion
measurements, it is difficult to ascertain the expansion speed of the
uniform emission component.  One can speculate that while the clumpy component
is accelerated by the thermonuclear outburst, the uniform component consists
of the infalling matter which has been pushed out by the action of radiation
pressure of the photons from the pseudosurface.  However this would require that
the observed radio spectrum of the clumpy and uniform component are different
as is possibly the source of the magnetic field.  This is a speculation which
needs to be investigated further with observational data. 

The variety of observed features of the different active nuclei has prompted
their classification into several types. 
Quasars are characterised by a luminous star-like nucleus which
overwhelms the galaxy emission, wide
and highly redshifted emission lines of permitted high ionization transitions and
wide absorption lines at redshifts lower than the wide emission lines.
BLRG are radio galaxies characterised by a bright nucleus (fainter than quasars),
wide emission lines of permitted transitions and narrow forbidden lines.  BLRG show
a FR II radio morphology.  NLRG are radio galaxies characterised by faint nuclear
emission and narrow lines of both permitted and forbidden transitions.
NLRG show a FR I radio morphology.  Seyfert 1 host a bright nucleus, wide permitted emission
lines and narrow forbidden emission lines while Seyfert 2 show a faint nucleus
and only narrow emission lines.  A radio core is commonly detected in Seyferts. 
While radio galaxies are radio bright, a large fraction of
quasars and Seyferts are radio faint.  Blazars are radio bright compact active nuclei and
comprises BL Lac objects and flat spectrum radio quasars (FSRQ).  
While BL Lac objects generally show a featureless continuum,
FSRQ show broad emission lines superposed on the continuum.

It appears that the observed range of properties of active nuclei arise
due to a combination of important physical differences and orientation to the sightline. 
We try to identify the differences here. 
The higher continuum luminosity and quanta of gravitational redshifts in quasars are
signatures of the proximity of the pseudosurface, BLR to the black hole as compared 
to other active nuclei.  This property has a physical origin in the non-polar
regions which differentiates quasars
from other active nuclei.  Considering that the radio structure of quasars, when
detected, is similar to
FR II radio galaxies, it indicates similarity of physical processes in the polar regions of
quasars and the active nucleus in FR II galaxies but differences in the non-polar
regions.  It means that the polar ejections happen close to a fast spinning black hole 
in quasars and FR II galaxies but the non-polar BLR in quasars is located closer to the 
black hole than in the latter. 
Many quasars are radio-quiet
like Seyferts and show the presence of Fe II lines as do several Seyfert 1. On the
other hand, BLRG are radio bright but notoriously deficient in Fe II lines.   
This aspect argues for the same physical process being responsible
for iron enrichment in active nuclei which could be nucleosynthesis in
thermonuclear reactions.  This could
occur in the BLR and/or in the accreted matter in the polar regions which is not
expelled.  The lack of iron lines in BLRG argues for their production in the
polar region in that in BLRG, the thermonuclear energy always expels matter
so that iron is seldom synthesised from the hydrogen-rich fuel whereas in the
radio-quiet quasars and Seyferts, ejection is infrequent and hence iron could
be ultimately synthesised at the poles.   However the prevalence of iron lines in 
Seyfert 1 as compared to Seyfert 2 argues for its synthesis in the BLR.  Thus
the ambiguity on where the iron lines are formed - in the BLR or the polar
region or both.  This needs further investigation.  Currently we can infer that
the observations provide strong independent support to the 
occurence of thermonuclear explosions in the accreted matter near the
supermassive black hole.
The orientation-based unification theory and radio observations have long
suggested that quasars and blazars are the same objects observed from different
angles.  The broad lines from blazars 
contain a large gravitational redshift component like quasars 
\citep{2016arXiv160901593K} which lends independent support to their unification.
The orientation-dependent
unification works for quasars and blazars with BL Lac objects being viewed exactly
pole-on so that the sightline misses the BLR explaining the lack of broad lines and 
FSRQ being viewed at a small angle to the jet axis so that the broad lines from the
BLR are detectable.
Quasars viewed from any other sightline will always
detect the BLR explaining the ubiquitous presence of broad lines in their spectra.
The ubiquity of radio emission from blazars which are observed along the
jet axis and its rarity in most quasars
might be due to the combination of Doppler beaming and inherent faintness due to
extremely narrow opening angles. 
Only quasars (includes blazars) amongst active nuclei show large gravitational 
redshift components in their spectral lines highlighting the 
important physical difference between quasars and other active nuclei of differing
separation of the BLR from the black hole. 
Since most FR II are BLRG while most FR I
are NLRG, these highlight the differences in the spin of the black hole, 
covering factor of the BLR and its separation from the black hole and the jet opening angle. 
The classes ranging from Seyfert 1 to 2 seem to signify the increasing separation
of the BLR from the black hole and only a subset of Seyfert 2 might host
a hidden BLR close to the black hole.  Since the radio properties of
Seyfert 1 and 2 are similar, it would indicate that their spins are comparable.
It appears that the radio galaxies referred to as FR I and NLRG are the same 
as are FR II and BLRG.  The classification of FR I/FR II has been made from 
polar properties (radio synchrotron) whereas
NLRG/BLRG is from non-polar (spectral lines from BLR) properties of the same objects.  
This means that in FR II/BLRG, the BLR is located close to the black hole and
its covering factor is large whereas in FR I/NLRG, the BLR is located relatively
far from the black hole and its covering factor is smaller. 

{\it To summarize: The black hole spin is the single-most important physical parameter 
responsible for several observed properties of active nuclei.
Only rotating black holes can launch ballistic relativistically expanding
bipolar jets and optical line forming NLR and higher the spin rate, smaller
is the jet opening angle.  Rotating black holes form a BLR in the non-polar
regions from which gravitationally redshifted and broadened lines arise in addition
to an accretion disk beyond the BLR. 
Infalling matter densities determine the maximum possible accretion rates while mass
of the black hole determines the linear size of the region over which the black hole
exerts its gravitational influence. }

Since the nuclei of all active galaxies are observed to be a multi-band compact
source, it supports the formation of a hot pseudosurface in all active nuclei. 
The simplest scenario for the location of the pseudosurface is 
at the event horizon of the rotating black hole.  Since the event horizon of
a rotating black hole is prolate-shaped, this means that the pseudosurface of
rotating black holes will be prolate-shaped if a thin layer is deposited at
the event horizon.  The main condition for the formation of a pseudosurface
is that matter enters a degenerate state and the degeneracy pressure is
sufficient to prevent the gravitational collapse of matter into the black hole.  
Matter can enter the degenerate state when densities exceed $10^5$ gm-cm$^{-3}$. 
Since volume matter densities will increase as $1/r^3$, it is possible that 
in dense environs where the infalling matter densities are high, such densities are reached
at a larger separation $r$ from
the black hole compared to when the infalling matter densities are small.
This means that the pseudosurface need not always form at the event horizon and
instead its separation 
from the black hole can be a variable which depends on the global infalling matter densities
such that when the density is low, the pseudosurface (and BLR) forms close to the black hole
and when the density is high, the pseudosurface (and BLR) forms far from the black hole.
The same argument will also hold in the polar regions so that the 
pseudosurface and hence matter ejection will happen far from the black hole 
if the infalling matter density is high and from closer to the black hole 
if the infalling matter density is low.  
This means that the radial separation of the pseudosurface 
from the black hole is likely to be dependent on the infalling matter densities. 

We can understand this further 
by using the gravitational redshift $z_g$ of the broad emission line to quantify 
the separation of the non-polar pseudosurface from the black hole and the bulk
relativistic velocity of the polar jet $v_{bulk}$ equated to the escape velocity to 
quantify the separation of the pseudosurface from the black hole in the polar regions. 
We consider the example of the quasar 3C 273 which is discussed later on wherein
its distance has been revised to 20 Mpc.  For this case,
the broad emission line from the BLR appears to include a component $z_g \le 0.131$ 
which indicates that the pseudosurface 
is located at a separation $\sim 3.8 R_s$ while the jet bulk velocity of $0.16c$ 
is derived from the proper motion of the blobs which
corresponds to a launch site being separated from the black hole by about
$39 R_s$.  These values indicate a prolate-shaped
pseudosurface in 3C 273.  
The varying separation of the BLR and jet launch site from
the black hole can be explained either by a prolate-shaped thin pseudosurface
which has formed far from the event horizon or an asymmetric pseudosurface 
of varying thickness at the event horizon such that it is thick at the poles and
thin at the equator which is expected from the latitude-dependent accretion rates.  
Although we refer to the two parts as polar and non-polar
for simplicity, we are cognisant that there will be variation of properties within 
the non-polar parts of the pseudosurface as a function of latitude.
This simplified reference is resorted to keeping in mind that only if 
we can fully grasp the distinction in properties 
between the polar and non-polar parts can we hope to understand the 
variation within the non-polar parts.  
The range of large gravitational redshifts that are detected in the quasar
spectra indicate that the BLR is radially extended and consists of emission line
and multiple absorption line zones located in close proximity 
to the black hole supporting the formation of the non-polar pseudosurface 
at the event horizon.  In fact, observations show that the emission line zone of
the BLR in quasars is located at a separation $\le 32 R_s$. 
Thus, it can be inferred that the pseudosurface forms close to the event horizon 
in quasars.  From the above discussion it might be possible to
infer that the non-polar accretion rates are low owing to lower 
infalling matter densities and/or high spin of the black hole in quasars.  
Only a small contribution of $z_g$ appears included in the 
broad line redshifts of BLRG and Seyfert 1 nuclei indicating a large separation
of the line forming zone in the BLR and the black hole.  Since in BLRG/FR II,
relativistically expanding jets are launched from close to the black hole, the
low observed values of gravitational redshifts would mean that the BLR is
located far from the black hole while the jet launching site is close to the black hole.
This can be translated to indicate a oblate-shaped pseudosurface.
The pseudosurface of the black hole in BLRG could show a latitude-dependent
thickness with the first layer of degenerate matter having deposited close 
to the event horizon. 
Gradually the non-polar pseudosurface has thicknened as an increasing quantity 
of matter has been accreted, compressed, heavier elements synthesised and
finally compressed to degenerate matter densities.
In Seyfert~1 nuclei, the radio emission is generally compact and faint
and jets are seldom observed to be expanding relativistically.  
Since gravitational redshifts are also negligible, it appears that both the
BLR and the jet launching site in Seyfert~1 are located far from the 
black hole.  The infalling matter densities in Seyfert nuclei can be large 
as their spiral hosts are gas-rich and it is
likely that the large infalling matter densities have led to the formation
of a pseudosurface far from the event horizon. 
However it is likely to be closer to the black hole and hotter in Seyfert~1 
than in Seyfert~2 galaxies as can be surmised from the luminous
core and broadening of the lines in the former. 

It was pointed out by \citet{1926MNRAS..87..114F} that matter can exist in the 
degenerate state if it has sufficient energy so that the electrons are not bound 
to any atom but free.  These electrons have energy to escape from 
any nucleus that they happen to venture close to during their journeys.   
Thus when matter enters the degenerate state, its energy content 
will be high which will be reflected in the high temperature which will
radiatively decline if there is no further energy input.  
As \citet{1926MNRAS..87..114F} also pointed out, the radiative losses will lead the
temperature of the degenerate matter to gradually approach $0^\circ$K 
while the energy content will continue
to remain finite which naturally emerges from quantum statistical mechanics.
To quote from \citet{1926MNRAS..87..114F} - 
"Temperature then ceases to have any meaning, for the
star is strictly analogous to one gigantic molecule in its lowest quantum
state. We may call the temperature then zero."
This giant molecule will continue to have its ground state energy but will
stop radiating since its temperature will be zero. 

The hot pseudosurface formed around the black hole provides 
the ionizing photons for the non-polar BLR and polar NLR. 
The degenerate matter will cool as it radiates and unless there is continuous
energy input or formation of a new degenerate matter layer, the pseudosurface
will cool.  The lower temperature of the
pseudosurface would lead to a softer radiation field which will reduce the quanta of
hard photons that can excite high-ionization lines in the BLR and/or NLR.  
This would eventually lead to the quenching of broad emission lines from the BLR so
that the existence of a BLR in systems with a low temperature pseudosurface is ambiguous. 
It is to be noted that in Seyfert~1, a luminous nucleus and wide lines are observed
whereas the Seyfert~2 nucleus is fainter and wide lines are absent.  Range of
line widths and nuclear continuum are observed for Seyfert nuclei which are then
placed between types 1 and 2. 
These observations tend to support the scenario in which the pseudosurface in Seyfert 1
and 2 are of different temperatures  which lead to varying nuclear strengths and 
different physical conditions in the BLR. 
Since the temperature of the pseudosurface can vary it could also lead to
varying physical conditions in the BLR.   For example, if the pseudosurface
was heated and brightened then the BLR will also suddenly be heated and ionized
and start emitting wide lines.  This, then, can explain the changing look
active nuclei (e.g. quasars and Seyferts) in which the properties flip-flop 
between the two types.  This hypothesis appears highly plausible 
since it explains both the changes in the continuum and line
properties that occur when a type 2 transforms to a type 1 nucleus or vice
versa.  Additionally, the
above can also explain the existence of the intermediate types between 1 and 2 as
simply a result of the range in the temperature of the pseudosurface.

{\it To summarise: A hot pseudosurface composed of degenerate matter is formed
around all black holes in active nuclei.  This quasi-spherical surface is
formed near the event horizon in quasars but much further in Seyfert nuclei. 
The radiating pseudosurface will cool and the temperature changes can
explain the range of properties observed for type 1 and 2 Seyferts and quasars.
A hotter pseudosurface will ionize the BLR so that wide lines arise from it
whereas the BLR near a cooler pseudosurface will be undetectable.
Thus, the bright nucleus and broad lines observed from quasars, 
Seyfert~1 and BLRG indicate the presence of a hot pseudosurface whereas the
faint nucleus and absence of broad lines in Seyfert~2 and NLRG can be explained
by a cooler pseudosurface.  The changing look active nuclei can be explained
by a change in the temperature of the pseudosurface.}

In the remaining part of the section, we discuss the 
wideband SED; BLR; thermonuclear explosions and variability; redshift components;
jets, lobes, hotspots;  magnetic field around a black hole; dust formation
in NLR; gas for accretion and radio faint active nuclei. 

\begin{figure}
\centering
\includegraphics[width=7cm]{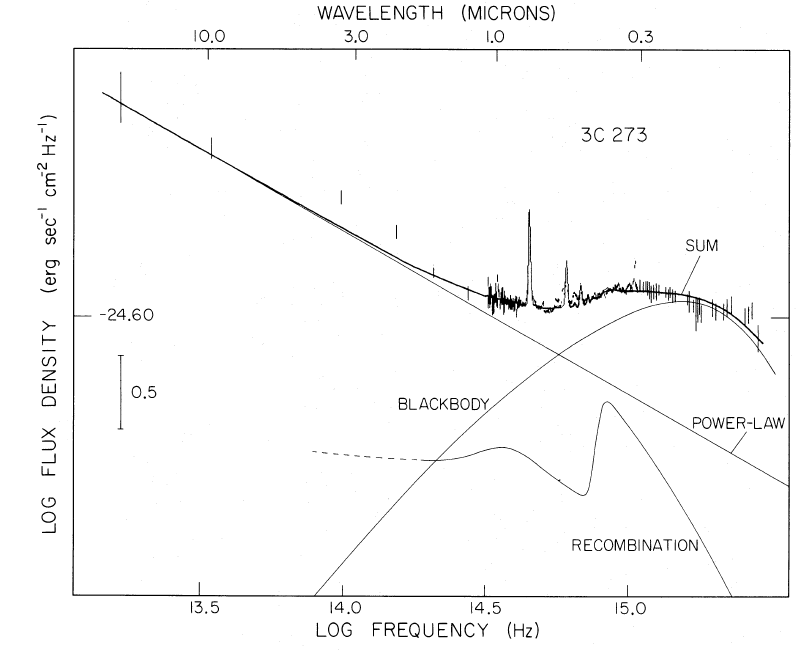}
\caption{\small Figure copied from \citep{1982ApJ...254...22M} showing the three component
fit to the SED of 3C~273 from infrared to ultraviolet bands.  The three components
are black body of temperature 26000 K, a Balmer continuum of temperature between
13000-15000 K and a power law of index 1.1.}
\label{fig26}
\end{figure}

\noindent
{\bf Wideband SED:}
\label{sed}
The wideband spectral energy distribution of several active nuclei are often well-fit with
three components - a black body component, a power law component and a Balmer continuum
component (such as thermal free-free) as was demonstrated for Seyfert 1 and quasar nuclei
by \citet{1982ApJ...254...22M} (see Figure \ref{fig26}).  
In the framework of the model presented here, the origin of all three components
is explained.  The black body 
component should arise from the hot pseudosurface of the black hole, 
the power law component should arise
from the synchrotron emitting relativistic plasma in the polar regions that
is energised in a thermonuclear explosion and the Balmer
continuum should arise in the photoionized BLR around the hot pseudosurface in the non-polar
regions.  The contribution of the three components can vary amongst active nuclei
as also their dominance at different wavelengths. 
For example, the black body component often dominates at soft X-rays, ultraviolet, 
optical  wavelengths
while the synchrotron component dominates at hard X-rays, radio and infrared wavelengths. 

We first discuss the black body component and start with the example of the quasar 3C 273.
An emitting area of about $3\times10^{33}$ cm$^2$ was
required to explain the 26000 K black body component that was surmised
for 3C 273 from observations \citep{1982ApJ...254...22M}.  
As suggested here, the emitting area should indicate the area of 
the quasi-spherical pseudosurface which for simplicity we consider to be
spherical and located at the event horizon of the black hole.  The radius 
of this pseudosurface would be
$1.5\times10^{16}$ cm which if equal to the Schwarzchild radius of a black hole
would require its mass to be close to $10^{11}$ M$_{\odot}$.   
However it is to be noted that the distance to
the quasar 3C 273 has been overestimated due to the presence of the $z_g$ component in the 
broad emission line redshift and this is discussed more in the next section.  
3C 273 is more likely to be located at $\sim 20$ Mpc (instead of $\sim 650$ Mpc) 
which will reduce its emitting area 
by the factor $(20/650)^2$ i.e. the area will be $2.8\times10^{30}$ cm$^{2}$. 
This corresponds to a spherical surface of radius $4.7 \times 10^{14}$ cm 
which is equal to the Schwarzchild radius of a black hole of mass 
$1.6 \times 10^9$ M$_\odot$.  The mass of the black hole in 3C 273 has been
estimated to be around a billion solar masses which then 
justifies the origin of the black body
component in the (quasi)spherical pseudosurface formed near the event horizon.
The physical parameters have emerged from observations without requiring any
contrived explanations.

It was found that a single temperature black body provided a good fit
to the observed SED and the best fit temperatures for several nuclei were estimated to lie
between 20000 and 30000 K \citep{1982ApJ...254...22M}.  The pseudosurface is
composed of degenerate matter e.g. degenerate electrons which have high conductivity.
The high conductivity ensures that the 
temperature is rapidly uniformised in the degenerate matter and hence the single temperature
fit that is observationally found to be the best fit, is 
actually expected if the emission was from degenerate matter.  From their 
fitting, \citet{1982ApJ...254...22M} found that the Balmer continuum had a 
temperature $\le 15000$ K.  This then supports its origin in
the BLR which is photoionized by the much hotter black body radiating pseudosurface.   
This physical system bears resemblance to the photoionized HII region around a 
hot star although the detailed physical parameters such as size, temperatures,
hardness of the radiation field etc are widely different. 

\begin{table}
\centering
\caption{The table lists the expected bolometric and V band magnitudes estimated from
the Stephen-Boltzmann law for different mass black holes assuming that the black body
radiation is from a spherical pseudosurface at the event horizon of the black hole. 
The estimates are shown for two values of temperature and the 
bolometric correction (BC) for $T=16000$ K is $-1.3$ and for $T=40000$ K is $-3.8$.
$M_V = M_b - BC$.  Bolometric absolute magnitude of the sun is taken as $-4.75$ magnitudes
and solar luminosity as $3.828\times10^{26}$ Watts.  
% For T=40000 K, $L = 1.641 \times10^{13} \times R^2/9}$
}
\begin{tabular}{c|c|c|c|c|c}
\hline
$\bf M_{BH}$ & $\bf R_{s,BH}$ & \multicolumn{2}{c}{$\bf T=16000 K$} &
\multicolumn{2}{|c}{$\bf T = 40000 K$} \\
$\bf M_\odot$  &  $\bf m$ & $\bf M_{b}$ & $\bf M_V$  & $\bf M_{b}$ & $\bf M_V$  \\
& & mag & mag & mag & mag \\
%& & (BC = -1.3)  & (BC = -3.8) \\
\hline
%$10^{11}$ & $3\times10^{14}$ & $-27.5$ & $-26.2$ & $ -31.8$ & $-28$ \\
$10^{10}$ & $3\times10^{13}$ & $-22.5$ & $-21.2$ & $ -26.8$ & $-23$ \\
$10^{9}$  & $3\times10^{12}$ & $-17.5$ & $-16.2$ & $-21.8$ & $-18$ \\
$10^{8}$  & $3\times10^{11}$ & $-12.5$ & $-11.2$ & $-16.8$ & $-13$ \\
$10^{6}$  & $3\times10^9$&  $-2.5$  &   $1.2$ & $-6.8$ & $-3$ \\
100       & $3\times10^5$ & $17.5$  &  18.8 & $13.2$ & $17$  \\
10        & $3\times10^4$ & $22.5$  & 23.8 & $18.2$ & $22$\\
\hline
\end{tabular}
\label{tab1}
\end{table}

If we assume that the pseudosurface forms at the event horizon, is a spherical 
surface and the entire surface radiates then
we can estimate the expected black body luminosity for a few different black hole 
masses and temperatures as listed in Table \ref{tab1}. 
The estimates demonstrate the steep dependence
of the absolute magnitude/luminosity on the black hole mass and temperature such that 
an order of magnitude change in the black hole mass (and hence $R_s$) 
will lead to a change of 5 magnitudes in the luminosity while  
a change in the temperature by a factor of 2.5 leads to a change of
4.3 magnitudes in the bolometric luminosity.  Importantly,
the estimated values cover the range of luminosities that are 
normally recorded for active nuclei with 
M$_{BH}$ ranging from $10^6$ to $10^{10}$ M$_\odot$.  
While all these values are determined assuming the entire pseudosurface
emits, it is possible that at times the emission we record is from part
of the pseudosurface because of obscuration and hence could be dimmer. 
These estimates also show that quasars which are the most luminous objects,
%and generally show the bump in the ultraviolet,
%Table \ref{tab1} shows that they 
host the most massive black holes with the hottest pseudosurfaces.
Seyfert 1 and BLRG/`N' type galaxies although less luminous than quasars, are also likely
to host massive black holes and/or hot pseudosurfaces explaining
their bright nuclear continuum.  The relatively faint nucleus in Seyfert 2 and FR I 
galaxies is likely due to a smaller radiating surface owing to a
low mass black hole and/or a cooler pseudosurface. 
The temperature of the pseudosurface can be estimated 
from the bump in the wideband nuclear spectra since the black body 
peak should move to longer wavelengths
in a cooler black body.  The measured temperature alongwith the luminosity can then
be used to estimate the area of the emitting surface and hence the mass
of the black hole assuming the entire pseudosurface emits black body radiation.
The effectiveness of such a method has already been demonstrated 
by \citet{1982ApJ...254...22M} and others who have determined the emitting surface
area and temperature of the black body component.
In literature, the origin of this black body component is often attributed to a hot 
accretion disk.  This appears to be the case since a black hole does not have a 
surface unlike white dwarfs and neutron stars. 
However with the formation of a hot pseudosurface composed of degenerate matter
around all accreting black holes, it has to be the source of black body radiation.
It rules out the origin in an accretion disk which is much further from the black
hole and saves us the trouble of having to hunt for models to explain the
heating of the radially extended accretion disk. 

The remaining two continuum emission components are also expected from the
structure around an accreting rotating black hole. 
The Balmer continuum component of the SED arises in the hot photoionized
non-polar BLR as mentioned earlier.  In addition to the hot pseudosurface and
BLR, there also exists the polar region where matter is accreted, ignited
and ejected at relativistic velocities.  Synchrotron emission which
shows a power law spectrum is observed
from the ejected jets.  This would suggest that the power law component of the SED would be 
due to synchrotron emission radiated by the population of relativistic electrons that is  
energised in the thermonuclear explosions frequently occuring at the poles of
an accreting black hole.   The electron energies will follow a normal distribution 
(see Section \ref{cosrays}) which readily explains the flatter or curved low radio frequency 
spectrum (due to lower energy electrons) as
compared to the higher frequency emission (due to higher energy electrons)
which is approximated by a power law.  
{\it To summarise: the origin of the three components which comprise the continuum emission 
from active nuclei are (1) black body emission
from the hot (quasi)spherical pseudosurface around the black hole 
(2) Balmer continuum from the hot emitting zone of the non-polar BLR (3) power law
synchrotron from the relativistic plasma energised and ejected from the poles. }
These origins imply that the SED of an active nucleus viewed pole-on should not include
any contribution from the Balmer component arising in the BLR while the SED of an object
viewed edge-on will not include any contribution from the synchrotron emission at the poles.

The three radiation components responsible for the wideband continuum spectrum that
are described above, will contribute varying amounts to emission at different wavelengths.
Soft X-rays and ultraviolet bands will contain a significant contribution from
the black body emission while emission at radio and hard X-rays will be
dominated by synchrotron emission from the radio jets.
The emission at optical and infrared wavelengths are likely to contain 
significant contributions
from black body radiation, synchrotron emission and Balmer continuum.
The energetic $\gamma-$rays that are often detected from flaring active nuclei can
be due to inverse Compton scattering of $\gamma-$ray photons that
are released in thermonuclear outbursts by relativistic electrons. 
It follows then that the variability detected in the continuum emission
at optical, ultraviolet, soft X-ray band
will be due to changes in the physical properties of the 
pseudosurface of the black hole, whereas the variability in the radio bands, hard 
X-rays, infrared bands will be associated with the polar changes 
such as new relativistic ejections, magnetic field inhomogeneities etc.

\noindent
{\bf Broad line region (BLR):}
The BLR, observationally characterised by broad permitted spectral lines, 
is formed on the pseudosurface in the non-polar regions from matter that
flows in from the accretion disk. It consists of emission and absorption zones
and is photoionized by the radiation field of the pseudosurface.
The matter in the BLR which is located next to the pseudosurface composed of
dense degenerate matter will also be of high densities. 
Densities of $10^8-10^{10}$ cm$^{-3}$ have been derived for the BLR
from photoionization arguments and lack of wide forbidden lines 
%{\bf reference?}.
The recombination (Balmer) component of the continuum emission has to arise in
the emission zone of the BLR as discussed above.  
The BLR is formed so close to the black hole that 
the spectral lines arising in it are often shifted by a measurable gravitational redshift. 
The main difference in the gravitational redshift component present in 
the wide emission and absorption
lines that are detected in the core spectra of quasars, BLRG, and Seyfert 1  galaxies is
in the quantum of $z_g$.  Moreover for a given active nucleus, 
the emission line redshifts are always greater
than the absorption line redshifts indicating that the emission line zone forms closer to the
black hole than the absorption line zone.  It then follows that the photoionizing source lies 
between the black hole and the BLR.  This gives further evidence to the existence
of the hot pseudosurface of the black hole.  In fact, noting that the emission line 
$z_g$ measured for some quasars is as large as 0.5 i.e. the line photons 
arise at a distance of $R_s$ from the black hole,  conclusively rules out the
accretion disk as the photoionizing source.  If the accretion disk was the ionizing
source, it would have to be located within $R_s$ which is impossible.  
In literature the BLR is often
considered to be located between the accretion disk and the dust torus since the
accretion disk is believed to be the photoionizing source.  This is now conclusively
ruled out as discussed here.   {\it The BLR
lies between the pseudosurface i.e. black hole and the accretion disk and is
photoionized by the radiation field of the hot pseudosurface.}
Most FR II radio galaxies are BLRG and it is found that even the FR II cores which
show the presence of only narrow lines in total flux show strong wide lines in 
polarised flux revealing the existence of a BLR \citep[e.g.][]{1999AJ....118.1963C}. 
Broad lines are generally not detected from the nuclei of Seyfert 2 and 
FR I galaxies.  However 
the existence of Seyferts with types between 1 and 2 in addition to changing 
look Seyfert galaxies which dynamically transition between the two types in addition
to existence of the rare FR I galaxy which show broad Balmer emission lines in their
nuclear spectrum indicate that there has to exist a continuum of properties for the BLR.
and that a BLR exists in all active nuclei.
The rarity of Seyfert 2 and FR I with wide lines prompts an interpretation that
a fraction of the detected narrow lines in these cores can arise in a BLR
which is located so far from the black hole that there is a negligible 
contribution from $z_g$.  Observations do find a range
of excitations in the narrow lines detected from Seyfert 2 which could be indicative
of origin of a fraction of these high excitation lines in the BLR but with widths which
are similar to the lines arising in the NLR making it difficult to differentiate
the two origins.  

If we assume that the many classes of active objects share the same space 
distribution since there
appears no a priori reason to assume distinct distributions, then   
the observed variation in the emission line redshifts of the different 
classes of active nuclei can be attributed to the varying contribution of 
the intrinsic component of the redshift i.e.  gravitational redshift.
This is expected from the differing separation of the emission line
zone in the BLR from the black hole in the different types of active nuclei.
We can, hence, use the observed emission line redshift which contains a measurable
component of gravitational redshift to estimate the separation of the
emission zone of the BLR from the black hole. 
The BLR is located at a separation $\le 30 R_s$ in quasars 
(as deduced from the lowest gravitational redshift
being $\sim 0.016$) and is within $2 R_s$ for half the quasars (as deduced from
the median gravitational redshift of 0.25).  Since the gravitational redshift
component in other active nuclei is generally smaller, 
it is assumed that the observed lowest redshift for a given
class of active nuclei is dominated by gravitational redshift so that a rough 
estimate of the separation of the BLR from the black hole can be obtained.  
The observed emission line redshifts of FR II/BLRG are generally greater than 0.003
(see Figure \ref{fig23}) indicating that the BLR is separated from the
black hole by $\le 170 R_s$ while those of FR~I and Seyferts is around 0.001
(see Figure \ref{fig23}) indicating that the BLR is located $\le 500 R_s$ from the 
black hole.  The GPS and CSS radio sources are detected at redshifts 
greater than 0.1 (see Figure \ref{fig21})
which would mean that the BLR is located at a separation $\le 5 R_s$.  
These GPS, CSS sources include objects classified as quasars and galaxies. 
When an entire class of objects systematically occur at high redshifts 
(e.g. quasars, GPS/CSS), it is important to first exhaustively explore intrinsic reasons 
before searching for exotic cosmological models which try to
explain their absence in the local universe.  This could prevent us from
embarking on a wild goose chase. 

The lines that are attributed to the BLR are very wide with FWZM ranging from 5000 to
30000 kms$^{-1}$ or so.  The BLR is located close to the event horizon of the black hole as 
conclusively demonstrated by the measurable gravitational redshift component
in the broad spectral lines.  In this case, the large widths of the lines from the BLR have 
to include a significant component due to the radially changing gravitational 
potential within the line forming zones of finite thickness.  In other words, 
line photons arising from
a range of radial separations from the black hole will be shifted by different
gravitational redshifts with the net effect being a broadened line. 
Additionally the higher ionization emission lines appear from the innermost zones
of the BLR and show larger gravitational redshifts and broadening while the
lower ionization emission lines and absorption lines appear from the outer
parts of the BLR and hence show lower gravitational redshifts and broadening. 
Thus the width of the broad lines and their gravitational redshift carries 
information on the radial extent and location of the line forming zone.  
The gravitational width ($\Delta \lambda$ expressed in the same units as wavelength 
$\lambda$) is related to the fractional thickness of the line zone 
$\Delta R/ R = \Delta \lambda / \lambda$ \citep{1964ApJ...140....1G}. 
The high ionization line C~IV 1550 A is typically observed with widths $\ge 50$A 
(i.e. a velocity width $\ge 10000$ kms$^{-1}$) from the BLR in quasars, Seyfert~1, BLRG. 
To make some quantitative estimates,  we assume that the pseudosurface and BLR in these
three classes of active nuclei typically form at a separation $\sim R$ from 
the black hole such that $R=2R_s$ ($z_g=0.25$) in quasars, $R=50 R_s$ ($z_g=0.01$) 
in BLRG and $R=250 R_s$ ($z_g=0.002$) for Seyfert~1.  For a linewidth of 50~A,
the thickness of the C~IV emitting
zones in the BLR will be $\Delta R = 0.0322 R $ i.e. $0.0644 R_s$, $1.6 R_s$ and $8 R_s$ 
respectively for quasars, BLRG and Seyfert~1.  
This tells us that for the comparable line widths that are observed, 
the emission zone in the BLR is much more radially compact in quasars than in Seyfert~1
for a given black hole mass. 
For a black hole of mass $10^9$ M$_\odot$, the thickness of the C IV zone will be 
$1.9\times10^8$ km, $4.8\times10^9$ km and $2.4\times10^{10}$ km - all within 
a light day ($2.59\times10^{10}$ km) for the three cases.  
For a $10^8$ M$_\odot$ black hole, the thickness of the C IV zone will be $1.9\times10^7$ km,
$4.8\times10^8$ km and $2.4\times10^{9}$ km.   All these explain the linewidth of
10000 kms$^{-1}$.  
If broader lines are detected, then the thickness of the region will be larger.  For example,
C IV lines of width 30000 kms$^{-1}$ ($\sim 150$ A) have been detected from NGC 4151 
which gives $\Delta R/R = 150/1550 = 0.0968$.  For this Seyfert galaxy for which the
black hole mass has been estimated to be $\sim 10^8$ M$_\odot$, the thickness
of the C IV zone would be about $24 R_s = 7.2\times10^9$ km. 
High ionization lines of such widths are commonly detected in quasars where black hole
masses are commonly estimated to be $\sim 10^9$ M$_\odot$.  This would result in 
the thickness of the C IV zone being $7.2\times10^{10}$ kms i.e. about 3 light days. 
Similar exercise can be done on emission and absorption lines of different
elements and different ionizations.
This method, then, allows us to unravel the radial structure of the BLR from the observed
gravitationally redshifted and broadened lines.  The results should
be comparable to those obtained from reverberation studies since the BLR spectral line 
variations are triggered by the photoionizing continuum of the pseudosurface. 

BLR is formed from the matter that is accreted through the accretion 
disk onto the pseudosurface of
the rotating black hole.  To recall, the event horizon around a rotating black hole
will be prolate-shaped whose semi-major axis (polar axis) is always equal to $R_s$ whereas
the semi-minor axis (equatorial extent) is $< R_s$ with its exact value 
depending on the spin. 
Noting that the local accretion rates are the lowest at the equator and 
gradually increase towards the poles means that the radial extent of
the BLR will be minimum at the equator 
and will increase towards the poles (see Figure \ref{fig25}).  
The poles are devoid of an accretion disk and a BLR. 
Thus, the BLR can be visualised to be arranged
in the form of an equatorially pinched, hourglass-shaped structure with 
small openings at the poles whose extent is determined by the spin of the black hole
(see Figure \ref{fig25}).
On the other hand, the accretion disk will accumulate maximum matter at the
equator and matter accumulation will decrease with latitude.  Hence 
the maximum radial extent of the accretion disk will be at the equator. 
The latitude coverage of the accretion disk will be similar to the BLR
and its inner contour will be similar to the outer contour of the BLR 
so that if the accretion disk was to move in towards the BLR, the two surfaces would
fit into each other (see Figure \ref{fig25}).

The BLR of a high spin black hole will have a larger covering factor by which 
we refer to the fraction of the event horizon/pseudosurface covered by the BLR as 
compared to low spin black holes.  The larger covering factor of the BLR 
of high spin black holes should make them detectable over most sightlines
except maybe directly polar. 
Since the BLR around a low spin black hole will have a smaller covering
factor and be confined to a smaller range of latitudes
about the equator, detection of wide lines from these BLR will be more sightline-dependent.

The emission from a latitude-thin BLR which is not in our sightline,
can be scattered by dust formed in the NLR or accretion disk into the sightline
making it detectable.  
The distribution of dust in the accretion disk should be constant over long timescales
while the distribution of the dust formed in the polar regions can be 
dynamic over shorter timescales.  If the main scatterers are
polar dust grains, then it provides another reason for the changing look active
nuclei.  The changing distribution of dust can change the properties of the
scattered radiation and the BLR can alternate between detection and non-detection.   
All this supports the existence of a BLR in all active black holes
with a spin-dependent covering factor.   
The covering factor of the BLR will approach unity in quasars and
BLRG which will also explain a small opening angle for the jet and NLR while it should
be small in Seyferts explaining the large jet angles when detected. 

Spectral observations of Seyfert 1 nuclei have found that broad emission lines generally 
respond immediately or within a few days to any change in the optical/ultraviolet
continuum emission.  It is also noted that the higher ionization lines vary more and
respond to the continuum changes before the lower ionization lines.  
The above is expected from the model presented here since the photoionizing 
is by the photons from the hot pseudosurface which lies between the black hole
and the BLR.   

{\it To summarise:  The matter accreted on the pseudosurface through the
accretion disk forms the dense BLR consisting of emitting and absorbing zones.  
Observationally, this region is characterised by gravitationally
redshifted and widened permitted spectral lines and Balmer continuum.  The BLR
should be arranged in a hourglass-like shape with conical openings at the poles
whose opening angle is determined by the spin of the black hole.  }

\noindent
{\bf Thermonuclear explosions and variability:}
In literature, the possible sources of the vast energy released in active nuclei
are believed to be gravitational and/or rotational and/or magnetic in origin. 
The above three sources of energy are, no doubt, very important in the evolution
of the active nucleus - gravitational contributing to the attraction and
arrangement of matter around the black hole, rotational contributing to 
the arrangement and rotation of accreted matter and magnetic contributing 
to providing a magnetic field for synchrotron processes.  These demonstrate
that the above three energy sources feed persistent processes.  However it 
is difficult to envisage these 
energy sources being responsible for energetic transient events which lead to 
rapid acceleration of electrons and launching of relativistic jets.  
These perplexing issues have been extensively explored in literature 
and processes that might be active far from the black hole are promoted 
such as shock acceleration of electrons and magnetic collimation of jets. 
However, a water-tight physical explanation for
basic jet-related issues like launching and collimation remain elusive and
hence continue to perplex us with explanations in literature ranging from launching
by the accretion disk to magnetic launching.  We point out a 
simple physical process, that should be active close to the accreting
black hole but has somehow been overlooked.  This mechanism
can provide energy to launch jets, accelerate electrons to relativistic velocities
and consistently explain several observational results on active nuclei. 
This physical process is widely observed in the universe and hence requires no validation.
In fact, it is the most obvious solution to the energy problem considering the 
physical conditions that should prevail in the accreted matter on the black hole
and it is somewhat of a surprise that it has not been a favoured mechanism. 

This important transient source of the vast energy released in active nuclei has to
be catastrophic thermonuclear reactions.  
This follows from the fact that active nuclei consist of an accreting black hole
wherein matter accumulates on the hot pseudosurface and should be 
progressively compressed and 
heated setting up physical conditions conducive to thermonuclear reactions.  
That such reactions can be highly explosive was convincingly demonstrated when
it was shown that simultaneous ignition of $0.1$ M$_\odot$ of accreted matter 
will release $\sim 10^{50}$ ergs of energy in $1-100$ seconds 
\citep{1960ApJ...132..565H}.  Such an explosion can be expected 
to occur in the accreted matter on the pseudosurface - leading
to acceleration of matter and launching of jets from the poles and  
nucleosynthesis which enriches the matter and variability in the non-polar BLR.
Observations have reported higher metallicity in the BLR which also supports 
nucleosynthesis in thermonuclear reactions.  
The ejected matter at the poles should consist of accelerated atoms/ions 
and relativistic electrons and observationally these are 
identified as the narrow line region (NLR) 
and radio synchrotron jets respectively.  Since the energy source for the entire ejecta 
is the same thermonuclear outburst, this explains the observational results which have 
long inferred a common energy source powering jets and NLR. 
There might also be explosive episodes at the poles wherein the energy released
is insufficient for matter to escape 
so that the enriched matter accumulates on the pseudosurface at the poles.  In
such active nuclei, the accumulated matter can show higher metallicity 
and the pseudosurface at the poles can thicken as more matter is retained. 

Correlated variability is noted in the ultraviolet/optical continuum
strengths and amplitude/linewidths of broad emission lines such that when the
continuum emission increases, the amplitude and linewidths of the broad lines also increase
either instantaneously or with a delay.  We explore if a thermonuclear explosion 
in the accreted matter in the non-polar regions i.e. in the BLR can explain
such behaviour.  Energy released in the thermonuclear explosion
in the BLR should heat the BLR and/or the pseudosurface which should increase the continuum
emission - Balmer continuum or black body radiation.  Such changes in the continuum
should trigger physical changes in the BLR which can consist of exciting a larger
number of atoms of high ionization lines thus increasing the amplitude
of the observed lines.  Such a change can also increase the radial
extent of the line forming zone in the BLR which can widen the line as the 
change in gravitational potential within the line forming zone increases. 
Thus, the energy input from the thermonuclear outburst can explain both the observables.
This means that the source of the correlated continuum and line variability 
in the BLR can be the energy released in a thermonuclear explosion in the BLR. 

The origin of the observed multi-band variability whether in polar or
non-polar regions of the active black hole lies in thermonuclear energy - this
could be either by direct heating or through acceleration of relativistic electrons.  
It needs to be appreciated
that in the currently hydrogen-dominated universe, thermonuclear energy 
has to be one of the most important driving forces of transient energetic events
especially in systems where there is ongoing accretion of normal matter. 
It will cease to be an energy injection mechanism when all the light elements 
have fused to form iron and the universe has become iron-dominated. 
The nature of the universe would then change and if transient events are seen to persist
then we would be forced to look for an alternative energy source.
{\it To summarise, thermonuclear explosions provide energy to fuel transient
events in active black holes - launching of jets and NLR from the poles; variability
and enrichment in the BLR.} 

\noindent
{\bf Redshift components:} 
The polar matter which is energised in a thermonuclear explosion has to be accelerated
to at least the escape velocity to enable it to leave the system. 
The escape velocity that is required will depend on the separation of the 
ejection site from the black hole.  For example, if matter is ejected from a site 
separated from the black hole by $2 R_s$, then it has to have a bulk velocity of $0.707c$ 
i.e. bulk Lorentz factor $\gamma_{bulk} = 1/\sqrt{1-v^2/c^2} = 1.414$ to escape the 
black hole.  The frequently deduced relativistic bulk velocities from knot expansion in the
jets and Doppler boosting leading to detection of solitary jets in
powerful radio sources, support ejection from regions located close to the black hole. 
It appears reasonable to assume that matter is efficiently expelled from the system
when the energy distribution of the particles, which will be normal 
(Gaussian) as discussed earlier, is centred at least on the escape 
velocity.  If the distribution is centred at a lower velocity, the matter ejection is
likely to be inefficient.

If any spectral line was
detectable just as the matter was ejected as it has been possible in the case of
the Galactic microquasar SS 433, then this line should contain three 
distinct redshift components: (1) Lorentz factor for bulk expansion
(2) gravitational redshift  and (3) cosmological redshift.  The emission lines detected 
with varying redshift from SS 433 contains contributions from (1) and (2) as discussed
in the section on microquasars.  Since SS 433 is a Galactic source there was no
contribution from (3).  Moreover as was observed in SS 433, the bipolar expansion of
the NLR will be observed as a blueshifted and a redshifted component about the
cosmological redshift.  If the bipolar ejections occur from sites of comparable separation
from the black hole then the gravitational redshifts that the blue and redshifted 
lines show will be similar.  This means that both the redshifted and 
blueshifted components have to sit on a common redshift pedestal which will 
be a combination of the gravitational redshift
and the cosmological redshift.  This behaviour has also been demonstrated by the
emission lines from SS 433. 

Both the bulk Lorentz factor i.e. escape velocity and the gravitational 
redshift depend on the separation of the launch site from the black hole 
and contribute independent redshift components to the spectral line. 
The bulk velocity is actual motion of the line forming region to escape
the gravity of the black hole whereas the 
gravitational redshift is the wavelength shift a photon suffers due to its formation 
in a deep gravitational potential.  For example, a line photon that is emitted 
from $2R_s$ will be subject to a gravitational potential of $z_g=0.25$ 
which will contribute an intrinsic redshift $z_{in}=0.333$ while the redshift due to
the bulk velocity (escape velocity) will be 1.414.  Excluding the contribution 
due to cosmological
redshift, the redshifted emission line from the poles should be detected at a redshift of
$(1+z) = (1+z_{in})(1+z_{bulk}) = 2.218$ i.e. $z=1.218$ whereas the blueshifted
component will appear at a widely different velocity.  
As the ejected matter expands away from the black hole it forms the distributed NLR.
If the bulk expansion velocity of the ejecta remains constant over sub-parsec scales,
then the contribution of $z_{bulk}$ to the spectral line velocity will remain unchanged
for lines arising from that region.  However $z_{in}$ will 
rapidly decrease as the photon in the
expanding NLR emerges progressively farther from the black hole.
$z_{in}$ would have declined to 0.005 for a line photon emerging at $100R_s$. 
The observational implication for this is that the spectral line should appear
at progressively smaller velocity shifts along the jet as $z_g$ approaches zero.  
Once the contribution of $z_g$ has disappeared then the emission lines from 
NLR should show two components centred at $\pm z_{bulk}=1.414$ about the
cosmological redshift i.e. a blueshifted and redshifted component for the bipolar NLR.
If the bulk velocity of the NLR declines over parsec scales
then $z_{bulk}$ will accordingly decrease and eventually approach the 
commonly observed few hundred kms$^{-1}$ 
about the systemic velocity of the galaxy on parsec and kpc scales.  
Decrease in the velocity of the NLR decreases as it expands away from the black hole
is expected.   This can also be deduced from the observational results on 
supernovae explosions which tell us that the initially recorded Balmer line velocities 
of 30000-40000 kms$^{-1}$ decline over
several weeks to $\sim5000$ kms$^{-1}$.  Since these velocities signify the
expansion rates of the supernova ejecta, it shows that the expansion slows down as the
ejecta expands.   A similar evolution is likely followed
by the NLR so that it is ejected from the poles at relativistic velocities 
required to escape the gravity of the black hole but decelerates over parsec
scales to a few hundred to thousand kms$^{-1}$ which is the commonly detected 
expansion speeds in the NLR as
surmised from the blueshifted and redshifted components of the narrow lines. 
Since no spectral lines attributed to the NLR have ever been detected with
relativistic expansion velocities from an active nucleus; it tends to bias
us towards their non-existence while the detection of high velocity lines
in SS~433 supports the model suggested here wherein the NLR should always
contain such lines.  There remains the strong 
possibility that observations have missed the relativistically
expanding NLR close to the core.  This could be due 
an observational bias in that we have not really probed the NLR at such high redshifts
or that the emission line gas rapidly slows down after ejection.  
It is interesting to note that in quasars, wide permitted emission lines 
and narrow forbidden lines are observed at similar redshifts.  Since it was established
that the wide lines contain a non-trivial contribution from gravitational redshift and
arise close to the black hole, the narrow forbidden lines were suggested to be arising
in small low density pockets in the same region (Kantharia 2016).  However it now
appears that the possibility
of these forbidden lines arising in the matter that has just been ejected from the
poles i.e. NLR cannot be ruled out and hence their origin requires further 
investigation using observational data.  It is also important to note that 
the number of quasars which show the presence of
forbidden lines decreases as the emission line redshift of the quasar increases. 
The reason for this needs to be investigated.  One can expect
the forbidden lines from the NLR to show extremely large redshifts when 
at the launch site (even larger than shown by lines from the BLR) and then consistently
decrease along the radio jet direction
as both the gravitational redshift and bulk velocity components
decline.  Considering that in several broad line galaxies, 
a narrow feature is seen riding the broad emission line prompts the question
if the narrow feature can arise in the polar NLR.  This point also requires
further investigation since its validity hinges on the similar net
redshift of the lines arising at the poles and in the BLR in some quasars
which, as explained above can happen since the lines from the NLR can also appear
at a range of redshifts except these can also include forbidden lines
and will be narrower than those from the BLR
and dependent on how often matter is ejected from the poles. 

There exists some evidence regarding the presence of high velocity narrow line gas 
in compact steep spectrum sources.  GPS and CSS are compact radio sources 
with typical sizes of GPS sources being a kpc or so and CSS extending to $\le 20$ kpc. 
Broad lines are detected from these sources.  The lowest redshifts at which these  
objects are detected is fairly high $\sim 0.1$ (see Figure \ref{fig21}).
Since there seems to be no compelling reason to believe that GPS and CSS sources
should avoid the local universe, we go with the assumption that these sources
share the same distribution as other active galaxies and the absence of lower
redshift objects is because the lower redshift limit is dictated by $z_g$
which is large.  This indicates that the BLR in
GPS/CSS sources lies within $5 R_s$ of the black hole. 
What is of interest here is that spectral lines from the NLR, extended along 
the radio jets, are detected at the relatively large emission line redshifts of these 
sources and in several cases is observed to extend beyond the radio jet
\citep{2000AJ....120.2284A} (see Figure \ref{fig20}).  
This is significant, since the observed emission line redshifts of these sources contain
a sizeable contribution from $z_g$ which the extended NLR should not include.
The large redshift of the NLR lines then indicate gas expanding at high velocities 
i.e. the origin of the large redshift of the lines from the NLR and of the broad 
lines are distinct but can be comparable at some stage. 
The optical line emission distribution from the NLR is abruptly cut off and 
hence appears boxy in a few CSS (see Figure \ref{fig21}) 
e.g. 3C 147, 3C 48 \citep{2000AJ....120.2284A}.  Such a morphology makes one
wonder if the NLR does extend beyond the boxy boundary but at a lower redshift
and hence has been missed.  In other words, if the NLR gas is slowing down along
its length then it will show a gradient in
expansion velocity till it is close to 
the cosmological redshift of the host galaxy and the observations centred
at high redshifts will miss the lower redshift extended components of the NLR. 
Further observational studies for quantifying gravitational
redshift components and bulk velocities for these sources are required. 
The spectrum of the host galaxy, when detected, should appear at a much lower 
redshift - devoid of both gravitational redshift and bulk polar motions.

{\it To summarise, the spectral lines from the BLR should contain two main redshift
components: (1) cosmological redshift (2) gravitational redshift.  The lines
from the compact NLR can contain three redshift components: (1) cosmological redshift
(2) gravitational redshift (3) redshift due to expansion.  
The distributed NLR will only contain two components i.e. 1 and 3.
The lines from the NLR in CSS sources seem to show very high expansion
velocities.  }

\noindent
{\bf Radio jets, lobes, hotspots:}
As mentioned above, following the thermonuclear explosion in the small 
polar region of the pseudosurface close to the black hole, matter is energised
to escape velocities and then ballistically shot out 
at relativistic velocities.  The radially ejected matter is observed in the form of
synchrotron emitting jets and the optical line emitting NLR extended in
the general direction of the jet.   
The large forward momentum of the synchrotron jet propels it and since it is
launched from a small polar region, it is highly collimated.   
Only when the forward momentum starts declining that  
the jet starts decelerating and losing its collimation.  The deceleration should
set in sooner if the ambient medium is dense or if the ejection velocities are low.
The jets in FR I sources lose their collimation sooner than in FR II sources
which sometimes show collimated jets upto Mpc scales. 
As the forward momentum declines, the synchrotron jet diffuses and thickens into a lobe. 
This explanation is generally accepted as the lobe formation mechanism in 
FR I radio sources.
However in FR II radio sources, a narrow collimated jet is detected all 
the way to the hotspot which is generally the extremum point of the entire radio 
structure.  This would mean that the forward momentum of the jet dominates 
upto the hotspot.  The lobe in FR II sources forms between
the core and the hotspot and literature attributes its formation to a backflow from
the hotspot due to the formation of a reverse shock at the hotspot.  
The hotspots are generally formed at the edge of the soft
X-ray distribution in the radio galaxy (e.g. Figure \ref{fig33}). The hotspots  
typically subtend angles of a degree or less at the core      
which supports small jet opening angles and collimation of the
jet upto hotspots in FR II sources \citep[e.g.][]{1984ARA&A..22..319B}.
Magnetic field in lobes is observed to be circumferential for both FR I and FR II galaxies.
A circumferential field can be understood since the
jet diffuses to form the lobe so that the frozen-in field is 
stretched and eventually becomes circumferential.

\begin{figure}
\centering
\includegraphics[width=7cm]{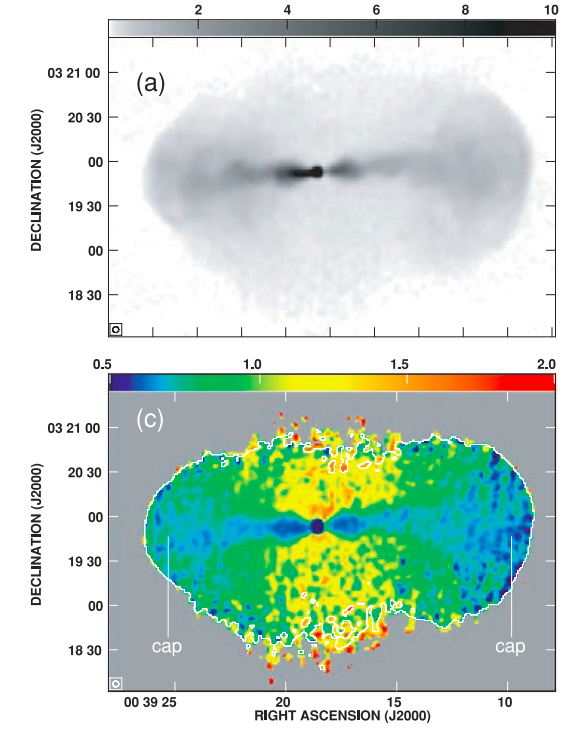} (a)
\includegraphics[width=5cm]{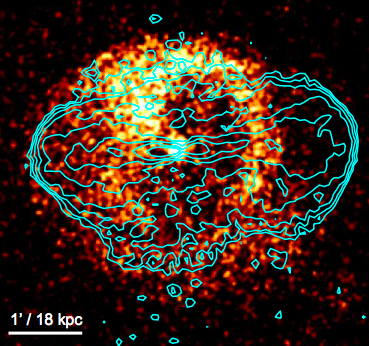} (b)
\caption{\small (a) Radio image of the FR~I galaxy NGC~193 at 4.9 GHz (top) and 
spectral index between 4.9 and 1.4 GHz at 4'' resolution 
reproduced from \citet{2011MNRAS.417.2789L}.
Notice the flaring jets and the halo-like radio emission surrounding the active nucleus.  The
spectral index is flattest for the core and jets and is steeper in the lobes.
The emission in the north-south of the core shows the steepest spectrum. 
Fragments of flat spectrum emission are detected at the boundary of the lobes. 
(b) Figure showing
X-ray emission in colour superposed on the radio contours around NGC~193 reproduced from
\citet{2014ASInC..13..259O}. The radio extent of the galaxy at 60 Mpc is about 80 kpc.
Notice that the radio and X-ray extents are comparable except  
in the west.    }
\label{fig27}
\end{figure}

In some FR I sources, the jet appears to be abruptly halted at some distance
from the core so that the radio lobe emission at all radio frequencies is 
confined within a well defined boundary (e.g. NGC~193 in Figure \ref{fig27}).  
The spectrum of emission from this interface region is observed to be similar to 
the jet i.e. a relatively flat spectrum as compared to the lobe emission 
(e.g. NGC~193 in Figure \ref{fig27}).  East-west directed jets are observed in
NGC~193 while radio emission is observed from a cylindrical region centred on the core. 
The radio jet in NGC~193 is likely to be precessing which explains the nature of
the lobe emission and the series of flat spectrum hotspots that are identifiable
in the interface region in the lower panel of Figure \ref{fig27}(a).
The radio emission in the north-south of the core in NGC 193 is faint and
is not confined within a well defined boundary.  
Such behaviour although not typical of FR I sources, is noticed in several FR I sources.
Such sources appear to be confined due to some external force.
The X-ray emitting thermal gas is of comparable extent as the diffuse radio emission 
in NGC 193 except in the western part (see Figure \ref{fig27}b).  
 
While double-sided jets are detected in FR I sources
with the jet-counterjet intensity ratios varying along the jet,
only single-sided jets are detected in FR II sources.  Literature attributes this
behaviour in FR II sources 
to Doppler boosting of the jet directed towards us and Doppler fading of the jet directed
away from us and observations have amply justifed this reason. 
Studies determine that the jet-counterjet asymmetry 
can be explained by Doppler boosting for relativistic jet velocities
between 0.65c and 0.8c and jet viewing angles between $45^\circ$ and $75^\circ$ 
\citep[e.g.][]{1997MNRAS.286..425W}.  
This means that the jet bulk velocities in FR II sources are $\ge 0.65c$ indicating
the jet launch site is separated from the black hole by $\le 2.4 R_s$. 
Although a single jet is detectable, 
the existence of double-sided jets in FR II sources, 
is surmised from the existence of symmetric lobes and hotspots located on
either side of the core and from the detection of both jets in the 
rare closeby FR II galaxy like Cygnus~A.  
In FR I sources, the jet-counterjet  ratio is large near the core i.e. the counterjet is
faint and the ratio  decreases along the jet i.e. the counterjet gets brighter.
This is explained in literature by the Doppler beaming argument 
and interpreted to mean that the jets in FR I galaxies are relativistic close to
the core but soon slow down so that the Doppler beaming effects are maximum
near the core but decline along the jet
\citep[e.g.][]{2011MNRAS.417.2789L}.  Using the Doppler beaming argument on
the non-detection of the counterjet in FR II galaxies leads to the
inference that the jets remain relativistic over their entire length i.e. 
upto the hotspot \citep{1989LNP...327...27L}.  These explanations
have remained well-supported by observations.  

In FR~II sources, a pair of hotspots can always be identified which are collinear 
with the core.  In literature, the hotspots are often considered as sites of
electron acceleration.   In many FR~II sources, multiple pairs of hotspots are detected 
with all pairs showing collinearity with the core.  The presence of multiple
pairs of hotspots provide
evidence to the existence of a precessing jet with the separation between the farthest
hotspots on the same side giving a measure of the precession angle which is generally 
rather small.  In FR~I sources, the precession angle appears to be larger than in
FR~II sources.   Since no hotspots are detected in FR~I sources, the precession 
angle in several FR~I sources is surmised from the thick lobes which straddle 
the jet such that the radio structure forms
an halo around the source e.g. M~87 and NGC~193 (Figure \ref{fig27}) and from
wiggles in the jets. 
Thus both the jet opening angle and precession angle in FR~I sources appear
to be larger than in FR~II sources. 
Till the magnetic field is observed to be aligned along the radio jet, 
the electrons whose relativistic random velocity is predominantly
along the radio jet will not radiate but travel freely \citep{1979ApJ...232L...7S} 
which should lead to lower energy losses and hence lengthen their lifetimes.  
We can extrapolate this argument
so that, at the hotspots where the magnetic field is oriented
perpendicular to the radio jet, all the electrons with the relativistic
velocity component along the jet will furiously begin to radiate 
synchrotron emission.  This can be an additional reason for the 
bright multi-band nature of hotspots where the magnetic field is enhanced compared
to the jets and lobes.  The higher magnetic field will also favour emission from
lower energy electrons which should be larger in number and have longer lifetimes.   
Basically there does not appear to be a compelling reason for attributing
electron acceleration at hotspots.   For example, if the hotspots are formed about
100 kpc from the core, this means that the electrons accelerated at the core and
travelling outwards at say a velocity of 0.8c 
should have sufficient energy to radiate at the hotspots.  For a mean magnetic
field of $\sim 30 \mu$G from the core to the hotspots, the electrons 
radiating at 1.4 GHz will have
a lifetime of 5.45 million years.  At a velocity of 0.8c, these electrons in
the jet can survive upto a distance of about 450 kpc while at a velocity
of 0.5c, the jet can survive upto about 250 kpc.  Thus, the electrons
accelerated at the core can explain the hotspots and there is no
compelling reason to require reacceleration of electrons at the hotspots. 

X-ray synchrotron jets are often detected close to the core in active galaxies 
and which indicates the existence of electrons with TeV energies in the jet.
We can infer from this that the central thermonuclear engine is 
capable of accelerating particles to these energies
before they are ejected.  This also provides direct evidence to the 
origin of high energy cosmic rays from active nuclei.  The early dousing of X-ray 
emission i.e. the shorter X-ray jet
compared to radio jets is expected since the TeV electrons should quickly lose
energy in magnetic fields of few tens of $\mu$G in the jets.
In literature, the physical mechanism that is surmised for the
X-ray emission from hotspots in FR~II sources
is inverse Compton emission or Self-Synchrotron Compton process. 
Synchrotron emission at X-rays, which requires exceptionally
high energy electrons, is seldom cited as the process for the X-ray emission 
from hotspots which also supports the inference that we arrive at namely 
that all the major electron acceleration is completed in the core and
hotspots are not sites of reacceleration.  

The radio spectra of the different components of a radio galaxy show distinct
behaviours.  The core spectrum is generally flat or peaked at high radio frequencies, 
the jet spectrum is a power law with typical
spectral index $\sim 0.6$, the lobe spectrum is also a power law but generally steeper than 
jets while the hotspot spectrum peaks at low radio frequencies and
shows a gentle break at a higher radio frequency like 10 GHz beyond which 
the spectrum steepens to a spectral index $\sim 1$.  As noted earlier in the paper, 
the explosion will result in  
a normal distribution of electron energies and the peak of the distribution
should move to lower energies and the dispersion of the distribution should
reduce as the population ages.  Depending on the magnetic 
field, electrons in a particular range of energies will radiate within a given range 
of frequencies.  Since the magnetic field in the hotspots is
enhanced, the electron energies which give rise to the radio emission in a 
given band of frequencies are lower than in regions with lower magnetic field.  
The curved low radio frequency spectrum
can be understood if we visualise the lower energy electrons close to the peak
of the aged distribution radiating in the higher magnetic fields.  
The steeper spectrum at higher radio frequencies would indicate enhanced losses due to
both synchrotron and synchrotron self-Compton which is often the physical process
invoked to explain the observed X-ray emission from hotspots. 
Although literature suggests electron reacceleration in the hotspots, it does
not seem to be warranted by observations.  There could be some reacceleration due
to the Compton process but is trivial compared to the acceleration in the core. 

Several radio jets are observed to show small wiggles along their length.
These wiggles in jets can be explained by precession of the
spin axis of the black hole as was conclusively shown for the microquasar SS~433. 
%{\bf reference}.
The presence of symmetric wiggles in jets on either side of the core in active nuclei 
lend strong support to their origin being related to
the central object and not the ambient medium.
The easiest model to explain a precessing spin axis is the presence of a companion
in a binary system.  However while this readily explains the precession in 
microquasars, it is difficult to imagine a quasi-stable binary system of two
supermassive black holes.  A binary of a supermassive black hole with a stellar
mass black hole is unlikely to result in a detectable precession of the spin axis
of the supermassive black hole and hence can be ruled out.  Since formation of 
supermassive black holes remains an ill understood topic, we avoid supporting 
binary supermassive black holes.  Instead other possible scenarios need to 
be examined using our current knowledge of the environment of the supermassive 
black hole.  Precession of the spin axis can be triggered due to the 
effect of a gravitational torque.  In Table 1 of \citet{2016arXiv160604242K},
the effect of a gravitational torque that is exerted on the matter in
two galaxies due to their close approach has been quantified in a simple
way.  The close approach of the two galaxies enhances the gravitational force
between them which is not symmetric and 
hence results in a torque i.e. induces a change in the angular momentum
of the galaxy.  The two components of angular momentum are the moment of inertia
and angular velocity and the change can be reflected in either component 
as was described in \citet{2016arXiv160604242K}.  From Table 1 in that
paper, we note that when two massive galaxies of mass $10^{11}$ M$_\odot$
approach within 100 kpc, then a mass of $10^9$ M$_\odot$ within one
of the galaxies can be torqued such
that it results in an equivalent velocity of about 200 kms$^{-1}$. 
In this simplified calculation, no adjustment for the finite sizes of the galaxies
or the fact that matter is distributed over a region comparable to the separation
was made.  The aim of that paper was to demonstrate that a gravitational torque
due to the approach of another galaxy 
can disturb the kinematics and spatial distribution of gas in a galaxy 
even for a closest approach which was as large as a Mpc. 
A lower mass packet can experience a stronger torque.  If we assume that the
central black hole is also subject to the torque then it can result in
precession of the spin axis i.e. change in the angular velocity of the black hole.
Change in moment of inertia of the black hole is much more difficult to
implement due to its enormous gravity.  However the distribution of gas in the
galaxy can change.  The detailed mechanism of how the black hole spin is changed 
remains a mystery as does the mechanism which actually implements the changed 
angular momentum in the gas in a torqued galaxy.
Moreover it should be mentioned that the gravitational torque which
was successful in explaining the velocity or spatial displacement of the stellar
and gaseous components, is observed to mostly affect the gaseous component
and not the stellar component.  So how does it affect the central black hole
which hosts the highest concentration of matter in the entire galaxy ?
Maybe when we discuss the effect of the torque, it has to be on the central
object which is then transmitted to the gaseous component through viscosity. 
This needs to be further investigated.  From existing data and knowledge,
it appears that the precession of the spin axis
of the black hole is likely to be connected to the passage of a massive
galaxy in the vicinity.  That several active galaxies are observed to have 
companion galaxies lends support to the significant role played by gravitational
torque in the evolution of the central black hole and the galaxy. 

\begin{figure}[t]
\includegraphics[width=8cm]{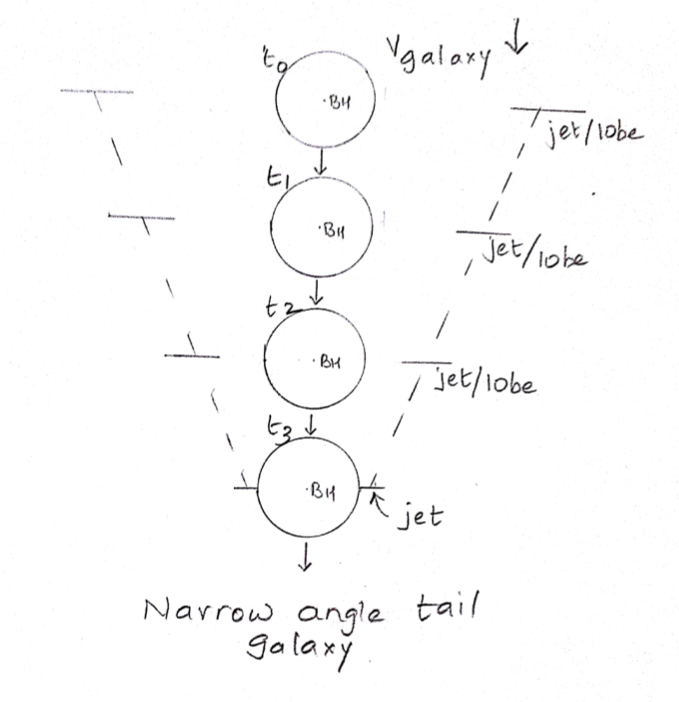}
\caption{\small The schematic explains the formation of the radio structure of
a narrow angle tail (NAT) galaxy.  The galaxy is moving at high spatial velocity from 
the top to the bottom.
At each position, the galaxy ejects plasma which ballistically expands away from
the nucleus while the galaxy continues its downward motion. 
At time $t_3$, the plasma from the previous outbursts
has expanded away from the position of the black hole at times $t_1$ and $t_2$,
so that the radio structure shows a V-shaped or
U-shaped structure.  The older plasma loses its ballistic nature
and diffuses.  If the central galaxy is moving slowly or sideways then
it would lead to different radio structures explaining the formation of NATs and WATs.  
This schematic shows that enhanced space motion of the galaxy under the influence
of the gravitational potential of the cluster can explain the observed structures.}
\label{fig28}
\end{figure}

Many low power radio sources show bent jets and such radio sources are referred
to as head-tail, narrow-angle tail (see Figure \ref{fig31}), 
wide-angle tail or S-shaped sources depending on their observed morphology.
Such sources commonly occur in cluster environments and hence their origin
has been attributed to the action of ram pressure exerted by the intracluster
medium on the jets launched from the galaxy moving in the cluster medium 
\citep{1979Natur.279..770B}.  
While this explanation is likely, it is noteworthy that the observed morphologies
can be explained simply by motion of the galaxy in the cluster. 
The schematic in Figure \ref{fig28} summarises the
picture.  The nucleus of the moving galaxy ejects matter from different positions
in space at different instants $t_0$, $t_1$...as shown in Figure \ref{fig28}.  
The jet knot that is ballistically ejected from the black hole at $t_0$ 
will keep expanding radially away from the position of the nucleus at $t_0$.
The same will be true for later times $t_1$, $t_2$...  The jet keeps expanding away
from the spatial position of the core when it was ejected and will be oblivious to 
the changed position of the core at later times. 
Assuming all knots are ejected with comparable velocities, the matter
ejected at an earlier epoch will have expanded further out than the later
eruptions (see Figure \ref{fig31}).
The earlier eruptions will also lose their ballistic nature and form lobes.  
The overall effect will be to lend a curved structure to the jets.  
The detailed morphology will depend on the velocity of the 
galaxy and the angle made by the sightline to the galaxy trajectory.  
A fast moving galaxy in the sky plane would result in a head-tail or narrow angle 
tail morphology as in Figure \ref{fig28} (e.g. NGC 1265 shown in 
Figure \ref{fig31}) whereas a galaxy with a larger component of motion 
along the sightline and smaller component in the sky plane would 
result in a wide-angle tail radio structure.
Thus a moving radio galaxy will show bent jets since the jet launching 
position is moving in space although the ejections are ballistic when launched. 
On the other hand, a stationary galaxy will always show 
straight jets since the jet launching position is constant in space.  
This suggests that the straight
jets of FR~II sources indicate very small space motion of the galaxy, if at all.
Most FR~II are field objects and hence lack of space motion is expected
since most of the space motion of FR~I is because they are members of clusters
and hence moving under the influence of the cluster potential. 
In fact, the larger frequency of cluster membership in case of FR~I can explain
a number of the differences between them and FR~II radio sources.  For example,
since FR~I radio sources are located in a crowded environment, the 
larger precession angle is expected from a gravitational torque induced by the
passage of another massive galaxy;
moreover it is also possible that the crowded environment has led to the slowing
down of the black hole and hence the active nucleus has evolved to a FR~I 
radio source instead of a FR~II source.

\noindent
{\bf Composition of jets:}
In literature, several studies are devoted to unravelling the composition
of the synchrotron jet - whether it is proton-electron or positron-electron.  
In other words, whether the radio jet is heavy or light.  Neither origin
has garnered convincing evidence from observations and the debate 
continues. 

Observations have provided ample evidence to the ejection of both a synchrotron-emitting
jet and emission line gas from the active nucleus -  both components are distributed along
the same axis, synchrotron blobs are observed to be expanding away at relativistic
velocities from the
nucleus, emission line gas has been observed to be expanding away from the
nucleus but with much lower velocities.  Such eruptions  
are expected from a thermonuclear explosion in the accreted matter at the poles. 
This means that the
synchrotron emission can arise from the relativistic electrons that are coincident
with protons/ions that give rise to the emission lines i.e. a proton-electron
plasma or it could also be that the synchrotron emission arises in a separate 
component comprised of electrons and positrons.  Thus while the thermonuclear 
explosion can explain the energising and ejection of matter from the black hole
and also provide an intense source of positrons,
it does not provide a water-tight resolution to the jet composition problem.  
It should be mentioned that in case of stars, the positrons generated 
in thermonuclear reactions are assumed to be immediately destroyed releasing
the annihilation photons around 511 keV.  However countering this assumption are
the observations which have detected the annihilation photons from the sun indicating
that all positrons are not destroyed on formation but some fraction does make
it to the solar surface.  Here we suggest that a large fraction of
the energetic positrons generated in the thermonuclear explosion
are ejected alongwith relativistic electrons instead of being destroyed.
At this point it is useful to mention that the annihilation line at 511 keV has
been detected from microquasars e.g. 1E 1740.7-2942 
\citep{1991ApJ...383L..45B,1991ApJ...383L..49S} which
lends support to the positron-electron plasma component of the ejecta. 
It is important to appreciate that the fate of positrons i.e. destruction
or survival is a studied assumption and we need to use observations to validate
the assumption.   
In fact the annihilation line was observed from the sun during solar flares
in 2002 and 2003 (i.e. near solar maximum) \citep{2004ApJ...615L.169S}.
A possible origin for the positrons involved in this could be those generated
in the thermonuclear fusion in the solar core which have managed to
reach the solar surface.  It is generally believed that the positrons
generated in the core are immediately absorbed - however no empirical evidence
for the same exists and hence a fraction of these emerging on the solar
surface cannot be ruled out.   

We might be able to obtain inputs on jet composition from observations of 
supernova explosions which are also 
thermonuclear eruptions and where the observed radiation also consists of synchrotron and
line emission.  Two separate radio synchrotron emission components from 
supernovae are recorded - the prompt radio emission following the supernova
explosion which has to be from a fast light ejecta i.e. positron-electron plasma
facilitating high forward velocities and the delayed radio synchrotron 
emission from the supernova remnant which is coincident with
the slower moving line emitting matter and hence arising in a proton-electron plasma.
In the microquasar SS 433, both the radio jet and the emission line gas are
observed to be expanding at a velocity of 0.26c - if in analogy to supernovae,
we assume that the radio jet is composed of fast positron-electron plasma then
the observations of SS~433 indicate that both the plasma are ejected with
high velocities which is expected since they are ejected from close to the black hole
where the escape velocities are relativistic.  As in supernovae, it is likely
that the heavy ejecta (i.e. line forming gas) from active nuclei rapidly slow 
down after ejection.  One would expect the situation in active nuclei to be similar 
to SS~433 i.e. both the synchrotron emitting plasma and line emitting gas ejected
at relativistic velocities especially since the matter is launched from a
few $R_s$ from the black hole.  However it needs to be investigated 
since there could be a range of possible
outcomes especially since no emission lines expanding with relativistic velocities
have been identified from an active nucleus.  While there exists a strong likelihood 
of all matter being ejected simultaneously, it could be that a fraction of the lighter
plasma acquires an energy distribution centred on higher 
velocities than the heavier matter and hence shoots out with higher expansion
velocities while the heavy matter fails to escape or it could be that all 
matter is ejected at relativistic velocities but the lighter plasma keeps ballistically
expanding at the same velocity while
while the proton-electron plasma quickly decelerates.  In some active nuclei,
it could be that the electron-proton plasma is the main synchrotron emitter.  
With the range in physical properties of the active nuclei that are possible, it
is likely that all the above scenarios are possible and manifested in different
active nuclei.  The lack of any relativistic line velocities from the NLR in
active nuclei support a distinct origin for the radio synchrotron emission. 
If the synchrotron plasma and NLR were associated then the latter should have
exhibited the large expansion velocities that are observed for the radio blobs. 
This, then, strongly supports the origin of the radio synchrotron emission in
active nuclei especially in jets from a positron-electron plasma.  
{\it To summarise, the composition of the radio synchrotron jet in active nuclei
is positron-electron plasma with a large pool of positrons being generated
in the thermonuclear explosion. }

\noindent
{\bf Formation of hotspots and lobes in FR~II:}
In FR II radio sources, the jet seems to be stopped at the hotspots although the ballistic
nature of the narrow jet indicates that the bulk velocity had continued to be relativistic
till the hotspot before the jet at all radio frequencies
suddenly lost its identity.  It appears as if the forward propagation of
the jet was suddenly braked at the hotspot where the relativistic plasma 
accumulated and radiated before being pushed backwards towards the core and forming
the lobe.  Since the medium beyond the hotspots is tenuous compared to
the medium closer to the host galaxy which consists of
X-ray emitting gas, emission line gas and radio plasma, the jet should have continued
unhindered in the low density medium.  What, then, stops the relativistic jet at the 
hotspots which is the interface between the dense medium around the active galaxy
and the tenuous medium beyond ?
In literature,  the abrupt nature of the jet termination, formation of hotspots 
and lobes is attributed to the effect of a reverse or termination shock travelling
towards the core from the interface. 
This explanation is vague and riddled with difficulty leaving it open to
individual interpretation.  We, hence,
revisit the problem in light of the increased understanding of active nuclei. 
All features at the hotspots point towards the influence of an external pressure
i.e. the jet is halted and the plasma starts to flow back towards the core.
The halting mechanism cannot be ram pressure   
since there is paucity of matter beyond the hotspots.  The other known halting
mechanism which is found to be active in several astrophysical systems
is radiation pressure.  For this mechanism, there needs to be a source of
hard photons beyond the jets which can halt the jets and form hotspots, lobes.  
Keeping in mind that the jet has a positron-electron composition, we 
investigate if there can be a source of photons beyond the detectable radio jet.
In the electron-positron jet, pair annihilation at some rate should be ongoing so that 
$\gamma-$ray photons of energy $\sim 511$ keV are continuously generated along the entire
radio jet.  Thus there should be soft $\gamma-$ray photons along the entire length
of the radio jet.  Since the jet is ploughing through a relatively dense medium 
as can be surmised from the soft X-ray emission and 
the optical emission lines, these $\gamma-$ray photons are likely to be scattered
to lower energies making detection of the annihilation photons difficult.  
However when the jet propagates beyond the boundary of the dense medium i.e. beyond hotspots 
and into rarified ambient densities, the $\gamma-$ray photons generated in the pair 
annihilation will no longer
be immediately absorbed and can instead retain their identity over a longer period due
to longer mean free paths.  This, then, provides a source of hard photons beyond
the hotspots, which can exert a radiation pressure on the jet!  
A positron-electron pair which annihilates will give 
rise to at least two $\gamma-$ray photons of energy 511 keV
moving in opposite directions.  The photons which move back towards the radio jet  
can exert radiation pressure on the jet and halt their forward motion so
that the plasma starts diffusing backwards and forms the lobes.
The radiation pressure can also compress the matter at the interface with the
intergalactic medium so that the plasma densities and magnetic field are 
enhanced leading to brighter radio emission i.e. hotspots.  This process can 
explain the formation
of hotspots, lobes in FR II and relics in clusters which form at the edge of the 
X-ray source.  A fraction of positron-electron plasma that continues to escape along 
the jet to the near-vacuum beyond the hotspots
will produce $\gamma-$rays of energy 511 keV which will continue providing the 
radiation pressure till the jet changes direction due to precession of the spin
axis of the black hole.  The old hotspot will gradually fade and a new hotspot will
form.  This explains the formation of multiple hotspots, their short lifetimes,
gradual formation and shift in their position alongwith the precessing jet.  
%{\bf Have gamma rays ever been detected from near hotspots?}
The annihilation photon can also provide the seed photons for inverse Compton
process so that
much higher energy photons can also be generated.  These photons can also
lose energy to relativistic electrons through the Compton process. 

The above strongly reiterates that there occurs no reacceleration of electrons
except through the Compton process at the hotspots.  The above explanation 
comprehensively describes the observed nature of FR~II radio sources without
requiring any further assumptions and lends convincing support to the positron-electron
composition of the jet and the implications thereof.   
{\it To summarise: Radiation pressure due to the positron-electron annihilation 
photons is responsible for halting the jet and forming hotspots, lobes in
FR~II radio sources.  }

\noindent
{\bf Formation of X-ray beams along radio jets:}
In several FR~II sources, two-sided thick X-ray beams coincident with the finer radio
jet(s) have been observed e.g. Cygnus~A and Pictor~A (see Figures \ref{fig33}, \ref{fig34}). 
Even if the FR~II galaxies show a single-sided radio jet, the X-ray beams are
observed to be double-sided.   We find that these X-ray beams can be understood
as the Compton-scattered annihilation photons that are continuously
generated along and around the radio jet thus leading to the thickening of
the X-ray emission beams. Since the X-ray beams are not moving outwards at the
rate at which the radio jet is expanding, no Doppler effect is expected and
hence X-ray beams are observed on both sides of the core.  

To demonstrate the feasibility of the above, we estimate the change in the wavelength of the
511 keV photon expected from Compton scattering by electrons
using Equation \ref{comp}.  If the scattering angle is $90^{\circ}$ then
the wavelength shift expected in the 511 keV photon
(i.e. photon of frequency $=1.2\times10^{20}$ Hz
or wavelength $=2.4\times10^{-12}$ m) by collision with an electron of $\gamma=10$
is $0.243 \times 10^{-12}$ m whereas scattering by an electron of
$\gamma=1$ will result in a wavelength shift of $2.43\times10^{-12}$ m.  If the
scattering angle is $180^{\circ}$ then the shift will be twice the above.
The wavelength of a 2 keV X-ray photon is $6.2 \times 10^{-10}$m.
Few Compton scatterings of the 511 keV photon are required to
downconvert the $\gamma-$ray photons to X-ray energies.  This physical explanation
can hence explain the X-ray beams coincident with the radio jet
without having to resort to any contrived arguments.
Since 511 keV corresponds to the rest mass energy of an electron, Compton scattering
by electrons is likely to be an important physical process that drain
photons of this energy and in turn accelerate the electrons.
If each Compton scattering event removes 50 keV from the photon then
it means that the involved electron gains this energy which although small
is non-zero.  This happens throughout the jet length.  This then has some
observable implications:  (1) deep observations might detect a
faint 511 keV annihilation line along the jet unless all photons are Compton
scattered to X-ray wavelengths  (2) X-ray photons should be detectable along
the jet which will not suffer Doppler beaming effects unlike synchrotron
radiation since these are produced by Compton scattering of the pair annihilation
photon.  This has already been demonstrated by the X-ray beams detected in a
few FR II sources all the way from the core to the hotspots.

\noindent
{\bf Magnetic field around a black hole:}
In a star, a magnetic field that exists within
the star can be ejected along with matter as in a supernova explosion.  However
a  magnetic field generated within a black hole will never be detectable.  However
a magnetic field whose source is located outside the black hole 
can thread the accreted matter on the pseudosurface. 
If we assume that the BLR is the source of the polar magnetic field, then
this field can be frozen in the accreted matter at the poles and ejected 
alongwith matter following episodic explosions.  
Since the ejecta is expanding away from the black hole, the magnetic field lines should
also be stretched along the jet.  This, then explains the observed magnetic field
in radio jets which is always observed to be aligned along the jet near the core in
FR~I and along its entire length in FR~II radio sources.
In other words, the magnetic field observed to be parallel to the jet 
strongly supports ejection of matter with a frozen-in magnetic field from
the poles of the black hole.  Since the ejection from the poles consists of
both positron-electron plasma which forms the radio jet and a proton-electron
plasma (line forming gas) which is the bulk of the accreted matter that is ejected,
the question that arises is whether the magnetic field is frozen in both plasma
or predominantly in the proton-electron plasma.  In supernova explosions,
observations support the frozen magnetic field in the massive
ejecta and not in the positron-electron ejecta.  It is likely that the same
exists for active nuclei so that the field is frozen in the proton-electron
plasma.  This would then require both the plasma to be expanding at relativistic 
velocities as long as the field is detected along the jet.  However if the
proton-electron plasma slows down soon after ejection, then the field has
to be frozen in the positron-electron jet.  If lack of high velocity
emission line gas is taken as proof of its absence along the jet then clearly
the field has to be frozen in the positron-electron jet.  However the
entire velocity space has not been investigated for the emission line gas and
hence it is premature to accept its absence along the radio jet.   
These are all possible, physically
allowed situations and although not presented here, it might be possible to
obtain convincing arguments one way or other from observations.     

Literature often suggests the presence of a helical field around the jet
and which is also used as a collimating agent. 
If the magnetic field was not frozen in the ejected material then it can
be generated by a net current flow in the jet.  Such a current
should lead to a helical field around the jet.   

The magnetic field is seen to become circumferential in the lobes and perpendicular
to the jet in the hotspots.  All this can also be explained by the diffusion
of matter and hence the frozen magnetic field. 

{\it To summarise: the magnetic field is observed along the radio jet and
suggests that it is frozen in the matter that was ballistically ejected from the poles
of the pseudosurface of the black hole.  Till further observations provide another
explanation, it appears that the field is frozen in the positron-electron jet
in active nuclei.}

\noindent
{\bf Dust formation in NLR:}
Fast ejection of normal matter from the poles of the pseudosurface of the
black hole should lead to atomic mass-based
segregation in the ejecta with the lighter elements leading and the
heavier elements lagging, in analogy with novae and supernovae. 
Since the line forming gas ejected from the poles forms the NLR, the
atomic mass-based segregation 
should lead to the formation of metal-rich dense clumps in the NLR.  
The inner parts of these optically thick metal-rich clumps will be 
shielded from the hard radiation
field of the hot pseudosurface of the black hole thus facilitating dust formation.
The dust formed in this process will be distributed along the polar axis.
High resolution inteferometric studies of active nuclei at mid-infrared wavelengths 
have indeed shown that dust emission elongated along the polar axis 
dominates the total mid-IR emission
e.g.  NGC 424 \citep{2012ApJ...755..149H}, NGC 3783 \citep{2013ApJ...771...87H},
Circinus \citep{2014A&A...563A..82T}, NGC 1068 \citep{2014A&A...565A..71L}.
These results are expected in the model suggested here but were surprising when 
first obtained since in the existing model of an active nucleus,
dust is expected to be distributed in an equatorial torus around
the black hole. 
Broad lines from some Seyfert 2 nuclei are often detected in polarised light 
from a biconical region about the core and are explained as being the
nuclear spectrum scattered into our sightline by dust in the biconical reflection nebula.
%{\bf reference}.  
These observations lend strong support to the formation of dust
in the polar biconical ejecta which then acts as a reflection nebula.

{\it Atomic mass-based segregation in the polar ejecta which forms the NLR,
leads to formation of 
metal-rich optically thick clumps within which dust forms.  This process 
succeeds in explaining the observed biconical dust distribution and properties in 
active nuclei.}

Dust could also form in the mid-to-outer parts of the accretion disk if 
densities are sufficiently large, metallicity is large and 
the radiation field of the pseudosurface is not able to penetrate far into
the accretion disk.  This could lead to the presence of dust in the 
parts of the accretion disk which are shielded from the radiation field of
the pseudosurface. 

\noindent
{\bf Gas for accretion:}
The cycle of accretion, compression, heating, explosion, ejection from the poles
of the supermassive black hole of the active nucleus will continue as long 
as there is matter for the black hole to accrete.  This will lead to the formation of
extended radio jets and NLR with the furthermost parts of the jet and NLR having
been ejected in the early days of the active nucleus. 
Since radio galaxies show the largest and hence oldest structures along 
the polar axis, it means that they have been going through the cycle much longer 
and more efficiently than Seyfert galaxies which show faint and compact
radio emission.  For the cycle to continue,  the most important
condition seems to be the existence of a pool of gas from which 
the supermassive black hole can accrete onto its pseudosurface.  
Once the black hole can accrete gas,
rest of the steps in the polar cycle have to follow with differences being
reflected in the strengths of the processes.  If accretion is halted, the cycle
stops due to lack of matter on the poles.  The prolonged active nature of the nucleus in
elliptical hosts of radio galaxies means that the required pool of gas has
always existed in their central parts.  While early shallow observations 
had failed to detect interstellar medium in early type galaxies, later
observations have detected gas and dust in the central parts of a large fraction 
of early type galaxies with masses being estimated to be upto $10^8$ M$_\odot$ or so.
%{\bf reference}.  
This gas is the fuel that powers the active nucleus.
If the black hole accretes at a rate of 1 M$_\odot$ per year then it means that
the nucleus can remain active for another 100 million years when the ambient
gas will be exhausted.  It is also interesting to note that several of the galaxies
classifed as LINERs \citep{1980A&A....87..152H} 
show distribution of ionized gas in their central few kpcs with excitation
properties between the NLR and HII regions.  
It is possible that the LINER signatures mark the gas that is falling in 
towards the black hole.  
The galaxies which have exhausted this pool of
gas will stop ejecting matter from the poles and only fossil lobes
and NLR will be left behind. The core will become radio-faint i.e. the synchrotron
continuum at the jet base will be extinguished.
The non-polar pseudosurface can possibly keep radiating black body emission and
the BLR can keep forming wide lines long after the galaxy has lost all accretable
gas since once formed, these regions do not have a critical connection to accretion
unlike the polar processes.  In fact the hot pseudosurface 
can keep shining till the degenerate matter is cooled to the ground state quantum
energy.  Moreover accretion in the non-polar regions through the accretion disk 
can continue long after matter has stopped falling in i.e. the polar accretion has stopped. 
Thus by all counts, the non-polar regions of the active nucleus will remain
active long after accretion has stopped i.e. the black body + recombination continuum 
of the active nucleus will be present long after the synchrotron 
component has disappeared.  In other words, the core of the active galaxy will
remain bright in X-ray to optical bands but will become undetectable at radio bands.
This, then, provides one possible explanation for the large fraction of active 
nuclei which are not detected in radio bands but are bright in higher frequency bands. 
Spiral galaxies host a large mass of interstellar matter ($10^9$ M$_\odot$
or more) distributed throughout the galaxy. 
If finite gas mass in the spiral galaxy keeps tunnelling inwards and feeding the black hole 
then the polar activity can continue over a long timescale which will eventually
deplete the interstellar medium of the galaxy.  
It can be speculated that the central black hole has been an important sink for
the interstellar medium in elliptical and lenticular galaxies.  If the black
hole has been active for a billion years and if it has been 
accreting at a rate of a M$_\odot$ per year than it would
have accreted about a billion solar mass of the interstellar medium of the
elliptical galaxy - some of this mass would have fallen into
the black hole, some arranged in a pseudosurface,  some used up in a BLR and 
accretion disk and some of it returned back as the NLR and jet plasma.  
No wonder elliptical galaxies have lost most of their gas!

{\it To summarise: Gas accretion is a very important step in the cyclic
process that signifies the radio-active black hole or nucleus.  The polar 
explosions and ejections are exclusively dependent on gas
accretion and stop soon after accretion is halted.  The non-polar activity of
a radiating pseudosurface and BLR, although
initiated by accretion, can continue a long time after accretion has stopped. 
These results can be interpreted to mean that the active core stops being
a radio core once accretion stops but continues to radiate in higher frequency
bands. In other words, the fraction of active cores which have stopped
accreting will be radio-quiet but not ultraviolet-quiet or X-ray quiet. 
}

\begin{figure}
\centering
\includegraphics[width=8cm]{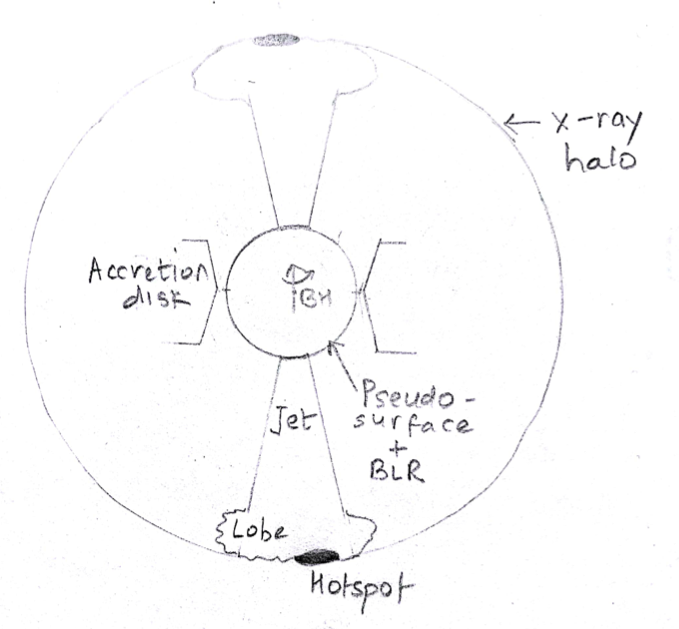} a
\includegraphics[width=8cm]{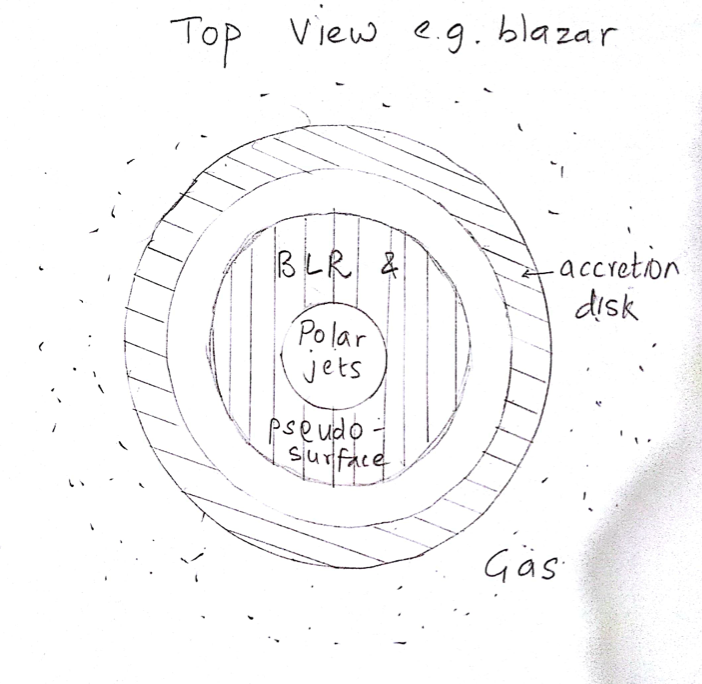} b
\caption{\small 
%{\bf modify (a) so that the shape of pseudosurface (prolate)
%and BLR (hourglass) is depicted + have the
%lobes flowing inwards towards the core instead of along the periphery} 
Schematic of the different components associated with an active 
nucleus/radio galaxy as described in the paper.  The figures are not to scale.  
(a) The observed structure of a radio galaxy - the hotspots are a feature of
FR~II sources.  The pseudosurface and BLR are approximated by the
circle but in fact as shown earlier, the BLR will be an hourglass-shaped
structure between the pseudosurface and accretion disk. The X-ray halo is
approximated by a circle here.  (b) Top view of an active nucleus (e.g. blazar)
showing the polar region surrounded by a BLR which, after a gap, is surrounded by an
accretion disk from which matter is accreted onto the BLR.  
%{\bf include a similar
%schematic but with a different polar region so that the two panels together
%summarise polar views of FR~I and FR~II.} 
}
\label{fig29}
\end{figure}

\paragraph{Summary}
In this section, we present schematics (Figure \ref{fig29}) 
which summarise the model and list a pointwise summary which encompasses it.

\noindent
{\it The core:}
The core of an active galaxy consists of an accreting rotating supermassive black hole.
Infalling matter on approaching the event horizon will be compressed to 
extremely high densities so that layers of degenerate matter deposit around the event
horizon and which are supported from collapse into the black hole by degeneracy pressure. 
These layers form the pseudosurface of the black hole and subsequent to its formation,
the infalling matter instead of falling into the black hole, deposits on its pseudosurface.  
In the polar regions which are free of any centrifugal influence, the accretion rate is
the largest and the infalling matter accumulates directly on the
pseudosurface.  This matter will be compressed and heated so that it will undergo
repeated thermonuclear outbursts which can energise and hurl matter outwards forming the
synchrotron jets and the NLR.  The accretion rates decrease on moving from the poles to 
the equator and while some matter accumulates on the pseudosurface and forms the BLR, the
excess matter collects beyond and forms the accretion disk. 
The BLR, heated and ionized by the hard radiation field of the pseudosurface, 
is confined between the pseudosurface and the accretion disk.  
It is dense and consists of emitting and absorbing zones from which 
gravitationally redshifted and broadened spectral lines are observed. 
Both these zones are stratified with properties like varying ionization, 
temperatures and densities with distance from the black hole. 
This configuration of the black hole, pseudosurface, BLR and accreted
matter at the poles would then define the structure of a compact core of 
the active nucleus or more generally an accreting black hole. 

\noindent
{\it Wideband continuum emission from the core:}  
The active core has a bright star-like appearance in quasars, Seyfert 1,  `N' type
radio galaxies/BLRG/FR II.
The observed spectral energy distribution is generally well accounted
by the combination of a black body spectrum, a Balmer continuum spectrum and a 
power law spectrum.  The hot pseudosurface contributes the
black body component - this is similar to a hot star but much larger and hotter.  
The continuum emission component of the BLR
which can be taken to be similar to a HII region
around a star would contribute a Balmer continuum spectrum.  Relativistic electrons
ejected from the poles and spiralling in a magnetic field will contribute
a synchrotron component to the spectrum.   
The active core in Seyfert 2, NLRG and FR I galaxies is fainter 
and the peak of the black body emission occurs at relatively longer wavelengths.  
The SED of all active nuclei is due to the combined contribution 
of the three components. 

\noindent
{\it Jet and narrow line gas launching from the core:}  
Infalling matter deposits at the poles at a faster rate than at the equator.
The situation at the polar region is similar to the accreting white dwarf in a nova so that
when favourable physical conditions of density and temperature are reached, 
a thermonuclear reaction can ignite with rapid release of energy
which instantly accelerates the overlying matter particles which are launched if their
forward velocity exceeds the escape velocity.  These episodic ejections 
will form the radio jet and the narrow line emitting region composed of matter with 
a range of excitation due to the photoionizing flux from the hot pseudosurface. 
The ejection episodes will keep recurring as long as the black hole keeps accreting
matter and this explains continuing injection of cosmic rays along the jets. 
Detection of X-ray jets requiring electron energies of TeV indicate the high efficiency
of the thermonuclear explosion in accelerating particles to such energies. 
The longer collinear jets in FR~II sources argue for a larger ejection velocity 
which ensures the ballistic nature of the jet to larger separations from the
black hole than in FR~I sources. 

\begin{figure}[t]
\centering
\includegraphics[width=7cm]{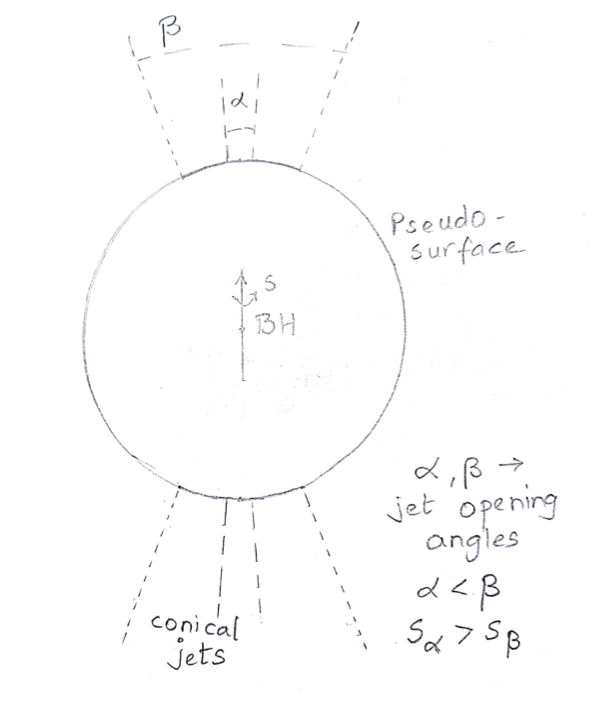}
\caption{\small Schematic representation of the jets radially launched from the polar
region of the spherical pseudosurface which has to lead to conical
jets.  A faster spinning black hole as in FR~II sources will have a smaller
opening angle ($\alpha$) while a slower spinning black hole as in FR~I sources will
have a larger opening angle ($\beta$).   The schematic can
also be used to understand the effect of a precessing spin axis of the black hole
so that at any instant, $\alpha$ can be taken to be the angle in which the jet and NLR
are launched and if the spin axis precesses in a cone of angle $\beta$ then over
time the entire region withing $\beta$ will contain radio plasma and NLR
as was the case for NGC~193 (Figure \ref{fig27}).}
%{\bf remove and include the conical jet detail in the previous figure.}
\label{fig30}
\end{figure}

\begin{figure}
\centering
\includegraphics[width=5cm]{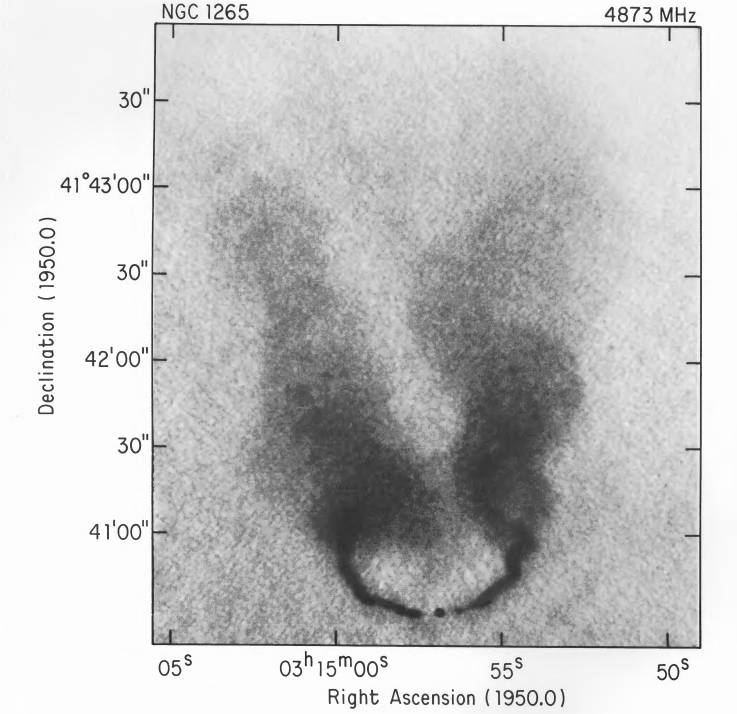}
\caption{\small A radio intensity map at 6 cm of the galaxy NGC 1265 
copied from \citet{1986ApJ...301..841O}.
The observed morphology of this narrow-angle tail galaxy can be explained simply
by motion of the galaxy in the sky as shown in the schematic in Figure \ref{fig28}.
Ram pressure might aid in its formation but it does not appear to be crucial.  } 
\label{fig31}
\end{figure}

\noindent
{\it Opening angles of conical jets:}
The jets (and NLR) are ejected from the polar region of the pseudosurface.
The circular extent of the polar region that determines the jet opening 
angle is dependent on the spin of the black hole.  A fast spinning black hole hosts 
a BLR with a large covering factor so that the jet opening angle is very small
(see Figure \ref{fig29}).  Since the pseudosurface is a quasi-spheroidal
structure, the radial ejection from the poles will happen from a convex-shaped
surface and hence the ejected matter will necessarily be confined to within 
a conical region whose apex will be the black hole (see Figure \ref{fig30}).   
As long as the ejected
matter retains its ballistic nature, the jet will propagate within the conical region. 
Thus, the observed conical jets are expected from this model. 
In other words, the ubiquity and persistence of the conical nature of the 
observed jets and NLR in active nuclei
provide irrefutable evidence to their ejection from a spherical pseudosurface while their
ballistic nature gives evidence to their energy budget being mainly determined at ejection. 

The conical angle recorded for the NLR, especially in Seyferts, 
is generally larger than that of the radio jet. 
The above can be explained to be due to precession which increases the
net conical angle over which persisting emission will be distributed.  
The narrow opening angle of the radio jet traces the current position 
of the jet whereas the NLR traces the entire precession region 
and hence is distributed over a larger conical angle.  Thus, it can be
inferred that in most cases, the jet opening angle
gives a measure of the spin of the black hole while the opening angle of
the NLR gives a measure of the precession angle. 

\noindent
{\it Edge-dimmed jets:}
The radio emission from jets which are resolved along their widths are observed to 
exhibit either uniform brightness or edge-brightened structure or edge-dimmed structure.
Some of this could be projection effects since we observe jets in the sky plane or
these could have a physical origin.  The edge-dimmed jets 
which have been noted in active nuclei with one of the first active galaxies in which it
was noticed being NGC 1265 \citep{1986ApJ...301..841O}.
The faint emission generally has a steeper radio spectrum and encloses the bright radio 
jet which has a flatter radio spectrum (e.g. Figure \ref{fig27}b).
The faint envelope, often referred to as the sheath, is detected around 
jets at several linear scales \citep[e.g.][]{1986ApJ...301..841O,1999ApJ...516..716K}. 
A possible origin for the sheath could be as follows. 
The collimated conical nature of the radio jet around which the sheath is
observed indicates dominance of the forward momentum i.e. the component
of the bulk velocity in the forward direction is much higher than in any 
other direction so that the jet retains its structure.  However,    
the boundary layers of the ballistic jet might undergo some erosion due to 
physical processes like turbulent viscous stripping \citep{1982MNRAS.198.1007N}.
This process can lead to stripping of 
the plasma from the edges of the jet which can then form a faint radio synchrotron 
sheath around the jet given an ambient magnetic field.   The sheath is from
the same electron population that is ejected
from the pseudosurface but the steeper spectrum and larger polarisation 
fraction that are recorded for the sheath 
\citep[e.g.][]{1999ApJ...516..716K} are suggestive of a distinct
magnetic field.  The differing magnetic fields in the jet and sheath could lead
to electrons of different energies emitting at the same radio frequency and 
hence explain the differing radio spectral index.  The differing orientation
and strength of the ordered magnetic fields could be due to the sheath shining
in the ambient magnetic field left behind by previous polar ejections while
the dense jet shines in the magnetic field frozen in itself.  The 
variations in the magnetic field and spectral index could be specific to the
radio source so that a range of plausible combinations can result. 
Such sheaths have been observed around jets in both FR I and FR II type sources
including Cygnus~A \citep{1994ApJ...426..116K,1996ASPC..100..233R,2000AAS...197.7512Y}.
It would be difficult to explain the existence of such
sheaths if jets were confined by an external medium which is often cited in literature
as a reason for their confinement and collimation.  In other words, the existence of 
sheaths indicate that jets are not confined by any external medium.
Sheaths support confinement of jets due to
their ballistic nature, after being ejected with relativistic velocities from the
polar pseudosurface, so that their lateral expansion is prevented although 
they are highly overpressured with respect to the ambient medium. 

Edge-brightened jets are detected when high resolution radio images which
can resolve emission close to the core in FR~I radio sources
like Virgo~A (3C~274) and Perseus~A (3C~84) are made.  Such edge-brightening
might be because the thermal gas in which the magnetic field is frozen is
compressed by the walls of the BLR which form a funnel at the poles enhancing
the magnetic field at the interface which enhances the radio emission at
the cone edges close to the launch site.  

\noindent
{\it Dust formation:}  
In the narrow optical line gas ejected from the poles, 
the heavier atoms will lag while the lighter atoms will lead in the ejecta. 
This will lead to segregation of heavier atoms due to enhanced mutual gravity which
will form dense metal-rich clumps whose insides will be shielded from the hot 
pseudosurface.  Dust will, hence, form inside these clumps and will eventually 
be distributed along the polar axis. 
Observations predominantly detect dust along the jet axis validating 
this origin scenario.

\noindent
{\it Magnetic field:}  
The magnetic field lines near the core in all radio galaxies, are aligned with the jet.
The structure of the magnetic field provides strong support to their being frozen  
in the jet which is radially expanding away from the black hole so that the lines
are stretched.  That the field lines change direction when the jets
start forming lobes in FR~I galaxies but continue to be aligned along the jet
upto the hotspots in FR~II sources also argue for an origin in the jet plasma.
In other words, the magnetic field lines are along the jet as long
as the jet retains its ballistic forward motion. 

\noindent
{\it Wiggles and bending of jets:}
The wiggles, observed to be symmetrically located on either side of the core 
in the jets are readily explained by precession of the spin axis of the black hole. 
Even the changing locations of hotspots on either side of the core in FR~II sources
such that all pairs remain collinear with the core are easiest 
to explain with a precessing jet axis.  This reason suffices to explain 
these observed features since matter is radially ejected
from the polar regions and when the polar axis precesses, the direction of radial
ejection of matter varies which results in symmetric wiggles commonly 
observed in FR~I jets and moving hotspot pairs in FR~II jets. 

In addition to wiggles, several FR~I jets show large scale symmetric or asymmetric
bending in the two jets.  There are cases wherein one of
the jets shows a sharp bend while the other continues uninterrupted in addition
to cases wherein both the jets change direction.  These features are difficult to
attribute to precession.  While the likely cause of 
asymmetric bending is inhomogeneities in the ambient
densities, symmetric bending of jets which give rise to NAT and WAT morphologies 
are best explained by rapid motion of the galaxy (i.e. of the launching site). 
For example, the radio morphology of the NAT source, NGC 1265 (Figure \ref{fig31}) 
can be explained by motion of the galaxy from the
north-east to the south-west and a non-trivial velocity component along the
sightline.   
A fast moving galaxy in the sky plane will be viewed as a head-tail or 
narrow angle tail galaxy while
a slow moving galaxy in the sky plane will be viewed as a wide-angle tail radio structure.
A stationary galaxy would show straight, oppositely directed jets 
as the jet launching position is constant.

\noindent
{\it Composition of jets:}
Radio jets in active nuclei are composed of positron-electron plasma.  
Several observations like highly relativistic propagation of the jet to large
distances, detection of the annihilation line from
a few microquasars,  X-ray emission coincident with the radio jets, abrupt
halting of jets where hotspots are formed etc provide convincing 
evidence to this composition.   A ready source of copious positrons is
the thermonuclear explosion at the poles which precedes an episode
of mass ejection in the form of a synchrotron jet and line forming gas. 
The positron-electron annihilation photons should be generated along the
entire radio jet and scattered by the thermal plasma.  

\noindent
{\it Radio lobes and hotspots:} 
The process of lobe formation is distinct in FR~I and FR~II radio sources 
as has been highlighted in literature.   In FR~I sources, the jet loses
its ballistic nature on kpc scales and subsequently diffuses out to form lobes. 
In case of FR~II sources, the jet retains its ballistic nature upto the hotspots.
As explained here, the forward motion of the jet is halted by radiation pressure exerted
by the positron-electron annihilation photons near energies of 511 keV produced
beyond the detectable radio jet.  This results in hotspots where the plasma is
halted which then diffuses back towards the core forming the lobes in FR~II sources.

\noindent
{\it Halo-like lobes in FR~I:}
The radio emission forms a halo around the active nucleus in FR~I galaxies like 
NGC 193 (see Figure \ref{fig27}), 0206+35, 0755+37 \citep{2011MNRAS.417.2789L} and 
M87 where the halo-like structure has been explained by a precessing jet 
\citep{1986NRAOW..16...79R}.
Such structures in all FR~I sources appears traceable to a precessing jet 
which sweeps over a large
region about the core so that over time, a halo-like structure seems to
form around the core.  Moreover the radio plasma appears to be confined within
a well-defined boundary around the active nucleus which is of similar extent as the
thermal X-ray emission.
The nature of the radio and X-ray structures are similar to those in centres of 
clusters of galaxies except for the different linear sizes.

\noindent
{\it Gravitational redshifts:} 
The broad emission lines from quasars, BLRG and Seyfert~1
include an intrinsic redshift component due to gravitational redshifts. 
Any line which contains
a gravitational redshift component will also be gravitationally broadened due to
the finite thickness of the line forming region. 
By determining the gravitational redshift of the broad emission lines, 
we can locate the BLR wrt to the black hole in terms of the Schwarzchild radius
and understand the structure of an active nucleus.
The lines forming in the NLR should also show a gravitational redshift component
if detected from close to the black hole as is seen in the case of SS~433.
The synchrotron jet and NLR are launched from within a few $R_s$ of the black hole
in radio sources so that any line photons emerging from there should show a significant
gravitational redshift component.   

\noindent
{\it Thermonuclear source of variability:} 
Variability is observed in the wide-band continuum emission 
at all wavebands.  The soft X-ray/ultraviolet/optical variability has to be associated 
with the black body component while radio variability will be related to the synchrotron
component.  The former will be predominantly in the non-polar regions whereas the latter
will be associated with activity in the polar regions.  It is interesting that
the variability in both regions can be powered by thermonuclear explosions which
are expected when matter is compressed to high densities and temperatures.
The variability in the non-polar continuum is also seen to
trigger variability in the properties of the broad lines.  

\begin{table*}[h]
\caption{The observed properties of FR~I and FR~II radio sources are summarised in the
first part of the table while the lower part lists
a few inferences that can be drawn from the model discussed in the paper.}
\centering
\begin{tabular}{l|c|c}
\hline
Parameter  & FR~I & FR~II \\
\hline
Bright nucleus & no  & yes$^1$ \\
Broad emission lines from BLR  &   no$^2$  & yes$^1$ \\
Broad absorption lines from BLR  &   no  & yes$^1$ \\
%Nuclear emission/absorption lines  &   yes & yes ? \\
%Nuclear absorption lines  & yes & yes \\
Emission lines from NLR & yes  & yes  \\
%Radio power {\bf include numbers from Fanaroff,Riley paper} & lower  & higher \\
Radio power  & lower  & higher \\
Radio jets  &  double-sided  & single-sided \\
Jet launching speeds (relativistic) & lower & higher \\
Radio jet opening angle & $\le 15^{\circ}$ & $\le 2^{\circ} $ \\
Jet magnetic field & $\parallel$, $\perp$ to jet & $\parallel$ to jet \\
X-ray jet  & single-sided  & single-sided \\
Radio lobes & double-sided & double-sided \\
Radio lobe magnetic field & circumferential   & circumferential \\
Radio hotspots & none  & double-sided  \\
Hotspot magnetic field & - & $\perp$ to jet \\
X-ray from lobes & faint or none & inverse Compton enhanced \\
Diffuse X-ray halo & yes & no \\
\hline
\multicolumn{3}{c}{Some Inferences} \\
\hline
Black hole mass$^3$ &  high & high \\
Black hole spin$^3$ &  low  & high \\
Accretion disk thickness$^3$ & thin  & thick \\
BLR latitude coverage$^3$ & small & large \\
Jet opening angle$^3$ & large & small \\
Jet launching speed$^4$ & $\le 0.65c$  & $\ge 0.65c$ \\
Jet launching region$^5$ & $>2.4 R_s$ &  $< 2.4 R_s$ \\
Gravitational redshift$^5$ at launch site & $<0.21$ & $>0.21$ \\
%Jet launching speed$^3$ & $\le 0.3c$  & $\ge 0.3c$ \\
%Jet launching region$^4$ & $>11 R_s$ &  $< 11 R_s$ \\
%Gravitational redshift$^4$ at launch site & $<0.045$ & $>0.045$ \\
\hline
\end{tabular}

{\small
$^1$ There do exist FR~II galaxies where the nucleus is not bright nor is there a BLR.
There also exist FR~II wherein the BLR is detected mainly in polarised light.
$^2$ In a few FR~I galaxies, broad lines from a BLR are detected in polarised light.
$^3$ These are comparisions between FR~I and FR~II.
$^4$ These numbers are based on studies in literature which give ballpark numbers for
relativistic speeds when Doppler beaming can explain the single-sidedness of the
jets in FR~II sources.
$^5$ These are estimated assuming the jet launching speed is equal to the escape
velocity from the black hole.
}
\label{fr1}
\end{table*}

\noindent
{\it LINER gas as the infalling gas:}
The low ionization lines which are often detected in active nuclei and from the
central regions of normal galaxies are believed to arise in the low ionization
narrow emission line region (LINER) \citep{1980A&A....87..152H}.  
This region whose excitation properties lie between HII regions and
the NLR is likely to be the pool of gas
from which the black hole accretes.  Many early type galaxies have been observed to
show LINER behaviour and the observed line strengths occupy a 
distinct place in the diagnostic diagram. 
If the LINER gas is indeed the infalling gas then once 
it is exhausted, the supermassive black hole will lose 
its source of infalling gas and stop accreting, thus entering a quiescent state
especially in case of its polar activity.    

\noindent
{\it Final state of an active nucleus:}  
When the galaxy runs out of gas, the black hole will stop accreting and
the processes, especially at the poles, which are put in action by 
accretion will be halted.
In particular, once accretion stops, the thermonuclear outbursts at the poles
will stop, no further
cosmic ray acceleration or ejection will occur and the already ejected radio plasma will
keep emitting till it synchrotron ages into oblivion.  Same will happen with the narrow
line gas which will diffuse into oblivion. 
The hot pseudosurface and BLR will exist around the black hole and
keep accreting from the accretion disk and radiating. 
If an instability leads to the
black hole gobbling up the pseudosurface and the BLR then the black hole will enter
a dark stage which can be forever unless gas becomes available to the black hole
and a fresh episode of accretion starts which reactivates the dormant nucleus. 

\noindent
{\it FR I and FR II radio sources:}
Lastly we compare and contrast the properties of FR I and FR II radio sources
in Table \ref{fr1} and also summarise the inferences that can be derived from the
model discussed in the paper.

\subsubsection{Case Studies}
In this section, we present a few case studies of active galaxies.

In several galaxies hosting an active nucleus, the luminosity is dominated by
the central source.  There remain uncertainties regarding the morphology
of the host galaxy of most quasars due to the extremely bright nature of the central
source.  However when a host galaxy is discerned, its morphology is surmised to be either 
elliptical or spiral.  The absolute magnitude of quasars which show a star-like  
appearance with excess ultraviolet emission is generally found to be $M < -23$ magnitudes
whereas for Seyfert 1 and `N' type radio galaxies the central luminosities are generally
$\ge -23$ magnitudes and it is much lower for Seyfert 2 and FR I where the
integrated luminosity is often dominated by the host galaxy.  
There appears to be a continuity in the ultraviolet/optical continuum properties of the nuclei in
quasars, radio and Seyfert galaxies.  However a word of caution, that the distances 
estimated from the  emission line redshifts of quasars are grossly overestimated
due to the large intrinsic redshift component that contributes to enhancing their
emission line redshifts.  This would mean that the actual distances to
quasars are lower and hence their luminosity will also be lower when the correct 
distances are used.  We follow the terminology in literature so that radio galaxies 
refer to radio doubles whose host is an elliptical galaxy whereas Seyferts 
refer to spiral galaxies hosting an active nucleus. 

A large fraction of the 
narrow emission lines detected in the spectra of most active nuclei are likely to
arise in the optical line forming gas that is ejected
from the poles which then indicates the presence of an ejected NLR in most active nuclei.
However in a fraction of active nuclei,  the narrow emission lines could arise in
a non-polar BLR that is far from the black hole so that the reduced gravitational broadening
leads to narrower lines.  It might be possible to differentiate the emission lines
of the two distinct origin from the presence of radio jet emission i.e. 
if the narrow emission lines arise in the NLR then these active nuclei should
also host a radio jet which can be compact or distributed whereas if the narrow emission
lines arise in a distant BLR then it would be missing a NLR and hence 
should also be radio-quiet.  Broad emission lines arising in the non-polar BLR
are observed from quasars, BLRG and Seyfert~1 nuclei which also show the presence of
narrow emission lines.  The similar morphology of radio-loud double sources 
consisting of bipolar radio jets, lobes (and hotspots in FR~II) and NLR around the jets
indicates that the physical processes occuring in the polar regions of all 
active nuclei are similar.  This is not surprising since all active nuclei
host an accreting supermassive rotating black hole and hence should also lead
to similar physical processes.  

There do exist interesting observational differences between active nuclei 
which should give us important insight into the range of implications of the 
physical processes.  Some of these differences are (1) iron lines are 
more frequently detected in radio-quiet quasars and Seyferts while rarely observed in
radio galaxies and radio-loud quasars. (2) The observed emission line redshifts    
of quasars are larger than radio galaxies and Seyferts (see Figure \ref{fig23}). 
The emission line redshifts of FR~II quasars are higher than FR~II galaxies 
which in turn are higher than FR~I galaxies.  Seyferts show the lowest emission line
redshifts amongst active nuclei. 
(3) The spectra of the majority of quasars and a few radio galaxies and 
Seyferts contain emission lines and absorption lines with the absorption lines 
blueshifted wrt to the emission lines.  While numerous
multi-redshifted absorption lines are detected in quasars, only a few 
such lines are discernible in the spectra of other active nuclei. 
In light of the model presented here, we can now understand these points.
The first point indicates that iron is more abundant in radio-quiet active nuclei.
Since the radio emission in an active nucleus is confined near the polar axis, this correlation
indicates that these iron lines arise in the matter accreted and/or ejected from the poles.
Such a correlation between the radio emitting plasma and
the iron lines arising in the BLR is not expected.  A thermonuclear blast which 
supplies the energy for energising the NLR and
radio plasma also forms heavier elements.  If the energy released in the explosion
is sufficient to accelerate and eject the accreted matter then this will 
lead to formation of a NLR and radio jets i.e. a radio-loud active nucleus.  
Assuming the infalling matter is hydrogen-rich, the thermonuclear
reaction will also synthesise helium so that the ejected
matter will consist of helium, hydrogen and other metals of say solar composition.   
The NLR of this active nucleus will not be rich in iron lines. 
However, if the energy released in the polar thermonuclear explosion in an active nucleus
is insufficient to acclerate the accreted matter to escape velocities, then
only nucleosynthesis will occur in the accreted matter at the poles with 
successive explosions leading to formation of increasing atomic mass elements
till iron is formed.  In this case, there will be no NLR or radio jets - in other
words, the active nucleus will be radio-quiet but the accreted matter at the
poles will be enriched in iron.  The observed correlation between the radio loudness
and presence of iron lines can thus be explained. 
The second point regarding the range of observed emission line redshifts of 
the different types of active nuclei can be
explained by the larger component of gravitational
redshift in the observed emission line redshift of quasars, followed by 
FR~II nuclei and then by FR I and Seyfert nuclei.    
If it is assumed that the emission line redshifts are derived from the lines arising
in the BLR, then the range of observed redshifts indicates the range of
separations of the BLR from the black hole.  The third point can be understood from
the proximity of the BLR to the black hole and hence the range of intrinsic
redshifts that can distinguishable.   In quasars the BLR is very close to the black hole
so that the range of intrinsic redshifts for lines forming in the BLR of finite
extent is large so that the emission and absorption lines of different redshifts
can be discerned.  As the separation of the BLR from the black hole increases,
the intrinsic contribution to the line redshifts decline.  All BLR are likely
to host an emission line and absorption line zone but the intrinsic redshift
component decreases from quasars to Seyferts so that the lines from the BLR
appear close to the systemic velocity of the host galaxy. 
 
We discuss the cases of radio-loud active nuclei namely
3C 273 (quasar), Cygnus A (FR II), Virgo A (FR I), 
NGC 4151 (Sy 1) in light of the model presented in the paper. 

\begin{figure}
\centering
\includegraphics[width=6.5cm]{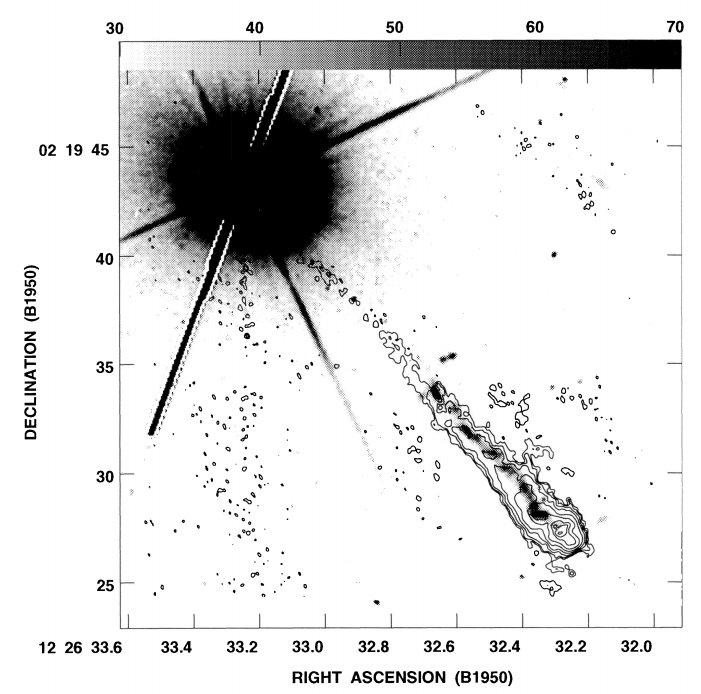}
\caption{\small Figure showing the radio (contours) and optical (grey) emission with 
a resolution of about 0.16'' and 0.1'' for 3C 273 reproduced from
\citet{1995ApJ...452L..91B}.  Notice that the active nucleus is so optically bright
that the host galaxy is not detected.  }
\label{fig32}
\end{figure}

\noindent
\paragraph{FR II quasar - 3C 273}:
3C~273 was the first quasar which was identified \citep{1963Natur.197.1040S}
and is also identified as the quasar with the lowest emission line redshift in the 3CR sample. 
The source consists of a bright core (see Figure \ref{fig32}) detected at multiple 
wavelengths, a one-sided jet detected from X-ray to radio wavelengths 
and a radio hot spot in which the radio jet terminates.
Its V band magnitude is measured to be about 12.8, its emission line redshift
is measured to be $z_{em}=0.158$ and the knots in the jet are observed to
expand at a rate of 0.76 mas/year.
These observed parameters have been interpreted to indicate that the quasar
is located at a distance of $\sim 650$ Mpc so that its absolute V band magnitude
is $-26.3$ magnitudes and the knots are expanding with superluminal 
velocities $\sim 8c$  \citep[e.g.][]{1990A&A...237....3K}.    
The radio spectrum of 3C~273 is observed to be variable
over timescales ranging from days to years \citep{1966ApJ...146..634P} with the behaviour
being qualitatively similar to microquasar in that the variability is
first detected at the higher radio frequencies. 
At the large distances that have been
inferred for 3C~273, the energy requirements to explain such repeated variability are 
huge but would be modest if the source was nearby \citep{1966ApJ...146..634P}.  
Radio polarisation studies show the magnetic field to be aligned with the radio jet
except in the hotspot where it becomes perpendicular to the jet 
\citep[e.g.][]{2017A&A...601A..35P}.  The jet is about 25'' long which 
for a distance of $\sim 650$ Mpc (corresponding to $z=0.158$) 
to the quasar would correspond to a length of about 79 kpc.  
An optical jet is also observed with comparable length but with a narrower 
transverse width than the radio jet \citep{1993Natur.365..133T}.
No counter-jet has been detected in any band.  
Broad emission lines at $z\sim 0.158$ and multi-redshifted absorption lines 
at redshifts $<0.158$ have been detected in the spectrum of 3C 273 which
is a property common to quasars.  

The lowest redshifts at which absorption lines have been detected in the spectrum of
3C~273 are $z=0.0053$ ($v\sim1600$ kms$^{-1}$) and $z=0.00337$  
\citep{2002ApJ...575..697T}.  We follow the reasoning in Kantharia (2016) and
assume that one of these is the actual cosmological redshift of the quasar which
means the quasar is closer and
the observed emission line redshift contains a gravitational redshift $z_g$ component.
We assume that $z_c=0.0053$ is the cosmological redshift of 3C~273 (it could be
0.00337 which would only make the arguments presented here stronger). 
Since $(1+z_{in}) = (1 + z_{em})/(1+z_{c}$),  
the intrinsic redshift suffered by the emission line will be $z_{in}=0.1519$.  
The gravitational redshift will be $z_g=z_{in}/(1+z_{in}) = 0.131$.   
The separation of the line forming zone from the black hole, $R$ can be
estimated from $ z_g = R_s/(2 R) = R_s / (2 n R_s)$ where $R = n R_s$.  
The lines from the BLR
which include a shift $z_g=0.131$ will be located at $R \sim 3.8 R_s$  and the
non-polar pseudosurface in 3C 273 will be formed within $3.8 R_s$ of the black hole. 

Considering 3C~273 is at $z_c=0.0053$ means it is located at a distance of about
20 Mpc so that $1'' \sim 0.1$ kpc.   The length of the observed $25''$ jet would then
be 2.5 kpc indicating that 3C~273 is a compact object similar to GPS and CSS sources. 
The central bright source in 3C~273 (Figure \ref{fig32}) which is the 
active nucleus is smaller than 50 pc.  Radio knots in the jet of 3C~273 
are observed to expand with angular rates of $\sim 0.76$ mas/year which translates to
an apparent velocity of about 8-10 times the speed of light at a redshift of 0.158
\citep{1981Natur.290..365P,1982IAUS...97..355P}.  However if  
3C 273 is at $z_c=0.0053$, then 0.76 mas/year would correspond to 
0.076 parsec/year = 0.24 light year/year.  
This velocity indicates the radial component of knot
expansion along the sightline i.e. $v_{knot} * cos\theta$ 
so that it approaches the actual knot
expansion velocity $v_{knot}$ for small angles between jet and sightline
$\theta$.  If we ignore the inclination of the jet with the sightline
for now then the knots are ejected at a velocity $0.24c$ which will in
the least be equal to the
escape velocity of the polar jet.  This, then translates to ejection of matter
from a region separated from the black hole by about $17 R_s$.  If the
jet is oriented at an angle of $30^\circ$ to the sightline, then the knots will be expanding
at a velocity of 0.277c and would have been ejected from $13 R_s$.  
Since Doppler boosting has led to the detection only a one-sided jet, the
inclination of the jet to the sightline cannot be too large. 
This discussion indicates that the inference of superluminal motions appears to
be wrong and has been misleading due to the incorrectly estimated cosmological redshifts of
quasars and other active nuclei.   As is shown here for 3C~273, once
we estimate and remove the intrinsic redshift component, the active nucleus is
found to be located close to us, thus reducing the linear motion of the knots 
estimated from their angular expansion. 

At a distance of 650 Mpc, the V band absolute magnitude of 3C 273 is
$M_V=-26.3$ magnitudes while if 3C~273 was separated by 20 Mpc, 
$M_V = -18.7$ magnitudes.  The latter luminosity is typical of bright 
normal galaxies except that while in those galaxies, light from the entire galaxy contributes
to the luminosity, in 3C 273 (and other quasars) the observed luminosity 
is from the compact active nucleus.  This, then, demonstrates that our
understanding of quasars can be flawed due to overestimated distances.   
Once the gravitational redshift component is removed from the quasar redshift,  
the distances to quasars will become comparable to other active nuclei
i.e. located much closer.  This will, amongst other properties, reduce the 
exceptionally large bolometric luminosities that have been attributed to quasars  
and make superluminal motions unnecessary. 

A temperature of 26000 K has been estimated for the black body component
in 3C 273 \citep{1982ApJ...254...22M}.
The emitting area required to explain the observed luminosity, if the source 
was closeby at 20 Mpc, is $2.8\times10^{30}$ cm$^{2}$ as discussed in Section \ref{sed}. 
If the emitting surface is spherical, then this area corresponds to a 
radius of $4.7 \times 10^{14}$ cm
which is equal to the Schwarzchild radius of a black hole of mass
$1.6 \times 10^9$ M$_\odot$ which is comparable to the mass that has been estimated
for the black hole in 3C~273.  For these values, the black body luminosity due
to a spherical emitting pseudosurface will be $-20.93$ magnitudes
and the V band magnitude will be $-18.43$ magnitudes.  Recall that the observed
V band luminosity, if 3C~273 was 20 Mpc away, is $\sim -18.7$ magnitudes.  
The matching of the estimates derived from distinct observed parameters 
tends to support the model presented here, in particular, the shorter distance
to 3C~273 and the black body component being from the hot quasi-spherical pseudosurface
of radius $\sim R_s$. 

The properties of 3C~273 determined for redshifts $0.158$ and 
$0.0053$ are summarised in Table \ref{c273}.  
\begin{table}
\caption{The table lists the change in properties of 3C~273 when the gravitational
redshift component is removed from the observed emission line redshift (0.158) which
leads to a revision in the distance estimate to the quasar (0.0053). }
\begin{tabular}{l|c|c}
\hline
Property & $z_c = 0.158$  & $z_c = 0.0053$ \\
\hline
Distance & $\sim 650$ Mpc & $\sim 20 $ Mpc \\
$M_V$ magnitudes & $-26.3$ & $-18.7$ \\
Emission line $z_g$ & none & 0.131 \\
Distance bet BLR, BH  & ?  & $\sim 3.8 R_s$ \\
$v_{esc}$ at BLR  & ?  & $\sim 0.5c$ \\
Length of $25''$ radio jet & 79 kpc  & 2.5 kpc \\
Knot expan 0.76 mas/yr & $\sim 8c$ &  $\ge 0.24c$ \\
Knot expansion rate & superluminal  & subluminal \\
Jet launching site$^1$ & ? & $\le 17 R_s$ \\
\hline
\end{tabular}

{$^1$ \small Assuming that the proper motion of the knot (i.e. 0.24c) indicates the
escape velocity from the black hole. }

\label{c273}
\end{table}
The changes listed in Table \ref{c273} underline the importance of correctly estimating
the cosmological redshift and hence distance to quasars since several important
properties of quasars depend on the distance.  Obtaining correct estimates of
physical properties of quasars is crucial to improving our overall understanding
else the discrepancy between observational results and our understanding will keep increasing.

{\it To summarise: 
(1) The corrected distance to 3C~273, as estimated from one of the lowest redshift
absorption lines in its spectrum is 20 Mpc. 
The V band absolute magnitude estimated from the observed apparent
magnitude for the corrected distance is $M_V=-18.7$ magnitudes (For a distance of 650 Mpc, 
$M_V = -26.3$ magnitudes).  The V band absolute magnitude expected for
black body radiation at 26000 K, as estimated from the Stefan-Boltzmann law using
parameters derived from observations ($R_s$ and $T_{blackbody}$) is $-18.43$ 
magnitudes.  The comparable values of $M_V$ strongly justify the reasoning. 
Some difference in the observed V band luminosity with that estimated from
black body luminosity is expected due to non-zero contribution by synchrotron emission
at V band and the quasi-spherical nature of the pseudosurface.  
(2) The emitting zone in the non-polar BLR is located at a separation of $3.8 R_s$ 
from the black hole.
(3) The radio jet is about 2.5 kpc long and the radio dimensions of 3C~273 are
similar to CSS/GPS quasars.
(4) The polar knots are ejected from a region separated from the black hole
by $\le 17 R_s$ at a velocity of $\ge 0.24c$.}

\paragraph{FR II galaxy - Cygnus~A}:
Cygnus~A is a powerful FR~II type radio galaxy at a redshift of 0.056 
(distance $\sim 230$ Mpc) located in a nearby cluster of galaxies \citep{1997ApJ...488L..15O}.
\citet{1954ApJ...119..206B} detected narrow emission lines in the spectrum of Cygnus~A
and identified the optical host galaxy of Cygnus~A. 
Broad lines were not immediately detected which prompted its classification as 
an NLRG FR~II source which is rare.  Generally FR~II radio sources are hosted by
BLRG nuclei while FR~I are hosted by NLRG nuclei.  An angle of 2 mas 
at the distance of Cygnus~A corresponds to a 
linear distance of 2.2 pc.  Some of the observed properties of Cygnus~A are 
listed below with comments in italics.

\begin{itemize}

\item The host galaxy is a faint but large cD galaxy. 
The radio structure extends out to about 70 kpc on either side of the optical
galaxy. 

\item Early observations detected faint optical nuclear continuum and
emission from strong forbidden lines ([O II],[O III]) at the galaxy 
redshift of $\sim 16800$ kms$^{-1}$ with widths $\sim 400$ kms$^{-1}$ with the forbidden
line emission extending beyond the nucleus \citep{1954ApJ...119..206B}.  
More recent observations have found the [O III] emission to be spread over the central 
$\sim 10''$ ($\sim 11$ kpc at the distance of Cygnus~A)
and elongated along the radio axis \citep{1988ApJS...68..643B}.  
The emission line spectrum consists of recombination lines and forbidden lines 
with a wide range of ionization  (Osterbrock and Miller 1975). 
The distribution of H$\alpha$ and [N II] lines also extends out from the
nucleus to about $10''$ \citep{1989AJ.....98..513C}.

\item Ionized gas distributed in the central 5 kpc region 
appears to be rotating about an axis aligned close to the radio axis of Cygnus~A 
\citep{1977ApJ...217...45S,1974MNRAS.166..305H}. 
{\it This could be the gas which the black hole is accreting. }

\begin{figure}
\centering
\includegraphics[width=5cm]{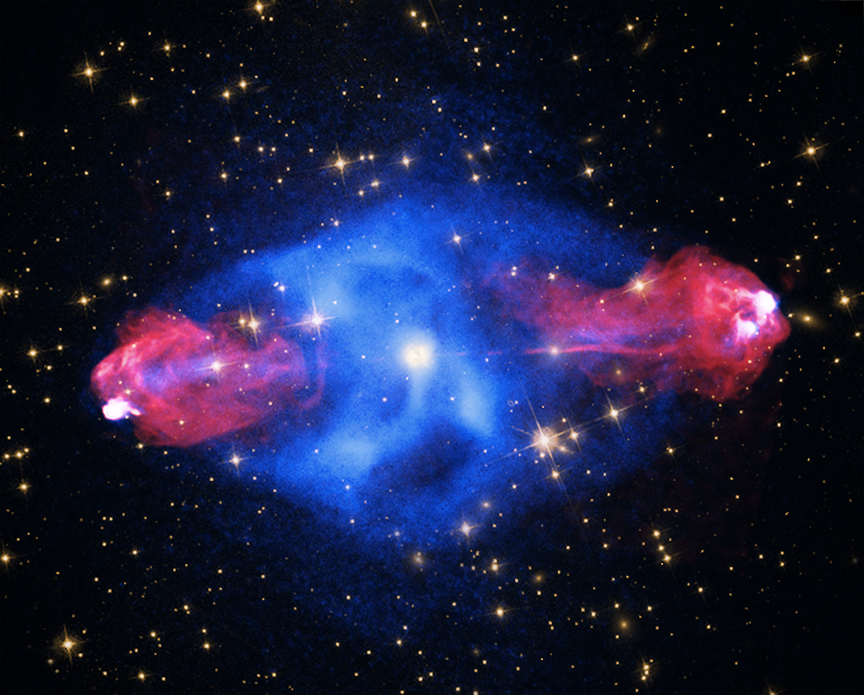} (a)
\includegraphics[width=5cm]{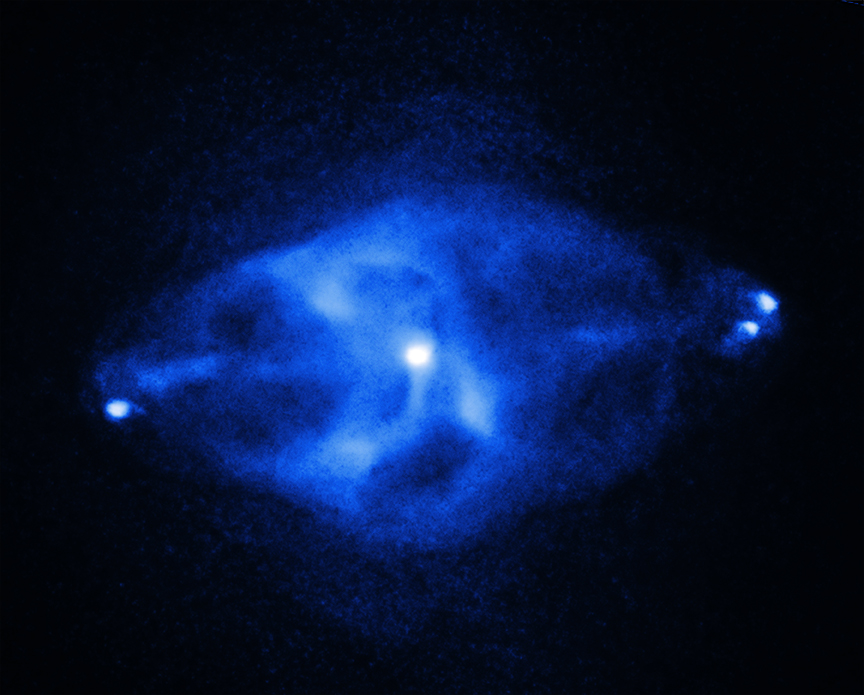} (b)
\includegraphics[width=8cm]{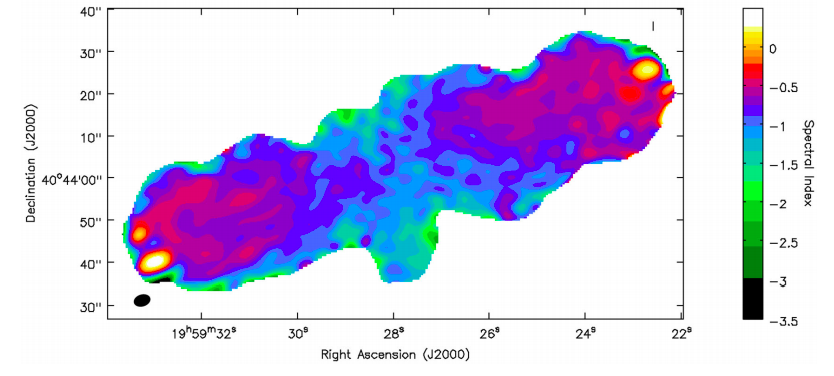}(c)
\caption{\small Multi-band image of FR~II galaxy Cygnus~A (a) and
X-ray image of Cygnus~A (b) downloaded from http://chandra.harvard.edu/photo/2015/iyl/.
(a) Blue is X-ray data from Chandra telescope, red is radio data from VLA and
yellow is optical data from the Hubble space
telescope and DSS.  The radio lobes and X-ray emission share a major axis 
and together define a oval cocoon although their distributions are 
anti-correlated.  The hot spot forms at the periphery of the  
X-ray cocoon which is common for all FR~II sources.  
(b) The X-ray image is shown to highlight the double-sided X-ray beams 
and their terminating pair of hotspots.  The X-ray beam in the east seems to terminate
above the hotspot. 
(c) Figure showing the distribution of spectral index between 109 and 183 MHz 
copied from \citet{2016MNRAS.463.3143M}.
Notice two pairs of hotspots - the bright hotspots A and D and another pair 
(north-east of D and south-west of A) which is not detected on total intensity maps.
The plume is also clearly detected with a steep spectrum.}
\label{fig33}
\end{figure}

\item Although Cygnus~A was originally classified as a NLRG due to detection
of only narrow lines, broad  Mg~II emission lines (FWHM $\sim 7500$ kms$^{-1}$) have 
subsequently been detected \citep{1994Natur.371..313A}. 
Broad Balmer lines (FWHM(H$\alpha$) $\sim 26000$ kms$^{-1}$) 
have also been detected in polarised light from a bipolar region centred on the core 
which have been interpreted as line emission from the 
BLR scattered into our sightline by polar dust \citep{1997ApJ...482L..37O}. 

\item No optical line emission is detected from the radio lobes.

\item The radio structure of Cygnus~A consists of a core, 
two jets with one being considerably brighter than the other which when 
explained due to Doppler boosting indicate a jet velocity 
$> 0.45c$ and the angle made by the jets with the sightline $\theta < 63^{\circ}$ 
\citep{1991AJ....102.1691C,1979Natur.277..182S}, two lobes and two bright 
hotspots \citep{1974MNRAS.166..305H}.  It is one of the brightest sources
in the radio sky and has been studied across the electromagnetic spectrum.  

\item X-ray emission detected from the radio hotspots cannot be explained
by the synchrotron process or thermal emission but well explains
the observations if due to synchrotron self Compton process as was surmised
from the hotspots in Cygnus~A which were the first to be detected in X-ray
bands \citep{1994Natur.367..713H}. 

\item Figure \ref{fig33}(a) shows a superposed image of 
the extended radio and X-ray emission in Cygnus~A. 
The radio and X-ray emission share a major axis and are of similar extent.
However the X-ray emission is faint at the location of the radio lobes. {\it Cygnus~A
is the rare FR~II which lies in a cluster of galaxies and is also the rare FR~II
wherein X-ray cavities are coincident with radio lobes.} 
Figure \ref{fig33}(b) shows only the X-ray emission which highlights the 
jet-like but laterally thick features which extend all the way
from the core to the hotspots and appear to be coincident with the expected radio jet
trajectories.  We refer to these as X-ray beams to distinguish them
from the radio jets which are significantly narrower.
Interestingly, the X-ray beams do not terminate in the 
bright hotspots A (west) and D (east) but in distinct hotspots. 
%with the western one being visible in Figure \ref{fig33}(b). 

\item High resolution images show that the two bright hotspots break up into two
smaller distinct hotspots. 
%{\bf reference}. 

\item Linear polarisation fractions which rise upto $70\%$ are detected in the lobes
\citep{1989AJ.....98..513C,1996cyga.book..168P}.

\item The linearly polarised emission from the jet shows that the magnetic field 
is aligned along the jet \citep{1996cyga.book..168P}.

\item The spectral index map between 109 and 183 MHz \citep{2016MNRAS.463.3143M}
shows two distinct pairs of hotspots with each pair appearing collinear with the core
(see Figure \ref{fig33}c).  One of these pairs is coincident with the bright 
hotspots A (west hotspot) and D (east hotspot) detected in
radio and X-ray images. The other pair of hotspots (south-west of A and north of D)
which show up on the spectral index map 
are not clearly discernible in the radio and X-ray images indicating their radio/X-ray 
intensities are currently comparable to the lobes although the low radio frequency 
spectrum is getting flatter.  Interestingly, it appears that the X-ray beams 
in Figure \ref{fig33}b terminate in this second pair of hotspots.  
{\it The radio jets should also terminate in this second pair of hotspots.
The second pair of hotspots till which the X-ray beams extend seem to
have just formed and should brighten over time.  The bright hotspots A and D indicate 
the location of the older pair of hotspots and hence the previous jet position.  
Multiple pairs of hotspots are identified in most FR~II and are characteristic 
of a precessing spin axis of the black hole. }

\item Like Cygnus~A, double X-ray beams have been detected in 
other FR II BLRG like Pictor A (see Figure \ref{fig34}) where a radio jet
is barely detectable.  The origin of the symmetric X-ray beams in Pictor~A has 
been suggested to be synchrotron \citep{2016MNRAS.455.3526H}.
An offset is noted between the location of the radio and X-ray hotspots 
%as has been noticed in other FR II sources 
and temporal variability in
the X-ray emission from hotspots has been noted over timescales of months 
to years which is inferred to be indicative of compact emitting regions
\citep{2016MNRAS.455.3526H}.
Existence of multiple hotspots, offset radio and X-ray hotspots, variable
hotspot emission, one-sided
radio jets but two-sided X-ray beams, similar radio and X-ray extents appear
to be common to several FR II sources. {\it As described in the previous sections,
fast jets are composed of a positron-electron plasma and hence positron-electron 
annihilation will proceed at some rate along the entire radio jet, releasing 
soft $\gamma-$ray photons of energy
$\sim 511$ keV.  These photons will be Compton-scattered by electrons to lower
wavelengths thus explaining the X-ray beams which extend from the core to the
hotspots.  The two-sided X-ray beams are detected since 
Doppler beaming is negligible unlike in radio jets.  }

\begin{figure}
\centering
\includegraphics[width=7cm]{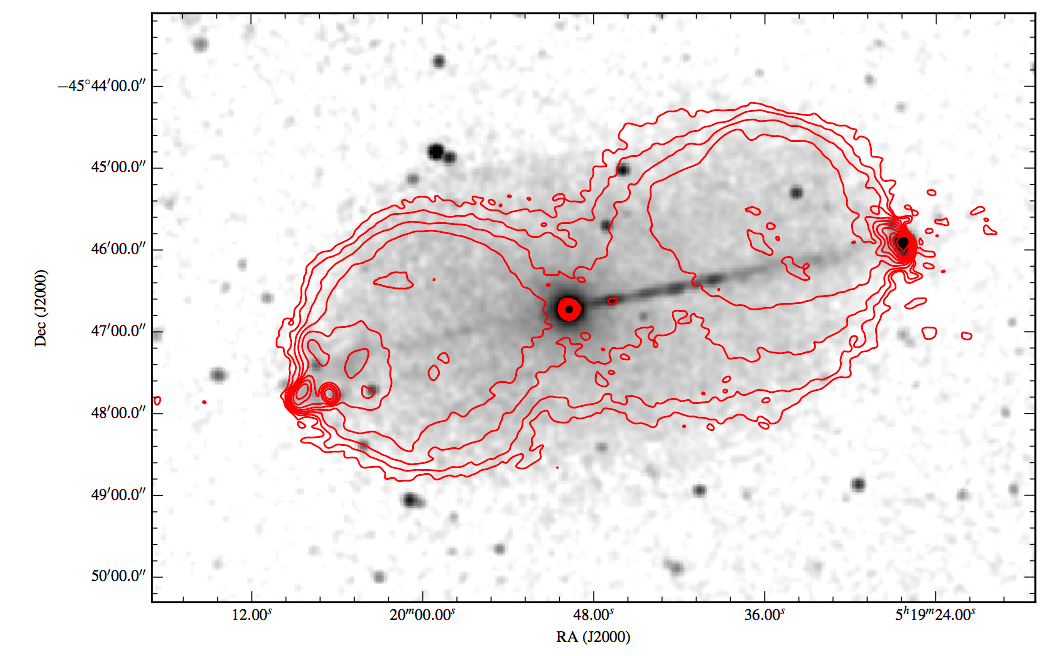}
\caption{\small X-ray and radio image of FR~II galaxy Pictor~A copied from
\citet{2016MNRAS.455.3526H}.  Grey scale shows the X-ray image between 0.5-5 keV
whereas the contours show the 5.5 GHz emission and their similar 
extents is striking.  Double-sided X-ray beams are detected
which extend from the core to the hotspots with the western one being brighter
than the eastern beam.  Radio jets are barely detectable.
The X-ray and radio hotspots show an offset especially noticeable in the eastern lobe.}
\label{fig34}
\end{figure}

\item The spectrum of the core is curved and peaks near 30 GHz as shown 
in Figure \ref{fig35}a \citep{1999AJ....118.2581C}.   
{\it The energy distribution of the electrons accelerated by the pulse of energy 
released in a thermonuclear outburst in the accreted matter on the polar pseudosurface 
of the black hole will be a 
normal distribution with a peak and dispersion determined by the expansion (escape) velocity 
and random motions respectively.  Thus, such a peaked SED with the peak at
high frequencies is the most probable distribution expected for
the instantaneous injection of energy to a large number of particles.}

\item The radio spectrum of the hotspot A (Figure \ref{fig35}b) and D
is a steep power law between 10 GHz and 1000 GHz ($\alpha \sim 1$)
but flattens at frequencies $<10$ GHz 
\citep{1999AJ....118.2581C} and $\alpha=0.34$ between 151 and 327 MHz
\citep{1991ApJ...383..554C}.  Low frequency LOFAR observations 
trace the turnover near 100 MHz in the hotspots A and D \citep{2016MNRAS.463.3143M}. 
This spectrum is very different from the core spectrum. 

\item Equipartition magnetic fields of 250 to 300 $\mu$G at the hotspots which decrease 
to $\sim 45 \mu$G near the core have been estimated \citep{1991ApJ...383..554C}.  

\item Radio variability in Cygnus~A is low as noted from the data at 43 GHz which
did not vary over a timescale of 1.3 years \citep{1999AJ....118.2581C}. 

\item The jet opening angle is estimated to be $<1.6^{\circ}$ on kpc scales
\citep{1996cyga.book...76C}.   A conical jet is not detected, instead the jet 
appears to be cylindrical at all scales - milliarcsec as shown
in \citet{2016A&A...585A..33B} 
to arcseconds. {\it This could indicate the fairly small opening angle of the jets
of FR~II sources which hence remain laterally unresolved upto the hotspots.}
%the observed size is the jet convolved by the resolution cell of the interferometer.  }

\begin{figure}[t]
\centering
\includegraphics[width=6cm]{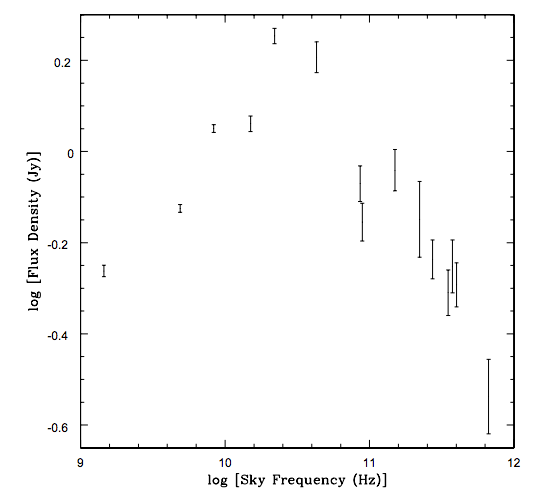}(a)
\includegraphics[width=6cm]{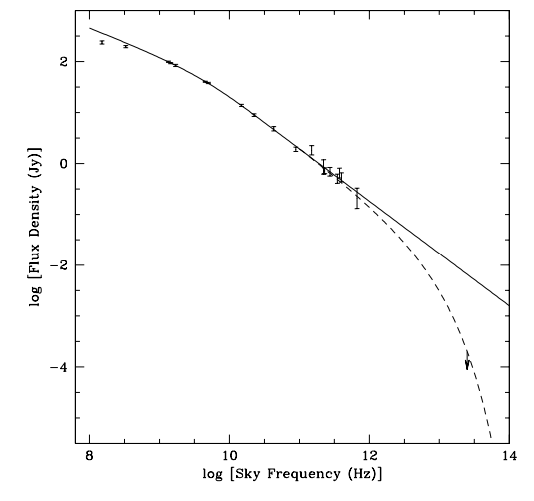}(b)
\caption{\small The spectra of the core (a) and hotspot A (b) in Cygnus~A reproduced from
\citet{1999AJ....118.2581C}.  Notice the turnover of the core spectrum around 30 GHz whereas
the hotspot spectrum shows a mild flattening below 10 GHz.   
On the other hand, the integrated spectrum
of Cygnus~A shown in Figure \ref{fig2} shows a turnover below 20 MHz with 
a convex shaped spectrum at higher frequencies. } 
%\label{cyg}
\label{fig35}
\end{figure}

\item The south-east radio lobe in Cygnus~A which is formed by the backflow
from the hotspot turns abruptly by an angle of $\sim 90^{\circ}$ near the core as
can be seen in the images at frequencies below 1.5 GHz or so  
\citep[e.g.][]{1989MNRAS.239..401L,2006ApJ...642L..33L,2016MNRAS.463.3143M}
and is often referred to as the `plume' in literature. 
No lobe emission is detected near the core at 5 GHz (see Figure \ref{fig33}).  
{\it The nature of the bending is similar to X-shaped galaxies indicating
that lobes bend in all FR~II galaxies when they reach the core.}

\end{itemize}

Having summarised the observations, we try to understand the observational
results on Cygnus~A using the model described in the paper. 

If the magnetic field in which the relativistic electrons emit synchrotron
radiation giving rise to the core spectrum is 45 $\mu$G, then the energy of the electrons 
emitting at the radio frequency of 30 GHz which is the peak of the SED (see
Figure \ref{fig35}) would be 6.5 GeV ($\gamma=1.2\times10^4$).  The energy of electrons 
emitting synchrotron at a X-ray frequency of $10^{18}$ Hz would
be $3.7\times10^4$ GeV i.e. 37 TeV ($\gamma=7.3\times10^7$).   
The core spectrum is the best available
representative of the injection spectrum of the relativistic electrons energised by 
the instantaneous energy released in the thermonuclear explosion.  The observed
core spectrum alongwith the estimated core magnetic field help infer that
the energy distribution of relativistic electrons injected into the jets in Cygnus~A
peaks around 6.5 GeV.  The peak energy could correspond to the escape velocity or likely
higher velocity expulsion of positron-electron plasma from the poles of the pseudosurface.  
The peak energy of 6.5 GeV corresponds to a ultrarelativistic velocity of 0.9999c 
for the electrons.  The observed energy distribution of all particles/cosmic rays should
peak at same energies which could be 6.5 GeV in Cygnus~A.  This energy would correspond
to peak velocities of about 36000 kms$^{-1}$ for the proton distribution.  
If this velocity is equal to the escape velocity from the black hole then the
line emitting matter will be ejected else it will remain trapped on the pseudosurface. 
Since launching of matter from the poles in FR~II cores happens from within a few $R_s$
of the black hole, it requires relativistic escape velocities. This suggests
that protons ejected with a velocity of 36000 kms$^{-1}$ might not be able to escape
the black hole gravity and will fall back on the pseudosurface.  However  
it should also be kept in mind that the peak energy estimates will be uncertain due to
the core spectrum being measured away from the launch site and due to inaccuracies
in the  magnetic field estimates so that the actual peak injection energy could be much larger
than what we record.  Since emission line gas is observed to be distributed 
in a region of $10''$ around the nucleus,
it indicates that there is a detectable fraction of line-forming gas that has been able to
escape from the black hole.  The large FR~II radio structure indicates that the
fast positron-electron plasma has been ejected in plenty from the polar pseudosurface. 

The equipartition magnetic field in the hotspots has been estimated to be 300 $\mu$G.
In this field, the energy of electrons predominantly radiating at the turnover
frequency of 100 MHz will be 0.14 GeV ($\gamma=280$) 
whereas those radiating at 10 GHz where the spectrum shows
a slight break will be 1.44 GeV ($\gamma=2813$).  
Since we assume that the peak energy is indicative of
the expansion velocity of matter, the slowing down of the jet 
will be reflected in the declining peak energy.  In other words, the number
of particles that are moving forward with the high injection momentum
keep declining along the jet.   This would mean
that the electrons which were injected into the jet with peak 
$\gamma \sim 1.2\times10^4$ has decelerated to peak $\gamma=280$ at the hotspots
where the jet is halted i.e. $\gamma \rightarrow 0$ and the plasma diffuses back. 
The integrated spectrum of Cygnus~A shows a turnover at frequencies below
20 MHz and the slight break at 10 GHz is also discernible (see Figure \ref{fig4}).  
If this low frequency turnover is due to the spectrum of the lobes and if
the magnetic field in the lobes is 45 $\mu$G then the electrons emitting
at 20 MHz in the lobes are of energy $0.17$ GeV which is similar to the peak electron 
energy in the hotspots.  Since the forward motion of the jet is abruptly arrested 
at the hotspot, it is interesting to think about the change in the nature of
the energy spectrum especially the peak energy.  Since there is no forward
motion, it should become centred on zero energy.  However on the other hand,
the abrupt stopping can also lead to the forward-directed energy being
randomised with little change in the energy spectrum.  The comparable peak
energies for the hotspot and lobe electron populations argues for the latter
wherein the electron energy spectrum remains the same after being stopped at 
the hotspot.  {\it Thus, the same electron population that flowed from
the core through the jets to the hotspots is flowing
into the lobes i.e. there is no electron reacceleration in the hotspots. }

Narrow lines were first detected from Cygnus~A and hence the 
emission line redshift was determined from the polar NLR which 
is generally detected at parsec and kiloparsec scales around active nuclei
and is typically seen to show expansion of a few hundred kms$^{-1}$ 
about the cosmological redshift.  Thus, 
the redshift of Cygnus~A ($z=0.056$) likely does not contain any contribution from 
gravitational redshift unlike the broad lines from the BLR located close to the
black hole.  In fact, the redshift of Cygnus~A lies towards the lower end of
the redshift distribution of FR~II type radio galaxies as shown in Figure \ref{fig23}. 
Since most FR~II sources host a BLRG nucleus, the wide emission lines which are observed
to appear at the highest redshifts in the spectrum and hence used to determine
the cosmological redshift of the host galaxy, contain a non-zero
contribution to the observed emission line redshift from a gravitational
redshift component.  The FR~II sources hence appear at relatively higher redshifts
than FR~I sources which are predominantly NLRG. 

As discussed earlier, the injection energy spectrum of electrons which will 
be launched along a jet will be a gaussian distribution
with a peak at an energy which
is equal to or larger than the required escape velocity from the black hole and a dispersion
indicating the random velocity component.  The energy spectrum will
evolve as the electrons proceed along the jet and lose energy - the forward velocity declines
to a lower value so that the peak will move to lower energies and as the electrons radiate and
suffer synchrotron losses, the high energy part of the energy spectrum will keep declining. 
In other words, the dispersion of the distribution will keep reducing i.e. the number
of high energy particles will keep declining which will cause the energy
distribution beyond the peak to get steeper.  This, then explains the steepening
of the observed radio spectrum.  Adhering to the hypothesis that
all electron accleration occurs in the core and applying the above
reasoning to the observed spectra of different components in Cygnus~A tells us that
the core spectrum is freshly injected into the jet and should peak 
at high radio frequencies as is observed (Figure \ref{fig35}).  In FR~II sources,
the jet is surmised to expand relativistically till the hotspot
so that the electron energy spectrum will continue to peak at relativistic energies
although much lower than at the core. 
Due to synchrotron losses, the energy spectrum will steepen beyond the peak 
so that it is steeper than at the core. 
As seen in Figure \ref{fig35}, the radio spectrum of the hotspot A (and also hotspot D) 
shows a gentle break at 10 GHz and is steep beyond it but flattens at lower
frequencies showing a turnover around 100 MHz.  The spectrum is
very different from the core spectrum which again argues against any
fresh episode of reacceleration. The steep spectrum beyond 10 GHz in the hotspots 
can be easily explained by the aged electron population so that 
the high energy electrons have lost energy and the dispersion of the
distribution has narrowed.  The varying magnetic field in the lobes and 
hotspots translate to distinct peak radio frequencies of the spectrum
(see Equation \ref{eqn2}) 
even if the underlying electron energy distribution peaked at the same energy.  
If the core frequently ejects matter then it can lead to several energy distributions
contributing to the observed radio spectrum which could result in 
a flat spectrum \citep{1980ApJ...238L.123C} as has been observed in several
active cores, especially if the peaks of the 
ejections are of similar intensity but varying energies.  

The X-ray emission from the hotspots in Cygnus~A has been explained by 
inverse Compton scattering of the radio photons generated in the synchrotron process 
by the same population of relativistic electrons i.e. synchrotron self-Compton (SSC).  
The frequency of the photons generated by the inverse Compton boosting of a low energy
photon can be estimated from Equation \ref{ic}.  In the hotspot where
the magnetic field is estimated to be 300 $\mu$G, electrons of energy 1.44 GeV
($\gamma\sim2813$) will radiate at 10 GHz.  In SSC,
the 10 GHz photons will be inverse Compton-scattered to higher energies by the 
same population of relativistic electrons which emit the radio photons e.g. 
the 10 GHz photon can be inverse Compton-scattered to a photon of frequency $10^{17}$ Hz
(X-ray photon of energy $\sim 1$ keV) by electrons of energy $\gamma=2813$. 
In this case, the energy losses in the hotspots below $\sim 10$ GHz will be predominantly due 
to the synchrotron process whereas above $\sim 10$ GHz, the losses in the hotspots will 
be due to both synchrotron and inverse Compton processes which will steepen the spectrum faster
as compared to synchrotron losses.  While some of the relativistic plasma is
trapped in the hotspots, most of it flows back and forms the lobes.  
The radio spectrum of the lobes of Cygnus~A 
shows a break near 5 GHz \citep{1991ApJ...383..554C}.  The synchrotron emission 
at this frequency would be due to electrons of energy 2.6 GeV ($\gamma=5080$) 
radiating in the measured magnetic field $\sim 45 \mu$G. 

The X-ray beams detected on either side of the core extending upto the hotspots
(see Figure \ref{fig33}) i.e. $\sim 70$ kpc from the core, are difficult 
to explain due to the synchrotron process since that would require the existence 
of TeV electrons upto the hotspots.  While TeV electrons are
generated in the polar regions of the pseudosurface energised by the thermonuclear
explosion, these would fade in 100-1000 years in the typical magnetic field of Cygnus~A indicating
that only the X-ray jet detected close to the core can be due to the synchrotron process.  
As suggested in the paper, the X-ray beams can be readily explained by the process of
Compton scattering of the soft $\gamma-$ray photons generated in the positron-electron
annihilation process to lower energies.  The larger transverse extent of the
X-ray beams compared to the radio jets indicates that the dense plasma around the jet
might also be involved in the Compton-scattering process.  Both the X-ray beams are detected
since the effect of Doppler boosting is diminished.  {\it To summarise: the X-ray beams
are a direct result of Compton-scattering of the positron-electron annihilation
photons, generated in the synchrotron jet, to lower energies.  This process can remain
active as long as the photons are emitted in a dense medium. }

The parsec scale jet-counterjet intensity ratio measured for Cygnus~A is between 
5 and 21 while the kiloparsec scale jet-counterjet intensity ratio is measured to be 
$2.6\pm1$ \citep{1996cyga.book...76C}.  This ratio is typically large in FR~ II galaxies 
since the counterjet is seldom detected.  
The detection of a faint counterjet in Cygnus~A has provided strong
support to the Doppler boosting hypothesis for the non-detection of the counterjet.
It also provides strong support to the existence of a counterjet in all FR~II 
radio galaxies even if not detected.  The existence of the
X-ray beams in Cygnus~A which, as explained above, owe their existence  
to positron-electron annihilation along the radio jets also supports 
the presence of both jets in FR~II galaxies.  These are additional pieces of
evidence - the existence of double lobes and pairs of hotspots on either side
of the core have already provided strong proof to the existence of double-sided 
jets even if the counter-jet is not detected. 

\paragraph{FR~I galaxy - Virgo~A}:

\begin{figure}[t]
\centering
\includegraphics[width=8cm]{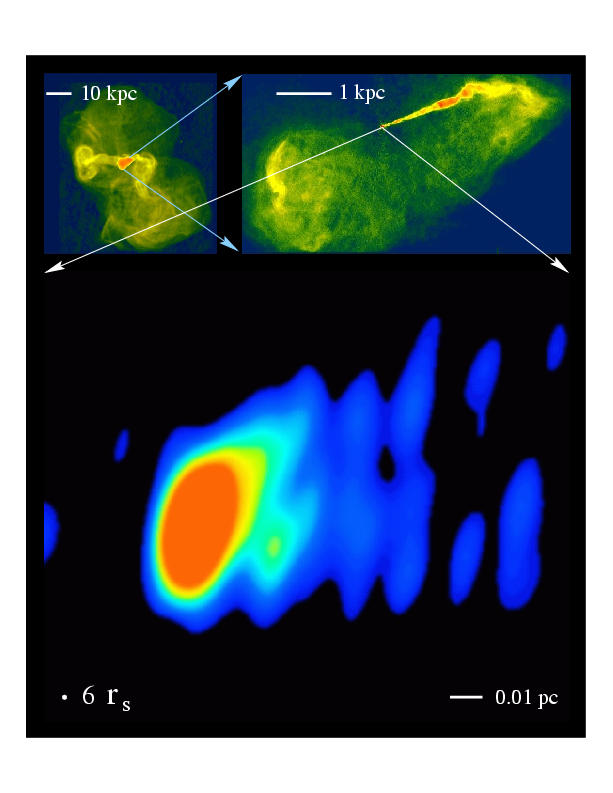}
\caption{\small Radio images of M87 - the above two are with VLA while the lowermost 
highest resolution image is at 43 GHz with the VLBA from \citet{1999Natur.401..891J}.
The image has been downloaded from https://www.nrao.edu/pr/1999/m87/.  Notice
how the conical jets are launched from the black hole as seen in the high resolution
image.  An opening angle of 60 degrees has been estimated near the core whereas
the opening angle on kpc scale is much smaller.  Notice how the triangular blob
in the centre of the top left panel is resolved into a double-lobed single-conical
jet structure which would have been difficult to surmise from the low resolution images. } 
\label{fig36}
\end{figure}

Virgo~A is the FR I radio source associated with the active nucleus of the
elliptical galaxy M 87 which is the central galaxy of the Virgo cluster. 
For a distance of 16.7 Mpc to M 87, 1 arcmin corresponds to about 4.95 kpc.
This galaxy has been extensively studied across the electromagnetic spectrum 
and at angular resolutions ranging from single dish to VLBI (see Figure \ref{fig36}).
Some of the observational results on this galaxy are summarised below: 

\begin{itemize}

\item M~87 is a large cD galaxy which is the host of the active radio source
Virgo~A with the radio structure enclosing a region of size $\sim 80$ kpc.  

\item M~87 has more than 12000 associated globular clusters.
%{\bf reference} 
About 2000 galaxies are believed to be members of the Virgo cluster.

\item A bright conical radio jet detected upto kpc scales extends towards the north-west 
(Figure \ref{fig36}) while a faint counter-jet is detected close to the core
as seen in VLBA images \citep{2018ApJ...855..128W}.  

\item Wiggles are noticeable in the radio jet in the north-west beyond a kpc or so
after which the jet bends to the south and diffuses out to a wider structure like
diffuse beams (see top left panel in Figure \ref{fig36}) and the
conical nature of the jet is no longer distinguishable. {\it This would indicate
that the jet remains ballistic upto a kpc or so till it bends, diffuses and loses
its identity.}  The counterjet in the east also seems to bend towards the south and diffuses.

\item The distribution of diffuse X-ray emission in the 0.6 to 2 keV band is
remarkably similar to the diffuse radio beams in Virgo~A
\citep{2010MNRAS.407.2046M}.  The X-ray emission from the beams is more intense
than the X-rays from the extended halo emission i.e. is similar to the 
distribution of radio emission shown in the top left panel of Figure \ref{fig36}.  
The X-ray emission from the diffuse X-ray halo is brightest at
2 keV whereas the emission from the X-ray beams peaks at 1 keV and 
is barely detectable at 3 keV
as can be surmised from the emission measure maps in \citet{2010MNRAS.407.2063W}.
This, then, could indicate that the X-ray emission coincident with the radio beams
includes a contribution from another process in addition to the thermal emission.  
A collimated X-ray jet coincident with the inner radio jet in the north-west is
also detected. 
%{\bf reference}.  
The X-ray jet is distinct from the X-ray beams.
{\it While the X-ray jet close to the core has a synchrotron origin like the radio jet, 
the X-ray beams which trace the diffuse radio beams will be due to 
Compton-scattering of the positron-electron annihilation photons
generated in the radio jets/beams.  A prerequisite for this explanation is the
coincidence of the radio synchrotron features due to the positron-electron plasma 
and the X-ray features due to Compton scattering of soft $\gamma-$ray photons. } 

\item A conical edge-brightened jet emerging with an opening angle of $\sim 60^{\circ}$ is
detected on sub-parsec scales \citep{1999Natur.401..891J} (lowermost panel in Figure \ref{fig36}).
The jet is observed to be collimated to less than $5^{\circ}$ on the scale of hundreds of 
parsecs \citep{1995ApJ...447..582B} 
(see top right panel in Figure \ref{fig36}). 
Such observations which suggest that the jet is ejected in a cone of a large 
angle and is collimated to a cone of a smaller angle at large separations from 
the active core, have prompted theories which can explain such collimation. 
{\it 
Before the widely different conical angles of the jet near the core and far from it
are taken to be indicative of the opening angle, possible effects on the sub-parsec
jet morphology due to finite angular resolution should be examined.  For example, 
such an effect is well demonstrated by the top panels in 
Figure \ref{fig36}.  In the top left panel which is at a lower resolution than the
top right panel, the entire radio structure of the top 
right panel is detected as a triangular region shown in orange.  
Large and asymmetric opening angles for the two jets would be surmised from
the top left image.  The higher resolution image of the central region shown
in the top right panel is required to measure the conical angle of the jet. 
A similar angular resolution effect could be afflicting the 
jet opening angle in the bottom panel of Figure \ref{fig36} which zooms in close to the core. 
It is important to obtain even higher resolution images before inferring different
opening angles at the base of the jet and at large separations.   In the model
presented in the paper, the conical jet has to be ejected within a small angle which 
will remain constant as long as the jet remains ballistic.  This is borne
out by the jet with a constant conical angle from several parsecs to kpc scales
as seen in the top right panel.  }

\item The base of the jet in Virgo A is resolved by interferometric observations at
230 GHz from which it is surmised that the emission which is fitted with a gaussian of 
FWHM $\sim 40 \mu$arcsec, corresponds to a linear size of $5.5 R_s$ \citep{2012Sci...338..355D}
at 16.7 Mpc. 
%which at a distance of 16.7 Mpc to Virgo~A 
This result indicates that the jet in Virgo~A is launched from close to the black hole.  
The black hole  mass in M~87 has been estimated to be $6.2\times10^9$ M$_\odot$ 
so that $R_s =1.9\times10^{15}$ cm.  
{\it If 40 $\mu$arcsec indicates the size of the pseudosurface i.e. the
separation between the two poles from which jets are launched then the jets will 
be launched from a region separated from the black hole by $\sim 2.75 R_s$.  These jets have
to be launched with a velocity $\ge 0.6c$ to be able to escape the gravity of the 
black hole.  If we could detect emission lines from the matter 
at $2.75R_s$ then their redshift would be observed to be large
due to the presence of an intrinsic component corresponding to a gravitational 
redshift of 0.18 and a Doppler component due to the expansion velocity of 0.6c.  Alternately,
40 $\mu$arcsec could correspond to the size of a blob in the jet close to
one of the poles in which its implications could be different.      }

\item The transverse extent of the optical jet is
narrower than the radio jet so that it appears to be confined to the centre of
the radio jet \citep{1994ApJS...90..909S} often referred to as the spine of the jet.  
The same is true for the X-ray jet and the correlation between the optical and
X-ray jets is good \citep[e.g.][]{2002ApJ...564..683M}. 

\item Knots in Virgo A have been deduced to be expanding with velocities which 
are generally
less than 0.5c although a few instances of superluminal expansion have also been
reported \citep{1995ApJ...447..582B}. 

\item The magnetic field lines are predominantly oriented along the jet as deduced
from radio polarisation measurements.
The percentage polarisation is typically 30\% in the bright parts 
of the jet (i.e. spine) and 50\% or more at the jet boundary (i.e. sheath) 
\citep{1989ApJ...340..698O}.
The magnetic field in the jet is estimated to be 200-300 $\mu$G 
\citep{1989ApJ...340..698O}.

\item The halo-like radio structure around M 87 is diffuse
and amorphous and is commensurate to having been generated by a precessing 
jet \citep[e.g.][]{1986NRAOW..16...79R}. 

\item The spectrum between 25 MHz and 10 GHz of the central part of Virgo~A 
is a power law with index $0.6$ while for rest of the radio source the index is
around $1.1$ (S $\propto \nu^{-\alpha}$).  Some steepening beyond 10 GHz is 
detectable from the diffuse regions
\citep{2012A&A...547A..56D}. They estimate equipartition magnetic fields which are on average
$30 \mu$G in the central part and $10-13 \mu$G in the extended diffuse parts. 
%{\bf also check Owen et al. 1989 for spectral index details} 

\item The radio source has well defined boundaries so emission is detected
from a region of the same extent at all radio frequencies. 
%and the lower frequencies do not detect a larger source.

\item VHE $\gamma-$ray ($E>100$ GeV) flares which last for only a day or so
are detected from Virgo~A and have been found to be accompanied by a rise
in the flux density at high radio frequencies of the core or of 
a jet component like HST-1 \citep{2012ApJ...746..151A}.

\item Optical filaments extending along the general direction of jets are
detected in H$\alpha$ and forbidden lines.  These filaments extend over 
about 3 kpc in the north-west and about 8 kpc in the south-east, 
are broader than the jets, patchy, show some lateral displacement
from the jet and are kinematically perturbed \citep{2018arXiv181009804B}.
Patchy dust is detected near the core and in patches around the core and in a
broader region in the general direction of the  jet
\citep{2018arXiv181009804B,2015ApJ...805..178B}. 
{\it All these signatures are typical of the NLR gas 
in Virgo~A which is ejected from the black hole along with the relativistic
plasma.  Dust has formed in the dense clumps within the NLR.  This explains why
the distribution of dust and H$\alpha$ emitting gas 
are oriented along the jet. }

\item VLBA images of Virgo~A taken at regular intervals indicate jet ejections on 
timescales of weeks \citep{2018ApJ...855..128W}. {\it This means that sufficient matter is
accreted on the polar regions of the pseudosurface and favourable physical conditions 
are established on timescales of weeks which
triggers a thermonuclear outburst.   A relativistically expanding positron-electron
jet in addition to line forming gas are then ejected. }

\begin{figure}
\centering
\includegraphics[width=7cm]{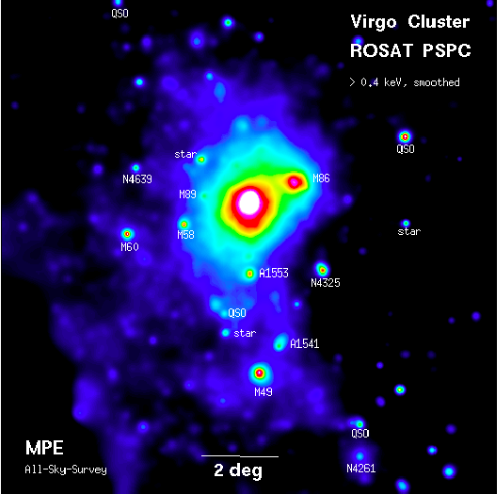}
\caption{ROSAT X-ray image of the Virgo cluster 
\citep{1994Natur.368..828B}.  While the X-ray emission
peaks at M~87, it extends to a large region and encloses several other
galaxies - both active and normal and shows several secondary peaks.  }
\label{fig37}
\end{figure}

\item Diffuse soft X-ray emission is detected over a large region extending
to several degrees and enclosing several member galaxies of the Virgo cluster
\citep{1994Natur.368..828B}.  This emission shows peaks associated
with several active galaxies with the brightest emission associated with
the brightest cluster galaxy M~87 (see Figure \ref{fig37}).  
The disk galaxies within the X-ray halo show truncated gas disks
\citep{2009AJ....138.1741C} which has been interpreted as being
due to the action of ram pressure on the interstellar medium of the galaxies. 
{\it The active nucleus in the central galaxy M~87 has led to the formation
of a central mini-halo (around 80 kpc) like region which is detected in radio synchrotron
and soft X-rays.  The large scale X-ray emission (Figure \ref{fig37})
in Virgo cluster over about 12$^{\circ}$ region (linear extent of 3.5 Mpc)
includes contributions from several active galaxies which
have lost gas either through mass loss induced by a gravitational torque or
in their active phase through polar ejections which would explain its patchy
nature.   Since the disk galaxies within the extended X-ray halo have either
evolved to lenticulars or ellipticals with small quantities of leftover
interstellar gas or have truncated gas disks would support the formation of
the extended halo from evaporation of the atomic gas.  Virgo cluster is believed
to host about 2000 galaxies while more than 12000 globular clusters have been
found to be associated with M~87.  If 1000 galaxies had lost their interstellar
medium of mass $10^9$ M$_\odot$ each which has evaporated to X-ray emitting
phase, then this can explain the origin of 
about $10^{12}$ M$_\odot$ of X-ray emitting intracluster medium. }

\end{itemize}

Only a single jet is detected on parsec scales from many FR~I radio galaxies while
both the jets are detected on kpc scales.  This behaviour 
has been interpreted as being due to the reducing effect
of Doppler beaming indicating declining bulk velocities along the jet in FR~I
sources.  Although no counterjet is detected even on kpc scales in Virgo~A a
diffuse radio beam which provides indirect evidence to the existence of a counterjet 
which transports the radio plasma from the core, is detected.

The observed large-scale radio structure of Virgo~A seems to require precession
and also other factors 
to explain its origin.  The symmetric halo around NGC~193 (Figure \ref{fig27}),
can be readily explained by
precession of the black hole spin axis and hence the changing direction of the
jets ejected from the poles.  However the observed radio structure of Virgo~A is
more complex and in addition to a precessing axis and episodic quasi-simultaneous 
ejections from the two poles, interaction with
the intracluster medium needs to be invoked, which has rerouted the radio plasma.   

High resolution VLBA radio images at 43 GHz have detected a faint
radio feature close to the core which is attributed to the counter-jet  
\citep{2018ApJ...855..128W}.  If the absence of
the counter-jet was due to Doppler beaming then it would not have been detected
on parsec scales where the bulk expansion velocities should be highly relativistic.
If the detected feature is indeed the counter-jet then we have to search for
reasons other than Doppler beaming for the non-detection of the counter-jet
on kpc scales in Virgo~A. 

The entire radio structure in Virgo~A is confined to a well-bounded region as is
observed in several FR~I sources (e.g. NGC~193).  The extent of the emitting region 
in such sources is same at all radio frequencies and also
comparable to the extent of the soft X-ray emission (e.g. NGC~193 and see Figure
\ref{fig27}).  
Good correlation is observed between the radio and X-ray beams in Virgo~A and supports
a connection between the emission processes.  In fact, the behaviour 
of such FR~I sources is comparable to
that of FR~II sources like Cygnus~A (see Figure \ref{fig33}) wherein,
the soft X-ray and radio emissions share a major axis and the extent of the soft 
X-ray and radio emissions appear to be similar.  All these argue for some
connection between the radio synchrotron and thermal X-ray emissions in active galaxies.  
In NGC~193 and Cygnus~A as in several other active galaxies, 
X-ray cavities are observed to be coincident with the radio lobes
\citep[NGC 193,][]{2011ApJ...732...95G}.  On a general note, it appears that the 
extremities of
the radio emission in many FR~I and FR~II radio sources is determined by the
thermal gas distribution associated with the galaxy.  In fact, the radio morphology of 
NGC 193 suggests that the precessing jet is forming multiple hotspot-like
features which trace the outermost boundary of the radio emission which is
well-defined in the east and west.  These hotspot-like features are 
identifiable on the spectral index image as the features where the spectrum is
flatter (see Figure \ref{fig27}). 
In the eastern part, the X-ray and radio are of similar extent while in the west, the
radio emission seems to extend further than the X-ray emission. 
In fact, it is interesting
to note on comparing the second panel in (a) with (b) that in the eastern lobe,
the hotspot-like features are confined to the outermost
boundary while in the west, these cover the entire region where soft X-ray is missing
from the lobe.  This means that these hotspot-like features have formed at the edge of
the X-ray distribution again arguing for a connection.  
To recall, in FR~II radio sources, the hotspots are formed just outside the soft
X-ray distribution (e.g. Cygnus~A in Figure \ref{fig33}) and the reason for the
same has been explained earlier.  The radio emission extended
along the equatorial region
has an especially steep spectrum and is coincident with faint soft X-ray emission.  

The jet consists of both knots and uniform emission -
if knots are ejected from the black hole than uniform emission might arise from
ablation of the knots - an argument already used to explain the faint sheath observed
around several bright jets.  Moreover if ejections are frequent, knots might be
too numerous to be resolved and could appear as uniform radiation. 
The radio movie at 43 GHz made from VLBA images showing the formation and expansion of 
jets from the
core of Virgo~A show the jet to be limb-brightened, dynamic and frequently
replenished by plasma from the poles \citep{2018ApJ...855..128W}. 

If for argument sake, we evolved the central active nucleus in 
FR~I radio sources of M~87 and NGC~193 to future epochs when accretion has
stopped so that no further radio ejections occurred from the source, then the 
jets and beams would diffuse and the large scale morphology of the fossil radio
emission would resemble a radio halo/mini-halo often detected in centres of clusters.  
The extent and major axis of the soft X-ray and radio halo in clusters
are comparable and relics are observed to form at the boundary of the soft X-ray emission.
The similarity to the radio/X-ray sources associated with an active nucleus is uncanny
and clearly argues for either a connection or similar physical processes being at work. 
This is discussed in the section on radio halos and relics.

\paragraph{Seyfert 1.5 - NGC 4151}:
The main observational results on NGC 4151 are described below with comments
included in italics. 
\begin{itemize}
\item NGC 4151 is an almost face-on barred spiral galaxy of type Sab with
HI distributed throughout the galaxy \citep{1973MNRAS.161P..25D}.
The emission line redshift of NGC 4151 is 0.0033 ($\sim 960$ kms$^{-1}$) 
which corresponds to a distance of $\sim 14$ Mpc.
At this distance, 1 arcsecond corresponds to a linear distance of 68 parsecs. 

\item The nuclear continuum emission is bright with the mean nuclear V band brightness
being $\sim 11.5$ magnitudes \citep{1977SvA....21..655L}.  {\it This corresponds to a
V band absolute magnitude of $\sim -19.2$ magnitudes for a distance of 14 Mpc.} 

\item The nuclear continuum emission in the ultraviolet bands 
consists of two components - a power law which fits wavelengths
$>2000$A while wavelengths $<2000$A include excess emission over 
the power law \citep{1981MNRAS.196..857P}.
{\it The two dominant components will be the synchrotron component from the
relativistic plasma at longer wavelengths and the black body emission from the
hot pseudosurface dominating at shorter wavelengths.}

\item The nuclear ultraviolet spectrum of NGC~4151 is characterised by wide-winged 
asymmetric emission lines (FWZM $\ge 10000$ kms$^{-1}$) of He II, C IV, Mg II, Balmer 
lines etc which led to its original classification
as a Seyfert 1 nucleus.  Absorption lines are also detected which are blue-shifted
wrt the emission lines.  Absorption lines are detected at velocities ranging from 
900 to $-100$ kms$^{-1}$ \citep{1979MNRAS.189P..45P,1981MNRAS.196..857P}. 
{\it The wide-winged emission lines arise in the BLR and the redshift contains
a gravitational redshift component as can also be surmised from the detected
absorption features which are blue-shifted wrt to the emission line redshift.  
This behaviour is typical of the broad spectral lines in quasar spectra although
the magnitude of redshifts is much larger in quasars.  
These results on NGC~4151 demonstrate that the BLR in all active nuclei
consist of emission and absorption line zones and the lines will be wide and
include a detectable gravitational redshift component if the BLR forms sufficiently 
close to the black hole. } 

\begin{figure}
\centering
\includegraphics[width=8cm]{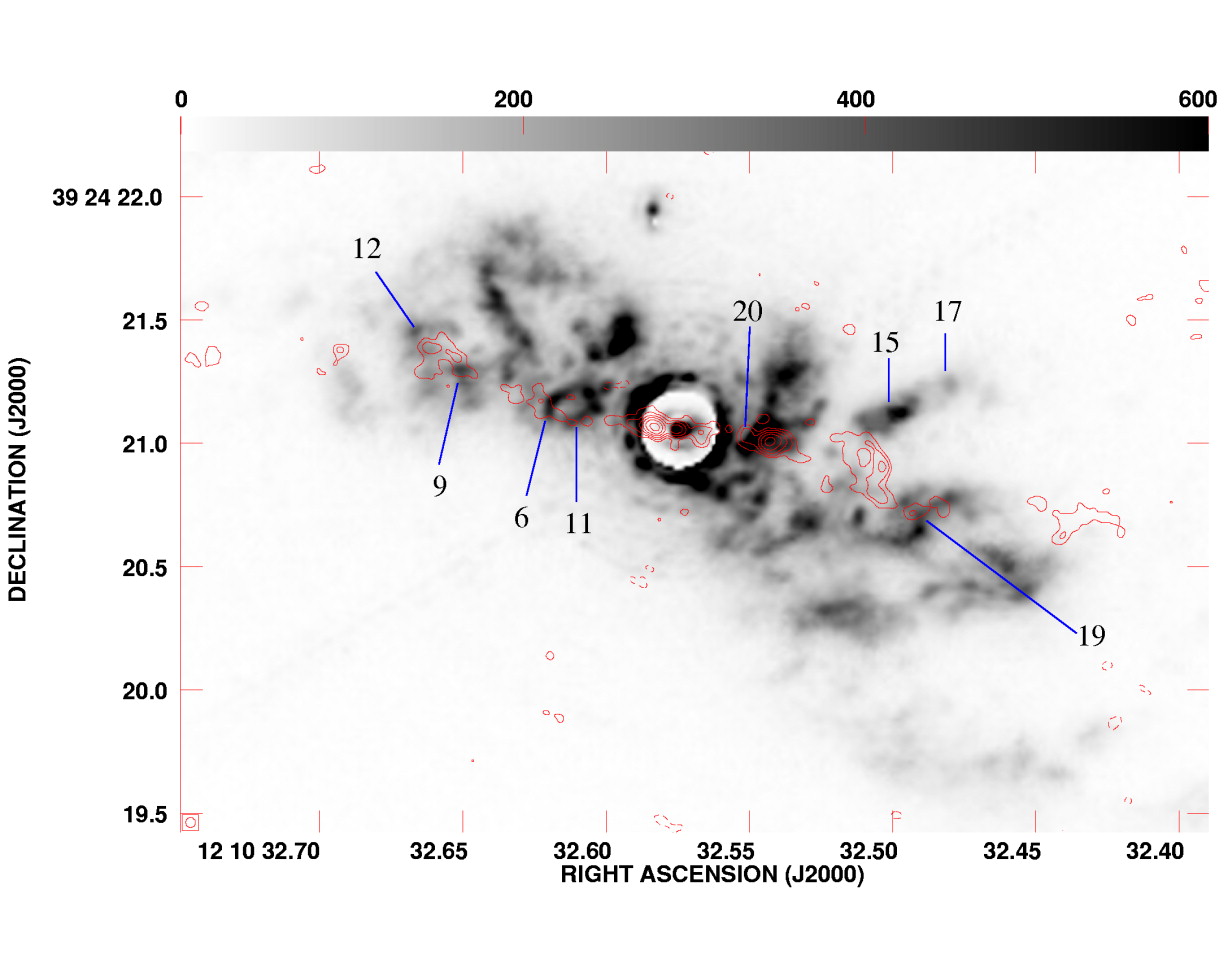}
\caption{\small Figure showing the superposition of the [O III] line emitting gas (grey) and 
the radio continuum emission (contours) from the jets in NGC 4151, copied from 
\citet{2003ApJ...583..192M}.  Notice the
optical line emitting gas is distributed over a larger conical region than the radio jet. }
\label{fig38}
\end{figure}

\item Correlated variability is detected in the continuum and the wide spectral lines. 
It is noted that when the ultraviolet continuum is stronger, the broad emission lines are
wider and stronger \citep[e.g.][]{1981MNRAS.196..857P}.  For example,
the FWZI of the C~IV and Si~IV lines were observed to vary from 
$\sim 30000$ kms$^{-1}$ in May 1978 to less than 20000 kms$^{-1}$ in January 1979  
which was correlated with the dimming of the ultraviolet continuum 
\citep{1981MNRAS.196..857P}.   Depending on the continuum strength and linewidths, the
source is in the `high' or `low' state.  
{\it The dominant continuum source at ultraviolet bands i.e. black body
radiation from the hot pseudosurface is the photoionizing
source for the BLR so that any change in the former leads to changes in the strength
and width of the emission lines from the BLR.  The correlated variability also
supports the radiating pseudosurface and the relative location of the BLR and the 
pseudosurface.  One of the original cause of variability could be  
the occasional thermonuclear outburst on the interface of pseudosurface/inner parts of the BLR. }

\item Deep absorption components are detected within the broad C~IV and Mg~II emission lines 
\citep[e.g.][]{1984MNRAS.206..221U}. 

\item In NGC~4151, the high ionization emission and absorption lines of C~IV, 
Si~IV, He~II are seen to vary with the ultraviolet continuum whereas the relatively 
narrow lines of C~III], Si~III] seem to remain unchanged \citep{1981MNRAS.196..857P}. 
In the Seyfert~1 galaxy NGC~5548, comparable variability was recorded for
C~IV and C~III] lines while the Mg~II feature remained unchanged 
\citep{1992ApJ...393..113C}.  {\it Such a range of behaviour is expected 
due to the range of BLR properties such as separation from the black hole,
extent, abundance, absorbing zone etc which will exist and be distinct for each
active nucleus.  Even in the same active nucleus, the observed BLR-related 
variablity can show a range of properties. }

\item The broad C~IV and Si~IV are the widest lines with FWZI reaching
30000 kms$^{-1}$ followed by He~I 5876 line observed to be in the range 
$12000-16000$ kms$^{-1}$ and then Balmer lines whose widths are
$\sim 8000$ kms$^{-1}$ \citep{1975MNRAS.173..381B}.  Broad
Fe~II lines are also detected from the nucleus of NGC~4151 \citep{1975MNRAS.173..381B}. 
NGC 4151 which was originally classified as a Seyfert 1 was later shifted to  
class 1.5 due to detection of the narrower Balmer features \citep{1976MNRAS.176P..61O}. 

\item While most of the variability in X-ray and ultraviolet bands in NGC~4151 is well 
correlated to within a few days, there exist the occasional ultraviolet flare
which is not 
accompanied by an X-ray flare \citep{1986ApJ...306..508P}.  A similar behaviour is
also recorded between the ultraviolet (1350 A) and X-ray (2-10 keV) emissions in NGC 5548
\citep{1992ApJ...393..113C}.  {\it This could be explained if the ultraviolet flare
arises due to variability in the continuum from the BLR.  The 
variability which arises on the hot pseudosurface will be detected in several wavebands
including X-ray and ultraviolet bands but this might not be the case for the relatively
cooler BLR. }

\item NGC~4151 is variable in all wavebands ranging from the infrared to X-rays. 
Soft X-ray emission extended over a 2 kpc region about the active nucleus
has been detected \citep{2010ApJ...719L.208W} which fills in the HI cavity.
No X-ray emission that extends around the galaxy in a halo has been detected.

\item The narrow line region in NGC~4151 has been inferred to extend by
$\ge 6''$  \citep{1973ApJ...181...51U}. 

\item Radio jets, extending to scales of about $10''$ (680 pc), are observed to emerge from
the active nucleus in NGC~4151 \citep{1981ApJ...247..419U,1982ApJ...262...61J}. 
The narrow line region detected in the optical 
and the radio jet are extended along similar directions 
\citep{1981ApJ...247..419U,1982ApJ...262...61J,1973ApJ...181...51U}. The optical
line emitting gas is spread over a larger conical region of opening angle
of $80^{\circ}$ \citep{1998ApJ...492L.115H} than the radio jet 
(see Figure \ref{fig38}) and shows
expansion with the northeast being redshifted and the southwest part being blueshifted
\citep[e.g.][]{1982ApJ...262...61J,1973ApJ...181...51U}.
{\it The differing transverse extent of the radio jet and NLR can be explained 
by a precessing spin axis of the black hole.
The radio jet composed of relativistic positron-electron plasma traces
the instantaneous position of the axis whereas the longer-living optical line emitting gas
traces the entire precession cone indicating that the angle of the precession cone
is close to $80^{\circ}$ while the ejection in each epoch is over a conical angle
determined from the radio jet.  } 

\item Linearly polarised optical continuum 
emission with a position angle along the radio jet is also detected in NGC~4151 
\citep{1980ApJ...240..759S}.

\item The optical lines form a biconical outflow with expansion velocity 
of $\sim 350$ kms$^{-1}$ and opening angle of $80^{\circ}$
\citep{1998ApJ...492L.115H}.  From the line ratios,
the authors infer that this narrow line gas is photoionized by the central continuum source
and they also infer a density gradient along the narrow line gas.  The lines close
to the core appear to be distributed along the radio axis and show higher expansion
velocities as compared to 
the gas located farther from the core which is also patchy but appears to be displaced from
the radio axis \citep[see Figure 4 in][]{1998ApJ...492L.115H}.  
{\it The photoionizing source for the NLR will be the ultraviolet photons from the hot
polar pseudosurface from which it is ejected. }

\item  Reverberation studies have deduced that the C~IV line arises 
in a region which is about 5 light days away from the continuum emitting region while the 
Mg~II variability shows a lag of about 12 days wrt to the continuum 
from which it is determined that it is formed
in a region about $\sim 12$ light days from the continuum emitting region 
\citep{1986ApJ...305..175G}.
Similar analysis with Balmer lines indicated that these lines arise in a region 
located about 10 times further than the region emitting C~IV and He~II lines 
\citep{1984MNRAS.206..221U}.  These results have led to the inference that the
BLR is stratified into at least three emission line zones with varying
ionization and densities \citep{1984MNRAS.206..221U}. 
A black hole of mass $0.5 - 1 \times10^8$ M$_\odot$ has been estimated in the centre of 
NGC 4151 from line variability studies  \citep{1984MNRAS.206..221U}. 
Another set of reverberation mapping of the broad emission lines has determined lags
of about 3 days and hence radius of the 
broad line region to be 3 light days $\sim 0.0025$ pc and the mass of the central black hole 
to be $1.2\times10^7$ M$_\odot$ \citep{1999ApJ...526..579W}. 
{\it Since the ionization structure
of the BLR can be dynamic depending on the variation in the photoionizing source,
the emitting zones will show radial shifts i.e. different lags. }

\item Extended narrow line gas (ENLR) is
detected around NGC~4151 such that the detected ratio [O~III] 5007/H$\beta$ is lower than
in the NLR and higher than in HII regions \citep{1987MNRAS.228..671U}. The ENLR 
has dimensions of several kpc 
%even beyond the radio jet 
whereas the NLR is extended to a kpc or so.  Moreover the linewidths
from the ENLR ($\le 45$ kms$^{-1}$) are smaller than in NLR ($\sim 300-500$ kms$^{-1}$)
and the velocity variation along the ENLR of about 112 kms$^{-1}$ suggests 
that the gas although extended along the inner milliarcsec ($\sim 2$ pc) radio jet 
(and not the larger scale radio jet) follows disk-like rotation 
\citep{1987MNRAS.228..671U}.
{\it  The ENLR could constitute the interstellar gas that has gotten channelled into
the central parts of the galaxy and from which the black hole is accreting.}

\end{itemize}

NGC~4151 which was originally classified as a Seyfert 1 was later shifted to  
class 1.5 due to detection of the narrower Balmer features \citep{1976MNRAS.176P..61O}. 
From the range of widths of the broad emission lines that were observed and
their ionization states, it was suggested by \citet{1984MNRAS.206..221U} that 
the broad line region around NGC~4151 was stratified into three regions such that 
BLR1 where the broadest lines of C~IV, Si~IV, He~II formed, was the closest line 
forming zone to the photoionizing source, had 
n$_e \ge 10^{10}$ cm$^{-3}$ and which recorded the highest line variabilities; 
BLR2 was the region where the narrower C~IV, Mg~II, He~I formed and which also had similar 
densities as BLR1; BLR3 was the region where the narrower lower ionization lines of 
Mg~II, C~III], Balmer lines formed and densities were lower than $10^{10}$ cm$^{-3}$ 
and variabiity was minimal. 
Balmer emission is faint or absent in BLR1 and BLR2 \citep{1984MNRAS.206..221U}.
Such a stratified behaviour of the BLR which lies immediately outside the pseudosurface
is readily supported by the model presented in the paper.  The hot pseudosurface is
the source of the photoionizing photons which ionize and excite the BLR that is formed in the
matter that is accreted on the pseudosurface.  
The BLR zones closest to the pseudosurface will show the highest ionization,
will show the broadest lines due to the largest gravitational redshift component, will
be the most dense of all the BLR zones and will show the highest correlated variability with
the continuum from the pseudosurface.  Beyond this there
will be a gradation in the ionization, temperature and densities of the regions i.e.
BLR2 and BLR3 zones.  
The stratified nature of the emission line region in the BLR into three zones  
based on the line widths is indicative of the differing 
separation of the respective zone from the pseudosurface of the black hole which provides
the black body continuum and the photoionizing photons.  
Beyond the emission line zones in BLR, there can form absorption line zones.
The gravitational redshift component in the emission lines will hence be larger
than in the absorption lines which explains why the absorption lines are blueshifted
wrt to the emission lines.  If all the stratified layers
of the BLR are of comparable linear thickness then the zones lying closer to
the black hole will form broader lines due to enhanced gravitational broadening. 
The line widths can be used to estimate the fractional width of the line forming region. 

A black hole of mass $0.5-1 \times 10^8$ M$_\odot$ has been estimated in the centre 
of NGC 4151 from line variability studies \citep{1984MNRAS.206..221U}.
If the pseudosurface in such a black hole has formed at a separation equal to $R_s$
then it will need to be very hot to explain the observed V band absolute magnitude of
$-19.2$ magnitudes which is comparable to the absolute magnitude of the quasar 3C~273.  
From Table \ref{tab1}, it can be deduced that the temperature will have to be $>> 40000$ K.
On the other hand, the pseudosurface and BLR are expected to be located farther than
$R_s$ in Seyferts, as can be inferred from the significantly lower quanta of the 
intrinsic redshift component in the wide emission lines.
The further the pseudosurface from the black hole,  larger should be its 
surface area i.e. emitting surface
which will bring down the requirement for an extremely high black body temperature.
%\citep{1984NYASA.422..291U}.  

We estimate a few typical physical parameters expected for the BLR which has
formed in the non-polar regions of the supermassive black hole in NGC 4151.  The galaxy 
(emission line) redshift is 0.0033 
%i.e. 990 kms$^{-1}$ 
so that the intrinsic redshift due to gravitational redshift
has to be $\le 0.0033$.  Lets assume the extreme case wherein the dominant contribution
to the observed redshift
is due to a gravitational redshift i.e.  $z_g=0.003$.  Since $z_g = R_s/2 R$, this 
photon will arise from a distance of about $167 R_s$ from the black hole which gives
a lower limit to the separation of the BLR from the black hole. If $z_g < 0.003$, then
the photon will arise from a larger separation from the black hole.   
For a black hole of mass $10^8$ M$_\odot$, $R_s = 3 \times 10^{11}$ m and hence
$167 R_s = 5\times10^{13}$ m $ \sim 2$ light days  which is the separation 
of the BLR from the black hole.  Thus, the high ionization broad emission lines in 
NGC~4151 would arise at a distance of 2 light days or more from the black hole.  For a 
gravitational width of 100A for the C~IV 1550A line (i.e. 20000 kms$^{-1}$), the ratio of
the thickness and radial separation of the line forming region will be 
$100/1550= 2/31$ and the thickness of the line forming
region at $167 R_s$ will be $2/31 (167 R_s) = 10.8 R_s$.  These estimates 
indicate the shortest possible separation of the C~IV zone in the BLR 
from the black hole and its thickness in NGC~4151 when
the line is gravitationally broadened to 100A.  
The C~IV zone will be farther from the black hole if there is a smaller contribution
from gravitational redshift while its thickness will be larger to allow for the required
change in the gravitational potential in the line forming zone. 
{\it To summarise: the main source of ultraviolet photons is the hot pseudosurface
which ionize and excite the BLR formed in the non-polar parts.  Lines broadened
and redshifted due to the influence of the strong gravitational potential of the black hole
are formed in the BLR around NGC~4151.  If the black hole is of mass $10^8$ M$_\odot$,
then from the maximum possible gravitational redshift, we can surmise
that the C~IV line forming zone will be located at a separation $\ge 2$ light days and
will form a line of width 100A if its thickness is $\ge 11 R_s$.}  

The physical processes active in the accreted matter in the polar regions
are different from those in the BLR as can be surmised from the presence of
bipolar radio jets and narrow line regions and we now discuss the polar properties in 
NGC~4151.  To recall, in a rotating black hole, matter will accrete on the polar 
pseudosurface over an area determined by the angular speed of the black hole
and hence the covering factor of the BLR. 
A fast rotating black hole will collect matter over a smaller polar region than a slower
black hole.  The accreted matter at the poles will be compressed and heated.
If the temperature in the bottom layer exceeds $10^8$ K, it will ignite in a
thermonuclear blast which will release energy depending on the mass that is ignited. 
This instantaneous release of energy will energise the accreted matter
and in many cases eject it.  This would lead to bipolar outflows but
since the two poles are separated by the BLR, the ignition and ejection at
the two poles should happen
independently.  This will lead to the detailed structure of the two radio jets and NLR 
being different.  The thermonuclear burst would eject relativistic 
positron-electron plasma and and heavier atoms/ions which would be ejected at high 
velocities from the black hole.   This, then, will constitute the 
radio jets and the narrow line gas which is found distributed along the jet.   
Faint radio jets and optical line emitting gas have been detected in NGC 4151 
\citep[e.g.][]{2003ApJ...583..192M} (see Figure \ref{fig38}) with the
optical gas being extended over a larger region compared to the radio jets.  
The slightly differing mean position angles of the radio jet axis and the axis
over which the optical line forming gas is distributed 
appears to be due to precession of the black hole so that the narrow line gas is spread
over the entire precession cone region while the radio jet is confined to the
instantaneous position of the spin axis which currently seems to be located at one edge 
of the line forming gas distribution (see Figure \ref{fig38}).  Over time,
the spin axis should precess so that the radio jet will be displaced in the clockwise
direction.  The line region and
radio jet are resolved into a knotty structure (see Figure \ref{fig38}) 
which supports the episodic ejection expected in an accretion dependent 
thermonuclear expulsion.  The low expansion velocity of 350 kms$^{-1}$
estimated for the narrow line gas indicates that the relativistic ejection from the
poles of the black hole rapidly slows down.  
However it can be noticed from Figure \ref{fig38} that the opening angle 
of the jet in NGC 4151 is large
and hence indicates that the black hole rotation is slower and the
covering factor of the BLR is smaller.  The small radial extent of the jet and
NLR indicate that the jet is not as energetically ejected as in FR~I or FR~II
sources.  Moreover the coexistence of the radio jet and optical line emitting
gas makes it possible for the radio jet plasma to radiate in the magnetic field 
frozen in the line forming gas.  As explained earlier, a conical flow is expected when 
matter is radially launched from a spherical region at the poles and the
biconical NLR in NGC 4151 strongly supports this scenario. 
NGC 4151 is classified
as a radio-quiet active nucleus due to the faint nature of the bipolar radio emission
compared to its optical continuum emission.  
However it clearly hosts radio jets supporting ongoing accretion, compression,
ignition and ejection from the poles.

% {\bf Seyfert 2: NGC 1068 \\}
%\citet{1976ApJ...206L...5A} found that the permitted emission lines (e.g. hydrogen,
%helium) from the nucleus of the Seyfert 2 NGC 1068 show similar polarisation fraction and position angle as the continuum emission outside the line whereas the forbidden lines are 
%weakly polarised
%with a different position angle.  They showed that a common dust scattering mechanism in a cloud
%of about 1'' size could explain the similar polarisation properties of the continuum and allowedemission lines with both arising in the nuclear region whereas the distinct 
%polarisation properties of forbidden lines indicated their
%origin outside the nuclear region. Antonucci and 
%Miller (1985) showed that broad permitted lines are present in the
%nuclear spectrum of Seyfert NGC 1068 but are only detected in polarised light.
%The position angle of the optical continuum and lines are perpendicular to the nuclear
%symmetry axes as defined by the radio jet axis.  They found
%that the broad lines (FWZI $\sim 7500$ kms$^{-1}$) were redshifted by
%600 kms$^{-1}$ wrt to the narrow lines.
%However detection of such broad permitted lines in polarised light is not true for all Seyfert 2.
{\it To summarise: the observed properties are well explained by the model
for active nuclei presented in the paper.  The coexistence of the radio jet
and optical line forming gas supports their ejection following a thermonuclear
blast in the accreted gas at the poles.     }  

\subsubsection{A short summary}
The main points of the model for active nuclei/galaxies can be summarised as follows:
\begin{itemize}
\item All active nuclei host an accreting supermassive rotating black hole. 
The density of the infalling matter increases as it nears the event horizon 
and at some separation, the high densities can push matter into the degenerate state. 
This will lead to layers of degenerate matter being deposited outside the 
event horizon in a quasi-stable configuration and forming
the pseudosurface of the black hole.  

\item Subsequent to the formation of the pseudosurface, infalling matter 
accumulates on the pseudosurface with a latitude-dependent accretion rate.  
Matter accretes directly on the pseudosurface
in the polar region and through an accretion disk in the non-polar regions 
of the rotating black hole. 
The matter accumulated on the non-polar parts of the pseudosurface forms 
the broad line region (BLR).

\item The separation of the pseudosurface and BLR from the black hole
determines the type of active nucleus i.e. observable properties.  
If the separation is $\sim R_s$, 
then the observed properties will be a bright compact continuum source, 
high emission line redshifts and multi-redshifted broad absorption lines 
(due to the effect of a large and varying gravitational redshift) resulting in
objects we identify as quasars;  if the separation is several tens of
$R_s$ then these objects show a bright nucleus and broad emission lines in their spectra 
(shifted and broadened by smaller gravitational redshifts) and we refer to them
as BLRG if in an elliptical host; for still larger separations but 
where the influence of the gravitational potential of the black hole
is still active in terms of broadening the lines and enhancing the nuclear
continuum, the active nuclei, if in spiral hosts, are referred to as Seyfert~1.  
When the separation between the black hole and the BLR is several hundred
$R_s$, then the observable effects of gravitational redshift are negligible
and the objects host a faint nucleus and narrow emission lines and 
are classed into a NLRG or Seyfert~2 nucleus.  The increasing separation of
the pseudosurface and BLR from the black hole leads to decreasing  
continuum strengths, 
decreasing variability, lower gravitational redshift contribution to the lines and 
narrower lines from the BLR, if at all.  This, then, forms a continuum of 
BLR properties so that there exist quasars found in both elliptical and spiral hosts  
and there will exist nuclei between extremes of BLRG and NLRG in elliptical hosts in 
addition to between type 1 and 2 Seyferts in spiral hosts.  

A fast spinning
black hole forms a latitude-thick BLR with a large covering factor which will 
be detectable from most sightlines except directly polar while a slow
spinning black hole forms a latitude-thin BLR with a small covering factor which
will be detectable only from sightlines which intercept the equatorial
regions, except in the odd case wherein
the lines from the non-polar BLR are scattered into our sightline by polar dust. 

It should be noted that narrow emission lines are observed in almost all active 
nuclei unlike broad lines.  In most active galaxies these lines are believed to
arise in the polar NLR but it is possible that a fraction of the lines
do arise in the non-polar distant BLR.  Thermonuclear explosions can occur in the 
inner parts of the BLR if sufficiently high temperatures are reached and this
can lead to enrichment of matter but expulsion of matter might be rare
due to the massive accretion disk that has formed outside the BLR.  
However radio synchrotron emission has been detected in the equatorial regions
of some active nuclei indicating that some relativistic plasma might be able to escape. 

\item Matter is directly accreted on the pseudosurface at the poles of a 
rotating black hole.  This matter will be compressed and heated
which will eventually lead to a thermonuclear outburst.  This explosion can 
energise and expel matter at relativistic speeds from the small region 
at the poles which lacks
an accretion disk.  This expelled matter will form the bipolar jets detectable 
from radio to X-ray wavelengths and the optical line emitting gas (NLR) 
distributed along the jets, where the narrow emission lines form.  The frequency 
of such ejections would depend on the accretion rates.  Thus large radio 
sources like FR I and FR II types would indicate a long duration of reasonably 
large accretion rates and frequent energetic expulsion of relativistic matter 
from the poles of the black hole.  The radio-quiet nature
of Seyfert nuclei would then indicate the short duration of accretion, lower accretion rates
and infrequent and low energy expulsion of relativistic matter from the poles 
of the black hole.  
In the active nuclei, where frequent explosions occur but are not sufficiently 
energetic to expel matter, matter will fall back on the poles and can get enriched 
through nucleosynthesis in the thermonuclear
outbursts - such active galaxies will lack large scale radio jets and optical line
forming gas.   The pre-dominant detection of iron lines in such radio-quiet active 
nuclei seems to suggest an origin at the poles and supports the importance 
of nucleosynthesis in the matter which has failed to be ejected. 

\end{itemize}

\subsection{Radio relics and halos in galaxy clusters}
Steep spectrum radio emission and soft X-ray thermal emission are
detected from a quasi-spherical region in the central parts of several galaxy clusters. 
These are referred to as the radio halo and X-ray halo
respectively and appear to occupy comparable volumes with the X-ray emitting gas
occupying a larger region, if at all.  The radio halo is sometimes accompanied
by steep spectrum radio arcs located on the outskirts of the X-ray halo
and which are referred to as radio relics in literature.
Deep, meticulous observations of several clusters have helped narrow down
the range of observed properties of radio halos and relics but 
the origin of these features remains a matter of debate.  
Radio emission from the halos and relics require the presence of relativistic
electrons and a magnetic field. 
A widely accepted model for their origin, in literature, concerns merging of two clusters
that generates turbulence and shocks which accelerate the 
electrons in the halo and relic respectively to relativistic velocities
enabling them to emit synchrotron radiation in the magnetic field frozen in the
thermal hot X-ray halo.  

In the rest of the section, we summarise some observed properties of these
enigmatic objects and then converge on a comprehensive model which 
observations and physics appear to dictate.  This model supports the
origin of magnetic field in the X-ray halo as has been suggested in literature, but
argues against requiring acceleration of electrons due to turbulence or shocks. 

\subsubsection{Summary of observed features}

\begin{itemize}

\item Most clusters of galaxies host an X-ray halo.  Some of these
clusters also host radio halos and relics and such clusters do not host
a bright radio galaxy in their centre.  

\item A radio halo is more commonly detected in clusters compared to radio relics.
The radio relics when detected are observed to be circumferential to the halo 
being either symmetrically located on diametrically opposite sides on the
longer axis of the radio/X-ray halo or only a single relic on the outskirts
of the X-ray halo is detected.  
Typical diameters of the X-ray and radio halos range from few hundred kpc to couple
of Mpc.  The length of radio relics is measured to be several hundred
kpc to Mpc or so.  

\item Clusters hosting radio halo and relics are also non-cooling flow clusters
which, observationally, are clusters which host a X-ray halo but lack 
a strong centrally located peak in X-ray emission.  
Cooling flow clusters generally host a bright radio galaxy in their centre and record peaked
X-ray emission of lower hardness ratio in the central 100 kpc or so.

\item The emission from the radio halo is unpolarised while that from the 
radio relics is highly polarised.
This means the radio halo is threaded by a randomly oriented magnetic field
whereas relics host a predominantly ordered magnetic field. 
The magnetic field in the relics is oriented along the length of the relic. 

\item Equipartition magnetic field in cluster halos and relics have been estimated 
to range from 0.1 $\mu$G to 10 $\mu$G 
\citep[e.g.][]{1996IAUS..175..333F,2019arXiv190104496V}.
More specifically, the equipartition magnetic fields estimated for halos appear
to range from 0.1 to 1 $\mu$G whereas in relics it is slightly higher ranging
from 0.5 to 2$\mu$G \citep[e.g.][]{2004IJMPD..13.1549G}.
Moreover the magnetic field is not constant throughout the cluster but declines
with the cluster radius \citep[e.g.][]{2004IJMPD..13.1549G}.

\item The spectra of many relics is
well-fit by a single power law between $\sim 100$ MHz to about 5 GHz 
with a typical spectral index between $1$ and $2$ ($S\propto\nu^{-\alpha}$). 
The radio spectrum of the radio halos is also similarly steep.
At frequencies below 100 MHz or so, the radio spectrum has been occasionally observed
to flatten while above $\sim 5$ GHz, the spectrum is observed to steepen. 
{\it This behaviour is reminiscent of the radio spectra of hotspots in FR~II sources.}

\begin{figure}
\centering
\includegraphics[width=6cm]{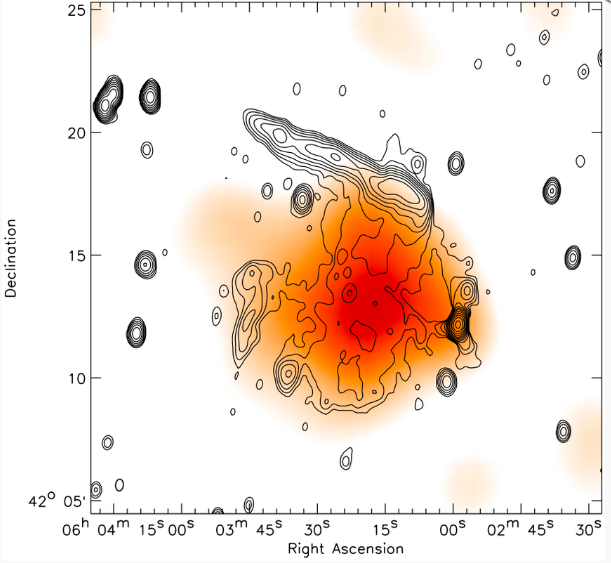}a
\includegraphics[width=6cm]{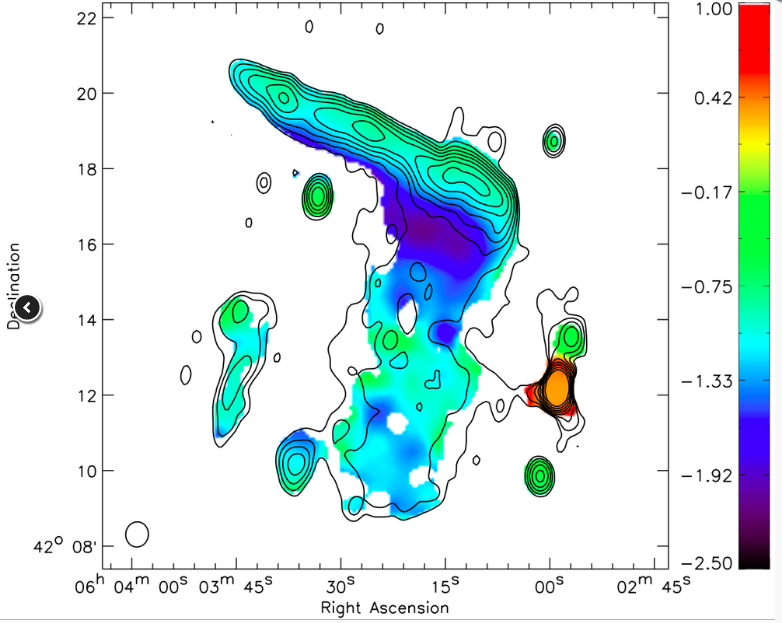}b
\includegraphics[width=6cm]{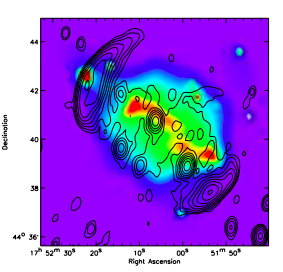}c
\caption{\small Figure panels (a) and (b) reproduced from 
\citet{2012A&A...546A.124V}.  The figures show
the X-ray and radio emission in a cluster at redshift $0.225$ 
which hosts the large relic referred to as the Toothbrush relic. 
(a) The radio emission in the band $1.2-1.8$ GHz from WSRT is in contours while the X-ray 
emission from ROSAT is in orange.  The $\sim 1.5$ Mpc long northern Toothbrush relic 
skirts the X-ray halo in the north. 
The radio halo emission is of similar extent as the X-ray halo. 
(b) The spectral index distribution between 147 MHz and 2273 MHz
made from maps with a resolution of 35'' is shown in blue.  Notice the extremely steep
spectrum on the lower part of the Toothbrush relic and in the region where it
meets the halo. 
(c) Figure showing the X-ray halo (colour) and radio relics, halo (contours) 
in the cluster MACS J1752.0+4440 
at $z = 0.366$ copied from \citet{2012MNRAS.425L..36V}.  Double relics are detected
at the edge of the X-ray halo.  The radio halo and X-ray halo are coincident and
share a major axis at the end of which relics are detected and this kind of
structure is commonly observed in clusters. }
\label{fig39}
\end{figure}

\item In the Toothbrush relic, the spectral index is constant along the length of 
the relic but shows a gradation along the minor axis of the relic 
(Figure \ref{fig39}b) \citep{2012A&A...546A.124V}. 
The spectrum is flattest ($\alpha \sim 0.6$) at the outer edge of the relic and
is observed to steepen towards the cluster centre 
($\alpha \sim 2$) \citep{2012A&A...546A.124V}.  
Such behaviour is typical of radio relics which are resolved along their minor axis. 

\item In the Toothbrush relic, $60\%$ polarisation is recorded at 5 GHz 
which declines
at lower frequencies indicating Faraday depolarisation and
an equipartition magnetic field of about $7-9 \mu G$  has been estimated
\citep{2012A&A...546A.124V}.

\item The Toothbrush and Sausage relics show a spectral break near 2.5 GHz such that
the spectrum below 2.5 GHz has $\alpha \sim -0.9$ while between 2.5 GHz to 30 GHz, 
the spectrum is steeper with $\alpha \sim -1.7$ \citep{2016MNRAS.455.2402S}.

\item The emission mechanism responsible for the X-rays from the cluster halo has been
deduced to be mainly thermal brehmstrahlung from 
hot gas \citep{1966ApJ...146..955F}  which is further supported
from the observation of a blend of highly ionized iron lines near 7 keV which
are detected from the intracluster gas.  This gas also has solar-like abundance of iron
which is inferred to suggest a stellar origin. 

\item \citet{2009A&A...507.1257G} find that $\sim 30\%$ of the clusters in their 
sample host both a halo and a relic radio source.   
The radio size and power are correlated for both small and giant radio halos while  
the cluster X-ray luminosity (indicative of cluster mass) is correlated with the 
radio halo power \citep{2009A&A...507.1257G}. 
{\it The correlation between the X-ray luminosity and radio halo power is 
reminiscent of the observed correlation between the 
luminosity of the optical line emitting gas that forms the NLR
and radio power of jets in radio galaxies.   The radio jet and NLR have
a common energy source in the thermonuclear energy.}

\item 
Radio power at 1.4 GHz ranges from $10^{23}$ to $10^{26}$ W Hz$^{-1}$ whereas
X-ray luminosity typically is in the range of few times $10^{43}$ erg s$^{-1}$
to few times $10^{45}$ erg s$^{-1}$ \citep[e.g.][]{2019arXiv190104496V}. 
{\it The X-ray energy released from the halo in a year will be $\sim 3 \times 10^{52}$ erg.
while over a billion years, about $3 \times 10^{61}$ ergs will be released in
the soft X-ray component.}
%{\bf think more} 

\item Mini-radio halos which are smaller in size than radio halos  
have been identified in clusters \citep[e.g.][]{1996IAUS..175..333F} with
the current number being about 10 mini-halos of size ranging from 50 to 300 kpc 
\citep[e.g.][]{2014ApJ...781....9G}.  Unlike halos, mini-halos are  detected 
in cool core clusters which host a central radio galaxy like the Perseus cluster.  
Mini-halos also show a quasi-spherical symmetry like halos but show more
small scale asymmetries than halos. 
{\it An evolutionary sequence seems to emerge on collating the properties 
of halos, mini-halos, central radio galaxy in clusters.  Central radio galaxy to
a radio mini-halo to a radio halo.  The first two commonly coexist
in clusters while there seem to be clusters in which the last two are seen to coexist.
However clusters in which a halo and central radio galaxy coexist are rare. 
The connection is chronological so that when a radio halo is formed, the
central radio galaxy has exhausted its fuel and entered a quiescent phase.  This can be
inferred to argue for an evolutionary connection between the three features.}

\item It is noted that in several radio galaxies, groups and clusters,
cavities in the diffuse X-ray emission coincide 
with extended radio emission like lobes.  The first such radio lobe - X-ray cavity correlation
was noted in the Perseus cluster by \citet{1993MNRAS.264L..25B} (see Figure \ref{fig40}c).
Perseus cluster has a central radio galaxy (NGC 1275 i.e. 3C 84), a radio mini-halo
of $\sim 300$ kpc extent \citep{1990MNRAS.246..477P} in addition
to a X-ray halo with cavities \citep{1993MNRAS.264L..25B}. 
Such correlations between X-ray cavities and radio emission have 
been widely observed. 
{\it Such correlations tend to argue for some connection between the
radio and X-ray processes.  Since the radio plasma 
radiates in the magnetic field frozen in the X-ray emitting intracluster 
medium, the presence of radio lobes at the X-ray cavities indicate the existence
of magnetised thermal gas although X-ray emission is significantly fainter.
The non-detection could be due to scattering of soft X-rays to different energies.}

\end{itemize}

\subsubsection{A comprehensive explanation}

The widely accepted model in literature used to explain 
the presence of relativistic electrons in the radio halo and relics invokes a cluster-cluster
merger which leads to turbulence in the central parts and sets up outwardly 
expanding shocks along the merger axis.  Turbulence
and shocks are believed to be the processes contributing to particle acceleration 
in the radio halo and relics
respectively.  In other words, the existence of radio halos and/or relics in a 
cluster is considered to be an indicator of two clusters undergoing a merger.  
The observational results on such clusters consist of an ellipsoidal distribution of the
X-ray and radio halos,  presence of relics at the edges of the major axis 
of the halos and the detection of some concentration of galaxies 
along the major axis of the halo.  These observations have been interpreted to mean that 
two clusters of galaxies are approaching along the major axis of the 
halos and merging.  The merger sets up turbulence in the central region of
the cluster explaining the radio halo while 
shocks propagate outwards on either side which are believed to result in the relic emission. 
Reacceleration is believed to be required since the assumed large energy losses that the 
relativistic electrons suffer means their lifetimes are fairly small.  

We searched for an origin which is
commensurate with one of the main results of this paper and \citet{2017arXiv170909400K}
that the dominant mechanism accelerating electrons to relativistic velocities 
is an explosive thermonuclear outburst on a compact object (supermassive 
black hole in this case) before the accelerated electrons are launched into space.
Any further acceleration as in Compton process that occurs post-ejection
leads to only a trivial increase in their velocities, if at all. 
Since the radio halo is quasi-symmetric over the centre of the cluster which 
has to host an active core although not accreting any more, 
it appears possible to start with the assumption that the relativistic electrons 
which populate the radio halo and relics were originally
energised in this active nucleus when it was in its accreting radio galaxy phase. 
It is possible that this pool of electrons is supplemented by electrons accelerated 
in other nearby radio galaxies within the cluster but this cannot be the dominant
contribution due to the central location of the halo.  
This means that we look for an explanation for formation of halos and relics with
the implicit understanding that the resident relativistic electron population has been 
energised in a thermonuclear blast and ejected along bipolar jets from the
pseudosurface of a rotating accreting supermassive black hole located in the
galaxy at the centre of the halo. 
Thermonuclear energy released in simultaneous ignition of a large mass of fuel is 
a source of enormous energy in a hydrogen-dominated universe 
- recall that simultaneous ignition of half a solar mass of fuel can release energy 
of $10^{51}$ ergs within 100s of seconds.  Since the clusters hosting a radio halo
and relic lack a powerful radio galaxy at the centre, it can mean that the 
active nucleus in the parent galaxy has long stopped accreting and ejecting
relativistic material.  
Whatever it ejected during its active phase 
has expanded along jets and diffused out to encompass a large
volume around the galaxy.  Soon after the bipolar ejections from the central
source stopped, the jets would have lost their identity as the last bursts
of plasma lost its ballistic nature and expanded into lobes.  This is not 
far-fetched as can be seen from the morphology of the extended radio emission
around Virgo~A at the centre of the Virgo cluster
and the halo-like radio emission in NGC~193 (see Figures \ref{fig36},\ref{fig27}).
If the jets were extinguished in both these galaxies,
then the extended radio emission would resemble a elliptical radio halo.  
The fossil plasma will continue to diffuse around 
the central black hole till the component electrons exhausted their energy and faded into 
oblivion.  If there was no ambient magnetic field i.e. no extended thermal
gas in which the field was frozen then the electrons will suffer only kinetic losses
and if a magnetic field was available in the extended thermal gas, then the
positron-electron plasma would radiate synchrotron emission and explain the existence
of the radio halo.  We examine the validity of 
this hypothesis on the basis of observational results, electron lifetimes etc. 
Since direct signatures of inverse Compton scattering which can contribute to
rapid draining of electron energies are yet to be observed, it is not included in
estimating the electron lifetimes. 

First of all, it is significant that the radio spectral index of the halo and
relic emission are similar and steep at $\alpha \ge 1$.  
Since the estimated  magnetic fields in the two regions differ by only a factor of few,
it appears safe to assume that similar energy electrons are 
radiating at a given frequency in both objects.  In this case, the similar spectral indices
argue for a common parent population of relativistic electrons and since we
suggest that the halo electrons arise from the central black hole, the relic electrons
also have the same origin.
The second line of support for the origin of the relativistic plasma
in the central supermassive black hole during its active phase comes from the
current large scale morphology of some FR~I sources like Virgo~A 
(top left panel in Figure \ref{fig36})
and NGC~193 (Figure \ref{fig27}) as mentioned above.  If we were to null out the central radio
galaxy, jets and structure traceable to recent activity,
then we would be left with diffuse elliptical structure
whose major axis of about 80 kpc or so is similar to that of the soft X-ray emission
and a radio mini-halo.  Literature often refers to the extended radio emission around
Virgo~A as a mini-halo and there appears little doubt that this extended
emission is fed by the central black hole and the observed sub-structure
would also include structure in the magnetic field.  In other words, the
electron population of the mini-halo around Virgo~A is certainly 
energised in the active nucleus. 
In fact, the sharp boundary of radio emission observed at all radio frequencies in
several such radio galaxies
is likely indicative of the drop in magnetic field intensity i.e. lack of X-ray
emitting gas beyond the observed radio emission.  In fact a physical process similar
to one responsible for hotspots is likely to be operational here but at a lower level so
that radiation pressure also contributes to confining the thermal and radio plasma
within the sharp boundary.  The X-ray gas is likely to slowly expand and encompass an
ever-increasing volume in which the positron-electron 
plasma will radiate in a declining magnetic field so that the 
radio mini-halo will keep expanding and the radio and X-ray sizes will remain comparable.  
The timescales over which the plasma (both positron-electron and thermal X-ray
emitting proton-electron plasma) diffuse to occupy a spherical region
of radius 1 Mpc which is typical of a radio halo will depend on the diffusion
velocities, region over which the jet retained some of its forward propelling velocity
etc.  However the paucity of clusters which host both a radio halo and a central
radio galaxy indicates that by the time a large Mpc-scale radio halo is detected 
the supermassive black hole has stopped accreting gas and entered a quiescent state. 
It is also possible that the positron-electron plasma has diffused beyond the
X-ray halo but is not emitting synchrotron radiation due to the absence of
an appropriate magnetic field.  The positron-electron plasma in this region
will suffer lower energy losses and hence enjoy longer lifetimes.  
There also exist other radio galaxies which show radio structures which are
several hundred kpc across and are unambiguously created by the central active nucleus 
such as Hydra~A.  This provides further
support to the positron-electron plasma succeeding in etching out such
large structures before fading.  

The origin of the hot thermal X-ray emitting gas in ejections from the central source is
supported by 
the estimated equipartition magnetic fields.  In active radio galaxies, the magnetic
fields range from several tens of $\mu$G in the jets and lobes to few hundred $\mu$G
in the hotspots while in radio halos and relics, the magnetic fields are
estimated to be of the order of a $\mu$G.  This kind of behaviour of the
magnetic field is expected if the field is frozen in the plasma ejected by the 
central source.  As the plasma keeps diffusing over
larger regions, the constant magnetic flux (BA = constant; where B is magnetic field
and A is the area through which the magnetic field is threaded) 
condition ensures that the magnetic field frozen in that plasma
keeps declining as the area it covers keeps increasing.  
New ejections will modify the magnetic field in the overlapping regions. 
We do not know the field strength close to the black hole from where 
matter is launched into space at relativistic velocities although we do know that 
in the lobes
of active galaxies, typical magnetic fields of few tens of $\mu$G are estimated.
These fields are frozen in the ionized thermal gas component i.e. either optical line
forming gas or the X-ray emitting gas.  If we
assume that the field threading a spherical surface of radius 
100 kpc around the active galaxy is 20 $\mu$G then the magnetic flux 
will be $BA = 4\pi \times 20 \times (100^2) = 8\pi \times 10^5 \mu$G-kpc$^2$.
For a spherical radio halo of radius 1 Mpc, such a flux would result in a magnetic
field of $ B = 8 \pi \times 10^5 / (4 \pi \times (1000^2)) = 0.2 \mu$G.
In the extended radio halos in clusters, typical magnetic fields have been estimated
to range from 0.1 to few $\mu$G with earlier studies systematically having determined
lower fields \citep[e.g.][]{1996IAUS..175..333F} as compared to 
more recent studies.  
This simple calculation demonstrates that the magnetic
field pervading the halo is the same as the field in which jets and
lobes associated with the central radio galaxy radiate and the simplest
explanation is as above wherein the field is frozen in the matter ejected
from the active nucleus of the galaxy at the centre of the cluster and
keeps declining as the matter occupies an ever-increasing region. 

Radio galaxies span a large
range in size from parsecs in FR~0 sources to few kpc in GPS, CSS sources to 
several tens to hundred kpcs in FR~I sources to several hundred kpc to Mpc 
in FR~II sources.  This tells us that the radio plasma energised and ejected from some
active cores is capable of traversing distances upto Mpc or so 
before fading i.e. electrons energised in the central engine and emitting
predominantly in the radio bands can survive upto
a Mpc or so from the central core when traversing along the jet wherein the
typical magnetic field can be upto a few tens of $\mu$G.  It also tells us
that the reason for the existence of small to large radio sources is the plasma energetics
i.e. the quantity of matter that is simultaneously ignited in the thermonuclear
burst is varied in different active nuclei leading to differences in the released
energy and quantity of matter energised, ejected and detected. 
If the mean magnetic field along the jets is assumed to average to about 
10 $\mu$G then the lifetime of the 
1.4 GHz emitting electron will be about 28 million years and it 
can traverse a distance of $\sim 5.3\times10^{22}$ m $=1.7$ Mpc at a 
bulk velocity of 0.2c before fading.  
If the mean magnetic field in which the electrons radiate synchrotron emission
is $2 \mu$G, then the lifetime of electrons dominantly emitting at 
1.4 GHz will be $3.2 \times 10^8$ years and these can travel a distance of 
$6 \times 10^{23}$ m at an average bulk velocity of 0.2c ($=60000$ kms$^{-1}$) i.e. 
%$2 \times 10^7$ pc 
$\sim 20$ Mpc!  
At larger bulk velocities as in FR~II sources, the electrons can traverse 
even longer distances.  
For comparison, the electrons predominantly emitting at 1.4 GHz in a field of $0.1 \mu$G 
can survive for about $28\times10^9$ years. 
The high energy electrons which lose energy will move
to lower energies and can contribute to the low radio frequency emissions. 
All this discussion is aimed at showing that the synchrotron lifetimes of
electrons are sufficiently long to explain the population in radio halos and relics. 
In a radio galaxy, the electrons along the jets can travel
ballistically to long separations which can be covered in shorter times.
However the electrons that make up the lobes can take longer to cover the
same distances since they are no longer moving ballistically at relativistic 
velocities but diffusing at much lower velocities.  If diffusion of the positron-electron
plasma from the jets is occurring at a velocity of about 1000 kms$^{-1}$, then
in a billion years, the plasma can diffuse over a distance of 1 Mpc. 
The electrons which started with high energies will move to progressively lower energies
so that the emission at high radio frequencies will fade. 
Since other processes which shorten the electron lifetime like inverse Compton, although
postulated, have not been unequivocally inferred from observations, these are
not included.  In future if other processes are found, the estimates can be revised.
It should be pointed out that instead of losses, the annihilation photons
near 511 keV are likely to be continuously generated in the positron-electron plasma
and if these are Compton scattered to X-ray wavelengths by the relativistic electrons
throughout the halo then it can result in small energy boosts to the electrons. 
An important reason for both relics and halos being detectable inspite of low
magnetic fields is likely to be their enhanced
synchrotron emissivity due to the long pathlengths in the cluster over
which the relativistic plasma is distributed.
Since relics are formed at the outer edge of the radio halo in clusters,
the population of electrons therein can also be explained with an origin in
the central black hole and there should be no lifetime problem.  
The steep spectra that are observed for both these and the absence of high
energy electrons as witnessed from the absence of synchrotron emission at even high
radio frequencies argues for an aged electron population. 
The rest of the observable properties of relics are distinct from the radio halo and
hence require a separate discussion.  {\it Here we limit ourselves to making the statement
that the electron population that was generated in the active
nucleus of the central radio galaxy can indeed survive in the low magnetic
fields of the radio halo and explain
synchrotron emission from the radio halo and relics and there is no need
to invoke an extensive reacceleration mechanism.  }

We now discuss the origin of radio relics in clusters.
The different observed properties of radio halos and relics in clusters 
indicate the distinct physical processes that
the plasma and magnetic field are subject to in the two features.  
Since the relics, when detected in a pair, are located on diametrically opposite
sides of the centre of the cluster, subtend a small angle there and are along the
major axis of the halo, 
we suggest that the axis connecting the relics tell us about the path followed
by the jets when the central nucleus was actively expelling matter.
The conical region defined by connecting the relic edges to the centre of the cluster defines 
the region over which the jet launched from the black hole would have traversed.   
While the low magnetic field in the radio halo region would have allowed electrons
to survive longer, another reason also exists for longer lifes for at least a fraction
of the electron population in the jets.  Along the jet, there would always be 
a fraction of electrons which do not suffer synchrotron losses due to their 
dominant relativistic random motion being along 
the magnetic field.  The bulk motion of the positron-electron plasma is ballistic
and radially outwards from the launch site.  However over and above this, all particles
have a random velocity component which for the jet plasma is relativistic and 
all electrons which have a velocity component which is not along the magnetic
field will execute circular motion about the field lines which is responsible
for the synchrotron emission whereas their bulk forward velocity will lead
to the motion about the magnetic field lines being helical.  
Once the random velocity component becomes aligned
to the radial magnetic field then no Lorentz force
will act on the particle and the particle will stop radiating synchrotron emission till
it regains a velocity component which is not along the magnetic field.
Such electrons along the field will only suffer kinetic energy losses as they expand
away from the active nucleus and it means that the lifetimes of such electrons 
can be enhanced so that they can traverse longer distances from the parent black hole.  
In reality the random velocity component of the relativistic electrons will 
define a range of pitch angles with the magnetic field and
hence experience different synchrotron losses with maximum loss being experienced
by the electron moving perpendicular to the field to minimum losses by the electron
moving along the field.  Quantifying this further, an electron with a pitch angle of
$10^\circ$ will suffer 3\% of the losses suffered by an electron moving perpendicular
to the magnetic field. 

The detection of strong polarised emission from relics (unlike halos)
from which magnetic field lines
stretched along the relic are surmised, gives strong evidence to the enhancement
of ordered magnetic fields at the periphery of the halo emission in clusters.  The 
relic and field lines in the relics are extended circumferential to the X-ray halo 
i.e. perpendicular to the direction in
which the jet would have traversed outwards from the central active nucleus and
their morphology is strongly indicative of having been subjected to compression. 
The magnetic field orientation in the relics facilitates synchrotron emission from the 
electrons which had a random velocity component which was predominantly 
along the field lines in the jet i.e. lower pitch angles and hence 
suffered lower synchrotron losses after being ejected from the central black hole.
In fact, there is similarity in the observational properties of the radio relics, which
appear to signify the edge of the fossil jets launched from
the central FR~I galaxy and radio hotspots observed at the end of jets in FR~II sources
except that the linear scales and hence radio intensities involved are very different. 
We, hence, explore if the formation scenarios could be due to a similar
physical process.  To recall, we pointed out that the formation of hotspots can
be attributed to radiation pressure exerted on the plasmas by the positron-electron 
annihilation photons near 511 keV, formed beyond the extended X-ray emission. 
If the axis connecting the relics does indicate the orientation of the historical jet 
including the precession cone then after the
positron-electron plasma of the halo leaves the X-ray emitting halo 
then the mean free path of the annihilation photons will increase which will lengthen their
lifetimes. These photons can exert radiation pressure back on the plasmas. 
This can result in further expansion of the plasma being stopped and the
plasma at the edge being compressed.  The magnetic field frozen in the thermal plasma 
will be enhanced and ordered due to the radial compression.  Thus it appears that the
formation of relics can also be explained by the action of radiation pressure on
the plasma.  
The radio spectrum of some relics is observed to show flattening at frequencies 
around 100 MHz which would indicate the peak of the normal distribution of the parent
electron population which has aged since it was ejected from the black hole.  
For a magnetic field of 5 $\mu$G in the relics, this indicates the peak of the
aged electron energy distribution is about 1 GeV.  

This explanation is comprehensive in that it explains the formation of radio halos,
radio relics and accounts for the location of relics at the periphery of
the X-ray halo of the cluster without invoking any exotic processes. 
The explanation starts with the same population of positrons-electrons injected along the
jets in an active nucleus and follows their evolution. 
In principle, such enhanced relic-like emission can surround the entire radio halo
since the positron-electron plasma will continuously diffuse outwards and if
it expands beyond the X-ray halo, the positron-electron annihilation photons
can exert radiation pressure on the halo plasma.  It is easier to envisage this
if the positron-electron plasma has a net outward motion which is larger
than that of the thermal plasma and hence it is easier to expect this along jets.

We list properties of hotspots and relics in Table \ref{relic} since
both occur in the transition zone from a dense to tenuous medium. 
Properties of the radio halo are also included for completion.  
There are several similarities between the hotspot and relic properties 
such as their location at the extremum of radio emission and X-ray halo,
high polarisation fractions, magnetic field orientation, steep
spectrum and their frequent occurence in symmetric pairs about the central source. 
The radio spectrum in both hotspots and relics is observed to be flattest
at the lowest observed radio frequencies (10-100 MHz), steepens at higher radio frequencies 
(upto 5 GHz) and steepens further beyond radio frequencies of 5 GHz or so. 
There is also similarity in the magnetic field behaviour.  
The magnetic field which had been predominantly along the jet 
changes orientation and becomes perpendicular to the jets at the hotspots in addition to
being enhanced.  The magnetic field in the radio halo which is predominantly randomly
oriented becomes highly ordered and along the relics in addition to being enhanced.
{\it To summarise:  the physical process responsible for the formation of hotspots
at the end of radio jets in FR~II sources and radio relics at the periphery of the X-ray
halo in clusters which would have hosted a powerful FR~I radio source in the past
are the same.  Positron-electron plasma fills up the radio halo and at least
in clusters which host relics, it has continued to move along the jets long 
after the matter ejection episodes from the central active nucleus have been quenched. 
This plasma will continue to generate the annihilation photons at some rate
which will be scattered/absorbed by the thermal electrons.  However the photons 
generated beyond the X-ray emitting thermal gas will survive
longer and can exert radiation pressure back on the thermal and non-thermal plasma.
This will result in compression of both the plasma - i.e. enhancement of
magnetic field in thermal plasma and enhancement of relativistic electron density
in the non-thermal plasma.  Both these factors can enhance the radio brightness
and we identify these regions as relics in clusters. }

\begin{table*}
\centering
\caption{\small Comparison of observed properties of hotspots in FR~II galaxies with 
relics and halos in clusters of galaxies.  The observed properties are listed on
the top and some inferences that are drawn in the paper are listed at the bottom.}
\begin{tabular}{l|c|c|c}
\hline
{\bf Property}  &  {\bf Hotspot} & {\bf Relic} & {\bf Halo} \\
\hline
Separation from centre in kpc & $\ge 100$  & $\sim 1000$ & centered \\
Location & Edge of radio jets/X-ray & Edge of radio/X-ray halo & centred \\
Radio polarisation  & $\sim 60 $\%  &  $\sim 70-80$ \% & unpolarised \\
Magnetic field direction &  $\perp$ to jet & $\perp$ to cluster radius & random \\
Magnetic field strength  &  $\ge 100 \mu$G  & few $\mu$G  &  $\le \mu$G \\    
Spectral index $\alpha$ ($S\propto\nu^\alpha$) & $\sim -1$ at $\nu\ge10$ GHz & $\le -1$  & $\le -1$ \\
Spectral turnover $\nu$  & $\sim 100$ MHz  & $\sim 100$ MHz  & ? \\
Symmetry  & diametric double  & single or diametric double & roughly spherical  \\
Multi-band emission & yes & only radio & only radio \\
\hline
\multicolumn{4}{c}{Some inferences}\\
\hline
Magnetic field B & $B_{ordered} \uparrow$ & $B_{random} \rightarrow B_{ordered}$ &
         $B_{ordered} \rightarrow B_{random}$ \\
Plasma expansion velocity & $v_{jet} \rightarrow -v_{random}$ & $v_{jet} \rightarrow
-v_{random}$ &  $v_{diffusion}$  \\
External pressure & yes & yes & no \\
\hline
\end{tabular}

$`\uparrow'$ means an increase in amplitude of the quantity.  \\
$`\rightarrow'$ means that the left hand side quantity is converted to the 
right hand side quantity.

\label{relic}
\end{table*}

\begin{figure}
\centering
\includegraphics[width=6.0cm]{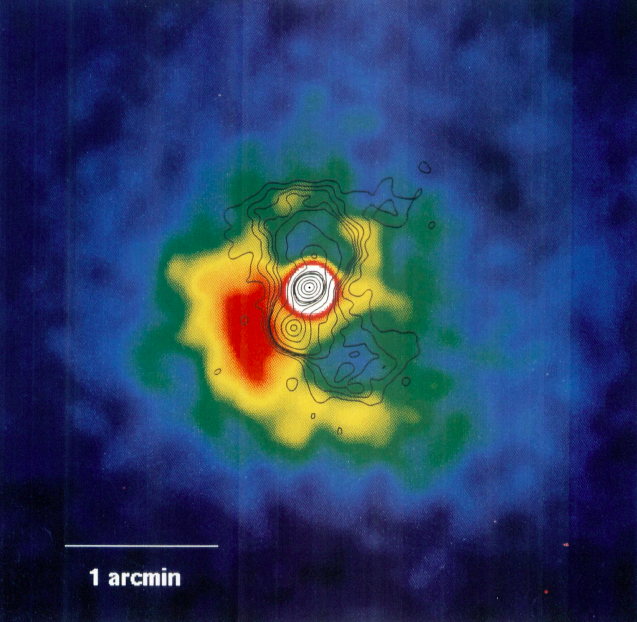}
\caption{\small 
NGC 1275/3C 84 from \citet{1993MNRAS.264L..25B}.  Notice how the radio (in contours) lobes are
located in X-ray (in colour scale) cavities.  }
\label{fig40}
\end{figure}

An important detection criterion for relic emission appears to be its location
i.e. the angle it makes with the sightline. 
All relics which have been detected are the ones whose long axis is in the sky plane 
and hence perpendicular to the sightline i.e. magnetic field is predominantly perpendicular
to the sightline which facilitates the synchrotron cones being directed towards us.
It might be that radio relics in the outskirts of the X-ray halo,
exist in all clusters which host a radio halo, but due to the highly
ordered magnetic field in the relic, the detection of relic emission is 
sightline-dependent.  
On the other hand, the magnetic field in the radio halo is randomly oriented 
indicating that it should be detectable along all sightlines.  
Radio halos are detected in only about 10\% of
clusters while a soft X-ray halo is detected in almost all clusters. 
It is interesting to note that radio-loud active nuclei are surmised to
constitute about 10\% of the entire population of active galaxies. 
%{\bf references?}
It could be inferred from this that only about $10\%$ of the population of
active galaxies are successful in launching detectable long-lived fast radio jets. 
The similar fraction of radio-loud active nuclei and radio halos lends
support to the direct connection between them.  Detection of synchrotron radio halos 
should be more common than detection of radio relics which is in line with the
detection of a relic source in about 30\% of the clusters which host a radio halo 
\citep{2009A&A...507.1257G}. 

The `Toothbrush' relic \citep{2012A&A...546A.124V} is located to the north of 
a $\sim 2$ Mpc large radio halo which is coincident with the X-ray emission 
from the cluster 1RXS J0603.3+4214 (see Figure \ref{fig39}a).  
The spectral index is flattest ($\sim 0.6$) on the outer edge with a gradient
along the minor axis so that the steepest spectrum is where the inner parts of
the relic which might be due to a backflow meets the radio halo ($\ge 2$).  
For the same parent electron population, the steepening
could indicate a varying magnetic field along the minor axis of the relic
such that it is maximum at the outer edge and decreases towards the halo. 
This is what we could expect from the field that is enhanced due to compression of gas
at the periphery.  The magnetic field gradient means 
relatively higher energy electrons will emit on the inner edge as compared
to the outer edge.  For a normal distribution of energies, this could indicate
that electrons near the peak energies emit at the outer edge and hence the spectrum
is relatively flatter.  
The Faraday depolarisation in the relic supports
the mixing of the X-ray emitting gas which provides the magnetic field
for the positron-electron plasma. 

Another interesting relic-halo cluster is MACS J1752.0+4440 which hosts a double relic 
of spectral index $\sim -1.1$ and lengths of 0.9 and 1.3 Mpc in addition to  
a radio halo with an extent of 1.6 Mpc \citep{2012MNRAS.425L..36V}. 
The relics are located outside and along the major axis of
the X-ray halo emission (see Figure \ref{fig39}c). 
The radio emission is highly polarised and the morphology of the relics
and magnetic field argue for a compression event.  

While the dominant contribution of relativistic electrons to the radio halo in 
clusters should be from the central active source which is often a massive cD
galaxy believed to have grown by mergers, other radio galaxies in the 
vicinity are also likely to contribute relativistic plasma and/or thermal
plasma to the halos as is the case for the cluster-scale X-ray halo in Virgo cluster
(see Figure \ref{fig37}).  

If the origin of the relativistic plasma populating the radio jets, lobes, mini-halos, halos
and relics and the thermal plasma which initially emits 
optical emission lines and graduates to emitting thermal X-rays, lie in the
ejected accreted gas around a black hole following a thermonuclear explosion
then a correlation between the radio power of mini-halos,
halos and X-ray luminosity should be observed.  The radio power of active galaxies
has been found to be correlated with the optical line power and the radio power
of halos is found to be correlated with the X-ray luminosity which supports the
above connection. The cases wherein the
X-ray emission is from several different galaxies while the mini-halo is
from the central galaxy, no correlation between the radio and X-ray luminosities
can be expected.  

{\it To summarise the discussion in this section, the relativistic electron population
in radio halos and relics in clusters has its origin in the energetic ejections
from the supermassive black hole in the central active galaxy which has since
been quenched.   The halo is formed by diffusion of the plasma over a large region,
similar to supernova remnants which distribute relativistic plasma throughout 
spiral galaxies.  Support for this comes from several active radio galaxies like Virgo~A,
Perseus~A which already show the existence of a radio mini-halo.  
It appears that ejections from the active nucleus in the central galaxy of the cluster
leads to the formation of radio mini-halo and radio halo as the relativistic
positron-electron plasma diffuse outwards and surround the entire region around
the central galaxy.  The optical line emitting gas ejected alongwith
the relativistic plasma leads to the formation of the tenuous hot X-ray emitting
thermal gas.  
The relics are formed at the edge of the X-ray halo and require
a compression from the outside to explain their shape, magnetic field
orientation, high polarisation and spectral index gradient.  It is suggested 
that this compression is provided by radiation pressure. }

\section{Formation of soft X-ray halos}
Gas at a million degrees which predominantly emits thermal brehmstrahlung and
highly ionized spectral lines is detected from several objects ranging from 
stars to clusters of galaxies.  For example, the sun is a strong soft X-ray emitter
with enhanced emission often associated with solar flares and other activity,
X-ray emission is detected from the hot ionized component of the 
interstellar medium while clusters
of galaxies are regularly identified from the presence of a large X-ray halo which 
encompasses several galaxy members of the cluster.  The X-ray emitting hot ($10^6-10^8$ K)
magnetised intracluster medium typically of mass $\sim 10^{13}-10^{14}$ M$_\odot$
constitutes a significant fraction of the cluster mass of $\sim 10^{15}$ M$_\odot$.  
The typical electron densities of the thermal intracluster medium are estimated to be
$\sim 10^{-3}-10^{-4}$ cm$^{-3}$.
X-ray emission is also detected from a halo of hot gas of temperatures of
10 million degrees K surrounding luminous early type galaxies
($M_B < -19$ magnitudes) and can
extend upto radial distances of 100 kpc from the centre of the galaxy 
with gas mass in the X-ray halo being estimated to range from a few times
$10^9$ M$_\odot$ to a few times $10^{10}$ M$_\odot$ i.e. upto $\sim 7\%$ of the 
stellar mass in the galaxy \citep{1985ApJ...293..102F} 
%$\sim 10^{10}$ M$_\odot$ 
and active galaxies (see Figure \ref{fig33}). 
Early studies which detected hot halo gas around normal ellipticals in the
Virgo cluster with X-ray luminosities of $0.5-7\times10^{40}$ ergs s$^{-1}$ prompted
the inference that the hot gas in the intracluster medium is the material 
lost by member galaxies \citep{1979ApJ...234L..27F}.
Subsequent inferences have also suggested an external accretion origin for
the hot X-ray gas especially on galactic scales.
All these observations point at thermal X-ray emission being common to several widely 
different extended astronomical systems indicating the ubiquity of low density thermal
plasma at temperatures in excess of million degrees.  
The origin of this important component of the universe is still debated.
Once the low density hot medium is formed, it is likely to be long-lived with
small energy inputs balancing the radiative losses. 
Here, we try to understand the origin of this hot gas on galactic and cluster levels. 

On a general note, it is possible to imagine that a large fraction of the diffuse
low density thermal free-free X-ray emitting gas around galaxies and
in clusters can be generated 
due to a process analogous to evaporation at the interface between
high density and low density regions like evaporation of water from the surface
of a water body into air as water vapour.  The water vapour has a lower density
than water and higher kinetic energy.  The same can occur on a galaxy level
wherein the cold dense gas is slowly evaporated and eventually forms hot tenous gas.
For an ideal gas $PV = nRT$ where $P$ is
the pressure, $V$ is the volume, $n$ is the molar density, $R$ is the
gas constant and $T$ is the temperature of the pressured gas.  Most of the gas in
galaxies except in self-gravitating molecular cores can be assumed to be 
ideal due to their low densities.  Since the gas mass is surrounded by
low density gas, the gas will expand i.e. $V$ will increase and $n$ will decrease.  
If the expanding gas takes away some energy from the cold gas mass then it
can increase its kinetic energy i.e. temperature. 
Since pressure of interstellar atomic gas and hot million degrees gas is observed 
to be nearly constant in
a disk galaxy, this can mean that the temperature of the gas increases according 
to the ideal gas law (i.e. $V_1/V_2 = n_1~T_1/(n_2~T_2)$).   
It is significant that the galaxies
which are surrounded by a hot gas halo seldom host large quantities of interstellar gas
within the galaxy 
whereas the galaxies which are not surrounded by a massive hot gas halo 
contain significant amount of cooler interstellar matter within the galaxy. 
This provides strong support to the origin of the soft X-ray emitting halo gas being
the interstellar medium of the galaxy itself and enhances the efficacy of
processes like evaporation. 
This scenario is opposite to the case wherein the density of matter keeps
increasing in a gravitational collapse so that eventually the atoms are stripped of 
their electrons and degenerate matter is formed as seen in compact objects.  

Three distinct mechanisms for the formation of soft X-ray halo around galaxies and in the
central parts of clusters appear to be operational as can be surmised from
observations: (1) The gaseous interstellar medium
of disk galaxies is lost due to the effect of a gravitational torque exerted
by the passage of another galaxy.  If the ideal gas expands isobarically,  the
densities decrease and the temperature increases as the gas occupies an ever-increasing
volume around the galaxy.  The galaxy will be depleted of its gas and evolve
to a gas-poor elliptical with the formation of a thermal soft X-ray halo.
(2) The gaseous interstellar medium in disk galaxies is channelled inwards
towards the supermassive black hole as a result of the gravitational torque
exerted by the passage of another galaxy.  The gas drives the black hole into
the active regime so that it repeatedly ejects matter from its poles
leading to the formation of a large radio double structure and optical
line emission gas distributed around it and the galaxy evolves to a gas-poor elliptical.  
The continued expansion of the optical line emitting gas especially if
isobaric can drive it into a low density phase such that its temperature 
increases and the gas evolves into a soft X-ray emitting thermal plasma. 
(3) The third mechanism involves the annihilation photons
near 511 keV that will be continuously generated in the relativistic positron-electron 
plasma ejected by the black hole.   Compton scattering of these photons to
X-ray energies might result in a X-ray halo like structure around the galaxy but
it remains difficult to ascertain this.  However 
this mechanism can be frequently identified with the thick X-ray beams that
are coincident with the radio jets in radio galaxies.   

We cite some examples of galaxies which appear to provide clues to the formation 
mechanism for the X-ray halo observed around them.  
NGC~6482 is an elliptical galaxy surrounded by a hot X-ray halo 
which hosts a weak active nucleus in its centre and no extended double
radio source is associated with it 
%{\bf reference}.  
NGC~4555 is another such case of 
an isolated elliptical galaxy with a X-ray halo \citep{2004MNRAS.354..935O},
no active nucleus and
no extended radio structure associated with the central black hole. 
NGC~4382, an S0 in the Virgo cluster shows a X-ray halo \citep{2006MNRAS.370.1541S}, 
is devoid of HI \citep{2003AJ....125..667H}, hosts no active nucleus and no
extended radio double structure is associated with it.
In all these galaxies, it appears that the galaxy has lost its interstellar medium 
which has subsequently formed a low density soft X-ray emitting halo around it 
while the galaxy has evolved from a gas-rich to a gas-poor galaxy.  The nuclear
black hole in these galaxies remain underfed and hence lack or host at most a faint
active nucleus.  In the pair NGC~7626 and NGC~7619, both of which are classified
as elliptical galaxies and have comparable optical sizes,
a more extended X-ray halo is detected around NGC~7619
in which radio emission is confined to the nucleus whereas the diffuse X-ray
emission around NGC~7626 is relatively compact but it shows bright radio jets and lobes
extending to several tens of kpcs beyond the optical galaxy and the
X-ray halo indicating the presence of a vigorous active nucleus \citep{2009ApJ...696.1431R}.
A cloud of gas of temperature of about 0.75 keV seems to be associated
with both the galaxies while the hot thermal gas that is observed to pervade the space
beyond the galaxies shows higher temperatures \citep{2009ApJ...696.1431R}.  
These observation can be used to infer the X-ray halo around NGC~7619 is 
formed when the galaxy lost most of its gas which evaporated
into the hot tenous thermal gas and only a small amount of the interstellar gas
was channelled inwards to
the supermassive black hole so that it hosts a low-activity nucleus which has
not resulted in a large radio structure.  
On the other hand, in NGC~7626 which hosts a large double radio source
but a smaller X-ray halo than NGC~7619, it could be that most of its interstellar medium 
was channeled inwards and fed the central black hole which in turn 
has ejected a large fraction of this in the form of positron-electron jets and
optical line emitting gas.  The observed X-ray halo in this galaxy might have
formed from the galaxy losing some of its interstellar medium or it could
have formed from isobaric expansion of the optical line emitting gas into
hot tenous X-ray emitting gas.  From the widely different sizes of the
soft X-ray halo around NGC~7626 and the radio double,
it appears more likely that the current X-ray halo
is formed from the interstellar medium and contribution from the ejected
matter from the black hole is yet to append to it.  The X-ray halo around the
host galaxy of the FR~II galaxy Cygnus~A whose extent is similar to the
radio extent would have formed from isobaric expansion of the optical line
forming gas ejected from the active nucleus.   This means that the host 
galaxy has channelled most of its gas to its centre which has fuelled the black hole.
In the process the galaxy has evolved to a gas-poor state. 
The accreting black hole has resulted in a vigorous active nucleus which has
experienced several episodes of mass ejections forming a large double radio source
and a matched X-ray halo.  It appears that this galaxy would not have formed
a X-ray halo around it prior to its active nucleus phase.  
These examples, although few, seem to support the suggestion that galaxies are often
subjected to a gravitational torque that changes their angular momentum which is primararily
mediated through the interstellar medium.  The change in the angular momentum 
results in changing the gas distribution in the galaxy so that in some cases, most
of the gas is gently ejected either as stellar winds or in form of rings which 
forms ring galaxies (whose main stellar body is often observed to be gas-poor)
as was suggested by \citet{2016arXiv160604242K}. In other cases, most of the 
gas in the galaxies is channelled inwards fueling the supermassive black hole 
in the nucleus triggering repeated polar ejections and extended radio bright 
structures.  The galaxies
which lose most of their gas can at most harbour a weak active nucleus 
while the galaxies which channel most of their gas inwards can harbour a vigorously
active nucleus.  This, then, also provides us with a prescription for the formation 
of early type galaxies wherein the gas is depleted in either of the ways
mentioned above.  

The third mechanism which can can result in soft X-ray emitting gas is
the Compton scattering of the positron annihilation photons near 511 keV that are
continuously generated in the positron-electron plasma that emits
the radio emission in jets and lobes in active galaxies.  
The soft $\gamma-$ray photons are down-converted
to X-ray energies which can contribute to the observed X-ray
emission.  The examples which almost certainly point to the existence of this 
process are (1) the double-sided thick X-ray beams detected along radio jets in FR~II 
galaxies like Cygnus~A and Pictor~A and (2) the X-ray emission which is coincident 
with the thick radio features extending on either side of the core in Virgo~A in
the general direction of the compact radio jets.

{\it To summarise:  The X-ray halo often observed surrounding gas-poor early
type galaxies
is formed from the interstellar medium that is lost by the originally gas-rich
galaxy.  The X-ray halo often observed around gas-poor active galaxies with extended
radio double structure with comparable radio and X-ray extents argue for 
an origin in the ejections from the active nucleus.  In the first case, 
interstellar gas is lost due to expansion out of the stellar body of the galaxy whereas 
in the second case, interstellar gas gas is used up in fuelling the 
central supermassive black hole.  There could also be some contribution from
the Compton scattering of positron-electron annihilation photons  to lower energy
X-rays.}

%{\bf include figures showing the different cases of Xray halo formation}

\section{Resolving outstanding problems; creating new ones}
\label{outstand}
We briefly summarise some of the outstanding questions that the work in this
paper has been able to address within the ambit of known physics and observational
constraints.  The questions that working on this paper has raised or not
explained and which need to be resolved with the help of solid observational 
evidence and physics are also mentioned.

\begin{itemize}

\item{\it Formation of S0/ellipticals and active nuclei:}
Since we are not in possession of any data that support the contrary, 
it appears appropriate to assume that all galaxies are formed with a 
central supermassive black hole surrounded by gaseous matter. 
In fact existing observations wherein gaseous matter is detected
either within or around galaxies, support this initial condition for all galaxies.
Star formation would have been triggered in the gaseous matter using 
up a large fraction of the gas with the leftover gas forming the interstellar medium
of the galaxy.  Taking a cue from the present day disk galaxies, the leftover
gas mass would have been at least $10^9-10^{10}$ M$_\odot$.  
This gas, which would be spread across the galaxy can be assumed to behave like a fluid
and hence should be able to change its global properties when a strong
gravitational torque is exerted by the passage of another massive galaxy. 
(Under other circumstances this gas can also be considered to be ideal due to
their extremely low densities.  The situation is somewhat like the wave-particle
duality of quantum particles so that the interstellar gas behaves like a fluid 
in some events and like an ideal collisionless gas for other events.  On the other hand,
the stellar distribution is always a collisionless system.) 
The torque will change the angular momentum of the galaxy which can be reflected 
in the gas component either as a change in its radial distribution or its kinetic
properties as was described in \citet{2016arXiv160604242K}  i.e.
torque $= dL/dt = d(I\omega)/dt$ where $L$ is the angular momentum, $I$ is 
the moment of inertia and $\omega$ is the angular speed. 
If we limit this discussion to the torque-induced radial rearrangement 
of the gaseous component,
then it can be inferred that the gas can either move outwards or inwards
towards the centre of the galaxy.  It should also be possible for
both processes to occur depending on the details of the gravitational interaction(s). 
This means that the gas distributed over the entire galaxy is gradually channeled
inwards towards the black hole providing fuel to drive it into the active state
and/or channeled outwards forming a ring around the galaxy.  
The galaxies which channel most of their gas inwards should trigger intense
activity in their nuclei
and large radio doubles whereas the galaxies which lose most of their
gas can only host weak short-lived active nuclei with faint radio structure, if at all. 
The observed results
that most large radio double sources are hosted by a gas-poor elliptical galaxies, most
ring (and polar ring) galaxies are gas-poor lenticular galaxies and 
most gas-rich spiral galaxies are seldom hosts to strong double radio sources and 
are seldom observed as ring galaxies strongly support the above explanation wherein
the dominant source of fuel for the central black hole and the gas in ring galaxies is
from the host galaxy itself and there exist more than one outcome for the
evolution of this gas.  This, then supports the present day elliptical and 
lenticulars as the first galaxies which were born as gas-rich disk galaxies. 

\item{\it Origin of large emission line redshifts of quasars:}
Quasars are active nuclei with the largest emission line redshifts. 
Since in literature, these emission line redshifts are considered to be 
cosmological in origin,
quasars have been placed at very large distances, their luminosities have
been estimated to be enormous and their radio jets have been observed to expand
with superluminal velocities.  However the study by \citet{2016arXiv160901593K}
pointed out the irrefutable evidence that exists in the redshift data of spectral lines
towards quasars which support a finite and large contribution from gravitational 
(intrinsic) redshift in the observed broad emission and absorption line redshifts. 
In their recipe, the lowest absorption line redshift detected in a quasar spectrum
is the cosmological redshift of the quasar so that the difference redshift for
the emission line provides an estimate of the largest intrinsic redshift present
in the quasar spectrum. 
Once this intrinsic component is removed from the emission line redshift of quasars
then then they seem to occupy a redshift space comparable to other active nuclei,
have comparable luminosities and their radio jets expand subluminally.
The large component of gravitational redshift in the wide spectral lines from
quasars tell us that the broad line region is located close to the event
horizon of the black hole.  In quasars, the emission 
lines are formed within $30 R_s$ of the black hole and in more than half
the quasars, the broad lines are formed within a separation of $2 R_s$
from the black hole.  The largest gravitational redshift component in other active
nuclei is smaller i.e. the emission lines are formed farther from the black hole. 
In fact, this is physical property which is responsible for the extreme
nature of quasars compared to other active nuclei. 

\item{\it Origin of multiply redshifted spectral lines in quasars:}
The multiple redshifts commonly detected in a quasar spectrum are due to 
the contribution of a varying gravitational redshift component in
the wide spectral lines of different species and ionizations, which arise in the BLR
from a range of separations from the black hole. 
Such behaviour but with redshifts of lower amplitude has also been observed in lower
redshift active nuclei.

\item{\it Source of transient energy in nova, supernovae, active nuclei:}
The source of the energy burst in all these transients can be traced to a
thermonuclear outburst.  This origin can explain both the
timescales and energy released in these events.
Moreover this means that all three objects should emit $\gamma-$rays
generated in the thermonuclear reaction - with energies ranging from tens of
MeV to several hundred MeV depending on whether the photons generated in the
reaction escape unchanged or are inverse Compton boosted to higher energies
by relativistic electrons.  It appears that
thermonuclear reaction is the source of energy which dominates transient
events in the universe in addition to powering the steady stellar radiative output.
This should continue to be the case till the universe
becomes iron-dominated when thermonuclear reactions will cease to generate
energy.  It could be that the transient
sky will become quiet in the iron-dominated era and the universe will enter
a quiescent phase of evolution.
 
\item{\it Origin of a conical, collimated radio jet:}
Radio jets and optical line forming gas are launched from the polar regions 
of the quasi-spherical pseudosurface around a black hole due to the energy input from
a thermonuclear explosion in the accreted gas.   
Since matter radially launched from a small spherical surface has to
be confined to a conical region with its apex at the centre of the sphere, the
jet and optical line forming gas describe a conical flow as they expand away from the
core. The conical angle (jet opening angle) is determined by the 
extent of the launch region which in turn will be determined by the
black hole spin.  In other words, the observed conical structure of the radio 
jets lends strong support to their being radially launched from a spherical surface 
i.e. pseudosurface of the black hole.  If the jet is
launched from close to the black hole which requires highly relativistic 
velocities to escape, then the jet can remain ballistic for long distances
and hence collimated.  

\item{\it Radio structure in active galaxies:}
The differences in the size of the radio structure associated with 
active nuclei in elliptical and spiral hosts is trivially explained by the 
active nucleus having switched on much earlier in the elliptical galaxy then
in the spiral galaxy i.e. accretion by the black hole would have commenced much earlier in
elliptical than in spiral galaxies.  Since accretion by the black hole is an important
sink for gas depletion, the gas-poor active ellipticals which host extended 
radio double sources have channeled most of its gaseous component inwards and
consistently fuelled the nuclear black hole. 
On the other hand, as indicated by the copious quantities
of gas in spiral galaxies, the nuclear black hole is yet to accrete a significant
amount of gas and subsequently eject it along radio jets
explaining the lack of extended radio doubles in spiral hosts.

Another contributory factor to the weak radio structure of active nuclei
in spiral hosts could be the lower spin of the black holes in spiral hosts
which leads to large jet opening angles and hence the jets will lose their
identity at short separations from the black hole.  
In the fast rotating black holes in elliptical hosts, the jet opening angles
are small and the jets launched from close to the black hole are launched with
relativistic ejection velocities which ensures collimation 
upto a long distance. 

The question remains as to why the black hole in elliptical hosts rotate
faster than in spiral hosts. 

\item{\it Most FR~II are BLRG while most FR~I are NLRG:}
FR~II radio sources are formed when the jet opening angle is extremely small
and the BLR covers most of the pseudosurface - both of which occur 
for fast-spinning black holes.  If the emission zone of the BLR lies close
to the black hole due to the diminished non-polar accretion rates then
lines will include a larger contribution from gravitational broadening 
and broader  emission lines will be more frequently detected
from such objects along most sightlines explaining the connection between
BLRG and FR~II radio sources.

In FR~I sources, the black hole spin is relatively lower, the jet opening
angle relatively larger and the covering factor of BLR smaller than in FR~II nuclei.
For the same infalling matter densities and hence comparable
polar accretion rates in the FR~I and FR~II cores, the 
accretion rates at a given latitude in FR~I cores can be larger than in FR~II 
cores due to different centrifugal contributions. 
This can cause the non-polar pseudosurface in FR~I cores to thicken and the 
BLR can form progressively farther from the black hole and hence the detected
emission lines are likely to be more frequently narrower due to the 
smaller gravitational broadening and the galaxy will be classified as NLRG.  
The smaller covering factor of the BLR in FR~I cores can also make detection of
lines from the BLR orientation-dependent.  

\item{\it Energy distribution of cosmic rays:}
\begin{figure}[t]
\centering
\includegraphics[width=8cm]{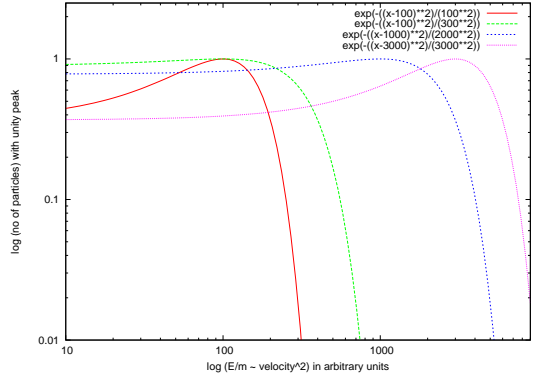}
\caption{The expected energy distribution for particles energised in an
explosive fast event like a thermonuclear outburst is shown on log-log.  Energy distributions  
expected for four different mass particles is shown with the peak normalised to unity
and the E/m in arbitrary units.  Neutrinos with their lowest mass will have
a distribution centred at the highest velocities followed by electrons/positrons,
protons and heavy ions.  The dispersion of the distributions will be as shown
in the plot.  Neutrinos have the highest dispersion followed by electrons/positrons,
protons and heavy ions.  The energy distribution for
protons and heavy ions is shown to peak at the same velocities and the difference in the
distributions are due to distinct dispersions.  }
\label{fig41}
\end{figure}
The energy input in supernovae and active nuclei which launches matter is from
an explosive thermonuclear outburst which instantaneously energises a large
quantity of matter.  This statistical event has to result in normal/gaussian distribution
of energies for all species of particles ranging from neutrinos to electrons/positrons
to protons and heavy cosmic rays.  The distributions can be assumed to
be centred on similar energies and can have different dispersions which depend on the
mass of the particle.  Thus if E/m is plotted as in Figure \ref{fig41},
neutrinos, due to their negligible mass, will show the distribution at the highest E/m
followed by electrons/positrons.  The distribution of protons and heavy ions are
shown to peak at the same E/m in Figure \ref{fig41}.   These plots are only
indicative but these already reproduce the observed particle energy distributions
fairly well.  The distribution beyond the peak  in all cases can be approximated
by power laws.  If the dispersion of the distribution is small then the curve
will show a convex shape whereas if the dispersion is large, the curve will show
a large flat part at lower energies.  Thus all particles should be injected with
such a energy spectrum which can be modified due to several physical processes 
after they leave the system.  Thus, a fraction of the ejected cosmic rays will
quickly leave the system and is detected by us.  Another fraction of the cosmic
rays especially protons and heavy ions can undergo collisions which
leads to thermalising of the plasma. 

\item{\it Composition of radio jets is positron-electron plasma:}
Several observations seem to point to a positron-electron plasma in the fast radio 
jets.  These are detection of the annihilation line from microquasars, 
detection of radio emission in two distinct phases in supernovae namely the prompt
radio emission which can be explained if attributed to a 
fast positron-electron plasma and the delayed supernova remnant emission
which can be attributed to a relativistic electron-proton plasma mixed with the
thermal plasma.  The formation of X-ray beams detected along radio jets and hotspots in 
several FR~II sources are readily explained by a positron-electron plasma in the jet
which emits positron-annihilation photons near 511 keV along the jet which
are Compton-scattered to X-ray energies or exert a radiation pressure.

Is there a relativistic electron-proton plasma that is mixed with the optical
line forming gas i.e. thermal gas ?  Are there any radio signatures 
which can distinctly be associated with synchrotron emission from the proton-electron
plasma ?  Or is it that the fast proton-dominated plasma also rushes out with
the positron-electron plasma and provides the magnetic field along the jet ? 
Since observations of the prompt radio emission in supernovae strongly
suggest that the magnetic field is not frozen in the positron-electron plasma,
we assume this is true for other astrophysical systems as well. 
Thus, we assume that in radio jets, the emitting plasma is positron-electron
but the magnetic field is provided by the proton-dominated plasma.  This is 
also supported by the comparable spatial extent of the radio plasma (relativistic
particles) and the thermal X-ray emitting or optical line emitting plasma 
(magnetic field). 

\item{\it Origin of the centrally located radio and X-ray halo in clusters:}
The origin of a large fraction of both the radio and X-ray halos detected in the
centres of clusters is likely to be due to energetic ejections from the active
nucleus of the central galaxy over long timescales.  If several galaxies
have either merged with the central galaxy or lost their interstellar medium
to the central galaxy in repeated encounters, then a large fraction of this matter
can fuel the central black hole and a large fraction can be ejected following
thermonuclear explosions at the poles.  The ejected matter will be composed of
two main detectable components: positron-electron plasma which gives rise to
radio synchrotron emission and the proton(ion)-electron plasma which gives rise
to the thermal free-free emission at X-ray wavelengths and provides the
magnetic field for the positron-electron plasma.

Other contributions to the radio and X-ray halo can include similar ejections
from other active galaxies in the cluster.   The X-ray halo can also include
contributions from galaxies which are losing their gas content via rings
and stellar winds.  For example, Virgo cluster is good example of a X-ray
halo which contains both a halo formed by the central active galaxy M~87 and
X-ray emission contributed by other galaxies in the cluster
as is obvious from the multiple peaks in the non-spherical X-ray distribution
(see Figure \ref{fig37}).

It is noticed that the X-ray luminosities of elliptical galaxies with the same
optical luminosity span more than an order of magnitude \citep{1997ASPC..116..375S}.
This could indicate the existence of two distinct origins for the thermal plasma
namely in the gas that is gently lost by the galaxy and in the gas that is
accreted onto the pseudosurface of the central black hole and then episodically
ejected.  Both the gas isobarically expand around the galaxy to more than
million degrees K.  The two different origins might be related to the formation of
cavities in the X-ray distribution that is observed to be associated with radio lobes 
in several low power FR~I radio galaxies especially in groups and clusters. 
In other words, the galaxy might have formed a X-ray halo as it gently lost gas 
before the black hole entered its active phase and started ejecting matter along
polar jets which formed the radio lobes.  

\item {\it Formation of hotspots and relics at the edge of the X-ray halo; 
formation of thick X-ray beams along radio jets:} 
In literature, formation of hotspots is understood as being due to a 
reverse/termination shock whereas formation of relics is attributed to 
shocks set up by the merging of two clusters.  These explanations are
incomplete and vague leaving them open to misinterpretation - for example, 
it is not clear in this scenario which physical process abruptly halts the still 
relativistic jet in FR~II sources at the hotspots nor is it clear
how such intense emission of hotspots can result from the jet encountering the lower density 
intergalactic medium?  In case of relics, the shock scenario fails to explain
the formation of relics at the edge of the X-ray halo. 
Since such crucial points lack clarity in the widely accepted models, they 
are interpreted by astronomers in a variety of ways. 
Concrete physical reasons can be identified for the formation of hotspots and relics as 
described in the paper.

The morphology and physical properties (e.g. magnetic field orientation)
of both the hotspots and relics argue for a halting mechanism like ram pressure
or radiation pressure both of which are found to be widely effective in the universe. 
Since the medium beyond the hotspot is more tenuous than the medium in which the
jet has already propagated, ram pressure can be ruled out. 
We hence examine the feasibility of radiation pressure being responsible for
hotspots and relics. 
Interestingly, both hotspots and relics are formed on the outskirts of the diffuse
extended thermal free-free X-ray emission.  
For radiation pressure to explain hotspots and relics, a
source of hard photons beyond the hotspots and relics is required.
The pair annihilation photons near 511 keV generated along the entire positron-electron
jet can be the photons which can exert a radiation pressure.
These photons will be promptly
scattered or absorbed by the dense gas around the jet and Compton scattering of these
photons to X-ray wavelengths explains the thick X-ray beams detected along radio jets.
However when the radio jet travels beyond the dense gas defined by the soft X-ray emission 
the annihilation photons near 511 keV can survive longer. 
and a fraction of these 
can exert radiation pressure on the matter at the periphery of the X-ray emission - both
radio plasma and thermal plasma.  This pressure can compress the thermal gas 
which enhances the magnetic field and orients it perpendicular to the jet 
direction.  The radiation pressure can also redirect a large fraction of the 
positron-electron plasma.  The main requirement for these being that the plasmas
are optically thick to the photons.   
This explains the formation of hotspot at the periphery of the X-ray distribution and
no reacceleration of electrons at the hotspots is required. 
The same population of electrons ejected from the active nucleus can explain the 
hotspot emission.  This also implies that deep observations
might detect the annihilation line from the region immediately beyond the hotspots. 
Since no radio synchrotron emission is detected beyond the hotspots although some of
the positron-electron plasma is certainly escaping beyond the hotspots supports  
the freezing of the magnetic field in the thermal hot gas and is hence not available
beyond the thermal gas boundary.  

It is possible to think that relics are formed from a similar process 
on the outskirts of the X-ray halo in clusters.  The X-ray emission is from the
thermal proton-electron plasma whereas the radio emission is from the relativistic
positron-electron plasma that radiates in the magnetic field frozen in the thermal plasma. 
The dominant source of both these plasma was the central active galaxy. 
The plasma in the radio halo co-exists with the X-ray halo and hence the annihilation
photons are rapidly absorbed or Compton scattered.  However beyond the X-ray halo, 
matter densities are extremely low and the 511 keV photons can survive longer. 
These photons can exert a radiation pressure on the relativistic plasma and the thermal 
plasma.  If the axis, which is generally the major axis of the radio and X-ray halos,
at the end of which relics are detected defines the original direction of the jets
from the nucleus of the central galaxy when it was in its active phase then it is
possible that the positron-electron plasma still retains a finite outward motion. 
When this plasma exits the dense region of the X-ray halo, 
the annihilation photons can survive longer and exert radiation pressure which can compress
the plasma in the interface region which compresses the thermal gas and 
enhances magnetic fields which will be oriented perpendicular to the radial axis of the
halo.  The compression of radio plasma and enhanced magnetic field can result in
enhanced radio emission i.e. relics.  The relic is seen to be 
spread over a large linear region
which when joined to the centre defines a larger opening angle than observed for hotspots.
This is not surprising since the central nucleus in its active phase would have been
a FR~I source with a larger jet opening angle which would been increased by a precessing spin
axis of the black hole. 
This explanation hence appears to be plausible and is consistent with rest of the discussion
on active nuclei.  However it needs to be further investigated and specific
observations that would provide targetted results need to be conducted.

\begin{figure}[t]
\includegraphics[width=7cm]{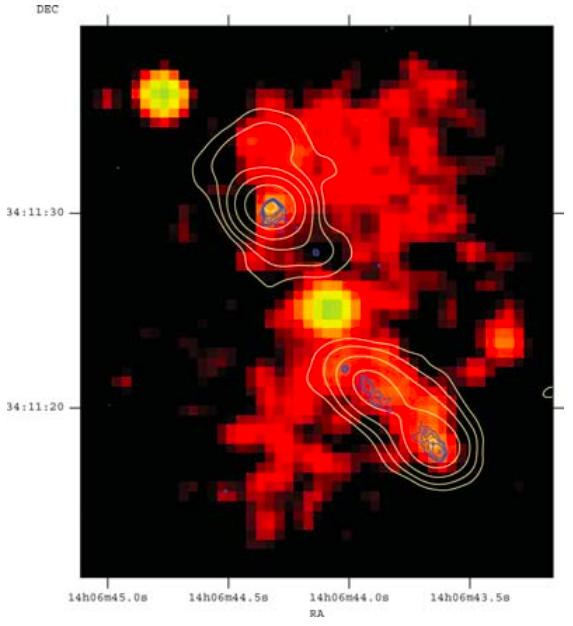}
\caption{The X-ray emission between 0.5-3 keV in colour overlaid with contours
of radio emission at 8.46 GHz (higher resolution) and 1.425 GHz (lower resolution)
of 3C~294 copied from \citet{2006MNRAS.371...29E}.
Notice the biconical morphology traced by the X-rays with the radio jets being confined
to one edge of the X-ray emission which is reminiscent of the ionization bicones
often imaged in forbidden lines of oxygen centred on the active core - for example, 
in the Seyfert galaxy Circinus as in Figure \ref{fig24}.  }
\label{fig42}
\end{figure}

\item{\it Evolution from radio galaxy to minihalo to halo ?} 
A correlation between the presence of a central X-ray peak which is inferred
to indicate a cooling flow 
in a cluster and the presence of a central radio galaxy has been noted.  
On the other hand, it is observed that the clusters which host both
a radio and X-ray halo often lack a radio galaxy at its centre.
This can be understood as follows.
The centrally located radio and X-ray halo in the cluster has an origin in the energetic
ejections from the active nucleus as discussed in the paper and 
hence has evolved from the nuclear ($R_s$)
scales to galactic (several 10s kpc) scales to cluster (several 100s kpc) scales.
In other words, the ejections from the central nucleus can lead to double radio
structure which evolves to a mini-halo and then a radio halo. 
Similarly the ejections will lead to a galaxy-size X-ray halo which evolves
to a mini-halo and halo-sized X-ray emissions.  It has been observed that
a central radio galaxy and mini-halo co-exist in several clusters while in
some clusters structures which can be dissociated into a radio mini-halo and 
halo have been observed.  Such observations justify an evolutionary connection.  It would 
also mean that by the time a uniform centralised cluster-level halo is formed, 
the active nucleus has exhausted its fuel and is no longer launching jets.

\item{\it X-ray cavities and radio lobes:}
In several radio galaxies and groups/clusters of galaxies, 
bright radio emission of the lobes is coincident with faint emission
in X-rays which is often referred to as X-ray cavities since these
regions appear dark as compared to rest of the X-ray halo.
Literature suggests that these cavities are a result of pressure differences
between the X-ray emitting gas and the synchrotron emitting plasma so
that the latter has displaced X-rays.  While this might be possible,
since the positron-electron plasma radiates in the 
magnetic field frozen in the thermal plasma which when at lower temperatures
is characterised by optical emission lines and when at higher temperature emits 
X-rays, it is possible that there is a transient stage when both the plasma
are mixed in the regions which emit synchrotron radiation e.g. lobes.  
This, then argues, against any displacement of the thermal gas. 
Optical line emitting gas i.e. thermal proton-electron gas is ejected 
alongwith the positron-electron plasma and 
is likely to eventually isobarically expand to a 
million degrees gas and contribute to X-ray emission.  Cases such as the one in
Figure \ref{fig42} wherein the X-ray distribution shows a biconical structure like
optical line emitting gas lends support to this scenario.  
There, then, exists the possibility that the X-ray cavities indicate the transition
regions wherein the cooler $10^4$ K gas is mixed with the hotter $>10^6$ K gas so that
the resultant temperature of the gas is between $10^4$ and $10^6$ K and does not
emit X-rays.  

There also exists another possible explanation for the X-ray cavities which is 
opposite of the above.  The relativistic electrons in the lobes
might inverse Compton scatter the existing X-ray photons to higher energies so that the
lobes appear fainter in the low energy X-ray photons.  
Observations should be used to disentangle the
two possible origins of the cavities.  The X-ray emission from most FR~II lobes 
are well explained by inverse Compton scattering of lower energy photons to 
X-ray energies.  Literature often supports the inverse Compton scattering of
the CMB photons as the source of the X-ray emission from the FR~II lobes.
However, it is also possible that inverse Compton scattering of the optical photons 
emitted by the $10^4$ K gas which was ejected alongwith the radio plasma from
the black hole and should be flooding the lobes is responsible for the X-ray photons.
Relativistic electrons of energy $\gamma\sim100$ or so can scatter the optical
photons to X-ray energies.  In analogy to this, if the X-ray photons (instead
of or alongwith optical photons) already present 
in the FR~I lobes are inverse Compton scattered to higher energies then the
lobes would shine at very high energies - hard X-ray or soft $\gamma-$ray energies
and be faint at soft X-ray energies as is observed in several such sources. 

It should be of relevance that diffuse X-ray halos are more often detected around FR~I 
sources compared to FR~II sources.  A possible explanation might lie in the
manner in which the host galaxy has lost its interstellar gas.  If in FR~I sources, 
the host galaxy has lost its gas through both
formation of external structures like rings and by channelling fraction 
of gas inwards to the black hole, it could lead to the formation of a soft
X-ray halo around FR~I hosts from the thermal gas evaporated from the ring 
by the time the central black hole in its vigorous active phase, ejects and 
distributes radio plasma along bipolar jets and lobes.   On the other
hand, it is possible that in FR~II hosts most of the gas is channelled inwards
towards the black hole and hence no soft X-ray halo is formed around the galaxy
prior to nuclear activity distributing radio plasma along jets and lobes.
This can consistently account for the distinct nature
of the diffuse X-ray emission around FR~I and FR~II sources, their
distinct power regimes and the distinct
observed nature of the X-ray emission from lobes of FR~I and FR~II sources. 
However it needs to be verified or ruled out by detailed examination of observations
such as evolution of ring galaxies into FR~I sources and non-ring galaxies into
FR~II sources.  For example, the observation that FR~I structures are commonly formed in
cluster-like environs while FR~II structures are commonly formed in isolated
environs does lend support to the above scenario.

\item{\it X-shaped radio galaxies:}
\begin{figure}
\centering
\includegraphics[width=6cm]{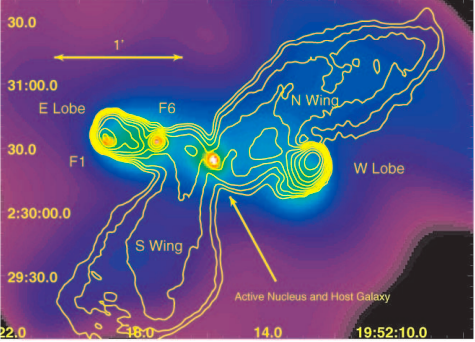} (a)
\includegraphics[width=6cm]{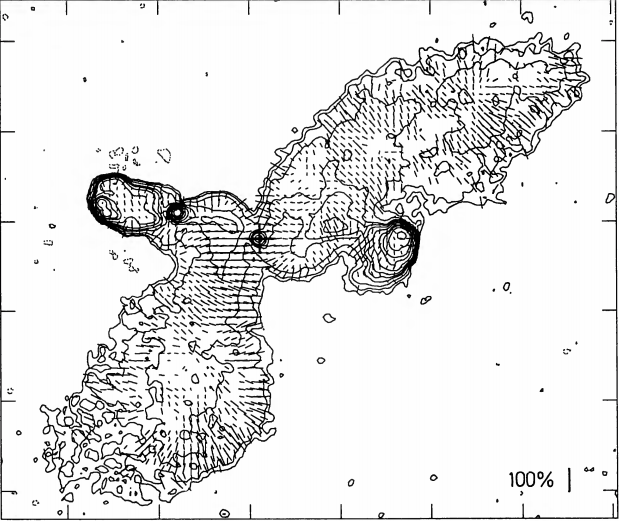} (b)
\caption{The X-shaped radio galaxy 3C~403.  (a) Figure
reproduced from \citet{2005ApJ...622..149K}. The contours denote the radio emission at
8.3 GHz overlaid on the X-ray emission between 0.5 and 2 keV.  Notice  
how the X-ray and radio lobes are coincident and the lobes are bent by right angles
near the core. 
(b) Figure reproduced from \citep{1992MNRAS.256..186B}. 
The contours depict the total intensity at 8.3 GHz whereas the dashes denote the
polarisation vectors.  Notice the enhanced polarised emission at the
edge of the wing near the core and indicates a magnetic field oriented
along the length of the wing.  It is suggested that such wings are formed due to
the backflow being stopped and diverted by the accretion disk of the black hole. }
\label{fig43}
\end{figure}
There exist radio galaxies which show a pair of intense active lobes along the
radio axis and a pair of diffuse lobes (often referred to as wings 
in literature) along an axis which is often displaced 
from the radio axis by an angle close to $90^{\circ}$, thus lending the 
radio structure a X-shaped morphology. These galaxies are referred
to as X-shaped radio galaxies in literature 
(see Figures \ref{fig43},\ref{fig44}).  The active radio lobes in
many of these galaxies terminate in a hotspot and hence most of these galaxies
have been classified as FR~II type sources although their radio power is 
generally estimated to be between the FR~I and FR~II types.   
The formation of the X-shaped radio structure in such 
galaxies has been a subject of debate in literature.

\begin{figure}
\centering
\includegraphics[width=7cm]{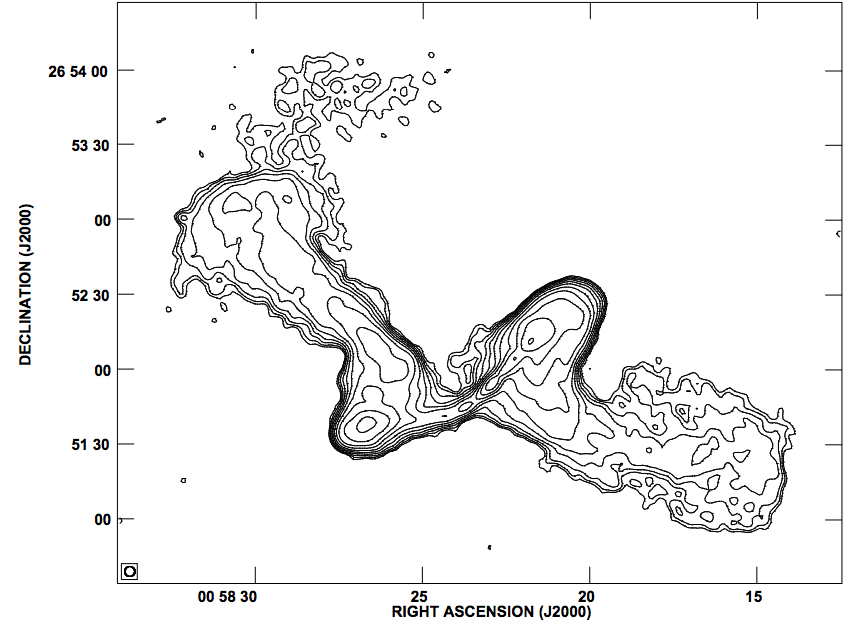}
\caption{The X-shaped radio galaxy NGC~326 at 1.4 GHz.  Figure copied from
\citet{2001A&A...380..102M}.  Notice how the wings are bent at right angles near the core.
The dimension of the linear side of the wing which continues would indicate
the size of the accretion disk around the supermassive black hole.   }
\label{fig44}
\end{figure}

A few inferences derived for X-shaped radio galaxies from a statistical
study \citep{2009ApJ...695..156S} are (1) The orientation of the
radio axis in X-shaped sources is within $\sim 50^{\circ}$ of the major axis
of the host galaxy whereas wings are oriented within 
$40^{\circ}$ of the minor axis of the host galaxy.  (2) The linear size of
radio galaxies in which the radio axis is oriented along the host 
minor axis, is larger than when the radio axis is 
closer to the host major axis with the latter more often showing
wings and other distortions.  They find that the largest linear 
dimension of the X-shaped galaxies is smaller than of linear FR~I and FR~II 
sources.  It was suggested that the X-shaped morphology could be explained 
if the backflow in FR~II sources
was deflected.  One of the deflecting agents is considered to be the thermal 
X-ray emitting halo around the host galaxy \citep{2009ApJ...695..156S}.  
Other origin scenarios for X-shaped radio galaxies in literature involve 
a change in the jet axis either due to precession or due to dual black holes in
the active galaxy. 
Magnetic field in the edge of the wings of X-shaped radio galaxies
is observed to be parallel to the length of the wings (see Figure \ref{fig43}b).
Such persisting nature of the ordered magnetic field in the jets and wings
can be inferred to indicate a net flow direction for the gas in which it
is frozen.  

The backflow explanation seems to well explain the observed morhpology
of X-shaped radio sources.  As pointed out in the paper, the
hotspots and backflow are due to the action of radiation pressure exerted
by the pair annihilation photons at 511 keV on the plasmas so that
these are redirected towards the core.   
In X-shaped radio galaxies, the backflow near the core is bent at right 
angles to the 
radio axis before the plasma diffuses out to form the wing-like structure
(see Figures \ref{fig43},\ref{fig44}).   
This behaviour argues for a restraining medium/structure near the core which is 
oriented perpendicular to the radio axis which first halts and
then deflects the plasma away from the core such that the plasma does
not cross the core towards the other jet.
In other words, both the positron-electron and thermal proton(ion)-electron plasma 
appear to be ramming into a dense medium near the core which deflects them.  
%While the deflecting agent could be the X-ray halo \citep{2009ApJ...695..156S}, 
The location and nature of the observed effect, especially the 
bending of the backflow by 90 degrees strongly supports bending
by the equatorial accretion disk that has formed around the 
central supermassive black hole.  The accretion disk will be formed close to
the core and be perpendicular to the bipolar jets.  If  
a line through the core is drawn perpendicular to the radio axis, then the
wings generally skirt this line in X-shaped sources.  This line would represents
the location of the accretion disk.  If precession
of the jet axis is included then the rest of the X-shaped sources wherein the
wings seem to flow a bit onto the other side can also be accounted for. 
This then suggests that the
accretion disk can extend to several kpc around the fast rotating
supermassive black hole as it keeps accreting matter over billions of years. 
The especially low non-polar accretion rates in the fast rotating black hole
in FR~II sources will lead to the formation of a radially extended
accretion disk. 
Winged sources appear to be more common when the radio axis is oriented
closer to the major axis instead of the minor axis \citep{2009ApJ...695..156S}.
This could simply be an effect of the larger resistance provided to the
polar ejections from the black hole by the galaxy when the jet is oriented closer to
its major axis so that the plasma turns back sooner and hence encounters the accretion disk
more frequently.  In the cases when the jet is along the minor axis, the jets 
extend much further before turning back.  This means the timescales required by 
the backflow to encounter the accretion disk and be deflected are much longer.  

The backflows on either side of the core in X-shaped galaxies are observed to
be bent in opposite directions giving it the appearance of a dual pair of lobes.  
There also exist sources in which the backflows on either side of the core are
bent on the same size instead of opposite sides 

FR~II radio galaxies which are believed to have restarted from the existence
of an inner pair of hotspots like A1425 \citep{1997ApJS..108...41O} (see Figure \ref{fig45}a) 
and J0116-473
\citep{2002ApJ...565..256S}, although not X-shaped show an abrupt turning of
the backflow at right angles to the radio axis in one of the lobes.   
The physical process at work is likely to be similar to X-shaped galaxies except that the
backflow on the opposite side is yet to reach the core.  The morphology of the
two radio galaxies is remarkably similar and the spin axis of the black hole
appears to be precessing as can be surmised from the displaced positions of the
pairs of hotspots which are collinear with the core.  Another interesting
inference that follows from the discussion so far is that the new radio axis
determined by connecting the new pair of hotspots
has to be displaced from the previous one such that the thermal gas along the new direction
is confined to a smaller region around the core thus enabling the formation of
the new pair of hotspots.  This means the new pair of jets and hotspots are only projected
on the old pair of lobes, but occupy an independent region of space along the sightline
if the explanation for the formation of hotspots presented in the paper is valid
which requires relatively empty space beyond the hotspots.  

The galaxy 0053+261A in Abell~115 is a variant of a X-shaped radio galaxy wherein 
both the lobes formed by the backflow from the hotspots are deflected towards the same side 
of the jet axis by the accretion disk (see Figure \ref{fig45}b).  
If the lobes had been deflected towards opposite
sides of the jet axis, the galaxy would have been classified as a X-shaped galaxy. 

\begin{figure}
\centering
\includegraphics[width=6cm]{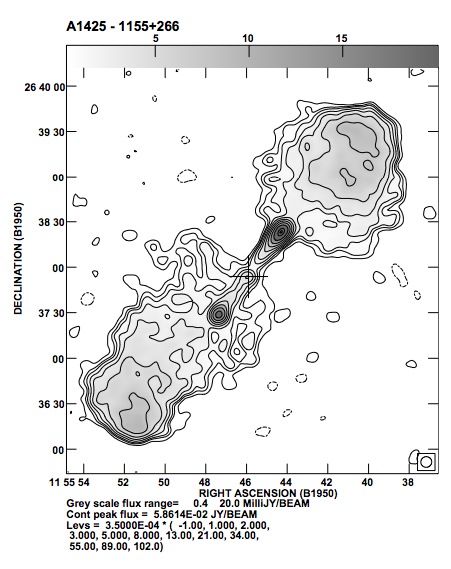} (a)
\includegraphics[width=6cm]{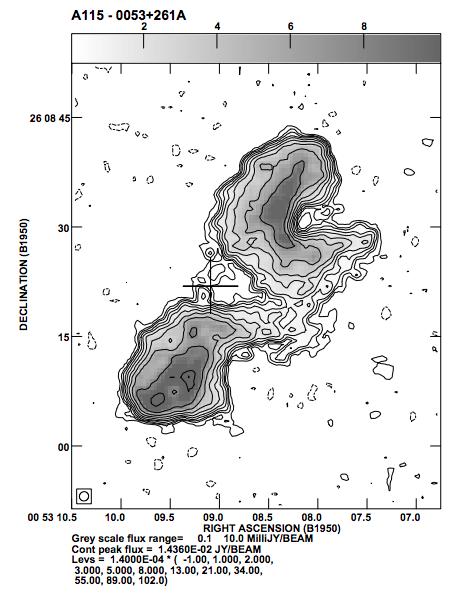} (b)
\caption{Figures showing two examples of radio galaxies wherein lobe(s) are bent by
right angles near the core. 
Figures reproduced from \citet{1997ApJS..108...41O}.  In (a), the galaxy 1155+266 in
Abell~1425 is shown wherein the backflow in the
south appears to have rammed into the accretion disk which has confined it to
the same side as the lobe, bending it by $90^{\circ}$ to the jet axis as defined
by the new pair of hotspots detected within the older lobes.   In (b), the galaxy 0053+261A
in Abell~115 is shown wherein lobes
formed by backflows on either side of the core have reached the accretion disk
around the black hole and have been deflected and confined
to the same side as the corresponding lobe.  In this case the deflection for both
lobes is towards the same side - if the lobes were deflected towards opposite sides,
it would have been classified as a X-shaped galaxy.  The formation mechanism appears 
to be the same. }
\label{fig45}
\end{figure}

{\it In summary, X-shaped sources are FR~II sources wherein the backflow on both sides
have encountered the accretion disk around the central black hole and hence 
deflected by $90^\circ$.  The extent of the perpendicular wing can be used to
obtain the radial extent of the accretion disk.  In principle, all FR~II sources 
will form X-shaped
sources when the plasma in the backflow reaches the core i.e. accretion disk. }

\item{\it Explaining radio spectra of blazars:}
The typical radio spectrum of a blazar (i.e. an active nucleus which
is viewed along or close to its polar axis) is flat over a large radio frequency
range and shows a convex curvature at low radio frequencies.  This can be understood 
from the discussion in the paper as follows.
Blazars are all radio-loud indicating repeated ejection of radio-emitting plasma,
each with a normal energy distribution.
It is possible that the peak of the energy distribution in the different
ejection episodes are at differing energies so that emission
from several such plasma blobs located along the sightline results in a flat
spectrum.  This composite spectrum will curve at the lowest radio
frequencies  - this could be postulated
to arise from the radio plasma which is closest to us along the sightline and
hence has aged the most.  This explanation is similar to that given by 
\citet{1980ApJ...238L.123C} to explain flat radio spectra of active nuclei
and only appends it with a physical reason which explains the spectral turnover.  

Several Fermi LAT $\gamma-$ray emitting sources have been found to be
blazars.  This supports the thermonuclear origin for the explosion which
emits $\gamma-$ray photons in addition to energising and ejecting relativistic plasma
as jets.  Since we view these objects along their polar axis, all the emission
will be Doppler-enhanced.  

\item{\it Origin of $\gamma-$ray bursts and their afterglow:}
Episodic thermonuclear ignition of accreted matter at the poles on the pseudosurface of
a black hole will release $\gamma-$rays and energy.   The thermonuclear explosion 
in which a large mass of fuel is ignited on the poles of a faint accreting black hole
can release energy that can heat the entire pseudosurface due to the high conductivity 
of degenerate matter which forms the pseudosurface.  A large number of 
$\gamma-$ray photons upto energies of 25 MeV or so will be released 
and can be inverse Compton-scattered to the higher energies that we detect. 
This burst of $\gamma-$ray photons can explain $\gamma-$ray bursts. 
It is also possible that the $\gamma-$ray burst is due to the $\gamma-$rays 
released in the thermonuclear explosion of a supernova which   
signals the formation of a black hole/neutron star or the annihilation of white dwarfs.  
The afterglow signatures for the two origins will be distinct due to
the different nature of the two physical systems.  
The typical afterglow signatures of a GRB consists 
of kilonova emission which can be explained by black body radiation, 
multi-band jet emission consistent with a synchrotron process and spectral lines
which can be narrow or wide, often multi-redshifted similar to the lines from the BLR in an
active nucleus.  Most supernova explosions are spherically symmetric and hence will
not result in jet-like structures giving rise to synchrotron emission.  
Hence the GRBs in which afterglow signatures include a synchrotron jet
cannot be associated with a supernova explosion and instead have to be associated
with an accreting rotating black hole.  
It does appear easier to explain the afterglow signatures for an accreting black hole.
The thermonuclear explosion in the accreted matter at the poles will heat and
hence brighten the entire pseudosurface due to the enhanced
black body radiation which explains the enhanced ultraviolet-optical
emission which is often referred to as kilonova emission in literature.  
The emergence of hard photons from the heated pseudosurface in the kilonova phase 
can excite the non-polar BLR lying dormant next to it so
that wide lines become detectable in the spectra of the afterglow emission.  
As the pseudosurface radiatively cools, the emission known as kilonova will fade and the 
broad lines will disappear.  The matter ejected at the poles will consist of
relativistic positron-electron plasma which explains the synchrotron emission 
observed from X-rays to radio bands and the 
narrow optical line emitting gas i.e. thermal proton(ion)-electron plasma.  
These plasmas will keep expanding along the jet so that jets are detectable
till the emission fades below sensitivity limits. 
GRB 170817A which was extensively observed demonstrated several of the above properties -
a $\gamma-$ray pulse followed by kilonova emission and synchrotron emission along jets. 
These would indicate that the burst was associated with an accreting rotating black hole.
A range of observational signatures are possible.  For example, the actual
$\gamma-$ray burst could be missed but afterglow signatures could be noted else
a fraction of the afterglow signatures are detected alongwith the GRB.

Thus, it appears that GRBs can be explained by transient events on
an accreting rotating black hole or supernovae.  

\item{\it Single or double-peaked light curves of supernovae:}
The light curves of type 1a supernova show a single peak especially in B 
and V bands and the typical $t_2$ i.e. time over which the peak magnitude 
drops by 2, is noted to be around 20 days.   
The light curves of the SN 1a at infrared and red bands often show a second peak.
On the other hand, light curves of type II supernovae are characterised by 
double peaks or plateaus and the typical $t_2$ is
longer.  As pointed out for novae \citep{2017arXiv170909400K}, the 
pre-maximum halt and final rise of light curve to maximum in addition
to the post-maximum plateau observed in light curves of some novae
could be understood as being
due to light contribution from a hot central source like the white
dwarf.  This contribution could be in form of radiation or
it could also lead to change in the temperature of 
the ejecta and hence its emission properties.  A similar explanation 
could explain the double peaks or plateau which are normally observed
in supernovae of type II.  The absence of double peaks and plateaus in the light curves
of type 1a supernovae in which the central object is believed to be destroyed 
lends support to the above model.  Since dust will be formed in the ejecta of
supernovae in a way similar to novae \citep{2017arXiv170909400K},
the second peak, often observed in the red and infrared lightcurves of SN 1a 
would signal the inclusion of radiation from newly formed dust. 
 
\item{\it Origin of the optical transient AT2018cow:}
This transient was first recorded on 16 June 2018 with an apparent magnitude of
$\sim 14.76$ magnitudes in ATLAS o-band with its position being inside
the galaxy CGCG 137-068 (distance of 60 Mpc)
but displaced from its centre \citep{2018ATel11727....1S}. 
At the distance to the galaxy, the maximum optical luminosity was
$-19.9$ magnitudes while the peak bolometric luminosity
was estimated to be  $\sim 4\times 10^{44}$ erg s$^{-1}$. 
The transient showed predominantly blue colours and a rapid rise followed by a rapid 
decay of the light curve with no secondary peaks 
\citep{2018arXiv180800969P,2018arXiv181010720M}.
The line spectrum was initially featureless but the spectra taken between days 4 and 8 
showed the presence of a very wide 
feature which was interpreted as an absorption feature of full width 1500 A 
centred around 4600 A arising in gas expanding with velocities $> 0.1c$ 
\citep{2018arXiv180800969P}.  Alongwith this feature, radio and sub-mm continuum emission
also became detectable while narrower (few thousand kms$^{-1}$), redshifted 
($\sim 3000$ kms$^{-1}$) asymmetric emission lines of hydrogen and helium 
appeared in the spectra after 10 days  \citep{2018arXiv180800969P,2018arXiv181010720M}.
The transient, first detected in the ultraviolet-optical-infrared
bands was detected from radio to X-ray bands. 
The ultraviolet-optical spectrum is well fit with a black body
spectrum whose temperature decreases from 30000 K to 15000 K
while the infrared and radio emission can be explained by 
a power law component \citep{2018arXiv180800969P}.
Two possible origins are suggested for the transient - supernova or a  
tidal disruption event by an intermediate mass black hole but both are found
to be inadequate in explaining all the observations \citep{2018arXiv180800969P}.
The V band light curve of AT2018cow drops by 2 magnitudes in $\le 7$ days 
as can be surmised from Figure 2 in \citet{2018arXiv180800969P}.
Long-lived soft X-ray emission and transient hard X-ray emission (between days
8 and 17) are detected from AT2018cow \citep{2018arXiv181010720M}.
It is inferred that less than $0.5$ M$_\odot$ of matter was ejected in the explosion
of AT2018cow with the observed ejecta velocities between 0.05 to 0.1c and the  
energy radiated in the optical/ultraviolet bands was  
$10^{50.5}-10^{51.5}$ ergs \citep{2018arXiv181010720M}. 
While the above estimates are model-dependent, it is useful to
mention that an energy of $10^{51}$ ergs can be released in a short timescale
from simultaneous thermonuclear ignition of about $0.5$ M$_\odot$ which is typical of
supernova explosions and explosions which eject matter in active nuclei. 

The well-known and widely observed objects in which highly energetic transient 
events are most certainly powered by a thermonuclear outburst are:
nova outbursts, supernova explosions, jet launching in microquasars (X-ray binaries)
and active nuclei.   $\gamma-$ray bursts can also be explained by the inverse
Compton boosted $\gamma-$rays generated in a thermonuclear explosion.  In fact, 
it appears that most
transient energetic events in the universe are connected to a thermonuclear explosion.  
It, hence, appears most reasonable to assume that AT2018cow is also a result
of the energy pulse from a thermonuclear outburst either in a supernova or
in a microquasar located in the galaxy CGCG 137-068.  AT2018cow cannot be associated
with the centre of the galaxy since its position is displaced.  The black body
emission is reminiscent of a hot pseudosurface around the black hole, the
absorption/emission lines can arise in the BLR or NLR while the
radio emission can be associated with polar jets launched from the pseudosurface.  
Only if the observational
data has fundamental problems with this simple explanation then other possible physical
scenarios need to be explored. 

\item{\it A possible origin of fast radio bursts (FRB):}
Fast radio bursts are energetic, isolated, highly dispersed radio pulses of millisecond 
duration.
The DM estimated from FRB signals appears
to have a mean of about 1000 cm$^{-3}$pc with the range being 200-2600 cm$^{-3}$pc 
with no dependence on Galactic latitude 
\citep[Figure 2 in][]{2018NatAs...2..860L}.  This feature 
has prompted an extragalactic origin for FRBs \citep{2007Sci...318..777L}. 
It is interesting to note that detection of highly dispersed pulses between
430 and 1420 MHz had been reported from the core of M~87 in 1980 
\citep{1980ApJ...236L.109L} which were not confirmed by further observations 
\citep{1981ApJ...244L..61H,1981ApJ...244L..65T,1981ApJ...245L..99M}.
Energetic, highly-dispered radio bursts 
had also been reported from the directions of Sgr A* and MXB1730-335 
\citep{1978BAAS...10R.657L}.  Several radio bursts from the frequent 
X-ray burster MXB1730-335 were reported at a frequency of 4.1 GHz 
\citep{1979IAUC.3347....1C,1980IAUC.3458....1C,1980IAUC.3467....1C} but
no radio bursts were detected at 327 MHz \citep{1980BASI....8...41R}. 
These radio detections were considered suspect since they were erratic and often
not repeatable.  However now with the detection of several
highly dispersed FRBs from which only a single highly dispersed millisecond
duration pulse is often detected at GHz frequencies,  it appears that
the early observations could have detected the same population of sources. 
The earlier targetted observations associate the origin of FRBs with accreting 
black holes i.e. active nuclei or microquasars (X-ray binaries). 

About 60 FRBs have now been detected out of which only one has been observed to be 
repetitive - FRB 121102 which is localised to within a dwarf galaxy at 
a redshift of  $\sim 0.192$ \citep{2017ApJ...834L...7T} for which a luminosity 
distance of 900 Mpc has been estimated. 
The number of theories put forward to explain FRBs 
exceeds the number of FRBs that have been detected. 
While this indicates the high level of interest 
amongst astronomers in the intriguing signals, it also warns us against letting 
free-wheeling imagination triumph over well-constrained imagination working 
within the confines of known physics and astrophyics.  Free-wheeling imagination
can make it almost impossible to ever understand the origin of FRB 
signals since we cannot unequivocally verify or rule out so many models.  
One can take the easy way out and say that each FRB
is due to a different physical process but this is obviously an incorrect
inference made to justify our favourite model which in the larger scheme of research
has no relevance beyond our individual frame of reference.  

\begin{figure}
\centering
\includegraphics[width=6cm]{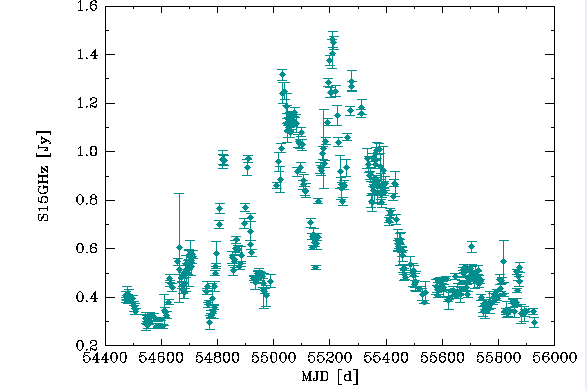}
\includegraphics[width=6cm]{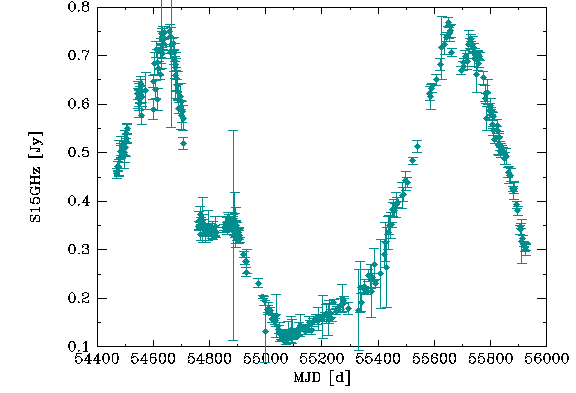}
\caption{\small Figure showing the 15 GHz light curves of blazars copied from
\citet{2014MNRAS.438.3058R}. Notice the amplitude and timscales of variability.
(a) The blazar CGRaBS J0128+4901 has an emission line redshift of 0.067 whereas in
(b) the FSRQ CGRaBS J0225+1846 has an emission line redshift of 2.69. }
\label{fig46}
\end{figure}

For the repeating FRB~121102, isotropic energies of individual
radio pulses have been estimated to be between $10^{37}$ and $10^{40}$ ergs
\citep[and references therein]{2018NatAs...2..860L}.
For comparison the solar luminosity is $3.8\times10^{33}$ erg s$^{-1}$ so that
the energy released in radio bands in FRB~121102 is equivalent to the energy that the 
sun releases over a week or so. 
If FRBs are confined to a jet then the energy released in each event will be reduced. 
For example, if the energy is released within an angular area of $25$ square degrees
then only 0.02\% of the energy will suffice i.e. $2\times 10^{33}-10^{36}$ ergs. 
The energy released in FRBs is small compared to supernovae ($\sim 10^{50}$ ergs), active 
nuclei (total energy $ >> 10^{50}$ ergs) and transient events 
like $\gamma-$ray bursts ($\ge 10^{50}$ ergs).  The short timescales over 
which FRBs are detected have helped infer that
the size of the emitting region is less than a few thousand kms.  

As demonstrated in the paper, thermonuclear outbursts which ignite a large mass of
fuel can release a huge quantity of energy in a short timescale and can successfully
explain the energetic transients.   We, hence, examine
if the source of energy in FRBs can be thermonuclear and are associated with explosive
ejection of radio plasma from stellar mass accreting rotating black holes in microquasars
or supermassive accreting rotating black holes in centres of galaxies.  
Large DM are measured for most FRBs which have been predominantly detected at GHz 
frequencies.  Only recently, detection of a few FRBs at frequencies between
400 and 800 MHz with CHIME have been reported \citep{2018ATel11901....1B}.
The radio emission from the repeating FRB~121102 shows varying high linear polarisation
and extremely high rotation measures ($10^4-10^5$ radians m$^{-2}$) 
at 4.5 GHz \citep{2018Natur.553..182M}.  The polarisation behaviour appears similar to
that noted in the vicinity of the supermassive black holes in the centres
of galaxies.  For example, the linearly polarised emission from
Sagittarius A* is detected at frequencies above 100 GHz \citep{2000ApJ...534L.173A} and is
found to vary  
over timescales of hours.  High rotation measures ($\ge 10^5$ radians m$^{-2}$) have 
been measured for Sagittarius A$^*$ at sub-mm wavelengths
\citep{2007ApJ...654L..57M}.  High rotation measures have also been measured for pulsars
located close to the Galactic centre 
\citep[e.g.][]{2013Natur.501..391E}.
High rotation measures of few thousand radians-m$^{-2}$ are commonly measured
at centimetre wavelengths and $\sim$mas resolution 
towards the core component of active nuclei which drops to a hundred or so 
radians-m$^{-2}$ over a few mas from the core
and a possible origin for such high rotation measures is suggested to be the
strong magnetic fields in the NLR 
\citep[e.g.][]{1998ASPC..144..113T,2012AJ....144..105H}.
Even higher RM $>10^5$ radians-m$^{-2}$ are recorded at mm and sub-mm wavelengths 
from the cores of radio galaxies 
and a possible origin is suggested to be the accretion flow onto the black hole
e.g. 3C~84 \citep{2014ApJ...797...66P}, M~87 \citep{2014ApJ...783L..33K}. 
Thus, the polarisation properties measured for the repeating FRB support an origin
close to the supermassive black hole in galaxies.  

It is relevant to note that several BL Lac objects show variability of 100\% or more
so that their radio flux densities double or halve or vary by various factors over
hours to days as shown for a couple blazars in Figure \ref{fig46} 
taken from \citet{2014MNRAS.438.3058R}.  In Figure \ref{fig46},
the intensity of the blazar in the top panel varies by a factor of four
whereas the one at the bottom varies by a factor of eight and the change is observed to
occur over timescales ranging from what appears to be a day to several tens of days.
This behaviour is typical of blazars which are known for their
high variability quotient.   The observed radio flux densities
of BL Lacs range from sub-Jy to few Jys which is also the typical range in which
FRB pulses are detected thus prompting us to suggest a possible association with
BL Lac-like active nuclei i.e. nuclei whose polar axis is along the
sightline.  FRBs can also arise in microquasars whose radio axis
is directed along the sightline.  The rapid quenching of the radio signal from
FRBs is intriguing and needs further investigation.  It is possible that FRBs
are predominantly detected in radio-faint accreting black holes. 

\item{\it Events referred to as TDE:}
A tidal disruption event (TDE) signifies an event in which a main sequence star
is fragmented due to its close approach to a black hole such that half of it
falls into the black hole and other half escapes from the black hole.  
In literature, TDEs are observationally identified with sources which show enhanced 
short-term brightening in the ultraviolet/blue bands which is well fit with a 
black body spectrum, wide emission lines of helium and/or hydrogen which
show correlated changes with the ultraviolet continuum \citep[e.g.][]{2012Natur.485..217G},
a constant black body temperature, lack of optical colour evolution 
and the events are not observed to repeat over the monitored timescales
of few hundred days.  This range of transient observational signatures that have 
been postulated to arise in a TDE near a black hole are similar to the 
observed signatures of microquasars, active nuclei and kilonova/afterglow signals
of a $\gamma-$ray burst which do not necessarily require a major accretion event. 
For example, sudden brightening, especially in the ultraviolet bands of
an apparently blank region, many times coincident with
the centres of galaxies and occasionally accompanied by brightening in the X-ray and
radio bands is attributed to a TDE.  Wide emission lines especially of hydrogen
and helium with FWZM of several thousand kms$^{-1}$ 
\citep{2014ApJ...793...38A,2017MNRAS.466.4904B} are detected in several 
of the events labelled as TDE, which evolve with the ultraviolet continuum such that the 
wide lines fade as the continuum fades.  The same observations could
also be associated with the afterglow signatures of a $\gamma-$ray burst or variability
in the non-polar pseudosurface and BLR of a Seyfert~1/BLRG/quasar nucleus.  
Reverberation studies are  
concerned with the correlation between the ultraviolet continuum and
wide lines from the BLR in Seyfert~1 nuclei.  This means that there seems
to exist some subjectivity in interpreting similar observational results. 
It would be useful if the observational
results are explicity shown to not be commensurate with, for example, emission from
the structure associated with a uniformly accreting black hole such as the
hot pseudosurface and BLR before it is attributed to
alternate explanations like a TDE.  Since few candidate TDEs have been detected in 
radio bands \citep[e.g.][]{2014ApJ...793...38A} which can be inferred to mean that
the variability identified as TDEs seldom launch radio plasma along jets, it 
might suggest that TDEs trace
variability in the non-polar regions around the accreting black hole. 

The debris of a TDE event can lead to enhancing the infall densities which can
increase the accretion rates at the poles and accretion rates onto the accretion disk
in the non-polar regions of the black hole,
but obviously require special circumstances to explain the ultraviolet brightening 
and occurrence of wide lines in the source which 
are, on the other hand, common signatures of the hot pseudosurface and BLR 
around the accreting black hole.  This is further supported by the detection
of variability in the broad lines which is correlated with the ultraviolet continuum emission. 
This means the transient signatures attributed to a TDE can be explained
by short-term variability (thermonuclear blast?) in the already present non-polar BLR 
and pseudosurface.  This needs to be investigated further and concretised 
by examining observational data on individual TDEs. 

\item{\it Double-peaked optical lines from active nuclei:}
Double-peaked optical lines are expected from the ionized line forming gas that is ejected
by the active nucleus alongwith the radio jets and which we identify
as the NLR.  Early investigations of resolved
radio galaxies often detected Lyman $\alpha$ line emission extended along the
bipolar radio jets which were blueshifted
and redshifted wrt to the galaxy redshift on the two sides.  Such lines would appear as
double-peaked lines in case of distant and unresolved radio galaxies about the
galaxy redshift and need not signify the presence of double active nuclei.  

\item {\it More questions:}
Is the pseudosurface formed around all accreting black holes or does it
require minimum infalling matter densities ?

The predominant source of synchrotron emission in radio galaxies appears to be
the positron-electron plasma - is there any synchrotron contribution 
from the proton-electron plasma in radio galaxies and cluster radio halos  
and can it be distinguished from synchrotron emission due to the positron-electron 
plasma ?  The magnetic field is observed to remain oriented along the 
straight jet as long as the radio jet appears to remain
ballistic i.e. forward momentum far exceeds the lateral momentum.  
Since the field is provided by the proton-electron plasma,
does this mean that this plasma remains ballistic till the hotspots
in FR~II radio sources and the positron-electron plasma which can have 
random relativistic motions comparable to the forward bulk motion
is merely dragged along the field lines and hence appears to be ballistic ?  
Is it that if the field was removed, the positron-electron plasma would diffuse in
all directions i.e. the perceived confinement of radio plasma in jets or in lobes is 
provided by the field frozen in the proton-electron plasma ?

Are the forbidden emission lines which are detected at redshifts similar to broad 
lines in the spectra of some quasars formed in low density pockets of the
BLR or in the NLR just ejected from the black hole ? 

What causes the spin axis of the supermassive black holes to precess ?  If 
due to an external gravitational torque, does it mean that the black hole does
exchange gravitational information with the outside world ?
How does the infalling matter external to the black hole surmise 
that the black hole is rotating and its rotational speed?  In other words,
how does information regarding the rotation state of the black hole 
leave the event horizon and influence the external infalling matter i.e.
how does the space-time outside the Schwarzchild radius get appropriately distorted 
and reflect the spin of the black hole ?  

The extent of the event horizon is determined by
the mass of the black hole while the
gravitational potential which gravitationally redshifts a photon,  
the escape velocity that a particle requires to leave the black hole etc are 
dependent on the separation from the event horizon (or black hole if it is assumed to
be a point source at the centre of a sphere of radius equal to the Schwarzchild
radius).  Recall that no matter or electromagnetic radiation can leave from within the
event horizon since they require velocities in excess of light then how is
the above information passed on to the world outside the event horizon ?  
Is it that the black hole can exchange gravitational information with the
outside world? 
This can be inferred to mean that the graviton, which if believed to be the
mediator of the gravitational interaction, travels with a speed which exceeds the
speed of light so that it can effortlessly escape the event horizon of the black hole
and transmit the gravitational information from within the
event horizon to the outside world.  Can there be any other explanation ? 

\end{itemize}

\section{Summary and Conclusions}

This paper has attempted to comprehensively explain observations of radio synchrotron 
objects within the framework of known physics.  It has also pointed out reasons
for the observed shapes of cosmic ray 
energy spectra, in particular the relativistic electron energy spectra and
the observed radio spectra of synchrotron sources.  The observed energy distribution  
adheres to what we would expect from thermodynamics and statistical physics.
Thermonuclear energy has emerged as the source of transient explosive
energy which is released in a myriad of phenomena near black holes and in supernovae. 
While conditions conducive for triggering thermonuclear outbursts rapidly converge due
to the immense gravity near black holes,  the actual source of energy 
is thermonuclear and is adiabatically transferred to matter in a short time. 
Jets and narrow line emitting gas in active galaxies, 
owe their existence to episodic radial eruptions from the polar regions of the 
quasi-spherical pseudosurface close to the event horizon of the rotating black hole
following a thermonuclear outburst.  
Observations favour a positron-electron composition for the fast radio synchrotron jets. 

The study can be summarised as follows:

\begin{itemize}
\item The observed cosmic ray energy spectrum and observed radio spectra 
which are approximate power laws at the high energy/frequency end are observed to
be convex-shaped at the lower energy/frequency side.  
The matter is energised in energetic transient events.  Once it is appreciated that
the instantaneous energising of a large number of particles is statistical in nature, it 
follows that it will lead to a normal/gaussian (log-normal) distribution of energies with a
centroid and a finite dispersion which is the largest entropy distribution and it follows
that the observed nature of the energy spectra is no longer a mystery.  
If the energetic event which energises
and expels matter (such as supernova explosions) adiabatically imparts comparable quanta of energy
to each particle (neutrinos, electrons, protons, ions etc)  which if predominantly 
kinetic in nature i.e. $E=m v^2 / 2 $ would lead to the centroid of
the resulting energy distribution for all particles being located at the energy
equal to or larger than the escape velocity from the system.  The centroid energy will
primararily be determined by the heavier particles that escape.  
This naturally accounts for the peak of the 
cosmic ray distribution which is observed to be near 1 GeV for various particle species.  The dispersion
of the E/m distribution for each particle species will determine  
the random velocity component for the species which will be largest for the 
lightest particles.  This will result in highly relativistic velocities for
neutrinos and electrons.  

As the forward bulk motion of the ejected matter declines, the centroid of 
the particle energy distributions should move to progressively lower energies.  
As the electrons suffer synchrotron losses, the dispersion of their energy
distribution will keep reducing
so that the post-peak distribution which we always approximate by a power law
of index $p$ related to the radio spectral index $\alpha$ by $p=2\alpha+1$
will appear to steepen. 

\item It appears beyond doubt that the composition of the fast radio synchrotron jet 
in accreting black holes (i.e. microquasars and active nuclei) and of the plasma
giving rise to the prompt radio synchrotron emission in supernovae consists
of positrons and electrons.

\item A supernova explosion of type II involves implosion of the core
of a massive star and thermonuclear explosion of half a solar mass or so of the
surrounding gas which releases more than $10^{51}$ ergs within seconds and explosively
ejects the outer parts of the star.  Supernova explosions of type 1a have been explained by
a thermonuclear explosion triggered when the mass of a white dwarf 
exceeds the 1.4 M$_\odot$ limit.  Observations support the ejection of
a fast light positron-electron plasma which is responsible for the prompt radio emission
from the supernova whereas the massive ejecta i.e. protron(ion)-plasma, which
slows down soon after expulsion, eventually forms the supernova remnant
and is responsible for the prompt optical light curve,
delayed radio emission and the long-lasting spectral line emission i.e. it
consists of a thermal plasma mixed with a non-thermal relativistic plasma. 
The same injected energy spectrum of the relativistic electrons results in
the synchrotron emission in the prompt supernova phase and delayed remnant phase 
although it might be possible for the two phases to contain relativistic electrons from 
different energy bins.  

Radio observations including polarisation studies indicate that the prompt synchrotron 
emission due to the fast positron-electron plasma requires an ambient magnetic field 
whereas the synchrotron emission from the supernova 
remnant is due to the  magnetic field frozen in the plasma.   
This, then, supports freezing of the stellar magnetic field in the heavy plasma.  

\item Hard X-ray emitting rims have been detected around the main body of
several supernova remnants 
like Kepler, Tycho, Cas~A.  Most of these supernova remnants (except Cas~A)
are explosions classified as type 1a from which prompt radio or X-ray emission has never been
detected and in literature this has been attributed to the lack of a dense circumstellar
medium.  While this study also supports this general explanation, the reasons are
different.  Since relativistic plasmas are ejected in all
supernova explosions, the reason for the lack of prompt radio synchrotron in
type 1a is the absence of an appropriate magnetic field due to the lack of a
circumstellar medium in which it would be frozen.  In the absence of the field, the 
ejected positron-electron plasma 
will expand with only kinetic losses and hence can survive longer. 
When this plasma encounters an appropriate magnetic field, it will start radiating synchrotron
emission.  This, then can explain the presence of hard X-ray synchrotron emitting rings 
that are detected mostly around supernova remnants
of type 1a.  The magnetic field configuration 
is predominantly radial within young remnants and shows a distinct
structure outside the main remnant extent when radio and/or hard X-ray synchrotron
emitting circumferential rings are detected.  
On the other hand, there is no difference in the synchrotron properties of 
the supernova remnant of types I and II and in both cases, the remnant consisting of
thermal and non-thermal plasma should 
become detectable once the thermal plasma becomes optically thin
to radio synchrotron emission as was the case with SN 1986J.
% with the morphology being determined by
%the interaction of the remnant material with the ambient material. 

\item Observations of type II SN 1987A support the fast propagation of a positron-electron
plasma ($\sim 25000$ kms$^{-1}$ till the equatorial ring) and relatively 
slow propagation of a thermal proton(ion)-electron plasma ($\sim 2000$ kms$^{-1}$).  The
presence of a mixed non-thermal proton-electron plasma is not yet clear. 
The rapid detection and quenching of prompt radio emission from 
SN 1987A can be understood as being due to paucity of distributed 
ambient matter and magnetic field due to episodic mass loss by the progenitor so
that most of the circumstellar matter is concentrated in the triple ring system.
The origin of the triple ring system 
can be understood if the progenitor of SN 1987A was in a loose binary
system and experienced episodes of rapid ejection of matter from its
equatorial regions at high velocities from three distinct locations in the
orbit.  

The brightening of the inner equatorial ring in optical would have happened 
when the fast positron-electron plasma exicted it, in the radio when
sufficient positron-electron plasma started radiating synchrotron emission in the
magnetic field of the ring.  
The optical/infrared source that has become visible at the centre of the inner ring
should be 
the heavy ejecta i.e. supernova remnant which was spherically symmetric in 1994 and 
has now evolved into a faint elliptical shell detected in spectral lines and
dust emission. 
Radio synchrotron emission has not been detected from this central remnant till now. 

\item  In general, accretion by a black hole should be spherical due to the 
small sizes of the black holes as compared to either a binary companion or the
scale height of the interstellar medium in centres of galaxies.  
Non-spherical accretion, if found to be the case for
a subset of black holes, can be studied as special cases. 
Irrespective of the black hole mass, accretion of matter by a rotating 
black hole should lead to formation of similar physical structures
around the black hole.  This should consist of (1) a hot quasi-spherical layer 
of degenerate matter that forms a pseudosurface of the black hole and emits 
black body radiation; (2) a broad line region formed from the matter deposited on 
the non-polar pseudosurface wherein broadening of emission and absorption
lines is dominated by the changing gravitational potential within the region; 
(3) bipolar conical radio synchrotron jets and optical line emitting
gas which is the accreted matter 
that is episodically ejected into space following energy input from 
thermonuclear outbursts in the hot accreted matter at the poles and (4)
an accretion disk in the non-polar regions which is
formed beyond the broad line region due to the accumulation of the excess infalling 
matter that is yet to be accreted due to the lower equatorial accretion rates 
for a rotating black hole.  Accreting black holes can show all or a subset of
the above properties since their detection will depend on the distinct 
physical parameters of each system and this can account for the observed
variety in the properties of active nuclei/galaxies and microquasars.

\item Observational results on microquasars can be understood with the above.
The Galactic microquasar SS~433 which is believed to have formed from
the supernova explosion which resulted in the remnant W~50, shows blueshifted and redshifted
optical emission lines with changing frequencies ($\sim 0.26c$ 
about a mean redshift of about 0.04) and helical twisting of 
the radio jet, both with a period of about 167 days.  These 
are consistently explained by precession of the black hole 
spin axis in literature and hence can be identified with the thermal proton(ion)-electron
and non-thermal positron-electron plasmas which are launched at escape 
velocities ($\sim 0.26c$) from the polar pseudosurface of the rotating black hole following
a thermonuclear explosion in the accreted matter.  The launching site has to be
separated from the black hole by about 14.8 Schwarzchild radii so that it needs to
be accelerated to at least 0.26c to escape the gravity of the black hole. 
The gravitational redshift component in the spectral
lines emerging from this location will be about $0.034$ (intrinsic redshift 
would be 0.035) which is comparable to the mean redshift over which the velocity 
of the spectral lines are observed to oscillate (about 0.04) lending further support
to the model. 
The occasional disappearance of the moving lines would support episodic
ejection of matter and/or indicate dust formation and obscuration in
the ejected optical line forming gas.  Thus several observations of SS~433 are
consistently explained.  The radio and $\gamma-$ray flares observed from
microquasars would be due to thermonuclear energising and ejection of matter from
the poles whereas the detection of the positron annihilation line near 511 keV
supports the positron-electron composition of the radio jets.   Positrons can
be generated in the thermonuclear blast.  

\item GRB 170817A (and several other GRBs) can be explained by the $\gamma-$ray
photons released in a thermonuclear explosion in the accreted matter deposited on the
polar pseudosurface of the black hole in a microquasar located in the galaxy NGC~4993.
These $\gamma-$rays could be inverse Compton scattered to the higher
energies by the relativistic electrons 
energised in the same explosion.   The width of the $\gamma-$ray pulse
could be indicative of duration of the thermonuclear outburst. 
Observationally, GRB 170817A was characterised by the $\gamma-$ray pulse,
thermal ultraviolet/optical emission and synchrotron emission. 
The energy released in the thermonuclear outburst
heats the entire pseudosurface of the black hole which enhances its 
black body emission at ultraviolet 
and optical bands i.e. the kilonova emission.  The outburst 
also energises and launches relativistic plasma from the poles i.e bipolar jets which
emit multi-band synchrotron radiation referred to as the afterglow. 
These emissions fade as the effect of the explosion wears off.
In general, it appears that a large fraction of GRBs can be explained
with an origin in energetic thermonuclear outbursts in
supernovae, microquasars and active nuclei.  Broad emission/absorption lines
arising in the BLR have been detected in several GRBs supporting their origin in
active nuclei.

\item The spin of the accreting black hole appears to be the most important parameter
determining the observable characteristics of active nuclei.  
High spin massive black holes result in radio galaxies and high spin massive
stellar mass black holes result in microquasars.  The spin of the black hole
determines the latitude coverage of the centrifugal barrier which in turn
determines the magnitude of the jet opening angles and 
the latitude thickness of the broad line region and accretion disk.
FR II radio sources appear to host the fastest spinning black holes
in the universe and hence should have the smallest jet opening angles, ballistic 
and collimated jets upto the largest distances, a broad line region
with the largest covering factor and an accretion disk with the largest radial
extent. 

\item The observed broad-band continuum emission from active nuclei which is known
in literature to consist of three emission components can be explained as follows:  
a black body component from the pseudosurface, 
a thermal continuum from the non-polar broad emission line region and a
synchrotron component from the relativistic positron-electron plasma ejected along the poles. 
The relativistic plasma is energised by the explosion energy before ejection.
The spectra of active nuclei consists of broad permitted emission and absorption 
lines from the non-polar broad line region (BLR) and narrow permitted plus forbidden 
emission lines from the polar narrow line region (NLR).  Narrow lines from a 
distant BLR could contribute lines to the spectra of type 2 active nuclei.

\item The observed emission line redshifts of active nuclei especially
quasars are large due to the inclusion of a significant gravitational redshift 
component and hence distances to quasars have been overestimated in literature. 
The gravitational redshift component needs to be
removed before using the redshift to estimate the distances to quasars.
This is demonstrated for the FR~II quasar 3C~273 whose distance is changed
to about 20 Mpc and it is shown that this
correction results in several observations converging on similar physical parameters. 
%observed characteristics match the expected values
%estimated from the measured physical properties.
For example, the expected area of the emitting black body of temperature
26000 K (temperature taken from literature) is comparable to the area of a spherical
pseudosurface for Schwarzchild radius of a billion solar mass black hole
which is the black hole mass that has been inferred for 3C~273 in literature, indicating
that the major source of the black body emission is indeed the pseudosurface formed at the
event horizon of the black hole; the expected V~band absolute magnitude 
estimated from the Stephen-Boltzmann
luminosity matches the observed V~band luminosity estimated for the corrected distance 
indicating that the major source of the V-band emission in 3C~273 is the hot pseudosurface
and finally the jet expansion speeds become subluminal for the corrected distance. 

\item In the FR~II galaxy Cygnus~A, thick X-ray beams are seen to extend
from the core to the hotspots along the finer radio synchrotron jets.  Even if
the radio jets are single-sided, the X-ray beams are observed to be two-sided.
These X-ray photons could be from the Compton 
scattering of the higher energy pair annihilation photons near 511 keV generated 
along the positron-electron jets and are not 
subject to any Doppler effects explaining
their two-sidedness.  A similar origin is suggested for 
the thick X-ray beams observed in Pictor~A, Virgo~A, 3C~294 and 3C~432 along 
the radio jets/lobes.  This origin follows once the composition of the
jets is narrowed down to positrons and electrons. Conversely the observed
X-ray beams support a positron-electron composition of the jets.

\item A more concrete origin for the hotspots in FR~II radio sources emerges 
from the positron-electron composition of the radio jets.
Most of the pair annihilation photons near 511 keV are likely to be Compton
scattered/absorbed by the  
thermal gas along the radio jet.  Once the pair plasma jet expands beyond the
periphery of the thermal gas and enters the highly tenuous intergalactic
medium, the annihilation photons will have longer mean free paths (and lifetimes) and
hence can exert radiation pressure back on the dense thermal plasma, halting it,
compressing it and enhancing the ordered magnetic field.  
The synchrotron jet will fade beyond the periphery of the thermal plasma due
to lack of a magnetic field.  However the pair annihilation photons can exert
a radiation pressure on the thermal and relativistic plasma pushing them
back towards the active nucleus explaining the formation of hotspots where
the jet is turning back, the backflow and formation
of radio lobes in FR~II sources.  If the backflow extends to the core then
it can be stopped and deflected by the radially extended, thick accretion disk 
around the supermassive black hole, thus forming X-shaped or winged radio sources.

\item Multi-band variability in microquasars and
active nuclei can be explained by temporal variations in the 
structure and accreted gas located close to the black hole.  
Thermonuclear outbursts which launch matter from the poles can account for the frequent
radio and $\gamma-$ray flares in blazars and microquasars, FRBs and also the
enormous instantaneous energy released in extreme events like GRBs.
The proximity of the accreted matter to the event horizon of the black hole makes it
inherently quasi-stable in nature so, for example, any change in the temperature
or thickness of
the non-polar pseudosurface and BLR can explain the variability in thermal
continuum and broad lines from the active nuclei. 

\item The aged relativistic plasma in radio halos and relics in clusters
is commensurate with an origin in the polar ejections from the pseudosurface
of the supermassive black hole in the central massive FR~I galaxy which has
long exhausted its gas supply and entered a quiescent phase.  Both, the low $\mu$Jy to
sub-$\mu$Jy magnetic fields in the halos and the
subset of electrons which have their relativistic motion along
the field lines and hence suffer low losses ensure longer lives for a large
pool of electrons so that the relativistic plasma can expand and survive so far from the 
black hole and diffuse forming a halo.  Such low magnetic fields are expected from
flux conservation
if they were originally frozen in the NLR which has expanded to halo dimensions.  The existence
of double relics which are located on diametrically opposite sides of the cluster
centre indicates the orientation of the jet axis when the central black hole was active
and along which the positron-electron plasma flowed outwards.  The relics
are always detected beyond the thermal X-ray halo (similar to hotspots in FR~II sources which
are formed beyond the thermal X-ray gas).  It is hence suggested that the relics are formed
due to the radiation pressure
exerted by pair annihilation photons near 511 keV generated beyond the thermal gas.  
The radiation pressure compresses the thermal plasma and circumferentially enhances 
the ordered magnetic field as observed in relics.  

\item Diffuse soft X-ray emitting gas is ubiquitous - it 
forms the hot ionized medium 
in spiral galaxies, a halo around elliptical galaxies and active galaxies and
a centrally located halo in clusters of galaxies.  
The kinematic and spatial correlation observed between the optical line
emitting gas and the X-ray emitting gas in active galaxies supports the
evaporation of the optical gas which is originally ejected from the poles of the 
active nucleus to the higher temperature tenuous X-ray gas. 
The presence of a X-ray halo around normal elliptical
galaxies argues for an origin in a gentler process such as through the formation
of a ring around the galaxy due to the action of a gravitational torque and 
eventual evaporation of the gaseous component to X-ray densities and temperatures
as it expands.   The X-ray halo in clusters could be a combination of both
the above as can be discerned in case of the Virgo cluster.

\end{itemize}

\section*{Acknowledgements}
I gratefully acknowledge using ADS abstracts, arXiv e-prints, AAVSO data, gnuplot, LaTeX,
Wikipedia, NASA Extragalactic Database, ViZier database and Google search engines enabled 
by the internet and the world wide web,
in this research.  {\bf If you happen to use any of the figures copied from literature, please
credit the original reference since I have only used these to better demonstrate a point.}
Although no effort is spared in studying literature, this paper addresses popular fields of research
so that literature is vast and it is practically impossible to study all of it - this again
underlines the need to be choosy in publishing our results to prevent ourselves from
getting lost in trivia.  
This work has benefitted from wide-ranging discussions with Prasad Subramanian.

\bibliography{novae2016,agn}

\end{document}